%% file: main.tex
\documentclass[
  journal=largetwo,
  manuscript=article-type,
  year=2020,
  volume=37,
]{cup-journal}

\usepackage{amsmath}
\usepackage{booktabs}

\usepackage{amsmath}	% Advanced maths commands
\usepackage{amssymb}	% Extra maths symbols
\usepackage{nameref}
\usepackage{varioref}
\newcommand{\twoPi}{$^{2}\Pi_{3/2},~J = 3/2$~}

\newcommand{\Tb}{$T_\mathrm{b}$}
\newcommand{\Tc}{$T_\mathrm{c}$}
\newcommand{\taunu}{$\tau_{\nu}$}
\newcommand{\Htwo}{H$_2$}

\newcommand{\amoeba}{\textsc{Amoeba}}
\newcommand{\HI}{H\textsc{i}}
\newcommand{\HII}{H\textsc{ii}}
\newcommand{\kms}{km s$^{-1}$}
\newcommand{\aap}{AAP}

\newcommand{\apjl}{ApJL}
\newcommand{\apjs}{ApJS}

\title{GNOMES II: Analysis of the Galactic diffuse molecular ISM in all four ground state hydroxyl transitions using \textsc{Amoeba}}

\author{Anita Petzler}
\affiliation{School of Mathematical and Physical Sciences and Research Centre in Astronomy, Astrophysics \& Astrophotonics, Macquarie University, Sydney, 2109, Australia}
\alsoaffiliation{Australia Telescope National Facility, CSIRO Space \& Astronomy, PO Box 76, Epping, NSW 1710, Australia}
\email[Anita Petzler]{anita.petzler@csiro.au}

\author{J. R. Dawson}
\affiliation{School of Mathematical and Physical Sciences and Research Centre in Astronomy, Astrophysics \& Astrophotonics, Macquarie University, Sydney, 2109, Australia}
\alsoaffiliation{Australia Telescope National Facility, CSIRO Space \& Astronomy, PO Box 76, Epping, NSW 1710, Australia}

\author{Hiep Nguyen}
\affiliation{Research School of Astronomy and Astrophysics, The Australian National University, Canberra, ACT 2611, Australia}

\author{Carl Heiles}
\affiliation{Department of Astronomy, University of California, Berkeley, 601 Campbell Hall 3411, Berkeley, CA, 94720-3411, USA}

\author{M. Wardle}
\affiliation{School of Mathematical and Physical Sciences and Research Centre in Astronomy, Astrophysics \& Astrophotonics, Macquarie University, Sydney, 2109, Australia}

\author{M.-Y. Lee}
\affiliation{Korea Astronomy \& Space Science Institute, 776 Daedeok-daero, Yuseong-gu, Daejeon 34055, Republic of Korea}
\alsoaffiliation{University of Science and Technology, 217 Gajeong-ro, Yuseong-gu, Daejeon 34113, Republic of Korea}

\author{Claire E. Murray}
\affiliation{Department of Physics \& Astronomy, Johns Hopkins University, 3400 N. Charles Street, Baltimore, MD 21218, USA, NSF Astronomy \& Astrophysics Postdoctoral Fellow}

\author{K. L. Thompson}
\affiliation{Davidson College, Davidson, NC, 28115, USA}

\author{Sne{\v{z}}ana Stanimirovi{\'c}}
\affiliation{Department of Astronomy, University of Wisconsin, Madison, WI 53706, USA}

\keywords{galaxies: ISM, ISM: molecules, radio lines: ISM}

\begin{document}

\begin{abstract}
    \input{0_Abstract}

\end{abstract}

%%%%%%%%%%%%%%%%% BODY OF PAPER %%%%%%%%%%%%%%%%%%

\section{Introduction}

\input{1_Intro}

\section{Observations}
    \input{2_Data}

\section{Method practicalities and limitations}
    \input{3_Methods}

\section{Results}
    \input{4_Results}

\section{Analysis}
    \input{5_Analysis}

\section{Conclusions and Future work}
    \input{6_Conclusions}

\begin{acknowledgement}
During the development of this work A.P. was the recipient of an Australian Government Research Training Program (RTP) stipend and tuition fee offset scholarship.
\end{acknowledgement}

%%%%%%%%%%%%%%%%%%%%%%%%%%%%%%%%%%%%%%%%%%%%%%%%%%

%%%%%%%%%%%%%%%%%%%% REFERENCES %%%%%%%%%%%%%%%%%%

% The best way to enter references is to use BibTeX:
\printbibliography
% \bibliographystyle{mnras}
% \bibliography{bibliography} % if your bibtex file is called example.bib

%%%%%%%%%%%%%%%%%%%%%%%%%%%%%%%%%%%%%%%%%%%%%%%%%%

%%%%%%%%%%%%%%%%% APPENDICES %%%%%%%%%%%%%%%%%%%%%

\appendix

    \input{7_Appendix}

%%%%%%%%%%%%%%%%%%%%%%%%%%%%%%%%%%%%%%%%%%%%%%%%%%

% Don't change these lines
% \bsp	% typesetting comment
% \label{lastpage}
\end{document}

%% file: 0_Abstract.tex
We present observations of the four $^2 \Pi _{3/2}\,J=3/2$~ground-rotational state transitions of the hydroxyl molecule (OH) along 107 lines of sight both in and out of the Galactic plane: 92 sets of observations from the Arecibo telescope and 15 sets of observations from the Australia Telescope Compact Array (ATCA). 
Our Arecibo observations included off-source pointings, allowing us to measure excitation temperature ($T_{\rm ex}$) and optical depth, while our ATCA observations give optical depth only. 
We perform Gaussian decomposition using the Automated Molecular Excitation Bayesian line-fitting Algorithm `\amoeba' \citep{Petzler2021a} fitting all four transitions simultaneously with shared centroid velocity and width. 
We identify 109 features across 38 sightlines (including 58 detections along 27 sightlines with excitation temperature measurements). 
While the main lines at 1665 and 1667\,MHz tend to have similar excitation temperatures (median $|\Delta T_{\rm ex}({\rm main})|=0.6\,$K, 84\% show $|\Delta T_{\rm ex}({\rm main})|<2\,$K), large differences in the 1612 and 1720\,MHz satellite line excitation temperatures show that the gas is generally not in LTE. 
For a selection of sightlines we compare our OH features to associated (on-sky and in velocity) H\textsc{i} cold gas components (CNM) identified by \citet{Nguyen2019} and find no strong correlations. 
We speculate that this may indicate an effective decoupling of the molecular gas from the CNM once it accumulates.

%% file: 1_Intro.tex
Molecular hydrogen (H$_2$) does not have readily observable transitions in the low densities and temperatures typical in the interstellar medium (ISM). Its presence must therefore be inferred from measurements of other `tracer' species. The most commonly used tracer of molecular hydrogen in the study of the interstellar medium is carbon monoxide (CO), through observations of its lower rotational transitions. 
% However, it has become increasingly apparent that CO fails to trace significant amounts of diffuse molecular gas \citep[e.g.][]{Blitz1990,Reach1994,Grenier2005,PlanckCollaboration2011,Paradis2012,Langer2014,Li2018}, as CO can be photodissociated in diffuse environments by external UV radiation \citep{Tielens1985a, Tielens1985b, vanDishoeck1988, Wolfire2010, Glover2011, Glover2016} even when hydrogen exists primarily as \Htwo. This is explained by the higher self-shielding threshold of CO compared to \Htwo: in the local ISM the extinction threshold for \Htwo~to form is $A_V \geq 0.14$\,mag, but CO requires $A_V \geq 0.8$\,mag \citep{Wolfire2010}, so CO is typically photo-dissociated by external UV radiation \citep{Tielens1985a,vanDishoeck1988,Wolfire2010,Glover2011,Glover2016}. 
% CO can therefore well-trace \Htwo~only in the inner part of molecular clouds, and overall it is not a perfect tracer of molecular hydrogen. Observations of CO emission alone then may miss a significant fraction of the molecular gas, especially in diffuse molecular clouds where the UV shielding is poor \citep{Reach1994,Meyerdierks1996}.
% New:
The abundance of \Htwo~can then be inferred from the integrated intensity of CO via the so-called `X-factor' \citep{Bolatto2013}. However, it has become increasingly apparent that this method fails to predictably trace significant amounts of molecular gas in more diffuse environments \citep[e.g.][]{Blitz1990,Reach1994,Grenier2005,PlanckCollaboration2011,Paradis2012,Langer2014,Li2018}. The primary reason for this limitation is the unreliable relationship between the integrated intensity of CO and the \Htwo~abundance in low extinction or low number density environments. CO can be photodissociated in low extinction environments by external UV radiation \citep{Tielens1985a, Tielens1985b, vanDishoeck1988, Wolfire2010, Glover2011, Glover2016} even when hydrogen exists primarily as \Htwo because of its higher self-shielding threshold compared to that of \Htwo. In the local ISM the extinction threshold for \Htwo~to form is $A_V \geq 0.14$\,mag, but CO requires $A_V \geq 0.8$\,mag \citep{Wolfire2010}, so CO is typically photo-dissociated by external UV radiation \citep{Tielens1985a,vanDishoeck1988,Wolfire2010,Glover2011,Glover2016}. On the other hand, in low number density molecular environments \citep[e.g. $n_{\rm H} \lesssim 10^{-2}\,{\rm cm}^{-3}$~as found by][in the region of Persius]{Busch2019} that do contain CO, the CO may not be sufficiently excited to be detectable due to its relatively high critical density.

This has motivated a resurgence of interest in hydroxyl (OH) as an alternative tracer of diffuse H$_2$ \citep[e.g.][]{Dawson2014, Allen2015, Engelke2018, Busch2021, Dawson2022}. OH has been demonstrated to trace `CO-dark' \Htwo\ in diffuse clouds \citep{Barriault2010, Cotten2012, Allen2015}, in the envelopes of GMCs \citep{Wannier1993}, in absorption sightlines scattered across the sky \citep{Li2015, Li2018}, and recently in a thick molecular disk of ultra-diffuse molecular gas in the outer Galaxy \citep{Busch2021}. Though there may be a weak relationship between the OH/\Htwo~abundance ratio $X_{\rm OH}$~and visual extinction $A_V$, $X_{\rm OH}$~appears relatively constant ($\approx 10^{-7}$) in a wide range of environments \citep[i.e. with $A_V=0.1-2.7$~and $n_{\rm H_2}>50\,{\rm cm}^{-3}$][and references therein]{Nguyen2018} including the CO-dark gas \citep{Black1977,Wannier1993,Weselak2009}. 

\begin{figure}
    \centering
    \includegraphics[trim={4cm 10cm 3cm 14cm},clip=true,width=\linewidth]{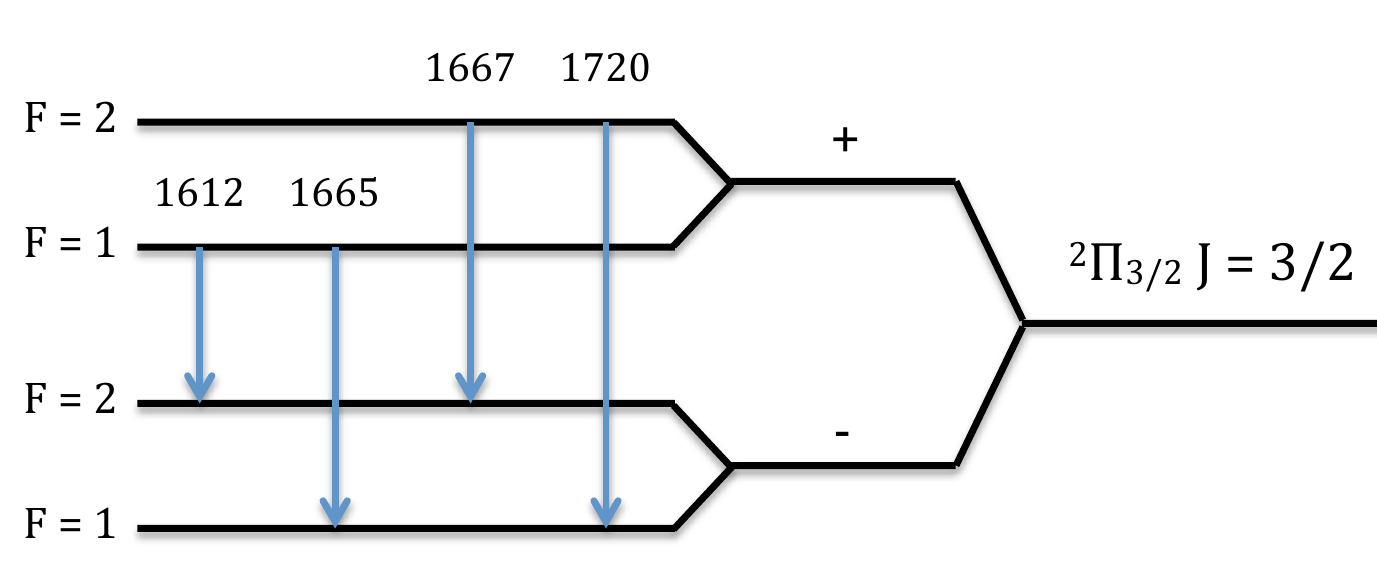}
    \caption{Energy level diagram of the \twoPi ground state of hydroxyl. The ground state is split into four levels due to $\Lambda$-doubling and hyperfine splitting, with 4 allowed transitions between these levels: the `main' lines at 1665.402 and 1667.359 MHZ, and the `satellite' lines at 1612.231 and 1720.530~MHz. Figure from \citet{Petzler2020}.}
    \label{fig:OHlevels}
\end{figure}

Most OH molecules in the diffuse ISM are expected to be found in the $^{2}\Pi_{3/2}\,J = 3/2$ ground state (see Fig. \ref{fig:OHlevels}) which is split into 4 levels via lambda doubling and hyperfine splitting. There are four allowed transitions between these levels: the `main' lines at 1665.402 and 1667.359\,MHz, and the `satellite' lines at 1612.231 and 1720.530\,MHz \citep[e.g.][]{Destombes1977}.

\subsection{Local thermodynamic equilibrium (LTE)}
\begin{figure}
	\begin{center}
	\includegraphics[trim={1cm 9cm 1cm 8cm},clip=true,width=\linewidth]{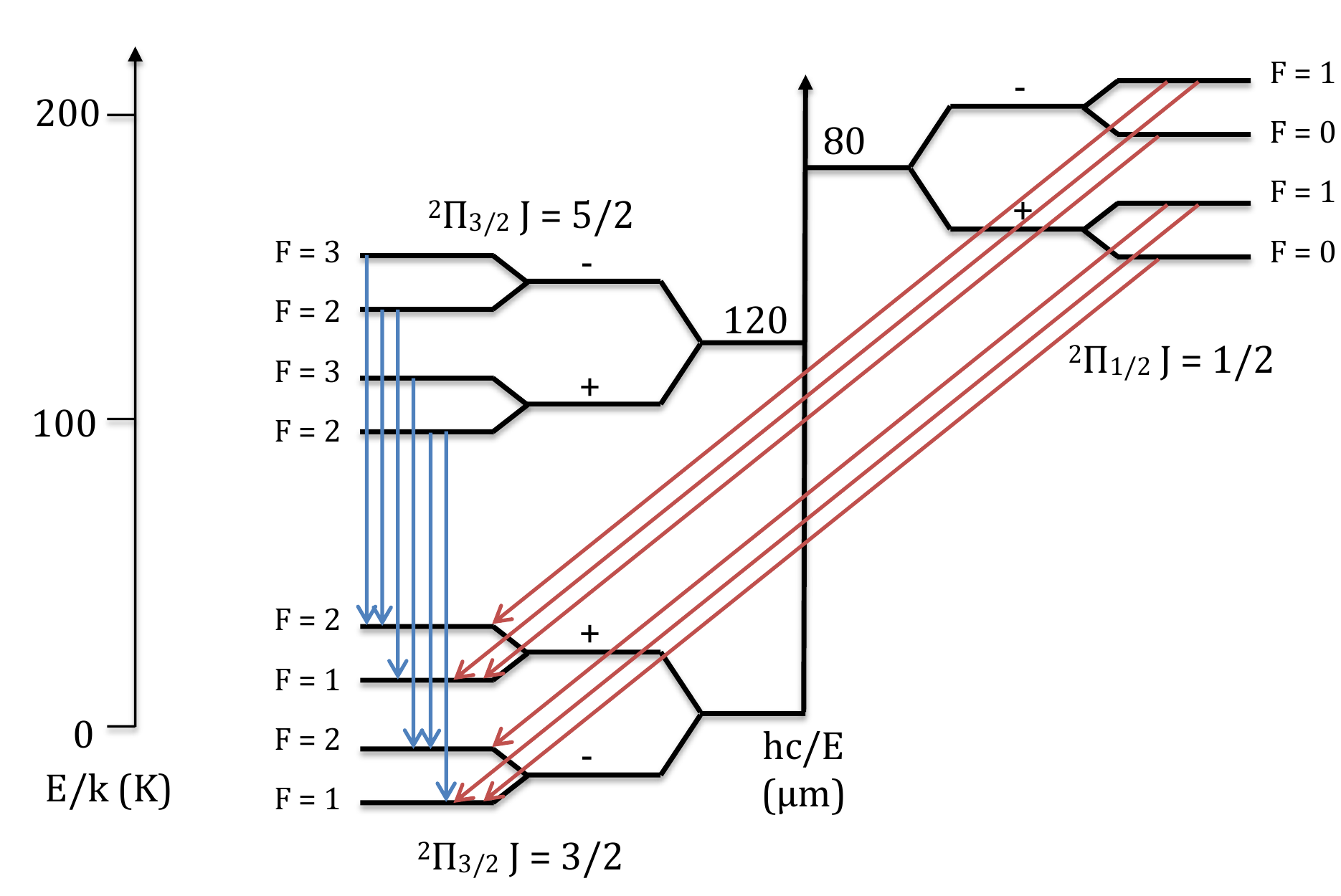}
	\end{center}
	\caption{Schematic of the three lowest rotational states of OH, indicating their $\Lambda$~and hyperfine splitting. Excitations above the \twoPi ground state will cascade back down to it via the $^2\Pi_{3/2},~J=5/2$~state, or the $^2\Pi_{1/2},~J=1/2$~state. Allowable transitions are those where parity is changed and $|\Delta F|$ = 1, 0; shown in blue at left and red on the schematic. The energy scale is given at left in kelvin, and the wavelengths of the IR transitions are shown at centre in $\mu$m. The splittings of the $\Lambda$~and hyperfine levels are greatly exaggerated for clarity. Figure from \citet{Petzler2020}.}
	\label{fig:rotational_ladder}
\end{figure}

OH excitation is complex. Significant departures from local thermodynamic equilibrium (LTE) are almost ubiquitous in the ISM, leading to anomalous excitation in all four of the ground state transitions \citep{Turner1979,Crutcher1977,Dawson2014,Li2018,Petzler2020}. The majority of this anomalous excitation is seen in the satellite lines and is due to asymmetries in the infrared (IR) de-excitation cascade pathways into the ground-rotational state from excited rotational states  \citep{Elitzur1976,Elitzur1976etal,Elitzur1978,Guibert1978}. All cascades into the ground-rotational state will pass through either the first-excited $^2 \Pi_{3/2}\,J=5/2$~rotational state or the second-excited $^2 \Pi_{1/2}\,J=1/2$~rotational state \citep{Elitzur1992}, and these and the ground-rotational state are shown in Fig. \ref{fig:rotational_ladder}. Radiative transitions between these states are subject to selection rules based on the parity and total angular momentum quantum number $F$~of the upper and lower levels: %of the IR transition: 
parity must change and $|\Delta F|$ = 1, 0. These allowed transitions are indicated in Fig. \ref{fig:rotational_ladder} by the blue and red arrows. The number of possible pathways into each level then introduces a natural asymmetry for intra-ladder (blue) or cross-ladder (red) cascades \citep{Elitzur1976}. Selective excitation into the first-excited $^2 \Pi_{3/2}\,J=5/2$~rotational state, for instance, will tend to cascade back into the ground state into its $F=2$~levels more often than its $F=1$~levels, while the opposite is true for cascades from the second-excited rotational level \citep{Elitzur1976etal}. In most cases these cascade mechanisms will be responsible for the majority of the divergence from equal populations seen in the levels of the ground-rotational state \citep{Elitzur1992}. This implies that the ground-rotational state transitions between levels with different $F$~quantum numbers (i.e. the satellite lines) will often have excitation temperatures that differ widely from one another and from those of the main lines. In contrast, the main lines -- which involve transitions between levels with the same $F$~quantum numbers -- will tend to have excitation temperatures similar to one another and to the kinetic temperature. 

However, the main lines are not fully immune from this anomalous excitation as noted observationally as early as the 1970s \citep[e.g.][]{NguyenQRieu1976,Crutcher1977,Crutcher1979}. The mechanism by which the main lines may diverge from LTE is an extension of the mechanism that leads to anomalies in the satellite lines: an additional imbalance in cascade pathways is introduced by an imbalance in the excitations into the upper and lower halves of the lambda-doublets. Briefly, this is caused by two key factors: transitions into the upper half of the lambda-doublet in the ground-rotational state originate from the upper half of the lambda-doublet in either the first- or second-excited rotational states (and vice-versa), and the energy difference between arms of these lambda-doublets increase moving up the rotational ladder. These factors imply that an imbalance can be introduced between pathways into the upper and lower level of the ground-rotational state lambda-doublet by a radiation field that diverges significantly from a Planck distribution \citep[i.e. from hot dust][]{Elitzur1978} or by collisional excitations from particles whose motions diverge significantly from a Maxwellian distribution \citep[i.e. from particle flows][]{Elitzur1979}. In general, since the main lines tend to be seen in their LTE ratio more often than the satellite lines, we may therefore conclude that the conditions required to create this imbalance in cascade pathways is less common in the ISM than those responsible for the satellite-line anomalies.

This, coupled with the fact that for practical reasons many researchers observe only the stronger main lines of OH \citep[e.g.][]{Li2018,Nguyen2018,Engelke2018}, has led researchers in the field of diffuse OH studies to describe the excitation of the OH via the idea of so-called `main-line LTE' -- where the main lines have excitation consistent with LTE -- as evidenced most often by the ratio of their optical depths ($\tau_{\rm peak}(1667)/\tau_{\rm peak}(1665)=1.8$~in LTE) or brightness temperatures ($T_{\rm b}(1667)/T_{\rm b}(1667)=1.8$~in the optically thin limit and $=1$~in the optically thin limit in LTE). Many works \citep[e.g.][]{Li2018,Rugel2018,Yan2017,Ebisawa2019,Engelke2019} then report the degree to which the main lines do or do not obey this relationship. 

It is often -- though not always -- the case (as these works clearly show) that the main-line optical depths or brightness temperatures have a ratio consistent with LTE within the observational uncertainties, and their excitation temperatures are often very similar. However, if the satellite lines are also observed, it is then quite clear that they do not exhibit the same `LTE-like' behaviour \citep[e.g.][]{Ebisawa2015,Ebisawa2019,Xu2016,Petzler2020,Dawson2014,Rugel2018,vanLangevelde1995,Frayer1998}. The nature of this divergence from LTE (i.e. the relationship between satellite-line excitation temperatures or the presence of population inversions) can then provide additional valuable information about the conditions of the gas that may otherwise not be apparent if only the main lines were considered \citep{Petzler2020}. In this work we examine all four ground-rotational transitions and explore the relationships between their optical depth ratios and differences in their excitation temperatures.

\subsection{Observing OH}\label{Sec:observing_OH}
The observed continuum-subtracted line brightness temperature $T_{\rm b}$~of an extended, homogeneous, isothermal ISM cloud towards a compact background continuum source of brightness temperature $T_{\rm c}$~and a diffuse continuum background of brightness temperature $T_{\rm bg}$~is related to the optical depth $\tau_{\nu}$~and excitation temperature $T_{\rm ex}$~of the transition via the solution to the radiative transfer equation: 

\begin{equation}\label{Eq:FullRadTran}
    T_{\rm b} = (T_{\rm ex}-T_{\rm c}-T_{\rm bg})(1-e^{-\tau_{\nu}}).
\end{equation}

We are interested in $\tau_{\nu}$~and $T_{\rm ex}$~because they allow us to characterise the excitation of the ground-rotational state. Excitation temperature is a re-parameterisation of the populations in the upper and lower levels of the transition, and can be described in terms of the column densities in the upper ($N_u$) and lower ($N_l$) levels as:
% Add all the stuff about why we want tau and Tex, how it lets us characterise the excitation
    \begin{equation}
    \frac{N_u}{N_l} = \frac{g_u}{g_l} \exp{\bigg[\frac{-h\nu_0}{k_{\rm B} T_{\rm ex}}\bigg]},
    \label{Tex}
    \end{equation}
\noindent where $g_u$~and $g_l$~are the degeneracies of the upper and lower levels of the transitions (determined by $g=2F+1$, see Fig. \ref{fig:OHlevels}), and $\nu_0$~is the rest frequency of the transition. Optical depth is defined by:
% If we can determine $T_{\rm ex}$ and $\tau_\nu$ for a given transition %\footnote{Technically only 3 excitation temperatures are needed as the fourth can be determined from the excitation temperature sum rule which arises because the four transitions share four levels:\\$\frac{\nu_{1612}}{T_{\rm ex}(1612)}+\frac{\nu_{1720}}{T_{\rm ex}(1720)}=\frac{\nu_{1665}}{T_{\rm ex}(1665)}+\frac{\nu_{1667}}{T_{\rm ex}(1667)}$.} 
% we can then find the absolute column densities in the relevant levels (as opposed to their ratios) using the definition of optical depth:
    \begin{equation}
    \tau_{\nu}=\frac{c^2}{8\pi \nu^2_0} \frac{g_u}{g_l}\,N_l\,A_{ul} \bigg(1-\exp{\bigg[\frac{-h\nu_0}{k_{\mathrm{B}} T_\mathrm{ex}}\bigg]}\bigg)\,\phi(\nu),
    \label{tau}
    \end{equation}
\noindent where $A_{ul}$~is the Einstein-A coefficient and $\phi(\nu)$~is the line profile. If both optical depth and excitation temperature can be determined for a given transition, we may then calculate the column densities in both the upper and lower levels of that transition. Since the four ground-rotational transitions share four levels, a minimum of two transitions are needed to fully characterise the excitation of the ground-rotational state. This excitation is a function of the local environment of the gas which may be parameterised through use of (or reference to) non-LTE molecular excitation modelling \citep[e.g.][]{Xu2016,Ebisawa2019,Petzler2020}. 

Unfortunately, Eq. \ref{Eq:FullRadTran} is insufficient to solve for both $\tau_{\nu}$~and $T_{\rm ex}$ uniquely, but several strategies exist to break this degeneracy. One such method is to make additional observations just off the compact background continuum source. These observations should not include any of the compact background continuum emission, but still point towards the same extended OH gas with the same $\tau_{\nu}$ and $T_{\rm ex}$, and include the same diffuse background $T_{\rm bg}$. In this case the average continuum-subtracted brightness temperature of these `off-source' positions will be described by:

\begin{equation}\label{Eq:off}
    T_{\rm b}^{\rm off} = (T_{\rm ex}-T_{\rm bg})(1-e^{-\tau_{\nu}}).
\end{equation}

Following \citet{Heiles2003}, we refer to this averaged off-source spectrum as the `expected brightness temperature' $T_{\rm exp}$~as it represents the spectrum we would expect to observe if we could turn off the compact background continuum source $T_{\rm c}$. We can then combine Eqs. \ref{Eq:FullRadTran} and \ref{Eq:off} to obtain the optical depth spectrum:

\begin{equation}
    \tau_{\nu}=-\ln \Bigg(\frac{T_{\rm b}-T_{\rm exp}}{T_{\rm c}}+1\Bigg).
\end{equation}

As we will describe further in the Observations section, we have observations of this type (which we refer to as `on-off' observations) from the Arecibo radio telescope toward 92 compact extragalactic continuum sources. These data include 8 spectra per sightline: one optical depth and one expected brightness temperature for each of the four ground-rotational transitions of OH.

The degeneracy between optical depth and excitation temperature can also be broken by observing bright compact background continuum sources with an interferometer -- and thus rendering the $T_{\rm ex}$~and $T_{\rm bg}$~terms in Eq. \ref{Eq:FullRadTran} insignificant. The reason for this is twofold: first, the emission from the extended OH in the intervening cloud and the diffuse background continuum are assumed to be smooth on the sky and large compared to the interference fringes of the interferometer, so that the flux detected from both will be negligible. Additionally, if $T_{\rm c}\gg |T_{\rm ex}|$~(which is likely to be the case if a bright compact background continuum source is targeted) then the $T_{\rm c}$~term will dominate Eq. \ref{Eq:FullRadTran}, and the observed brightness temperature will be well-described by $T_{\rm b}=T_{\rm c}(e^{-\tau_{\nu}}-1)$, even if some flux from the extended cloud is detected. We have observations of this type from the Australia Telescope Compact Array (ATCA) towards 15 bright compact continuum sources in the Galactic plane. These data only include 4 spectra per sightline; since the observing strategy rendered the $T_{\rm ex}$~and $T_{\rm bg}$~terms in Eq. \ref{Eq:FullRadTran} insignificant we are unable to construct an expected brightness temperature spectrum, and only have optical depth spectra for each transition. 

The individual features in these OH spectra will be broadened by mostly Gaussian processes \citep[turbulent or thermal broadening, e.g.][]{Leung1976, Liszt2001}. Other sources of broadening that are not Gaussian also contribute to the line profile (i.e. natural and collisional broadening -- both Lorentzian in shape) but are assumed to have negligible contribution to the feature shape. A single telescope pointing will tend to detect several blended Gaussian-shaped features arising from the same transition at different line-of-sight velocities. In our analysis these Gaussian-shaped profiles are interpreted as individual isothermal clouds along the line of sight: each cloud may then be expected to result in a feature with the same centroid velocity and full width at half-maximum (FWHM) in all the observed spectra (8 in the case of on-off observations, 4 if only optical depth spectra are obtained).

This work represents an unprecedented analysis of OH in the diffuse ISM due primarily to the Gaussian decomposition method used. The observed spectra were decomposed into individual Gaussian components using \amoeba\footnote{https://github.com/AnitaPetzler/AMOEBA} \citep{Petzler2021a}: an automated Bayesian line-fitting algorithm in Python. \amoeba's key advantage over other Gaussian decomposition methods is that it is able to simultaneously fit optical depth and expected brightness temperature spectra in all four ground-rotational transitions. Each Gaussian feature is parameterised by its centroid velocity, FWHM, $\log$~column density of the lowest ground-rotational state level ($\log N_1$), and inverse excitation temperatures of the 1612, 1665 and 1667\,MHz transitions. These parameters are then sufficient to fully characterise the associated peak optical depths and expected brightness temperatures in all four transitions. Alternatively, in the case of our ATCA data, \amoeba~can take a set of 4 optical depth velocity spectra, and each Gaussian component is then parameterised by its centroid velocity, FWHM and peak optical depths in each of the four ground-rotational state transitions. Further details about our usage of \amoeba~are given in the Method practicalities and limitations section.

\subsection{OH and H\textsc{i} cold neutral medium}
% background for comparison with CNM
In this work we will compare our OH data with published measurements of the atomic \HI\ gas. In pressure equilibrium, most of the \HI\ is expected to reside in two distinct thermal phases \citep{Field1969, McKee1977, Wolfire1995, Wolfire2003}: the warm neutral medium (WNM) at temperatures of several thousand kelvin, and the cold neutral medium (CNM) at temperatures at or below $\sim100\,$K for typical pressure ranges found in the Galaxy \citep[e.g.][]{Dickey1978, Heiles2003b, Jenkins2011,Murray2018,Nguyen2019,Murray2020}. It is generally accepted that, in cool regions of the ISM (like the CNM) molecular hydrogen forms primarily on dust grains \citep{McCrea1960, Gould1963, Hollenbach1971}, and can accumulate once it is sufficiently shielded from dissociating UV. This is not a uni-directional process, as matter can cycle back and forth from one stable phase to another \citep{Ostriker2010}, and the phases (WNM, CNM and H$_2$) are generally mixed \citep{Goldsmith2009}. Therefore, one might expect the properties of the molecular gas (as traced here by OH) to maintain some relationship to the CNM gas from which it presumably formed. 

72 of the 92 Arecibo sightlines with on-off observations examined in this work were simultaneously observed in H\textsc{i}. \citet{Nguyen2019} identified 327 individual CNM components along these sightlines (seen in absorption and emission, see \citet{Nguyen2019} for further details), and characterised their individual centroid velocities, FWHMs, peak optical depths, spin temperatures and column densities. As we describe in the Analysis section, we identify a total of 43 OH features along 20 of these sightlines, and we match these in velocity to their closest CNM feature. Some CNM features are matched with several OH features, for a total of 43 OH components matched with 26 CNM components that we then discuss.

%% file: 2_Data.tex
\begin{figure*}
    \centering
    \includegraphics[trim={1.5cm 1cm 2.5cm 3cm},clip=true,width=\linewidth]{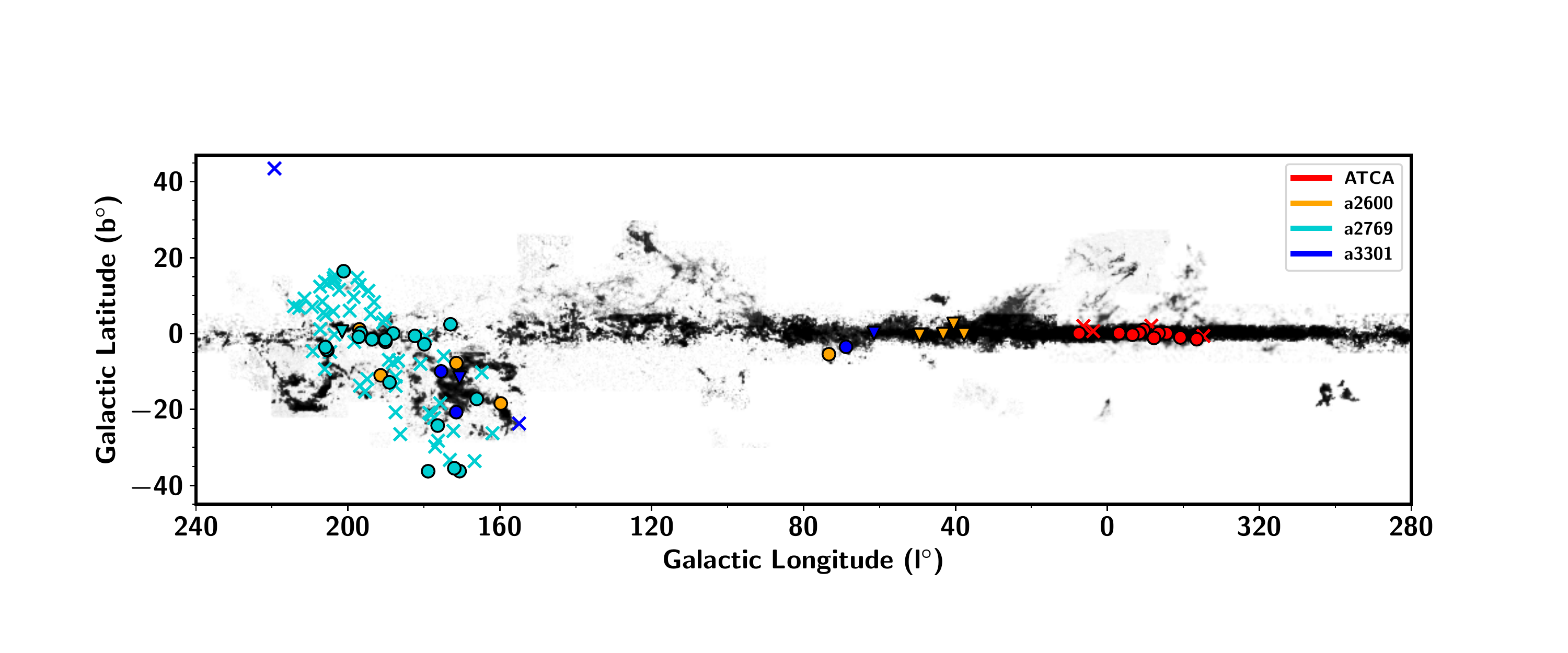}
    \caption{Positions of sightlines examined in this work from the Australia Telescope Compact Array (ATCA), and from the projects a2600, a2769 and a3301 from the Arecibo telescope. Sightlines with detections are indicated by filled circles, non-detections are indicated by crosses. Sightlines excluded from analysis are indicated by triangles. The grey-scale image is CO emission \citep{Dame2001} and is included for illustrative purposes only.}
    \label{fig:All_data}
\end{figure*}

This work utilises two distinct sets of OH observations. The first is a collection of observations toward 92 compact background continuum sources obtained through the GNOMES (Galactic Neutral Opacity and Molecular Excitation Survey) collaboration taken by the Arecibo telescope. The second set are observations towards 15 bright, compact continuum sources in the region of the Southern Parkes Large Area Survey in Hydroxyl \citep[SPLASH,][]{Dawson2014} made with the Australia Telescope Compact Array (ATCA). The locations of all sightlines examined in this work are shown in Fig. \ref{fig:All_data}. 

\subsection{Arecibo observations}
Our observations from the Arecibo telescope obtained through the GNOMES collaboration are comprised of data from three projects: a2600 \citep{Thompson2019}, a2769 \citep{Nguyen2019} and a3301. This data set consists of on-off spectra of the four OH ground-rotational transitions toward 92 sightlines in the Arecibo sky. These sightlines are listed in Table \ref{tab:Arecibo} along with their sensitivities in optical depth and expected brightness temperature (quantified by the rms noise of the individual spectra) for each transition. As can be seen in Fig. \ref{fig:All_data} the majority of these sightlines were out of the Galactic Plane. The angular resolution of the Arecibo telescope at the frequency of the OH ground-rotational transitions is $\sim 3'.5$.

\begin{figure}
    \centering
    \includegraphics[trim={0.5cm 0.5cm 0.1cm 0cm}, clip=true, width=\linewidth]{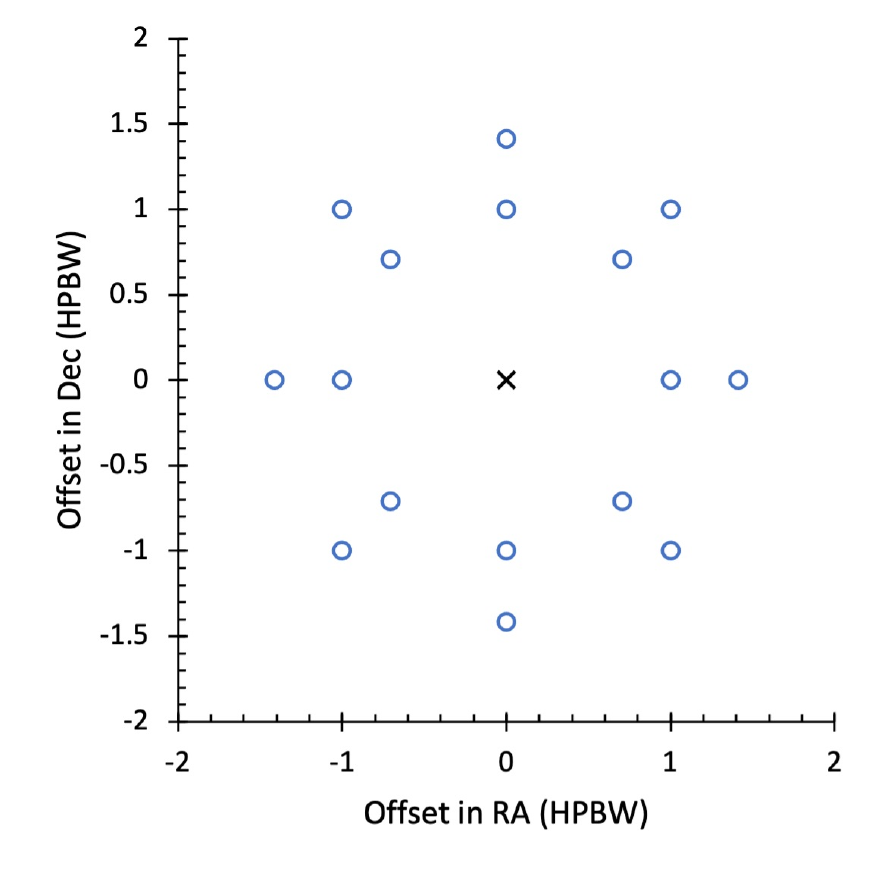}
    \caption{Offsets (in degrees) of off-source pointings (blue circles) in RA and Dec in terms of the telescope half-power beam width (HPBW) relative to the on-source pointing (black cross). The 16 off-source pointings are placed at distances of 1 and $\sqrt{2}$~times the HPBW in the four cardinal directions and in directions rotated 45$^{\circ}$~from these as shown.}
    \label{fig:off-source}
\end{figure}

The aim of project a2600 (PI Thompson) was to use Zeeman splitting of the OH ground-rotational state transitions to measure magnetic field strengths in the envelopes of molecular clouds. The targets for this project were compact extragalactic continuum sources chosen from the National Radio Astronomy Observatory Very Large Array Sky Survey \citep[NVSS][]{Condon1998} with brightness $S_{\nu}\gtrsim 0.5\,$Jy behind molecular clouds identified from CO emission maps \citep{Dame2001}. This project targeted regions of the inner and outer Galaxy, and includes sightlines passing through molecular clouds with low-mass star formation (e.g. Taurus) and high-mass star formation (e.g. Mon OB1) mostly near the Galactic plane. Observations were made both on- and off-source, allowing optical depth and expected brightness temperature spectra to be produced following the method of \citet{Heiles2003}. The 16 off-source pointings were arranged as illustrated in Fig. \ref{fig:off-source}. This pattern of off-source pointings was also used in the other projects outlined in this section. We have observations towards 12 sightlines from this project.

The aim of project a2769 (PI Stanimirovi{\'c}) was to explore the relationships between WNM, CNM and molecular gas in the Taurus and Gemini regions. Their observations also included on-off measurements, and targeted compact extragalactic continuum sources in the Taurus, California, Rosette, Mon OB1 and NGC 2264 giant molecular clouds. Their continuum sources were also selected from the NVSS catalog and have typical flux densities of $S_{\nu}\gtrsim 0.6\,$Jy at 1.4\,GHz. Our data include observations towards 73 sightlines from this project.

The aim of project a3301 (PI Petzler) was to follow up lines of sight observed in previous projects included in the GNOMES collaboration that showed `anomalous excitation': this generally involved interesting patterns of emission and absorption across the available transitions. Most of these were chosen because not all four transitions had been observed in the original project. These sightlines will therefore be biased towards anomalous excitation, but due to poor data quality in some of the 1720\,MHz spectra, only 6 of the 16 sightlines observed in that project were included in this work.

\subsection{ATCA observations}
Our ATCA data (taken under project code C2976) include sightlines towards 15 bright compact continuum sources selected from the 843 MHz Molongo Galactic Plane Survey catalogue \citep[MGPS,][]{Murphy2007}, the Southern Galactic Plane Survey \citep[SGPS,][]{Haverkorn2006} and the NVSS 1.4 GHz continuum images. All sources were cross-checked against the recombination line measurements of \citet{Caswell1987} in order to discriminate between H{\sc ii} regions and other source types, and were also examined for evidence of H{\sc i} absorption in SGPS datacubes in order to confirm near- or far-side Galactic distances where relevant. 
Bright, compact sources (unresolved or with sufficient unresolved structure at a beam size of $\sim30''$) were chosen, located between $332^{\circ} < l < 8^{\circ}$, $|b| < 2.1^{\circ}$ to match the region mapped in the Southern Parkes Large Area Survey of Hydroxyl \citep[SPLASH][]{Dawson2022}. Sources with a spectral flux density $\sim 1$~Jy at 1.6 GHz were preferred, which would result in brightness temperatures of $\sim 500$ K when observed with our array configuration (ATCA 1.5D, excluding antenna 6). Distant sources were considered preferable as they probe a larger number of absorbing components along the line of sight. However, the number of extragalactic and far-side Galactic sources with sufficient flux density and compact structure was small. Therefore the target criteria were expanded to include nearside \HII~regions with evidence for bright and compact substructure and intervening H{\sc i} absorption. 

The CFB 1M-0.5k mode on the ATCA Compact Array Broadband Backend (CABB) was used to simultaneously observe all four ground state OH lines in zoom bands centred on the line rest frequencies (a single zoom band was used for the main lines, centred at 1666 MHz). This provided a raw channel width of 0.09 km s$^{-1}$. The 1.5D array resulted in a synthesised beam size of $\sim30''$ at 1.6 GHz. The total observing time for all 15 sources was 50 hours. 

The raw visibility data from the ATCA (excluding antenna 6) was reduced using the \textsc{miriad}\footnote{http://www.atnf.csiro.au/computing/software/miriad} package \citep{Sault1995}. The main-line observations at 1666 MHz contained more radio frequency interference (RFI) than the satellite-line observations. Flagging this RFI resulted in systematically larger synthesised beams for the main-line observations, and hence lower continuum brightness temperatures in the main lines (see Table~\ref{Tab:ContSources}). This would not affect the peak optical depths measured in our analysis as they are derived from a ratio of $T_{\rm b}$~and $T_{\rm c}$~which are equally affected by this increase in synthesised beam. 
The visibilities were inverted using a Brigg's visibility weighting robustness parameter of 1 \citep{Briggs1995}, corresponding to roughly natural weighting. The velocity spectrum at the location of the brightest continuum pixel was selected for further analysis. A linear baseline was fit to these velocity spectra to determine the background continuum brightness temperature \Tc, which was subtracted to produce line brightness temperature (\Tb) spectra.  These were then converted to optical depth (\taunu) spectra, assuming that $T_{\rm b}=T_{\rm c}(e^{-\tau_{\nu}}-1)$. 
The rms noise levels of the optical depth spectra ranged from 0.006 to 0.023, and are outlined in Table \ref{Tab:ContSources}. %From the 15 sightlines observed with the ATCA, 11 had detections yielding a total of 51 individual features.

\begin{table*}
\centering
\setlength\tabcolsep{4pt}
\begin{tabular}{llrrccccccccccccc}
\hline
&&&&\multicolumn{4}{c}{$T_{\rm bg}$\,(K)}&\multicolumn{4}{c}{$\tau_{\sigma}~(10^{-3})$}&\multicolumn{4}{c}{$T_{{\rm exp}\,\sigma}$\,(10$^{-2}$\,K)}&\\
\cline{5-16}
Source$^a$&Project$^b$&$l^{\circ}$&$b^{\circ}$&1612&1665&1667&1720&1612&1665&1667&1720&1612&1665&1667&1720&Det.$^{c}$\\
\hline
$^*$B1858+0407&a2600&37.76&-0.21&9&10&10&10&3&3&3&3&4&5&5&4&N\\
$^*$B1853+0749&a2600&40.50&2.54&7&7&7&7&3&2&2&4&5&4&4&6&N\\
$^*$B190840+09&a2600&43.25&-0.18&8&9&9&9&1&1&1&1&5&6&5&5&N\\
$^*$B1919+1357&a2600&48.92&-0.28&12&12&12&13&3&2&3&3&10&9&9&8&N\\
$^*$B1920+1410&a2600&49.21&-0.34&11&12&12&12&2&2&2&2&9&8&8&8&N\\
$^*$B1921+1424&a2600&49.49&-0.38&11&12&12&12&1&1&1&1&12&11&12&11&N\\
$^*$PKS1944+251&a2600&61.47&0.09&5&5&5&5&2&2&2&3&4&3&7&5&N\\
SRC44&a3301&68.83&-3.49&214&234&235&257&1&1&1&1&3&3&3&3&Y\\
3C417&a2600&73.33&-5.45&14&24&26&8&4&4&4&6&3&3&3&4&Y\\
4C+30.04&a3301&154.92&-23.69&164&179&180&197&2&3&3&5&3&4&4&4&N\\
3C092&a2600&159.74&-18.41&9&32&33&9&3&2&2&2&2&1&2&2&Y\\
4C+24.06&a2769&161.92&-26.26&4&6&2&5&4&3&3&4&2&2&1&2&N\\
3C115&a2769&164.76&-10.24&10&5&6&5&6&4&4&6&4&3&3&4&N\\
4C+28.11&a2769&166.06&-17.22&4&7&11&4&5&3&4&6&3&2&2&3&Y\\
4C+16.09&a2769&166.64&-33.60&79&110&110&81&2&1&1&2&5&4&4&6&N\\
$^*$SRC10&a3301&170.58&-11.66&3&3&3&3&1&1&1&1&4&5&5&6&N\\
PKS0319+12&a2769&170.59&-36.24&18&24&29&19&2&1&1&2&2&2&2&2&Y\\
3C131&a2600&171.44&-7.80&36&51&52&23&1&1&1&1&2&1&1&2&Y\\
3C108&a3301&171.47&-20.70&166&182&182&199&2&2&2&2&3&3&3&4&Y\\
4C+11.15&a2769&171.98&-35.48&6&10&13&6&4&4&4&5&3&3&3&4&Y\\
4C+18.11&a2769&172.23&-25.66&7&3&4&5&4&3&3&4&3&2&2&3&N\\
4C+36.10&a2769&172.98&2.44&5&6&11&4&5&3&4&5&3&2&2&3&Y\\
3C090&a2769&173.15&-33.30&15&17&17&14&2&1&1&2&2&2&2&2&N\\
4C+29.16&a2769&174.77&-5.97&4&5&4&4&7&4&5&7&3&2&2&3&N\\
4C+27.14&a3301&175.46&-9.96&171&187&188&205&2&5&5&5&2&4&4&5&Y\\
4C+21.17&a2769&175.70&-18.36&6&7&8&10&3&2&2&3&3&2&2&3&N\\
3C096&a2769&176.27&-28.26&6&8&9&5&3&2&2&3&2&2&2&2&N\\
4C+17.23&a2769&176.36&-24.24&6&10&18&5&3&2&3&4&2&2&2&2&Y\\
J035613+130535&a2769&177.02&-29.78&5&4&5&4&3&2&2&3&2&1&1&2&N\\
3C114&a2769&177.30&-22.24&7&6&6&5&3&2&2&3&2&2&1&2&N\\
4C+17.25&a2769&178.11&-21.31&5&6&2&4&5&3&3&5&3&2&2&3&N\\
4C+17.26&a2769&178.56&-20.88&5&6&7&5&5&4&4&6&4&3&3&4&N\\
4C+07.13&a2769&178.87&-36.27&10&10&5&7&4&3&3&4&3&2&2&3&Y\\
4C+29.19&a2769&179.53&-0.59&3&4&3&4&5&4&4&6&3&2&2&3&N\\
B0531+2730&a2769&179.87&-2.83&7&15&17&5&4&4&4&5&3&2&2&3&Y\\
4C+23.14&a2769&180.86&-8.01&7&8&7&5&5&4&4&5&3&2&2&3&N\\
4C+26.18b&a2769&182.36&-0.62&7&7&15&5&6&4&5&7&4&3&3&4&Y\\
4C+08.15&a2769&186.21&-26.51&5&3&6&5&2&2&2&3&2&1&1&2&N\\
PKS0531+19&a2769&186.76&-7.11&85&74&84&84&2&2&2&2&6&4&4&6&N\\
3C138&a2769&187.41&-11.34&122&97&105&101&2&1&2&2&6&4&4&6&N\\
PKS0509+152&a2769&187.41&-13.79&4&4&1&4&7&5&4&6&3&2&2&3&N\\
PKS0446+11&a2769&187.43&-20.74&7&5&6&5&5&4&3&5&3&2&2&3&N\\
4C+22.12&a2769&188.07&0.04&23&22&32&21&2&2&2&3&3&2&3&4&Y\\
4C+17.33&a2769&188.22&-7.67&4&5&3&3&6&4&5&7&3&2&3&3&N\\
4C+14.14&a2769&189.04&-12.85&4&4&3&3&4&3&4&5&3&2&2&2&Y\\
\hline
\end{tabular}
\caption{Summary of sightlines observed by the Arecibo telescope included in this work. $^a$Source names are given along with the original Arecibo $^b$project designation and the galactic longitude and latitude. $^{c}$Sources with detections are indicated `Y' and those without are indicated `N'. Source names indicated with asterisks were excluded from analysis due to contamination of off-source pointings as described in the text. The brightness temperature of the background continuum $T_{\rm bg}$~at each of the four OH ground-rotational state transitions are given along with the rms noise of the optical depth $\tau_{\sigma}$~and expected brightness temperature spectra $T_{\rm exp\,\sigma}$.}
\label{tab:Arecibo}
\end{table*}

\begin{table*}
\centering
\setlength\tabcolsep{4pt}
\begin{tabular}{llrrccccccccccccc}
\hline
&&&&\multicolumn{4}{c}{$T_{\rm bg}$\,(K)}&\multicolumn{4}{c}{$\tau_{\sigma}~(10^{-3})$}&\multicolumn{4}{c}{$T_{{\rm exp}\,\sigma}$\,(10$^{-2}$\,K)}&\\
\cline{5-16}
Source$^a$&Project$^b$&$l^{\circ}$&$b^{\circ}$&1612&1665&1667&1720&1612&1665&1667&1720&1612&1665&1667&1720&Det.$^{c}$\\
\hline
4C+17.34&a2769&189.21&-6.93&11&7&9&8&4&3&3&4&4&2&2&4&N\\
4C+19.18&a2769&190.09&-2.17&7&6&13&5&5&3&3&5&3&2&2&3&Y\\
4C+19.19&a2769&190.13&-1.64&5&6&14&4&3&3&3&4&2&2&2&2&Y\\
4C+22.16&a2769&190.16&3.91&8&4&5&5&4&3&4&5&3&2&2&3&N\\
J062019+210229&a2769&190.74&2.94&9&6&10&5&4&2&3&4&3&2&2&3&N\\
PKS0528+134&a2600&191.37&-11.01&22&43&60&24&3&2&2&2&4&3&3&3&Y\\
3C166&a2769&193.12&8.30&16&15&20&14&2&2&2&2&3&2&2&3&N\\
4C+16.15b&a2769&193.64&-1.53&3&5&8&4&4&3&3&5&2&2&2&2&Y\\
J063451+190940&a2769&193.99&5.10&7&5&6&5&4&3&3&5&3&2&2&3&N\\
4C+21.22&a2769&194.63&11.26&3&3&2&3&6&5&5&7&2&2&2&3&N\\
4C+09.21&a2769&194.89&-11.98&5&5&5&5&6&4&4&6&4&3&3&4&N\\
4C+07.16&a2769&195.51&-15.35&5&5&4&5&5&4&4&5&3&2&2&3&N\\
3C158&a2769&196.64&0.17&14&27&10&14&3&2&2&3&4&3&3&4&Y\\
J053239+073243&a2769&196.84&-13.74&16&14&18&17&3&2&2&3&4&3&3&4&N\\
4C+19.26&a2769&196.91&12.80&4&3&5&3&4&4&4&5&2&2&2&2&N\\
4C+14.18&a2600&196.98&1.10&20&45&47&22&2&1&1&1&2&1&1&1&Y\\
4C+13.32&a2769&197.15&-0.85&4&4&4&4&2&2&2&3&3&2&2&3&Y\\
PKS0715+20&a2769&197.52&14.74&4&5&4&5&5&4&4&5&3&2&2&3&N\\
J061622+115553&a2769&198.33&-2.20&3&6&5&5&3&3&3&4&2&2&2&2&N\\
J070001+170922&a2769&198.47&9.58&9&11&8&8&3&2&2&3&3&2&2&3&N\\
4C+14.20&a2769&199.52&6.04&5&7&6&4&4&3&3&5&3&2&2&3&N\\
4C+17.41&a2769&201.13&16.42&2&4&8&4&5&4&3&5&2&2&2&2&Y\\
$^*$4C+10.20&a2769&201.53&0.51&4&4&4&4&4&3&3&4&5&4&4&5&N\\
3C175.1&a2769&202.29&11.53&10&18&14&12&2&2&2&3&3&2&2&3&N\\
4C+15.20&a2769&203.42&15.42&9&11&9&9&3&2&2&3&3&2&2&3&N\\
4C+08.21&a2769&203.54&-0.27&6&5&8&5&4&2&3&4&3&2&2&3&N\\
4C+14.23&a2769&203.64&13.91&5&5&5&6&4&3&3&4&3&2&2&3&N\\
3C181&a2769&203.75&14.63&12&8&11&14&2&2&2&2&3&2&2&3&N\\
4C+10.21&a2769&203.85&5.82&6&5&3&4&5&3&4&5&3&2&2&3&N\\
J061900+050630&a2769&204.66&-4.84&7&5&4&3&6&5&4&7&4&2&2&3&N\\
PKS0722+12&a2769&205.35&13.17&4&4&4&3&5&5&5&6&2&2&2&2&N\\
4C+04.22&a2769&205.41&-4.43&8&7&14&4&5&3&3&5&3&2&2&3&Y\\
J134217-040725&a2769&205.58&-4.14&4&4&4&4&5&3&4&5&3&2&2&3&Y\\
4C+08.23&a2769&205.81&4.91&3&2&3&4&6&4&4&6&3&2&2&3&N\\
4C+04.24&a2769&205.92&-3.57&4&4&4&4&5&3&4&5&3&2&2&3&Y\\
4C+01.17&a2769&206.08&-9.37&3&2&2&3&6&5&5&7&2&2&2&3&N\\
4C+12.30&a2769&206.09&13.67&6&9&3&4&6&4&4&6&4&3&2&4&N\\
J065917+081331&a2769&206.48&5.48&8&6&4&5&5&4&4&5&3&2&2&3&N\\
4C+09.27&a2769&206.72&8.44&8&5&6&5&4&2&3&4&2&2&2&2&N\\
3C167&a2769&207.31&1.15&9&7&8&7&5&4&3&5&4&3&3&4&N\\
4C+10.22&a2769&207.31&12.37&9&7&10&7&4&3&3&4&3&2&2&3&N\\
4C+01.19&a2769&209.24&-4.64&3&3&4&3&5&4&4&6&3&2&2&3&N\\
4C+06.28&a2769&209.43&7.00&2&3&3&3&6&5&5&7&3&2&2&3&N\\
PKS0719+056&a2769&211.43&9.23&2&6&6&3&5&3&3&5&3&2&2&3&N\\
4C+03.12&a2769&212.82&6.78&2&4&4&3&6&6&5&8&3&3&2&3&N\\
J071924+021035&a2769&214.18&7.22&2&4&3&3&8&6&5&8&3&2&2&3&N\\
SRC19&a3301&219.34&43.50&160&175&176&192&1&1&1&1&4&2&2&3&N\\
\hline
\multicolumn{17}{l}{Table \ref{tab:Arecibo} continued.}\\
\end{tabular}
% \caption{continued}
\end{table*}

\begin{table*}
\begin{tabular}{lccccccclc}
\hline
&\multicolumn{3}{c}{T$_{\rm c}$ (K)}&\multicolumn{4}{c}{$\tau_{\sigma}~(10^{-3})$}&&\\
\cline{2-8}
Source&1612$^{*}$&1666$^{*}$&1720$^{*}$&1612&1665&1667&1720&Notes&Det.\\
\hline
G334.72-0.65&503&470&498&54&56&55&58&2 (16 \kms)$^{a}$&N\\
G336.49-1.48&2339&2220&2374&15&15&15&16&1 (-25 \kms)$^{a}$&Y\\
G340.79-1.02&1462&1406&1401&20&21&21&23&1 (-25 \kms)$^{a}$&Y\\
G344.43+0.05&908&887&898&28&30&29&29&1 (-67 \kms)$^{a}$&Y\\
G346.52+0.08&428&405&417&59&60&59&61&2 (2 \kms)$^{a}$&Y\\
G347.75-1.14&1427&1313&1389&17&18&18&18&3$^{c}$&Y\\
G348.44+2.08&466&415&430&41&45&44&48&3$^{d}$&N\\
G350.50+0.96&1120&1056&1143&21&22&22&22&1 (-11 \kms)$^{b}$&Y\\
G351.56+0.20&975&1028&902&26&23&23&29&2 (-45 \kms)$^{b}$&Y\\
G351.61+0.17&935&867&905&27&27&27&30&2 (-45 \kms)$^{b}$&Y\\
G353.41-0.30&1286&1142&1139&18&23&22&25&1 (-16 \kms)$^{b}$&Y\\
G356.91+0.08&582&532&537&38&40&40&42&3&Y\\
G003.74+0.64&394&372&380&43&45&45&45&3$^{e}$, 4$^{f}$&N\\
G006.32+1.97&471&426&422&31&33&32&37&3$^{g,h}$&N\\
G007.47+0.06&394&366&398&42&44&42&42&2 (-18 \kms)$^{b}$&Y\\
\hline
\end{tabular}
\caption{Detailed information for the continuum sources coinciding with the sightlines observed by the ATCA examined in this work and their optical depth sensitivities. $^{*}$Central frequency of zoom band (MHz). The systematically lower brightness temperatures in the central band are a result of the slightly larger synthesized beam at this frequency (see text). 
\newline Notes: 
1. \HII~region near-side, radio recombination line in brackets, 
2. \HII~region far-side,  
3. Extragalactic,  
4. Nearby \HII~region. 
\newline References: 
$^{a}$\citet{Caswell1987}, 
$^{b}$\citet{Lockman1989}, 
$^{c}$\citet{Petrov2006}, 
$^{d}$\citet{Condon1998}, 
$^{e}$\citet{Gray1994}, 
$^{f}$\citet{Wink1982}, 
$^{g}$\citet{Helfand1989}, 
$^{h}$\citet{Griffith1993}. 
Sources with detections are indicated `Y' and those without are indicated `N'. 
}
\label{Tab:ContSources}
\end{table*}

%% file: 3_Methods.tex
In this section we discuss practical details and limitations of the methods used in this work. We will also discuss the process of Gaussian decomposition used to obtain our results. This will include details of our use of \textsc{Amoeba}, an automated Bayesian Gaussian decomposition algorithm developed primarily for this dataset. \textsc{Amoeba} is described extensively in \citet{Petzler2021a}, and this section will provide additional details on its use in this work.

Before being decomposed into individual Gaussian components using \textsc{Amoeba}, the OH data from our on-off observations from the Arecibo telescope were processed into sets of optical depth and expected brightness temperature spectra (see Observing OH subsection), following the method of \citet{Heiles2003}. This method included a step where the antenna temperatures $T_{\rm a}$~were converted to brightness temperatures $T_{\rm b}$~by considering the convolution of the antenna beam with the background continuum source through the following relation:
\begin{equation}
    T_{\rm a}=T_{\rm b} \epsilon_{\rm eff},
\end{equation}
\noindent where $\epsilon_{\rm eff}$~is an effective beam efficiency parameter. This parameter accounts for the efficiency of the main beam and the sidelobes as they overlap with the background continuum source. Previous surveys of H\textsc{i} \citep[GALFA-H\textsc{i}][]{Peek2011} apply a single value of $\epsilon_{\rm eff}$, found by averaging the convolution of the beam efficiency with continuum source size over the whole survey. The Millennium survey used a similar approach, adopting an effective beam efficiency of 0.9. Though the OH observed in our data from Arecibo is likely to be less smoothly distributed than the \HI\ of the Millennium survey, in the absence of exact information about that distribution we adopt the same effective beam efficiency of 0.9. This may lead to an underestimation of the brightness temperatures $T_{\rm b}$~and hence our derived excitation temperatures $T_{\rm ex}$, likely by no more than 10\%. Our derived optical depths would be unaffected.

Our method of generating the OH expected brightness temperature spectra differed slightly from the method used for \HI\ observations described by \citet{Heiles2003}, in that we did not interpolate between the off-source pointings to determine $T_{\rm exp}$, but rather simply averaged the off-source brightness temperature spectra. This choice was made because (for a majority of sightlines) there were not significant differences between the features seen in the individual off-source brightness temperature spectra.

As noted in the Observations section the on-off method assumes that the OH optical depths and excitation temperatures, and the diffuse background continuum brightness temperature are the same in both the on-source and all the off-source positions. If one or more of these assumptions is incorrect -- i.e. if the OH gas varies in optical depth or excitation temperature across the on- and off-source pointings or if there is additional continuum behind any of the off-source positions -- then the averaged off-source spectra will not be a good estimation of the expected brightness temperature spectrum of the on-source pointing. For the majority of sources presented in this work (for which the individual off-source pointings were available), there was little noticeable difference between the individual off-source spectra surrounding each on-source pointing before the background continuum $T_{\rm bg}$~had been subtracted. Any variation in the OH gas or continuum between the off-source pointings in these cases is therefore likely to be small. This is in contrast to the findings of \citet{Liszt1996} who note inconsistencies between the absorption (`on-source') and emission (`off-source') spectra of OH. 

We did, however, find a small number of sightlines (9, all indicated in Table \ref{tab:Arecibo} with asterisks) that did show differences in diffuse background continuum and/or off-source OH features. For a given transition, variations such as these affect both the derived optical depth and excitation temperature. In our data, this resulted in un-physical relationships between either the optical depth and the expected brightness temperature of the individual transitions (e.g. positive $\tau_{\nu}$~ but $T_{\rm exp}$~implies a negative $T_{\rm ex}$, or vice-versa), or between the four transitions (e.g. excitation temperatures that violate the excitation temperature sum rule $\frac{\nu_{1612}}{T_{\rm ex}(1612)}+\frac{\nu_{1720}}{T_{\rm ex}(1720)}=\frac{\nu_{1665}}{T_{\rm ex}(1665)}+\frac{\nu_{1667}}{T_{\rm ex}(1667)}$). \amoeba~was unable to construct a model to fit these un-physical features, which remained as significant residuals of the fits. Since the optical depth spectra tend to have higher signal-to-noise, these residuals were mostly seen in the expected brightness temperature spectra (i.e. \amoeba~fitted the optical depth spectra at the expense of residuals to the expected brightness temperature spectra). However, even if the optical depth spectra were well-fit, the resulting parameters from the entire sightline were suspect. Therefore, even if the original individual off-source pointings were not available to us we were still able to identify this problem in the data. Since sightlines with this problem represented a small minority of the overall dataset (9 of the 92 observed with Arecibo) the decision was made to exclude these sightlines from further analysis.

As outlined in the Observing OH subsection we assume that our observations from the ATCA do not contain any emission from the extended OH cloud or the diffuse background and are well-described by $T_{\rm b}=T_{\rm c}(e^{-\tau_{\nu}}-1)$ (i.e. there is no contribution from $T_{\rm ex}$~or $T_{\rm bg}$ in Eq. \ref{Eq:FullRadTran}). If there is contribution from the $T_{\rm ex}$~term our method will underestimate optical depth. If there is contribution from the $T_{\rm bg}$~term, optical depth will be overestimated if the actual optical depth is positive, and underestimated if it is negative. Across the four transitions this will change the line optical depth ratios, which in most cases (i.e. where $|T_{\rm ex}|\gg h\nu_0/k_{\rm B}=0.08$\,K) are expected to have the relation $\tau_{\rm peak}(1612) + \tau_{\rm peak}(1720) = \frac{\tau_{\rm peak}(1665)}{5} + \frac{\tau_{\rm peak}(1667)}{9}$, known as the optical depth sum rule. \amoeba~includes a weak prior that penalises deviations from this relation, but will still fit features that do not adhere to it. 

Another challenge that is more relevant for our ATCA observations is the presence of high-gain OH masers in the primary beam, whose sidelobes may coincide with our sources. Interferometric maser sidelobes manifest as either a positive or negative feature in a single transition (the maser transition), apparent as a feature in the residual of the sum rule. \amoeba~is hesitant to fit such features in a single transition, since the improvement to the likelihood gained by fitting the feature may not be able to overcome the penalty from the prior in violating the sum rule to such a degree. Therefore when we present our fits of our ATCA data in the Results section we include a plot of the sum rule residuals. 

More generally, our assumption that the foreground OH gas is uniform across the on- and off-source pointings (for both our on-off and our ATCA observations) is also limited by the fact that molecular gas is clumpy on sub-parsec scales (below the resolution of our observations). \citet{Engelke2019} addressed this issue, as well as the presence of unresolved structure in the bright background continuum source. This is a difficult problem to solve directly without higher resolution observations, but the overall consequence appears to be that our measurements of optical depth may represent lower limits rather than their true values.

%% file: 4_Results.tex
\begin{figure*}
    \centering
    \begin{tabular}{cc}
    \includegraphics[trim={.6cm 0.6cm 1cm 1cm}, clip=true, width=0.45\linewidth]{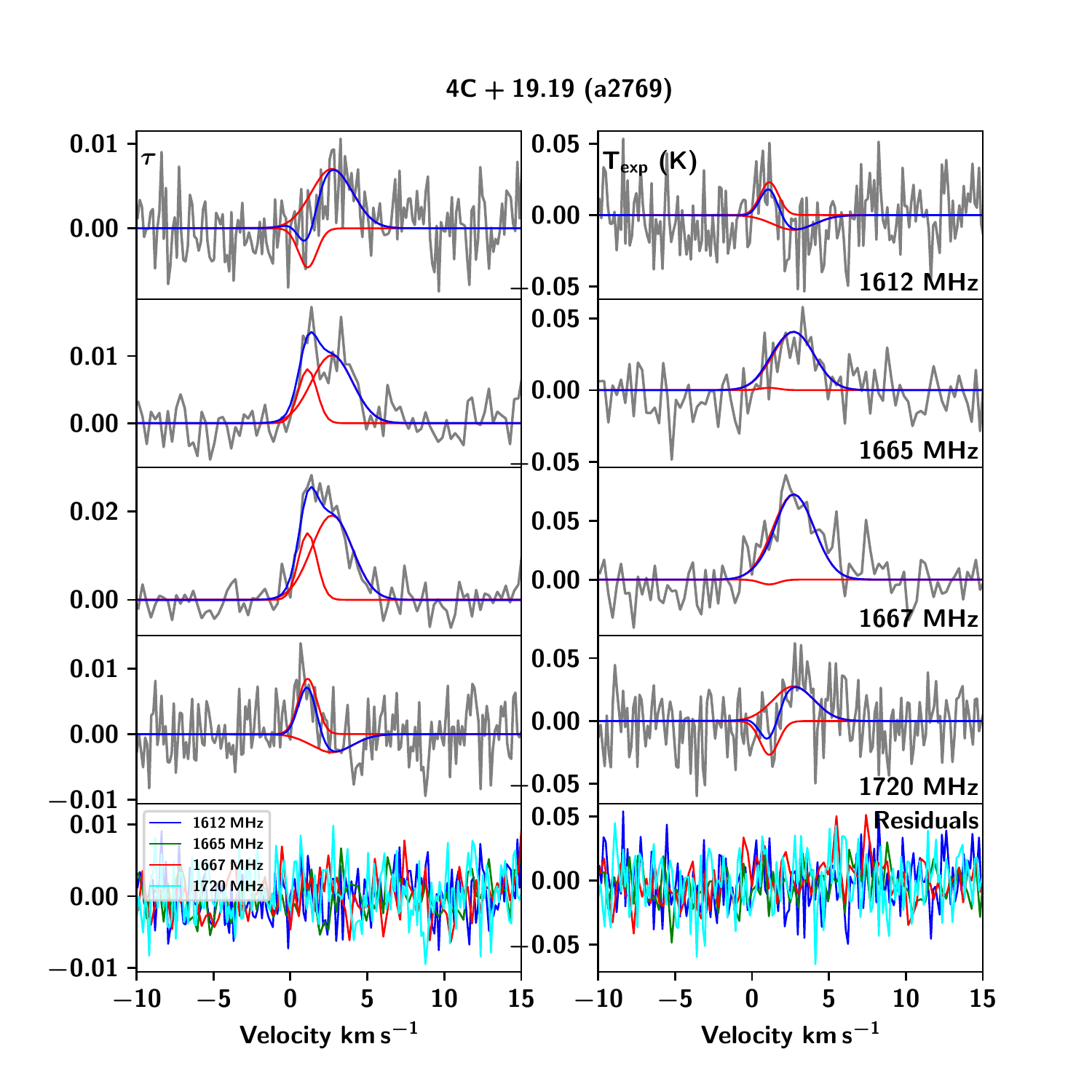}&\includegraphics[trim={0.6cm 0.6cm 1cm 1cm}, clip=true, width=0.45\linewidth]{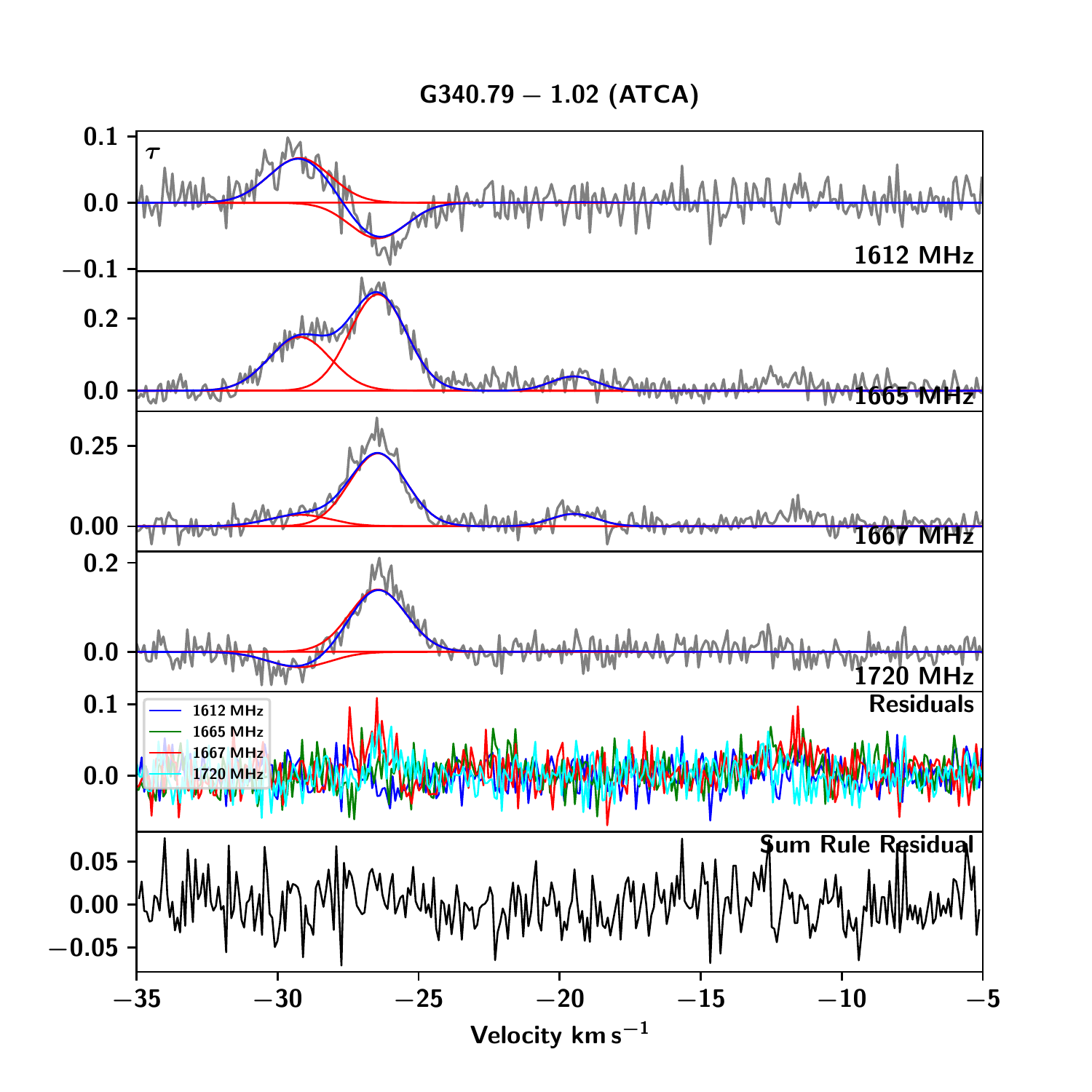}
    \end{tabular}
    
    \caption{Typical examples of data examined in this work from the Arecibo Radio Telescope (left towards 4C+19.19 from project a2769) and the Australia Telescope Compact Array (ATCA, right towards G340.79-1.02). Data from Arecibo (left) consist of 8 spectra plotted in grey: four optical depth ($\tau$) spectra (at 1612, 1665, 1667 and 1720 MHz) at left and four expected brightness temperature ($T_{\rm exp}$) spectra at right. Each identified Gaussian component is indicated in red and the total fit (the sum of Gaussian components) is shown in blue. The bottom panels then show the residuals of the total fit in each transition as described in the legend. Data from the ATCA (right) consist of four optical depth ($\tau$) spectra. In addition to the residuals of the total fit shown in the fourth panel, these plots also show the sum rule residual, as described by $\tau_{\rm peak}(1612)+\tau_{\rm peak}(1720)-\tau_{\rm peak}(1665)/5-\tau_{\rm peak}(1667)/9$.
    }
    \label{fig:example1}
\end{figure*}

Across the 107 sightlines examined in this work (92 with on-off observations from Arecibo, 15 with optical depth observations from the ATCA), 38 had detections (27 on-off, 11 optical depth only). We have identified a total of 109 features from these sightlines. 58 of these were from on-off observations, and therefore include excitation temperatures and column densities. Data toward 4C+19.19 from project a2769 from Arecibo and towards G340.79-1.02 from the ATCA are shown with their fitted features in Fig. \ref{fig:example1} as typical examples of the observations examined in this work. The peak optical depth values of these features are given in Table \ref{tab:tau}, excitation temperatures in the four ground-rotational state transitions of features identified from the on-off observations are shown in Table \ref{tab:Tex}, and the OH column densities in the ground-rotational state levels (as well as total OH column density) are shown in Table \ref{tab:N}. Data from sightlines with detections are plotted with their individual features and total fits (and residuals of those fits) in Figs. \ref{fig:results1} to \ref{fig:results7}. The sightlines are organised by Galactic longitude in all tables, and alphabetically by their background source name in all figures for easy reference.

As described in detail in \citet{Petzler2021a}, \textsc{Amoeba} parameterises individual Gaussian features in on-off spectra with a set of 6 parameters: $\boldsymbol{\theta}=[v,$ ~${\rm log}_{10}\Delta v,~{\rm log}_{10}N_{1},$ ~$T_{\rm ex}^{-1}(1612),$ ~$T_{\rm ex}^{-1}(1665),$ ~$T_{\rm ex}^{-1}(1667)]$. These are the centroid velocity, $\log$ FWHM, $\log$ column density of OH in the lowest level of the ground-rotational state, and inverse excitation temperatures of the 1612, 1665 and 1667\,MHz transitions, respectively. Alternatively (in the case of our ATCA data), if only optical depth spectra are available \textsc{Amoeba} parameterises an individual Gaussian feature with $\boldsymbol{\theta} =$ ~$[v,$ ~${\rm log}_{10}\Delta v,$ ~$\tau_{\rm peak~1}(1612),$ ~$\tau_{\rm peak~1}(1665),$ ~$\tau_{\rm peak~1}(1667),$ ~$\tau_{\rm peak~1}(1720)]$. These are the centroid velocity, $\log$ FWHM, and the peak optical depth in the 1612, 1665, 1667 and 1720\,MHz transitions, respectively. In both cases these are then sufficient to describe the features seen in the observed spectra, and the parameters given in Tables \ref{tab:tau} to \ref{tab:N}. Therefore, the 68\% credibility intervals quoted in these tables for centroid velocity, $\log$ FWHM, all four peak optical depths for our ATCA data and $\log$ column density in the lowest energy level for our Arecibo data are determined from the smallest volume in parameter space that contains 68\% of the converged Markov chains as found by \textsc{Amoeba}, thus representing a $1\sigma$~uncertainty assuming that those distributions are Gaussian. The remaining parameters and their associated credibility intervals in Tables \ref{tab:tau} to \ref{tab:N} are then derived from those fitted parameters and their credibility intervals. All Gaussian features identified in this work were accepted if their inclusion resulted in a Bayes factor of at least 10 compared to a model that did not include them, in keeping with the standard defined by \citet{Jeffreys1961}. We note again here (as discussed in previous sections) that our models assume (in the case of our on-off spectra) that the OH gas in the on-source position has the same optical depth and excitation temperature as the gas in the off-source positions. If this assumption is incorrect, \textsc{Amoeba} will fit a quasi-average model that best satisfies the available spectra, and any residual signal (relative to the noise) will decrease the value of the likelihood for that particular set of parameters, spreading out the model's posterior distribution in parameter space and lowering its Bayes factor compared to simpler models. Therefore both the noise level of the spectra and the validity of our assumptions will drive the detectability of features and the size of the 68\% credibility intervals of the fitted parameters.

\begin{table*}
\centering
\begin{tabular}{llrrrrrrrr}
\hline
Source&Project&\multicolumn{1}{c}{$l^{\circ}$}&\multicolumn{1}{c}{$b^{\circ}$}&\multicolumn{1}{c}{$v$}&\multicolumn{1}{c}{$\Delta v$}&\multicolumn{4}{c}{$\tau_{\rm peak}\,  (10^{-3})$}\\
\cline{7-10}
&&&&\multicolumn{2}{c}{km\,s$^{-1}$}&\multicolumn{1}{c}{1612}&\multicolumn{1}{c}{1665}&\multicolumn{1}{c}{1667}&\multicolumn{1}{c}{1720}\\
\hline
G007.47+0.06&ATCA&7.47&0.06&-17.86$^{+3.78}_{-31.30}$&0.49$^{+1.70}_{-0.25}$&4.2$^{+12.5}_{-7.3}$&43.2$^{+67.3}_{-96.9}$&8.3$^{+53.3}_{-9.8}$&0.5$^{+12.9}_{-6.9}$\\
G007.47+0.06&ATCA&7.47&0.06&-14.05$^{+0.10}_{-0.10}$&5.27$^{+0.82}_{-0.25}$&49.4$^{+5.0}_{-8.1}$&175.9$^{+10.7}_{-51.4}$&179.3$^{+8.7}_{-26.9}$&25.0$^{+5.2}_{-6.0}$\\
G007.47+0.06&ATCA&7.47&0.06&-1.91$^{+0.17}_{-0.16}$&2.68$^{+0.49}_{-0.40}$&26.9$^{+8.3}_{-8.4}$&66.1$^{+10.5}_{-10.1}$&65.8$^{+8.6}_{-7.3}$&-1.4$^{+2.7}_{-8.6}$\\
G007.47+0.06&ATCA&7.47&0.06&12.36$^{+0.14}_{-0.12}$&3.81$^{+0.37}_{-0.36}$&13.9$^{+7.2}_{-7.4}$&67.6$^{+8.8}_{-8.1}$&84.0$^{+5.8}_{-6.3}$&56.2$^{+7.0}_{-6.1}$\\
G007.47+0.06&ATCA&7.47&0.06&16.02$^{+0.22}_{-0.22}$&5.32$^{+0.74}_{-0.51}$&-0.7$^{+2.3}_{-6.1}$&86.5$^{+9.8}_{-7.4}$&68.7$^{+5.7}_{-4.9}$&43.1$^{+6.0}_{-5.8}$\\
G007.47+0.06&ATCA&7.47&0.06&122.27$^{+0.19}_{-0.16}$&4.99$^{+0.39}_{-0.43}$&2.1$^{+6.2}_{-2.7}$&76.0$^{+6.4}_{-7.3}$&71.9$^{+6.1}_{-5.4}$&17.4$^{+5.7}_{-6.1}$\\
SRC44&a3301&68.83&-3.49&6.17$^{+0.11}_{-0.11}$&0.80$^{+0.03}_{-0.03}$&-0.2$^{+0.3}_{-0.2}$&9.3$^{+6.0}_{-3.5}$&18.6$^{+11.9}_{-7.1}$&4.2$^{+2.3}_{-1.3}$\\
SRC44&a3301&68.83&-3.49&6.28$^{+0.11}_{-0.11}$&2.50$^{+0.11}_{-0.12}$&1.1$^{+0.2}_{-0.2}$&2.9$^{+0.4}_{-0.5}$&8.0$^{+1.2}_{-1.0}$&0.3$^{+0.0}_{-0.0}$\\
SRC44&a3301&68.83&-3.49&11.08$^{+0.11}_{-0.55}$&0.72$^{+0.08}_{-0.08}$&0.2$^{+0.4}_{-0.1}$&-2.9$^{+6.5}_{-1.1}$&-7.9$^{+15.4}_{-1.6}$&-1.4$^{+2.5}_{-0.1}$\\
SRC44&a3301&68.83&-3.49&11.20$^{+0.32}_{-0.11}$&1.29$^{+0.07}_{-0.25}$&-0.1$^{+0.2}_{-0.3}$&7.7$^{+3.9}_{-3.0}$&16.8$^{+9.1}_{-7.2}$&3.7$^{+1.7}_{-1.1}$\\
3C417&a2600&73.33&-5.45&9.51$^{+0.07}_{-0.07}$&0.88$^{+0.06}_{-0.07}$&4.2$^{+2.0}_{-1.8}$&40.4$^{+5.4}_{-5.0}$&58.9$^{+7.5}_{-7.1}$&10.5$^{+0.1}_{-0.1}$\\
3C417&a2600&73.33&-5.45&9.92$^{+0.07}_{-0.07}$&0.63$^{+0.02}_{-0.01}$&52.0$^{+8.6}_{-7.6}$&75.3$^{+8.3}_{-8.4}$&136.8$^{+15.6}_{-15.2}$&-20.0$^{+3.8}_{-4.7}$\\
3C417&a2600&73.33&-5.45&10.68$^{+0.09}_{-0.07}$&2.91$^{+0.10}_{-0.11}$&8.0$^{+0.8}_{-0.8}$&14.1$^{+1.1}_{-1.1}$&31.1$^{+1.5}_{-1.6}$&-1.6$^{+0.4}_{-0.4}$\\
3C092&a2600&159.74&-18.41&8.71$^{+0.07}_{-0.07}$&1.56$^{+0.07}_{-0.06}$&6.6$^{+1.0}_{-1.0}$&11.0$^{+1.4}_{-1.3}$&19.3$^{+1.6}_{-2.0}$&-2.2$^{+0.5}_{-0.5}$\\
3C092&a2600&159.74&-18.41&8.77$^{+0.07}_{-0.07}$&0.76$^{+0.01}_{-0.01}$&15.1$^{+2.7}_{-2.4}$&63.5$^{+9.2}_{-7.8}$&95.6$^{+13.3}_{-11.4}$&8.3$^{+0.7}_{-0.4}$\\
4C+28.11&a2769&166.06&-17.22&6.91$^{+0.21}_{-0.21}$&1.10$^{+0.05}_{-0.05}$&13.6$^{+1.5}_{-1.5}$&29.8$^{+1.9}_{-1.8}$&52.9$^{+2.1}_{-2.3}$&-1.7$^{+0.9}_{-0.9}$\\
PKS0319+12&a2769&170.59&-36.24&7.73$^{+0.21}_{-0.21}$&1.01$^{+0.16}_{-0.13}$&2.2$^{+1.5}_{-0.9}$&4.6$^{+2.6}_{-1.7}$&6.9$^{+3.7}_{-2.4}$&-0.5$^{+0.3}_{-0.5}$\\
3C131&a2600&171.44&-7.80&4.56$^{+0.07}_{-0.07}$&0.48$^{+0.02}_{-0.02}$&1.1$^{+0.5}_{-0.4}$&5.0$^{+0.7}_{-0.6}$&9.6$^{+1.1}_{-1.1}$&0.9$^{+0.2}_{-0.3}$\\
3C131&a2600&171.44&-7.80&5.71$^{+0.07}_{-0.08}$&3.21$^{+0.10}_{-0.10}$&2.3$^{+0.2}_{-0.3}$&4.1$^{+0.2}_{-0.2}$&7.1$^{+0.4}_{-0.4}$&-0.7$^{+0.2}_{-0.2}$\\
3C131&a2600&171.44&-7.80&6.59$^{+0.07}_{-0.07}$&0.44$^{+0.01}_{-0.01}$&2.8$^{+0.6}_{-0.6}$&16.3$^{+1.2}_{-1.1}$&25.8$^{+1.9}_{-1.8}$&3.4$^{+0.2}_{-0.2}$\\
3C131&a2600&171.44&-7.80&7.23$^{+0.07}_{-0.07}$&0.56$^{+0.01}_{-0.00}$&11.9$^{+0.6}_{-0.6}$&50.2$^{+1.0}_{-1.1}$&83.4$^{+1.6}_{-1.8}$&7.4$^{+0.2}_{-0.3}$\\
3C131&a2600&171.44&-7.80&7.48$^{+0.07}_{-0.07}$&1.93$^{+0.09}_{-0.08}$&5.7$^{+0.8}_{-0.8}$&6.4$^{+0.9}_{-0.9}$&10.7$^{+1.5}_{-1.4}$&-2.9$^{+0.4}_{-0.4}$\\
3C131&a2600&171.44&-7.80&7.79$^{+0.07}_{-0.07}$&0.57$^{+0.09}_{-0.07}$&0.1$^{+0.2}_{-0.2}$&-2.4$^{+0.3}_{-0.2}$&-3.1$^{+0.5}_{-0.4}$&-0.9$^{+0.1}_{-0.1}$\\
3C108&a3301&171.47&-20.70&9.42$^{+0.11}_{-0.11}$&1.19$^{+0.03}_{-0.03}$&8.0$^{+1.0}_{-0.9}$&29.9$^{+2.6}_{-2.3}$&31.1$^{+2.5}_{-2.4}$&1.5$^{+0.2}_{-0.2}$\\
3C108&a3301&171.47&-20.70&9.74$^{+0.11}_{-0.11}$&0.48$^{+0.03}_{-0.03}$&11.4$^{+10.5}_{-5.2}$&23.1$^{+17.8}_{-10.2}$&31.2$^{+24.3}_{-13.6}$&-2.9$^{+1.4}_{-3.4}$\\
4C+11.15&a2769&171.98&-35.48&7.18$^{+0.21}_{-0.21}$&0.65$^{+0.04}_{-0.03}$&8.5$^{+1.7}_{-1.9}$&36.2$^{+3.9}_{-3.7}$&64.9$^{+5.2}_{-5.3}$&6.0$^{+0.6}_{-0.3}$\\
4C+36.10&a2769&172.98&2.44&-16.74$^{+0.21}_{-0.21}$&2.89$^{+0.12}_{-0.11}$&5.3$^{+0.8}_{-0.8}$&17.2$^{+0.9}_{-0.9}$&29.0$^{+1.0}_{-1.1}$&1.3$^{+0.5}_{-0.5}$\\
4C+27.14&a3301&175.46&-9.96&7.19$^{+0.11}_{-0.11}$&1.62$^{+0.04}_{-0.05}$&0.8$^{+1.7}_{-1.0}$&16.0$^{+1.7}_{-1.7}$&36.5$^{+3.0}_{-1.9}$&6.5$^{+0.5}_{-1.1}$\\
4C+27.14&a3301&175.46&-9.96&7.89$^{+0.11}_{-0.11}$&0.84$^{+0.06}_{-0.04}$&30.0$^{+24.0}_{-20.4}$&1.1$^{+3.1}_{-1.7}$&15.3$^{+13.1}_{-10.4}$&-13.2$^{+8.0}_{-7.0}$\\
4C+17.23&a2769&176.36&-24.24&9.35$^{+0.21}_{-0.21}$&0.72$^{+0.04}_{-0.04}$&6.8$^{+1.3}_{-1.2}$&22.2$^{+2.0}_{-2.0}$&44.3$^{+3.3}_{-3.3}$&2.6$^{+0.4}_{-0.5}$\\
4C+17.23&a2769&176.36&-24.24&11.42$^{+0.21}_{-0.21}$&0.77$^{+0.03}_{-0.03}$&9.8$^{+1.5}_{-1.4}$&27.5$^{+2.6}_{-2.4}$&51.5$^{+4.1}_{-3.2}$&1.5$^{+0.5}_{-0.5}$\\
4C+07.13&a2769&178.87&-36.27&3.48$^{+0.21}_{-0.21}$&1.07$^{+0.13}_{-0.11}$&2.3$^{+1.7}_{-1.2}$&10.3$^{+3.5}_{-2.6}$&14.4$^{+4.1}_{-3.5}$&1.3$^{+0.3}_{-0.5}$\\
B0531+2730&a2769&179.87&-2.83&3.04$^{+0.21}_{-0.21}$&0.72$^{+0.03}_{-0.04}$&17.9$^{+14.6}_{-7.9}$&-21.3$^{+21.7}_{-20.4}$&92.8$^{+60.6}_{-34.2}$&-11.4$^{+2.9}_{-0.3}$\\
B0531+2730&a2769&179.87&-2.83&3.17$^{+0.21}_{-0.21}$&0.78$^{+0.03}_{-0.04}$&5.2$^{+7.0}_{-5.7}$&79.6$^{+63.6}_{-37.9}$&20.3$^{+21.8}_{-20.0}$&12.8$^{+7.9}_{-4.0}$\\
4C+26.18b&a2769&182.36&-0.62&-11.93$^{+0.21}_{-0.21}$&1.57$^{+0.11}_{-0.11}$&-8.4$^{+1.0}_{-0.6}$&23.3$^{+7.4}_{-6.2}$&41.1$^{+11.1}_{-10.2}$&18.3$^{+1.7}_{-1.7}$\\
4C+26.18b&a2769&182.36&-0.62&-9.99$^{+0.21}_{-0.28}$&2.66$^{+0.38}_{-0.35}$&1.1$^{+1.3}_{-1.2}$&5.5$^{+1.9}_{-1.6}$&14.8$^{+2.3}_{-1.8}$&1.6$^{+0.7}_{-0.7}$\\
4C+22.12&a2769&188.07&0.04&-1.62$^{+0.21}_{-0.21}$&0.66$^{+0.04}_{-0.04}$&0.4$^{+1.2}_{-0.7}$&15.2$^{+9.8}_{-6.2}$&31.0$^{+20.5}_{-12.3}$&6.1$^{+3.0}_{-1.9}$\\
4C+14.14&a2769&189.04&-12.85&2.60$^{+0.21}_{-0.21}$&3.82$^{+0.28}_{-0.29}$&1.3$^{+0.6}_{-0.6}$&8.4$^{+0.8}_{-0.9}$&16.3$^{+1.1}_{-0.9}$&2.2$^{+0.4}_{-0.3}$\\
4C+19.18&a2769&190.09&-2.17&-0.62$^{+0.21}_{-0.21}$&2.18$^{+0.23}_{-0.19}$&-4.2$^{+0.9}_{-0.9}$&4.7$^{+1.2}_{-1.2}$&13.6$^{+1.7}_{-1.7}$&6.9$^{+0.6}_{-0.6}$\\
4C+19.18&a2769&190.09&-2.17&2.39$^{+0.21}_{-0.21}$&1.47$^{+0.12}_{-0.11}$&4.3$^{+1.1}_{-1.1}$&13.6$^{+1.5}_{-1.4}$&29.6$^{+1.8}_{-1.8}$&1.7$^{+0.6}_{-0.6}$\\
4C+19.19&a2769&190.13&-1.64&1.12$^{+0.21}_{-0.21}$&1.42$^{+0.19}_{-0.16}$&-4.6$^{+1.2}_{-1.3}$&8.0$^{+6.6}_{-3.3}$&15.1$^{+13.0}_{-6.4}$&8.5$^{+1.4}_{-0.3}$\\
4C+19.19&a2769&190.13&-1.64&2.70$^{+0.21}_{-0.21}$&3.15$^{+0.20}_{-0.26}$&7.0$^{+0.9}_{-0.9}$&10.1$^{+1.0}_{-1.1}$&19.1$^{+1.4}_{-1.4}$&-2.8$^{+0.5}_{-0.5}$\\
PKS0528+134&a2600&191.37&-11.01&9.60$^{+0.07}_{-0.07}$&0.90$^{+0.01}_{-0.01}$&5.2$^{+0.6}_{-0.6}$&25.4$^{+1.6}_{-1.6}$&46.2$^{+2.8}_{-2.7}$&5.0$^{+0.0}_{-0.0}$\\
4C+16.15b&a2769&193.64&-1.53&11.88$^{+0.21}_{-0.21}$&0.87$^{+0.06}_{-0.05}$&5.6$^{+1.0}_{-1.1}$&24.0$^{+1.5}_{-1.5}$&43.2$^{+2.3}_{-2.3}$&4.0$^{+0.5}_{-0.5}$\\
3C158&a2769&196.64&0.17&3.14$^{+0.21}_{-0.21}$&0.98$^{+0.14}_{-0.15}$&0.5$^{+1.5}_{-0.7}$&7.6$^{+11.0}_{-4.5}$&8.4$^{+11.6}_{-4.9}$&1.9$^{+2.0}_{-0.7}$\\
\hline
\end{tabular}
\caption{Fitted centroid velocity, FWHM and peak optical depth of the features identified in this work. Columns give the targeted background source of each sightline, the project name, Galactic longitude and latitude, centroid velocity $v$, FWHM $\Delta v$, and peak optical depth (10$^{-3}$) at 1612, 1665, 1667 and 1720\,MHz. The uncertainties of all parameters are the 68\% credibility intervals, except in the case of centroid velocity, where this interval is replaced with the channel width if the channel width is greater than the 68\% credibility interval \citep{Petzler2021a}.}
\label{tab:tau}
\end{table*}

\begin{table*}
\centering
\begin{tabular}{llrrrrrrrr}
\hline
Source&Project&\multicolumn{1}{c}{$l^{\circ}$}&\multicolumn{1}{c}{$b^{\circ}$}&\multicolumn{1}{c}{$v$}&\multicolumn{1}{c}{$\Delta v$}&\multicolumn{4}{c}{$\tau_{\rm peak}\,  (10^{-3})$}\\
\cline{7-10}
&&&&\multicolumn{2}{c}{km\,s$^{-1}$}&\multicolumn{1}{c}{1612}&\multicolumn{1}{c}{1665}&\multicolumn{1}{c}{1667}&\multicolumn{1}{c}{1720}\\
\hline
4C+14.18&a2600&196.98&1.10&4.28$^{+0.07}_{-0.07}$&0.55$^{+0.05}_{-0.05}$&0.2$^{+0.4}_{-0.2}$&4.7$^{+5.6}_{-2.7}$&10.2$^{+13.0}_{-5.7}$&1.9$^{+2.3}_{-1.0}$\\
4C+14.18&a2600&196.98&1.10&4.94$^{+0.07}_{-0.07}$&1.84$^{+0.04}_{-0.03}$&3.5$^{+0.3}_{-0.3}$&11.9$^{+0.5}_{-0.5}$&16.5$^{+0.6}_{-0.7}$&0.7$^{+0.1}_{-0.1}$\\
4C+14.18&a2600&196.98&1.10&7.39$^{+0.07}_{-0.07}$&0.81$^{+0.04}_{-0.03}$&2.8$^{+0.6}_{-0.6}$&6.8$^{+1.1}_{-1.0}$&8.7$^{+1.3}_{-1.2}$&-0.4$^{+0.2}_{-0.3}$\\
4C+14.18&a2600&196.98&1.10&16.49$^{+0.07}_{-0.07}$&1.29$^{+0.09}_{-0.09}$&-0.5$^{+0.5}_{-0.4}$&5.1$^{+0.8}_{-0.7}$&10.6$^{+1.7}_{-1.5}$&2.7$^{+0.1}_{-0.2}$\\
4C+14.18&a2600&196.98&1.10&17.59$^{+0.07}_{-0.07}$&0.70$^{+0.06}_{-0.06}$&1.2$^{+0.7}_{-0.5}$&4.9$^{+1.7}_{-1.3}$&10.9$^{+3.7}_{-2.8}$&1.0$^{+0.1}_{-0.0}$\\
4C+14.18&a2600&196.98&1.10&18.40$^{+0.22}_{-0.18}$&3.76$^{+0.26}_{-0.36}$&2.3$^{+0.4}_{-0.4}$&1.9$^{+0.4}_{-0.3}$&3.4$^{+0.5}_{-0.4}$&-1.4$^{+0.3}_{-0.3}$\\
4C+14.18&a2600&196.98&1.10&31.98$^{+0.07}_{-0.07}$&0.42$^{+0.86}_{-0.01}$&4.3$^{+2.0}_{-1.0}$&21.6$^{+12.3}_{-5.6}$&37.2$^{+21.1}_{-9.9}$&4.2$^{+2.8}_{-1.2}$\\
4C+14.18&a2600&196.98&1.10&32.33$^{+0.07}_{-0.33}$&1.19$^{+0.08}_{-0.80}$&1.3$^{+0.5}_{-1.0}$&6.2$^{+1.9}_{-7.5}$&11.4$^{+2.6}_{-14.3}$&1.2$^{+0.2}_{-2.1}$\\
4C+13.32&a2769&197.15&-0.85&-5.65$^{+0.01}_{-0.01}$&6.01$^{+0.29}_{-0.25}$&4.9$^{+3.4}_{-2.0}$&12.8$^{+9.0}_{-5.1}$&24.0$^{+15.8}_{-9.7}$&0.4$^{+0.2}_{-0.2}$\\
4C+13.32&a2769&197.15&-0.85&-0.35$^{+0.11}_{-0.11}$&1.02$^{+0.11}_{-0.10}$&0.0$^{+0.1}_{-0.0}$&2.2$^{+5.8}_{-1.6}$&6.0$^{+16.1}_{-4.4}$&1.3$^{+4.4}_{-1.0}$\\
4C+13.32&a2769&197.15&-0.85&4.46$^{+0.11}_{-0.11}$&1.20$^{+0.07}_{-0.07}$&-0.4$^{+0.2}_{-0.1}$&3.6$^{+5.8}_{-2.2}$&8.6$^{+14.1}_{-5.3}$&2.1$^{+2.5}_{-0.9}$\\
4C+13.32&a2769&197.15&-0.85&6.99$^{+0.25}_{-0.20}$&2.21$^{+0.74}_{-0.45}$&1.6$^{+1.3}_{-0.8}$&1.0$^{+1.1}_{-0.6}$&2.4$^{+1.6}_{-1.2}$&-1.1$^{+0.5}_{-0.9}$\\
4C+13.32&a2769&197.15&-0.85&9.50$^{+0.21}_{-0.21}$&1.15$^{+0.08}_{-0.08}$&0.7$^{+0.7}_{-0.4}$&4.0$^{+4.4}_{-2.2}$&7.0$^{+7.4}_{-3.7}$&0.9$^{+1.0}_{-0.4}$\\
4C+17.41&a2769&201.13&16.42&0.23$^{+0.21}_{-0.21}$&1.38$^{+0.24}_{-0.17}$&3.4$^{+1.2}_{-1.1}$&9.1$^{+1.6}_{-1.7}$&18.9$^{+2.1}_{-2.2}$&0.6$^{+0.5}_{-0.6}$\\
4C+17.41&a2769&201.13&16.42&1.89$^{+0.21}_{-0.21}$&0.73$^{+0.06}_{-0.06}$&11.7$^{+1.8}_{-1.8}$&28.7$^{+2.5}_{-2.4}$&51.2$^{+2.9}_{-3.0}$&-0.2$^{+0.9}_{-0.9}$\\
4C+04.22&a2769&205.41&-4.43&11.92$^{+0.21}_{-0.21}$&0.71$^{+0.06}_{-0.07}$&7.7$^{+2.1}_{-1.7}$&17.0$^{+2.8}_{-2.7}$&38.1$^{+5.1}_{-4.4}$&0.0$^{+0.7}_{-0.9}$\\
4C+04.22&a2769&205.41&-4.43&13.33$^{+0.11}_{-0.11}$&0.92$^{+0.08}_{-0.07}$&10.6$^{+1.9}_{-2.1}$&17.8$^{+2.7}_{-2.7}$&38.8$^{+5.1}_{-5.1}$&-2.5$^{+1.0}_{-0.7}$\\
J134217-040725&a3301&205.58&-4.14&9.19$^{+0.09}_{-0.09}$&1.20$^{+0.04}_{-0.04}$&7.9$^{+1.2}_{-1.2}$&23.6$^{+2.4}_{-2.4}$&41.0$^{+3.6}_{-3.9}$&1.4$^{+0.3}_{-0.3}$\\
4C+04.24&a2769&205.92&-3.57&9.39$^{+0.03}_{-0.03}$&1.25$^{+0.07}_{-0.07}$&7.3$^{+1.3}_{-1.2}$&23.7$^{+2.1}_{-2.0}$&45.3$^{+3.3}_{-3.0}$&2.5$^{+0.4}_{-0.5}$\\
G336.49-1.48&ATCA&336.49&-1.48&-23.32$^{+0.16}_{-0.14}$&4.06$^{+0.21}_{-0.15}$&12.9$^{+5.0}_{-4.1}$&130.4$^{+3.7}_{-5.0}$&175.9$^{+5.0}_{-5.2}$&42.0$^{+5.2}_{-4.1}$\\
G336.49-1.48&ATCA&336.49&-1.48&-20.44$^{+0.09}_{-0.09}$&2.43$^{+0.27}_{-0.12}$&92.3$^{+3.7}_{-3.3}$&130.0$^{+5.5}_{-8.4}$&143.7$^{+11.2}_{-15.3}$&-57.4$^{+7.7}_{-9.4}$\\
G336.49-1.48&ATCA&336.49&-1.48&-14.25$^{+0.12}_{-0.12}$&2.37$^{+0.31}_{-0.23}$&11.1$^{+3.4}_{-3.0}$&29.8$^{+3.9}_{-3.4}$&34.8$^{+4.2}_{-3.5}$&-1.0$^{+1.7}_{-3.0}$\\
G340.79-1.02&ATCA&340.79&-1.02&-29.22$^{+0.09}_{-0.09}$&2.56$^{+0.11}_{-0.12}$&67.5$^{+3.9}_{-3.9}$&149.5$^{+5.1}_{-5.3}$&35.9$^{+5.1}_{-5.0}$&-34.5$^{+4.6}_{-4.2}$\\
G340.79-1.02&ATCA&340.79&-1.02&-26.44$^{+0.09}_{-0.09}$&2.37$^{+0.06}_{-0.04}$&-53.7$^{+3.9}_{-4.4}$&267.5$^{+6.1}_{-6.2}$&226.2$^{+5.1}_{-4.6}$&140.0$^{+4.6}_{-4.9}$\\
G340.79-1.02&ATCA&340.79&-1.02&-19.50$^{+0.10}_{-0.11}$&1.91$^{+0.50}_{-0.31}$&0.9$^{+2.8}_{-2.1}$&40.0$^{+6.0}_{-5.7}$&38.4$^{+5.7}_{-5.5}$&1.6$^{+4.5}_{-2.5}$\\
G344.43+0.05&ATCA&344.43&0.05&-67.57$^{+0.09}_{-0.09}$&5.41$^{+0.18}_{-0.18}$&-30.0$^{+3.7}_{-3.4}$&129.1$^{+4.7}_{-5.1}$&139.4$^{+4.6}_{-4.9}$&68.3$^{+4.0}_{-4.8}$\\
G344.43+0.05&ATCA&344.43&0.05&-62.70$^{+0.09}_{-0.09}$&1.55$^{+0.12}_{-0.10}$&-29.0$^{+13.1}_{-10.7}$&-65.6$^{+24.0}_{-29.7}$&133.9$^{+8.1}_{-8.5}$&22.0$^{+11.6}_{-11.9}$\\
G344.43+0.05&ATCA&344.43&0.05&-61.64$^{+0.09}_{-0.09}$&1.98$^{+0.16}_{-0.13}$&-80.0$^{+9.4}_{-7.6}$&228.9$^{+9.6}_{-12.2}$&3.5$^{+12.8}_{-5.4}$&93.7$^{+7.5}_{-8.2}$\\
G344.43+0.05&ATCA&344.43&0.05&-22.45$^{+0.09}_{-0.09}$&1.34$^{+0.12}_{-0.11}$&103.1$^{+8.9}_{-8.0}$&-18.8$^{+13.5}_{-13.3}$&24.1$^{+9.0}_{-7.7}$&1.4$^{+9.1}_{-3.4}$\\
G344.43+0.05&ATCA&344.43&0.05&-21.91$^{+0.12}_{-0.12}$&4.25$^{+0.34}_{-0.28}$&87.0$^{+7.5}_{-6.6}$&31.9$^{+7.3}_{-7.2}$&61.9$^{+5.6}_{-5.4}$&24.8$^{+4.5}_{-6.0}$\\
G344.43+0.05&ATCA&344.43&0.05&-17.90$^{+0.19}_{-0.23}$&3.09$^{+0.70}_{-0.52}$&48.4$^{+4.8}_{-4.8}$&15.2$^{+6.3}_{-5.6}$&29.5$^{+6.2}_{-5.6}$&8.5$^{+5.4}_{-7.2}$\\
G344.43+0.05&ATCA&344.43&0.05&-5.31$^{+0.09}_{-0.09}$&0.50$^{+0.06}_{-0.04}$&12.0$^{+9.6}_{-10.8}$&-209.9$^{+17.5}_{-15.2}$&-15.5$^{+13.0}_{-10.2}$&8.6$^{+10.5}_{-8.7}$\\
G344.43+0.05&ATCA&344.43&0.05&-3.79$^{+0.09}_{-0.09}$&0.80$^{+0.06}_{-0.06}$&22.0$^{+7.8}_{-9.9}$&-146.7$^{+13.6}_{-12.6}$&-80.3$^{+8.1}_{-8.1}$&13.7$^{+8.4}_{-10.3}$\\
G344.43+0.05&ATCA&344.43&0.05&-2.30$^{+0.09}_{-0.09}$&0.90$^{+0.05}_{-0.05}$&16.0$^{+8.1}_{-9.6}$&-399.5$^{+12.6}_{-15.6}$&-43.6$^{+8.6}_{-8.2}$&0.7$^{+7.0}_{-5.6}$\\
G344.43+0.05&ATCA&344.43&0.05&-1.55$^{+0.09}_{-0.09}$&0.50$^{+0.04}_{-0.05}$&2.4$^{+10.7}_{-3.6}$&300.7$^{+23.1}_{-22.3}$&20.4$^{+9.7}_{-10.6}$&0.1$^{+5.1}_{-8.7}$\\
G344.43+0.05&ATCA&344.43&0.05&-0.79$^{+0.09}_{-0.09}$&0.43$^{+0.06}_{-0.05}$&1.2$^{+6.7}_{-3.2}$&-220.8$^{+20.4}_{-22.4}$&-1.1$^{+5.0}_{-9.4}$&7.3$^{+8.8}_{-6.8}$\\
G344.43+0.05&ATCA&344.43&0.05&1.38$^{+0.16}_{-0.21}$&0.33$^{+0.15}_{-0.13}$&3.8$^{+12.0}_{-5.3}$&22.4$^{+17.3}_{-16.8}$&1.6$^{+8.8}_{-3.4}$&0.3$^{+6.0}_{-8.6}$\\
G344.43+0.05&ATCA&344.43&0.05&5.39$^{+0.09}_{-0.09}$&1.19$^{+0.20}_{-0.13}$&18.3$^{+7.6}_{-7.5}$&161.3$^{+13.3}_{-14.7}$&26.6$^{+6.8}_{-7.3}$&-1.5$^{+3.1}_{-9.7}$\\
G344.43+0.05&ATCA&344.43&0.05&14.79$^{+0.09}_{-0.09}$&3.11$^{+0.20}_{-0.21}$&130.2$^{+5.5}_{-5.0}$&-1.0$^{+7.1}_{-5.9}$&-0.9$^{+2.1}_{-4.6}$&-0.2$^{+2.1}_{-4.2}$\\
G346.52+0.08&ATCA&346.52&0.08&3.30$^{+0.25}_{-0.21}$&6.65$^{+0.57}_{-0.49}$&19.9$^{+6.2}_{-6.6}$&143.0$^{+9.6}_{-9.2}$&109.7$^{+6.3}_{-6.8}$&26.6$^{+7.9}_{-7.4}$\\
G346.52+0.08&ATCA&346.52&0.08&6.90$^{+0.16}_{-0.18}$&3.65$^{+0.51}_{-0.41}$&19.7$^{+8.6}_{-8.8}$&156.0$^{+11.9}_{-10.9}$&88.4$^{+8.3}_{-8.6}$&10.1$^{+10.4}_{-9.4}$\\
G347.75-1.14&ATCA&347.75&-1.14&-36.77$^{+73.74}_{-23.18}$&1.54$^{+0.53}_{-1.32}$&12.6$^{+6.7}_{-12.1}$&25.1$^{+9.8}_{-42.9}$&37.9$^{+8.9}_{-38.1}$&1.3$^{+7.1}_{-3.0}$\\
G350.50+0.96&ATCA&350.50&0.96&-10.74$^{+0.09}_{-4.37}$&2.34$^{+0.10}_{-1.44}$&-16.7$^{+17.1}_{-10.7}$&-53.0$^{+65.4}_{-48.6}$&166.0$^{+9.5}_{-161.8}$&81.2$^{+11.6}_{-69.3}$\\
G350.50+0.96&ATCA&350.50&0.96&-10.53$^{+0.09}_{-0.09}$&2.72$^{+0.05}_{-0.04}$&-19.3$^{+10.3}_{-10.4}$&356.3$^{+45.9}_{-43.8}$&177.3$^{+130.3}_{-11.7}$&90.6$^{+61.4}_{-11.1}$\\
G350.50+0.96&ATCA&350.50&0.96&-3.62$^{+0.09}_{-0.09}$&2.30$^{+0.20}_{-0.16}$&-0.9$^{+2.2}_{-5.2}$&59.9$^{+5.5}_{-5.3}$&59.7$^{+4.2}_{-4.4}$&18.6$^{+5.0}_{-5.2}$\\
G350.50+0.96&ATCA&350.50&0.96&6.69$^{+0.09}_{-0.09}$&1.67$^{+0.21}_{-0.21}$&24.0$^{+4.9}_{-4.9}$&36.1$^{+6.2}_{-5.7}$&43.1$^{+5.6}_{-5.2}$&0.6$^{+3.3}_{-2.4}$\\
\hline
\multicolumn{10}{l}{Table \ref{tab:tau} continued.}\\
\hline
\end{tabular}
\end{table*}

\begin{table*}
\centering
\begin{tabular}{llrrrrrrrr}
\hline
Source&Project&\multicolumn{1}{c}{$l^{\circ}$}&\multicolumn{1}{c}{$b^{\circ}$}&\multicolumn{1}{c}{$v$}&\multicolumn{1}{c}{$\Delta v$}&\multicolumn{4}{c}{$\tau_{\rm peak}\,  (10^{-3})$}\\
\cline{7-10}
&&&&\multicolumn{2}{c}{km\,s$^{-1}$}&\multicolumn{1}{c}{1612}&\multicolumn{1}{c}{1665}&\multicolumn{1}{c}{1667}&\multicolumn{1}{c}{1720}\\
\hline
G351.56+0.20&ATCA&351.56&0.20&-93.20$^{+0.49}_{-0.31}$&4.84$^{+1.01}_{-1.37}$&0.6$^{+2.9}_{-1.9}$&26.1$^{+6.8}_{-4.8}$&37.9$^{+6.4}_{-3.9}$&12.5$^{+3.9}_{-4.4}$\\
G351.56+0.20&ATCA&351.56&0.20&-45.18$^{+0.19}_{-0.20}$&4.20$^{+0.54}_{-0.50}$&22.4$^{+5.3}_{-5.7}$&47.5$^{+6.3}_{-7.4}$&55.1$^{+7.1}_{-6.4}$&-1.1$^{+3.3}_{-6.3}$\\
G351.56+0.20&ATCA&351.56&0.20&-40.62$^{+0.30}_{-0.30}$&8.24$^{+0.66}_{-0.59}$&-30.6$^{+4.9}_{-4.0}$&71.3$^{+4.5}_{-4.3}$&71.4$^{+4.8}_{-5.1}$&57.6$^{+4.3}_{-3.8}$\\
G351.56+0.20&ATCA&351.56&0.20&-36.77$^{+0.49}_{-0.52}$&6.75$^{+0.81}_{-0.90}$&-21.4$^{+4.2}_{-4.7}$&30.6$^{+3.9}_{-4.1}$&39.3$^{+3.8}_{-3.8}$&18.8$^{+4.6}_{-3.7}$\\
G351.56+0.20&ATCA&351.56&0.20&-7.63$^{+0.09}_{-0.09}$&0.76$^{+0.10}_{-0.08}$&0.1$^{+3.2}_{-4.2}$&125.0$^{+10.6}_{-10.7}$&3.8$^{+9.6}_{-4.1}$&1.4$^{+8.2}_{-3.3}$\\
G351.56+0.20&ATCA&351.56&0.20&0.09$^{+0.09}_{-0.09}$&0.71$^{+0.08}_{-0.07}$&8.2$^{+9.8}_{-8.1}$&-137.0$^{+12.3}_{-10.6}$&8.1$^{+10.2}_{-7.8}$&0.3$^{+5.1}_{-6.3}$\\
G351.56+0.20&ATCA&351.56&0.20&6.90$^{+0.09}_{-0.09}$&1.88$^{+0.20}_{-0.22}$&13.7$^{+5.8}_{-7.2}$&27.0$^{+5.9}_{-5.8}$&60.4$^{+5.6}_{-4.9}$&1.1$^{+5.7}_{-2.7}$\\
G351.61+0.17&ATCA&351.61&0.17&-43.39$^{+0.38}_{-0.40}$&8.51$^{+0.98}_{-1.21}$&6.5$^{+4.3}_{-5.0}$&55.4$^{+4.2}_{-4.5}$&62.3$^{+3.6}_{-4.2}$&20.6$^{+4.3}_{-4.0}$\\
G351.61+0.17&ATCA&351.61&0.17&-7.93$^{+0.09}_{-0.09}$&0.50$^{+0.03}_{-0.04}$&0.6$^{+4.2}_{-4.7}$&215.1$^{+14.9}_{-15.4}$&2.6$^{+10.4}_{-3.1}$&0.2$^{+5.7}_{-5.8}$\\
G351.61+0.17&ATCA&351.61&0.17&-0.07$^{+0.09}_{-0.09}$&0.38$^{+0.06}_{-0.05}$&0.7$^{+8.5}_{-4.6}$&-162.1$^{+18.5}_{-19.2}$&1.2$^{+8.7}_{-5.3}$&1.3$^{+8.5}_{-3.3}$\\
G353.41-0.30&ATCA&353.41&-0.30&-94.54$^{+0.09}_{-0.09}$&3.03$^{+0.16}_{-0.15}$&94.3$^{+3.8}_{-4.1}$&-0.4$^{+4.5}_{-4.8}$&0.7$^{+2.6}_{-1.6}$&-0.1$^{+2.2}_{-3.2}$\\
G353.41-0.30&ATCA&353.41&-0.30&-58.88$^{+13.22}_{-0.14}$&1.67$^{+0.27}_{-1.41}$&59.6$^{+9.1}_{-59.9}$&2.3$^{+22.9}_{-8.8}$&-0.4$^{+5.5}_{-4.7}$&0.6$^{+6.7}_{-5.8}$\\
G353.41-0.30&ATCA&353.41&-0.30&-19.10$^{+0.09}_{-0.09}$&1.55$^{+0.03}_{-0.03}$&26.8$^{+4.5}_{-5.5}$&-308.0$^{+7.0}_{-8.4}$&25.7$^{+5.3}_{-5.7}$&-290.4$^{+7.2}_{-5.8}$\\
G353.41-0.30&ATCA&353.41&-0.30&-14.40$^{+0.09}_{-0.09}$&5.31$^{+0.11}_{-0.10}$&-59.5$^{+2.7}_{-2.6}$&182.8$^{+4.1}_{-3.9}$&98.4$^{+3.4}_{-3.8}$&117.9$^{+3.4}_{-3.5}$\\
G356.91+0.08&ATCA&356.91&0.08&-75.58$^{+0.38}_{-0.33}$&5.76$^{+1.08}_{-0.76}$&0.2$^{+2.2}_{-3.0}$&34.8$^{+7.3}_{-5.5}$&60.8$^{+6.0}_{-5.7}$&12.4$^{+5.9}_{-8.6}$\\
G356.91+0.08&ATCA&356.91&0.08&-4.90$^{+0.10}_{-0.10}$&2.44$^{+0.25}_{-0.25}$&-11.0$^{+9.7}_{-7.7}$&59.0$^{+10.5}_{-10.4}$&83.4$^{+7.9}_{-6.8}$&41.3$^{+7.7}_{-7.2}$\\
G356.91+0.08&ATCA&356.91&0.08&4.71$^{+0.12}_{-0.11}$&2.83$^{+0.34}_{-0.28}$&43.7$^{+7.7}_{-6.8}$&58.6$^{+8.6}_{-7.9}$&71.6$^{+6.9}_{-7.4}$&2.8$^{+8.2}_{-3.2}$\\
G356.91+0.08&ATCA&356.91&0.08&9.81$^{+67.06}_{-0.29}$&2.23$^{+0.57}_{-2.00}$&1.5$^{+7.1}_{-4.9}$&42.7$^{+12.4}_{-97.8}$&53.1$^{+12.5}_{-54.2}$&24.2$^{+10.4}_{-23.2}$\\
\hline
\multicolumn{10}{l}{Table \ref{tab:tau} continued.}\\
\hline
\end{tabular}
\end{table*}

\begin{table*}
\centering
\begin{tabular}{llrrrrrrrr}
\hline
Source&Project&\multicolumn{1}{c}{$l^{\circ}$}&\multicolumn{1}{c}{$b^{\circ}$}&\multicolumn{1}{c}{$v$}&\multicolumn{1}{c}{$\Delta v$}&\multicolumn{4}{c}{$T_{\rm ex}$~(K)}\\
\cline{7-10}
&&&&\multicolumn{2}{c}{km\,s$^{-1}$}&\multicolumn{1}{c}{1612}&\multicolumn{1}{c}{1665}&\multicolumn{1}{c}{1667}&\multicolumn{1}{c}{1720}\\
\hline
SRC44&a3301&68.83&-3.49&6.17&0.80&-30.52$^{+88}_{-19.61}$&3.00$^{+0.93}_{-0.60}$&2.63$^{+0.87}_{-0.52}$&1.38$^{+0.26}_{-0.21}$\\
SRC44&a3301&68.83&-3.49&6.28&2.50&9.72$^{+1.94}_{-1.50}$&19.43$^{+2.80}_{-2.15}$&12.61$^{+1.18}_{-1.30}$&33.05$^{+3.19}_{-3.12}$\\
SRC44&a3301&68.83&-3.49&11.08&0.72&2.48$^{+10.79}_{-1.15}$&-0.56$^{+1.63}_{-0.34}$&-0.44$^{+1.35}_{-0.20}$&-0.23$^{+1.02}_{-0.12}$\\
SRC44&a3301&68.83&-3.49&11.20&1.29&-9.20$^{+28.20}_{-7.05}$&0.86$^{+0.21}_{-0.15}$&0.64$^{+0.20}_{-0.13}$&0.36$^{+0.04}_{-0.05}$\\
3C417&a2600&73.33&-5.45&9.51&0.88&7.94$^{+5.14}_{-2.15}$&4.30$^{+0.27}_{-0.24}$&5.27$^{+0.29}_{-0.26}$&3.43$^{+0.26}_{-0.25}$\\
3C417&a2600&73.33&-5.45&9.92&0.63&1.10$^{+0.08}_{-0.08}$&3.75$^{+0.13}_{-0.13}$&3.91$^{+0.12}_{-0.13}$&-2.92$^{+0.38}_{-0.40}$\\
3C417&a2600&73.33&-5.45&10.68&2.91&2.98$^{+0.25}_{-0.22}$&8.57$^{+0.51}_{-0.47}$&7.13$^{+0.22}_{-0.20}$&-15.18$^{+2.68}_{-4.23}$\\
3C092&a2600&159.74&-18.41&8.71&1.56&4.59$^{+0.71}_{-0.54}$&13.97$^{+1.53}_{-1.41}$&14.54$^{+1.35}_{-0.93}$&-14.68$^{+2.65}_{-4.11}$\\
3C092&a2600&159.74&-18.41&8.77&0.76&2.62$^{+0.27}_{-0.23}$&3.19$^{+0.18}_{-0.20}$&3.84$^{+0.20}_{-0.22}$&5.05$^{+0.04}_{-0.12}$\\
4C+28.11&a2769&166.06&-17.22&6.91&1.10&2.89$^{+0.34}_{-0.29}$&6.72$^{+0.41}_{-0.41}$&6.93$^{+0.26}_{-0.25}$&-24.55$^{+8.49}_{-27.01}$\\
PKS0319+12&a2769&170.59&-36.24&7.73&1.01&1.91$^{+0.77}_{-0.62}$&4.63$^{+1.44}_{-1.29}$&5.75$^{+1.66}_{-1.46}$&-8.89$^{+4.10}_{-8.15}$\\
3C131&a2600&171.44&-7.80&4.56&0.48&3.93$^{+2.00}_{-1.13}$&4.53$^{+0.37}_{-0.36}$&4.29$^{+0.35}_{-0.28}$&4.98$^{+2.21}_{-0.95}$\\
3C131&a2600&171.44&-7.80&5.71&3.21&3.02$^{+0.33}_{-0.26}$&8.83$^{+0.46}_{-0.43}$&9.25$^{+0.40}_{-0.35}$&-10.47$^{+1.77}_{-2.94}$\\
3C131&a2600&171.44&-7.80&6.59&0.44&4.29$^{+1.02}_{-0.66}$&3.77$^{+0.16}_{-0.16}$&4.29$^{+0.17}_{-0.17}$&3.79$^{+0.30}_{-0.31}$\\
3C131&a2600&171.44&-7.80&7.23&0.56&3.69$^{+0.18}_{-0.17}$&4.49$^{+0.07}_{-0.07}$&4.89$^{+0.07}_{-0.06}$&6.27$^{+0.26}_{-0.23}$\\
3C131&a2600&171.44&-7.80&7.48&1.93&0.79$^{+0.07}_{-0.06}$&3.33$^{+0.28}_{-0.26}$&3.90$^{+0.29}_{-0.27}$&-1.54$^{+0.11}_{-0.13}$\\
3C131&a2600&171.44&-7.80&7.79&0.57&8.26$^{+4.94}_{-26.70}$&-1.93$^{+0.38}_{-0.55}$&-2.90$^{+0.66}_{-0.89}$&-1.05$^{+0.06}_{-0.02}$\\
3C108&a3301&171.47&-20.7&9.42&1.19&4.07$^{+0.40}_{-0.34}$&5.60$^{+0.29}_{-0.30}$&9.72$^{+0.48}_{-0.46}$&23.62$^{+4.42}_{-3.64}$\\
3C108&a3301&171.47&-20.7&9.74&0.48&0.85$^{+0.25}_{-0.27}$&2.06$^{+0.59}_{-0.56}$&2.88$^{+0.73}_{-0.74}$&-3.40$^{+1.41}_{-1.25}$\\
4C+11.15&a2769&171.98&-35.48&7.18&0.65&3.65$^{+0.98}_{-0.56}$&4.38$^{+0.40}_{-0.37}$&4.42$^{+0.29}_{-0.26}$&5.46$^{+0.40}_{-0.59}$\\
4C+36.10&a2769&172.98&2.44&-16.74&2.89&5.40$^{+0.92}_{-0.72}$&8.62$^{+0.51}_{-0.47}$&9.28$^{+0.38}_{-0.35}$&23.13$^{+13.61}_{-5.90}$\\
4C+27.14&a3301&175.46&-9.96&7.19&1.62&75.92$^{+52.70}_{-292}$&18.63$^{+2.24}_{-1.74}$&14.69$^{+0.80}_{-1.04}$&9.47$^{+1.90}_{-0.64}$\\
4C+27.14&a3301&175.46&-9.96&7.89&0.84&0.10$^{+0.04}_{-0.02}$&6.60$^{+4.33}_{-13.36}$&1.86$^{+0.65}_{-0.37}$&-0.12$^{+0.02}_{-0.04}$\\
4C+17.23&a2769&176.36&-24.24&9.35&0.72&3.10$^{+0.58}_{-0.46}$&4.86$^{+0.39}_{-0.37}$&4.41$^{+0.26}_{-0.27}$&8.58$^{+2.00}_{-1.30}$\\
4C+17.23&a2769&176.36&-24.24&11.42&0.77&2.40$^{+0.34}_{-0.26}$&4.34$^{+0.34}_{-0.28}$&4.24$^{+0.19}_{-0.21}$&16.53$^{+8.07}_{-4.52}$\\
4C+07.13&a2769&178.87&-36.27&3.48&1.07&3.25$^{+3.00}_{-1.23}$&3.75$^{+0.83}_{-0.79}$&4.85$^{+0.91}_{-0.80}$&5.91$^{+3.93}_{-1.56}$\\
B0531+2730&a2769&179.87&-2.83&3.04&0.72&1.53$^{+0.54}_{-0.47}$&-6.25$^{+396}_{-3.84}$&2.75$^{+0.56}_{-0.66}$&-2.41$^{+0.02}_{-0.63}$\\
B0531+2730&a2769&179.87&-2.83&3.17&0.78&4.08$^{+1.87}_{-36.41}$&1.44$^{+0.57}_{-0.42}$&9.80$^{+491}_{-3.82}$&1.84$^{+0.10}_{-0.39}$\\
4C+26.18b&a2769&182.36&-0.62&-11.93&1.57&-2.38$^{+0.54}_{-0.56}$&4.69$^{+0.65}_{-0.82}$&4.55$^{+0.46}_{-0.64}$&1.23$^{+0.01}_{-0.10}$\\
4C+26.18b&a2769&182.36&-0.62&-9.99&2.66&14.70$^{+7.80}_{-151}$&15.18$^{+6.19}_{-3.62}$&10.16$^{+1.20}_{-1.06}$&10.48$^{+7.93}_{-3.21}$\\
4C+22.12&a2769&188.07&0.04&-1.62&0.66&15.72$^{+11.00}_{-32.70}$&2.36$^{+0.65}_{-0.68}$&2.03$^{+0.50}_{-0.60}$&1.21$^{+0.11}_{-0.26}$\\
4C+14.14&a2769&189.04&-12.85&2.60&3.82&7.86$^{+7.47}_{-2.47}$&6.20$^{+0.71}_{-0.55}$&5.75$^{+0.30}_{-0.35}$&4.86$^{+0.80}_{-0.68}$\\
4C+19.18&a2769&190.09&-2.17&-0.62&2.18&-2.37$^{+0.48}_{-0.72}$&11.44$^{+3.61}_{-2.30}$&6.87$^{+0.76}_{-0.67}$&1.61$^{+0.16}_{-0.16}$\\
4C+19.18&a2769&190.09&-2.17&2.39&1.47&4.41$^{+1.41}_{-0.87}$&7.13$^{+0.82}_{-0.71}$&5.93$^{+0.33}_{-0.33}$&11.52$^{+5.33}_{-2.86}$\\
4C+19.19&a2769&190.13&-1.64&1.12&1.42&-1.25$^{+0.55}_{-1.01}$&4.05$^{+1.04}_{-1.12}$&3.57$^{+0.83}_{-1.00}$&0.79$^{+0.10}_{-0.22}$\\
4C+19.19&a2769&190.13&-1.64&2.70&3.15&2.25$^{+0.27}_{-0.21}$&7.88$^{+0.76}_{-0.57}$&7.71$^{+0.45}_{-0.39}$&-5.94$^{+0.84}_{-1.11}$\\
PKS0528+134&a2600&191.37&-11.01&9.60&0.90&4.25$^{+0.48}_{-0.38}$&4.47$^{+0.19}_{-0.17}$&4.44$^{+0.16}_{-0.16}$&4.67$^{+0.11}_{-0.13}$\\
4C+16.15b&a2769&193.64&-1.53&11.88&0.87&4.40$^{+1.07}_{-0.69}$&5.25$^{+0.35}_{-0.33}$&5.26$^{+0.27}_{-0.26}$&6.42$^{+0.84}_{-0.77}$\\
3C158&a2769&196.64&0.17&3.14&0.98&8.07$^{+5.06}_{-23.59}$&3.08$^{+1.75}_{-1.36}$&4.93$^{+2.56}_{-2.11}$&2.53$^{+0.03}_{-0.85}$\\
4C+14.18&a2600&196.98&1.10&4.28&0.55&3.04$^{+1.57}_{-24.37}$&0.80$^{+0.40}_{-0.30}$&0.62$^{+0.29}_{-0.26}$&0.40$^{+0.12}_{-0.16}$\\
4C+14.18&a2600&196.98&1.10&4.94&1.84&3.30$^{+0.24}_{-0.24}$&4.99$^{+0.15}_{-0.16}$&6.54$^{+0.20}_{-0.20}$&17.12$^{+4.04}_{-2.33}$\\
4C+14.18&a2600&196.98&1.10&7.39&0.81&1.53$^{+0.26}_{-0.22}$&3.20$^{+0.31}_{-0.31}$&4.62$^{+0.39}_{-0.39}$&-10.11$^{+3.42}_{-8.92}$\\
4C+14.18&a2600&196.98&1.10&16.49&1.29&-11.67$^{+536}_{-5.72}$&5.77$^{+0.46}_{-0.47}$&4.88$^{+0.39}_{-0.37}$&2.24$^{+0.29}_{-0.23}$\\
4C+14.18&a2600&196.98&1.10&17.59&0.70&2.97$^{+1.69}_{-0.84}$&3.71$^{+0.62}_{-0.59}$&3.05$^{+0.50}_{-0.46}$&3.80$^{+0.28}_{-0.42}$\\
4C+14.18&a2600&196.98&1.10&18.40&3.76&1.46$^{+0.20}_{-0.16}$&8.49$^{+1.14}_{-0.92}$&9.09$^{+0.76}_{-0.74}$&-2.38$^{+0.29}_{-0.36}$\\
4C+14.18&a2600&196.98&1.10&31.98&0.42&3.57$^{+0.93}_{-0.84}$&3.65$^{+1.06}_{-0.26}$&3.83$^{+1.16}_{-0.28}$&3.91$^{+0.75}_{-1.31}$\\
4C+14.18&a2600&196.98&1.10&32.33&1.19&3.52$^{+1.55}_{-0.84}$&3.79$^{+0.58}_{-0.77}$&3.69$^{+0.48}_{-0.56}$&3.97$^{+0.40}_{-0.26}$\\
\hline
\end{tabular}
\caption{Fitted excitation temperatures of features identified in this work. Columns give the background source, project name, Galactic longitude and latitude, centroid velocity $v$, FWHM $\Delta v$, and excitation temperatures at 1612, 1665, 1667 and 1720\,MHz. The uncertainties are 68\% credibility intervals.}
\label{tab:Tex}
\end{table*}

\begin{table*}
\centering
\begin{tabular}{llrrrrrrrr}
\hline
Source&Project&\multicolumn{1}{c}{$l^{\circ}$}&\multicolumn{1}{c}{$b^{\circ}$}&\multicolumn{1}{c}{$v$}&\multicolumn{1}{c}{$\Delta v$}&\multicolumn{4}{c}{$T_{\rm ex}$~(K)}\\
\cline{7-10}
&&&&\multicolumn{2}{c}{km\,s$^{-1}$}&\multicolumn{1}{c}{1612}&\multicolumn{1}{c}{1665}&\multicolumn{1}{c}{1667}&\multicolumn{1}{c}{1720}\\
\hline
4C+13.32&a2769&197.15&-0.85&-5.65&0.78&1.21$^{+0.33}_{-0.32}$&2.31$^{+0.61}_{-0.62}$&2.29$^{+0.63}_{-0.56}$&15.25$^{+4.03}_{-2.45}$\\
4C+13.32&a2769&197.15&-0.85&-0.35&1.02&2.54$^{+24.45}_{-1.31}$&0.23$^{+0.31}_{-0.10}$&0.11$^{+0.20}_{-0.06}$&0.08$^{+0.13}_{-0.04}$\\
4C+13.32&a2769&197.15&-0.85&4.46&1.20&-6.94$^{+3.64}_{-15.72}$&4.42$^{+2.35}_{-1.98}$&3.24$^{+1.79}_{-1.49}$&1.53$^{+0.08}_{-0.52}$\\
4C+13.32&a2769&197.15&-0.85&6.99&2.21&2.35$^{+0.93}_{-0.92}$&18.97$^{+12.75}_{-8.94}$&14.59$^{+6.17}_{-4.89}$&-3.56$^{+1.37}_{-1.23}$\\
4C+13.32&a2769&197.15&-0.85&9.50&1.15&6.45$^{+5.22}_{-2.61}$&5.56$^{+2.37}_{-2.23}$&5.68$^{+2.08}_{-2.22}$&5.02$^{+0.98}_{-1.96}$\\
4C+17.41&a2769&201.13&16.42&0.23&1.38&2.48$^{+1.17}_{-0.64}$&4.68$^{+0.97}_{-0.75}$&4.14$^{+0.44}_{-0.42}$&15.81$^{+509}_{-7.87}$\\
4C+17.41&a2769&201.13&16.42&1.89&0.73&1.86$^{+0.31}_{-0.24}$&3.86$^{+0.32}_{-0.31}$&3.98$^{+0.20}_{-0.20}$&-120.31$^{+99}_{-150}$\\
4C+04.22&a2769&205.41&-4.43&11.92&0.71&2.37$^{+0.58}_{-0.45}$&5.43$^{+0.81}_{-0.63}$&4.45$^{+0.40}_{-0.39}$&2751.44$^{+2773}_{-2725}$\\
4C+04.22&a2769&205.41&-4.43&13.33&0.92&1.43$^{+0.26}_{-0.19}$&4.27$^{+0.49}_{-0.50}$&3.65$^{+0.31}_{-0.35}$&-6.20$^{+1.29}_{-3.23}$\\
J134217-040725&a3301&205.58&-4.14&9.19&1.20&2.55$^{+0.35}_{-0.28}$&4.36$^{+0.30}_{-0.30}$&4.58$^{+0.27}_{-0.26}$&15.01$^{+4.68}_{-3.13}$\\
4C+04.24&a2769&205.92&-3.57&9.39&1.25&2.72$^{+0.48}_{-0.38}$&4.31$^{+0.35}_{-0.31}$&4.10$^{+0.23}_{-0.23}$&8.59$^{+2.22}_{-1.38}$\\
\hline
\multicolumn{10}{l}{Table \ref{tab:Tex} continued.}\\
\hline
\end{tabular}
% \caption{Fitted excitation temperatures of features identified in this work. Columns give the background source, project name, Galactic longitude and latitude, centroid velocity $v$, FWHM $\Delta v$, and excitation temperatures at 1612, 1665, 1667 and 1720\,MHz. The uncertainties are 68\% credibility intervals.}
% \label{tab:Tex}
\end{table*}

\begin{table*}
\centering
\begin{tabular}{llrrrrrrrrrr}
\hline
Source&Project&\multicolumn{1}{c}{$l^{\circ}$}&\multicolumn{1}{c}{$b^{\circ}$}&\multicolumn{1}{c}{$v$}&\multicolumn{1}{c}{$\Delta v$}&\multicolumn{1}{c}{N$_{1}$}&\multicolumn{1}{c}{N$_{2}$}&\multicolumn{1}{c}{N$_{3}$}&\multicolumn{1}{c}{N$_{4}$}&\multicolumn{1}{c}{N$_{\rm OH}$}\\
\cline{7-11}
&&&&\multicolumn{2}{c}{km\,s$^{-1}$}&\multicolumn{5}{c}{log$_{10}$\,cm$^{-2}$}\\
\hline
SRC44&a3301&68.83&-3.49&6.17&0.80&12.25$^{+0.13}_{-0.10}$&12.46$^{+0.13}_{-0.10}$&12.24$^{+0.13}_{-0.10}$&12.45$^{+0.13}_{-0.10}$&12.96$^{+0.13}_{-0.10}$\\
SRC44&a3301&68.83&-3.49&6.28&2.50&13.05$^{+0.03}_{-0.04}$&13.27$^{+0.03}_{-0.04}$&13.04$^{+0.03}_{-0.04}$&13.27$^{+0.03}_{-0.04}$&13.77$^{+0.03}_{-0.04}$\\
SRC44&a3301&68.83&-3.49&11.08&0.72&10.97$^{+0.42}_{-0.33}$&11.27$^{+0.34}_{-0.24}$&11.03$^{+0.33}_{-0.22}$&11.35$^{+0.22}_{-0.17}$&11.79$^{+0.31}_{-0.22}$\\
SRC44&a3301&68.83&-3.49&11.20&1.29&11.83$^{+0.12}_{-0.21}$&12.01$^{+0.11}_{-0.21}$&11.79$^{+0.11}_{-0.21}$&11.95$^{+0.10}_{-0.20}$&12.51$^{+0.11}_{-0.21}$\\
3C417&a2600&73.33&-5.45&9.51&0.88&13.09$^{+0.06}_{-0.07}$&13.30$^{+0.06}_{-0.07}$&13.08$^{+0.06}_{-0.07}$&13.30$^{+0.06}_{-0.07}$&13.81$^{+0.06}_{-0.07}$\\
3C417&a2600&73.33&-5.45&9.92&0.63&13.15$^{+0.04}_{-0.05}$&13.40$^{+0.05}_{-0.05}$&13.14$^{+0.04}_{-0.05}$&13.39$^{+0.05}_{-0.05}$&13.89$^{+0.04}_{-0.05}$\\
3C417&a2600&73.33&-5.45&10.68&2.91&13.45$^{+0.02}_{-0.03}$&13.68$^{+0.02}_{-0.03}$&13.45$^{+0.02}_{-0.03}$&13.67$^{+0.02}_{-0.03}$&14.18$^{+0.02}_{-0.03}$\\
3C092&a2600&159.74&-18.41&8.71&1.56&13.28$^{+0.02}_{-0.03}$&13.51$^{+0.03}_{-0.03}$&13.28$^{+0.02}_{-0.03}$&13.51$^{+0.03}_{-0.03}$&14.01$^{+0.02}_{-0.03}$\\
3C092&a2600&159.74&-18.41&8.77&0.76&13.09$^{+0.04}_{-0.04}$&13.32$^{+0.04}_{-0.04}$&13.08$^{+0.04}_{-0.04}$&13.31$^{+0.04}_{-0.04}$&13.82$^{+0.04}_{-0.04}$\\
4C+28.11&a2769&166.06&-17.22&6.91&1.10&13.24$^{+0.02}_{-0.02}$&13.47$^{+0.02}_{-0.02}$&13.24$^{+0.02}_{-0.02}$&13.47$^{+0.02}_{-0.02}$&13.97$^{+0.02}_{-0.02}$\\
PKS0319+12&a2769&170.59&-36.24&7.73&1.01&12.24$^{+0.12}_{-0.14}$&12.47$^{+0.12}_{-0.14}$&12.23$^{+0.12}_{-0.14}$&12.46$^{+0.12}_{-0.14}$&12.97$^{+0.12}_{-0.14}$\\
3C131&a2600&171.44&-7.80&4.56&0.48&11.94$^{+0.03}_{-0.04}$&12.17$^{+0.04}_{-0.04}$&11.94$^{+0.03}_{-0.04}$&12.16$^{+0.04}_{-0.04}$&12.67$^{+0.04}_{-0.04}$\\
3C131&a2600&171.44&-7.80&5.71&3.21&12.96$^{+0.02}_{-0.02}$&13.19$^{+0.02}_{-0.02}$&12.96$^{+0.02}_{-0.02}$&13.19$^{+0.02}_{-0.02}$&13.69$^{+0.02}_{-0.02}$\\
3C131&a2600&171.44&-7.80&6.59&0.44&12.34$^{+0.02}_{-0.02}$&12.56$^{+0.03}_{-0.03}$&12.33$^{+0.02}_{-0.02}$&12.55$^{+0.03}_{-0.03}$&13.06$^{+0.03}_{-0.03}$\\
3C131&a2600&171.44&-7.80&7.23&0.56&13.01$^{+0.01}_{-0.01}$&13.23$^{+0.01}_{-0.01}$&13.00$^{+0.01}_{-0.00}$&13.22$^{+0.01}_{-0.01}$&13.73$^{+0.01}_{-0.01}$\\
3C131&a2600&171.44&-7.80&7.48&1.93&12.52$^{+0.05}_{-0.05}$&12.77$^{+0.05}_{-0.05}$&12.51$^{+0.04}_{-0.05}$&12.76$^{+0.05}_{-0.05}$&13.26$^{+0.05}_{-0.05}$\\
3C131&a2600&171.44&-7.80&7.79&0.57&11.33$^{+0.10}_{-0.12}$&11.58$^{+0.10}_{-0.12}$&11.35$^{+0.10}_{-0.11}$&11.59$^{+0.10}_{-0.12}$&12.08$^{+0.10}_{-0.12}$\\
3C108&a3301&171.47&-20.70&9.42&1.19&13.20$^{+0.03}_{-0.03}$&13.43$^{+0.03}_{-0.03}$&13.20$^{+0.03}_{-0.03}$&13.42$^{+0.03}_{-0.03}$&13.93$^{+0.03}_{-0.03}$\\
3C108&a3301&171.47&-20.70&9.74&0.48&12.26$^{+0.13}_{-0.17}$&12.50$^{+0.15}_{-0.18}$&12.24$^{+0.13}_{-0.17}$&12.49$^{+0.14}_{-0.17}$&12.99$^{+0.14}_{-0.17}$\\
4C+11.15&a2769&171.98&-35.48&7.18&0.65&12.91$^{+0.03}_{-0.03}$&13.14$^{+0.03}_{-0.03}$&12.91$^{+0.03}_{-0.03}$&13.13$^{+0.03}_{-0.03}$&13.64$^{+0.03}_{-0.03}$\\
4C+36.1&a2769&172.98&2.44&-16.74&2.89&13.54$^{+0.02}_{-0.02}$&13.76$^{+0.02}_{-0.02}$&13.53$^{+0.01}_{-0.02}$&13.76$^{+0.02}_{-0.02}$&14.26$^{+0.02}_{-0.02}$\\
4C+27.14&a3301&175.46&-9.96&7.19&1.62&13.59$^{+0.01}_{-0.01}$&13.81$^{+0.01}_{-0.01}$&13.59$^{+0.01}_{-0.01}$&13.81$^{+0.01}_{-0.01}$&14.31$^{+0.01}_{-0.01}$\\
4C+27.14&a3301&175.46&-9.96&7.89&0.84&11.70$^{+0.14}_{-0.31}$&12.25$^{+0.20}_{-0.39}$&11.69$^{+0.13}_{-0.30}$&12.23$^{+0.20}_{-0.38}$&12.65$^{+0.18}_{-0.37}$\\
4C+17.23&a2769&176.36&-24.24&9.35&0.72&12.79$^{+0.03}_{-0.03}$&13.02$^{+0.03}_{-0.03}$&12.79$^{+0.03}_{-0.03}$&13.01$^{+0.03}_{-0.03}$&13.52$^{+0.03}_{-0.03}$\\
4C+17.23&a2769&176.36&-24.24&11.42&0.77&12.87$^{+0.03}_{-0.02}$&13.09$^{+0.03}_{-0.03}$&12.86$^{+0.03}_{-0.02}$&13.08$^{+0.03}_{-0.03}$&13.59$^{+0.03}_{-0.02}$\\
4C+7.13&a2769&178.87&-36.27&3.48&1.07&12.52$^{+0.08}_{-0.09}$&12.74$^{+0.08}_{-0.09}$&12.51$^{+0.08}_{-0.09}$&12.74$^{+0.08}_{-0.09}$&13.24$^{+0.08}_{-0.09}$\\
B0531+2730&a2769&179.87&-2.83&3.04&0.72&12.89$^{+0.11}_{-0.14}$&13.14$^{+0.12}_{-0.14}$&12.89$^{+0.11}_{-0.14}$&13.12$^{+0.11}_{-0.14}$&13.63$^{+0.11}_{-0.14}$\\
B0531+2730&a2769&179.87&-2.83&3.17&0.78&12.85$^{+0.12}_{-0.15}$&13.06$^{+0.12}_{-0.16}$&12.83$^{+0.11}_{-0.15}$&13.05$^{+0.11}_{-0.15}$&13.56$^{+0.11}_{-0.15}$\\
4C+26.18b&a2769&182.36&-0.62&-11.93&1.57&13.14$^{+0.07}_{-0.11}$&13.34$^{+0.07}_{-0.11}$&13.13$^{+0.06}_{-0.11}$&13.33$^{+0.07}_{-0.11}$&13.85$^{+0.07}_{-0.11}$\\
4C+26.18b&a2769&182.36&-0.62&-9.99&2.66&13.25$^{+0.07}_{-0.07}$&13.47$^{+0.07}_{-0.07}$&13.25$^{+0.07}_{-0.07}$&13.47$^{+0.07}_{-0.07}$&13.97$^{+0.07}_{-0.07}$\\
4C+22.12&a2769&188.07&0.04&-1.62&0.66&12.28$^{+0.10}_{-0.15}$&12.49$^{+0.10}_{-0.15}$&12.26$^{+0.09}_{-0.15}$&12.47$^{+0.09}_{-0.15}$&12.99$^{+0.09}_{-0.15}$\\
4C+14.14&a2769&189.04&-12.85&2.60&3.82&13.20$^{+0.03}_{-0.04}$&13.42$^{+0.03}_{-0.04}$&13.19$^{+0.03}_{-0.04}$&13.41$^{+0.03}_{-0.04}$&13.92$^{+0.03}_{-0.04}$\\
4C+19.18&a2769&190.09&-2.17&-0.62&2.18&12.97$^{+0.05}_{-0.05}$&13.18$^{+0.05}_{-0.05}$&12.97$^{+0.05}_{-0.05}$&13.17$^{+0.05}_{-0.05}$&13.69$^{+0.05}_{-0.05}$\\
4C+19.18&a2769&190.09&-2.17&2.39&1.47&13.06$^{+0.03}_{-0.04}$&13.28$^{+0.04}_{-0.04}$&13.05$^{+0.03}_{-0.04}$&13.27$^{+0.03}_{-0.04}$&13.78$^{+0.03}_{-0.04}$\\
4C+19.19&a2769&190.13&-1.64&1.12&1.42&12.57$^{+0.17}_{-0.18}$&12.75$^{+0.18}_{-0.20}$&12.56$^{+0.17}_{-0.17}$&12.74$^{+0.18}_{-0.19}$&13.27$^{+0.18}_{-0.19}$\\
4C+19.19&a2769&190.13&-1.64&2.70&3.15&13.30$^{+0.03}_{-0.05}$&13.54$^{+0.03}_{-0.05}$&13.30$^{+0.03}_{-0.05}$&13.53$^{+0.03}_{-0.05}$&14.03$^{+0.03}_{-0.05}$\\
PKS0528+134&a2600&191.37&-11.01&9.60&0.90&12.91$^{+0.02}_{-0.02}$&13.14$^{+0.02}_{-0.02}$&12.91$^{+0.02}_{-0.02}$&13.13$^{+0.02}_{-0.02}$&13.64$^{+0.02}_{-0.02}$\\
4C+16.15b&a2769&193.64&-1.53&11.88&0.87&12.94$^{+0.03}_{-0.03}$&13.17$^{+0.03}_{-0.03}$&12.94$^{+0.03}_{-0.03}$&13.16$^{+0.03}_{-0.03}$&13.67$^{+0.03}_{-0.03}$\\
3C158&a2769&196.64&0.17&3.14&0.98&12.26$^{+0.20}_{-0.27}$&12.48$^{+0.19}_{-0.27}$&12.25$^{+0.19}_{-0.26}$&12.47$^{+0.19}_{-0.27}$&12.98$^{+0.19}_{-0.27}$\\
4C+14.18&a2600&196.98&1.10&4.28&0.55&11.22$^{+0.17}_{-0.23}$&11.41$^{+0.16}_{-0.22}$&11.18$^{+0.15}_{-0.21}$&11.35$^{+0.12}_{-0.21}$&11.90$^{+0.15}_{-0.21}$\\
4C+14.18&a2600&196.98&1.10&4.94&1.84&12.94$^{+0.01}_{-0.01}$&13.17$^{+0.01}_{-0.01}$&12.93$^{+0.01}_{-0.01}$&13.16$^{+0.01}_{-0.01}$&13.67$^{+0.01}_{-0.01}$\\
4C+14.18&a2600&196.98&1.10&7.39&0.81&12.15$^{+0.04}_{-0.05}$&12.38$^{+0.04}_{-0.05}$&12.14$^{+0.04}_{-0.04}$&12.37$^{+0.04}_{-0.05}$&12.88$^{+0.04}_{-0.05}$\\
4C+14.18&a2600&196.98&1.10&16.49&1.29&12.48$^{+0.06}_{-0.06}$&12.69$^{+0.06}_{-0.07}$&12.47$^{+0.06}_{-0.06}$&12.68$^{+0.06}_{-0.07}$&13.20$^{+0.06}_{-0.07}$\\
4C+14.18&a2600&196.98&1.10&17.59&0.70&12.01$^{+0.09}_{-0.10}$&12.23$^{+0.09}_{-0.10}$&12.00$^{+0.09}_{-0.10}$&12.22$^{+0.09}_{-0.10}$&12.73$^{+0.09}_{-0.10}$\\
4C+14.18&a2600&196.98&1.10&18.40&3.76&12.70$^{+0.05}_{-0.07}$&12.94$^{+0.05}_{-0.07}$&12.69$^{+0.05}_{-0.07}$&12.93$^{+0.05}_{-0.07}$&13.43$^{+0.05}_{-0.07}$\\
4C+14.18&a2600&196.98&1.10&31.98&0.42&12.42$^{+0.09}_{-0.04}$&12.64$^{+0.09}_{-0.03}$&12.41$^{+0.09}_{-0.03}$&12.64$^{+0.09}_{-0.03}$&13.15$^{+0.09}_{-0.03}$\\
4C+14.18&a2600&196.98&1.10&32.33&1.19&12.35$^{+0.05}_{-0.07}$&12.57$^{+0.05}_{-0.07}$&12.34$^{+0.05}_{-0.07}$&12.56$^{+0.05}_{-0.07}$&13.07$^{+0.05}_{-0.07}$\\
\hline
\end{tabular}
\caption{Fitted column densities of features identified in this work. Columns give the background source of each sightline, project name, Galactic longitude and latitude, centroid velocity $v$, FWHM $\Delta v$, and column densities of the hyperfine levels of the OH ground-rotational state (where N$_1$~is the lowest level) and the total OH column density. The uncertainties are the 68\% credibility intervals.}
\label{tab:N}
\end{table*}

\begin{table*}
\centering
\begin{tabular}{llrrrrrrrrrr}
\hline
Source&Project&\multicolumn{1}{c}{$l^{\circ}$}&\multicolumn{1}{c}{$b^{\circ}$}&\multicolumn{1}{c}{$v$}&\multicolumn{1}{c}{$\Delta v$}&\multicolumn{1}{c}{N$_{1}$}&\multicolumn{1}{c}{N$_{2}$}&\multicolumn{1}{c}{N$_{3}$}&\multicolumn{1}{c}{N$_{4}$}&\multicolumn{1}{c}{N$_{\rm OH}$}\\
\cline{7-11}
&&&&\multicolumn{2}{c}{km\,s$^{-1}$}&\multicolumn{5}{c}{log$_{10}$\,cm$^{-2}$}\\
\hline
4C+13.32&a2769&197.15&-0.85&-5.65&0.78&12.27$^{+0.11}_{-0.12}$&12.50$^{+0.11}_{-0.13}$&12.25$^{+0.10}_{-0.12}$&12.49$^{+0.10}_{-0.12}$&12.99$^{+0.11}_{-0.12}$\\
4C+13.32&a2769&197.15&-0.85&-0.35&1.02&10.62$^{+0.36}_{-0.26}$&10.71$^{+0.26}_{-0.19}$&10.47$^{+0.24}_{-0.17}$&10.40$^{+0.01}_{-0.14}$&11.17$^{+0.24}_{-0.16}$\\
4C+13.32&a2769&197.15&-0.85&4.46&1.20&12.18$^{+0.18}_{-0.25}$&12.39$^{+0.18}_{-0.26}$&12.18$^{+0.18}_{-0.25}$&12.38$^{+0.17}_{-0.25}$&12.90$^{+0.18}_{-0.25}$\\
4C+13.32&a2769&197.15&-0.85&6.99&2.21&12.52$^{+0.17}_{-0.24}$&12.75$^{+0.18}_{-0.24}$&12.52$^{+0.17}_{-0.24}$&12.75$^{+0.17}_{-0.24}$&13.25$^{+0.17}_{-0.24}$\\
4C+13.32&a2769&197.15&-0.85&9.50&1.15&12.31$^{+0.13}_{-0.22}$&12.53$^{+0.13}_{-0.22}$&12.30$^{+0.12}_{-0.22}$&12.52$^{+0.12}_{-0.22}$&13.03$^{+0.13}_{-0.22}$\\
4C+17.41&a2769&201.13&16.42&0.23&1.38&12.67$^{+0.06}_{-0.07}$&12.90$^{+0.07}_{-0.07}$&12.67$^{+0.06}_{-0.07}$&12.89$^{+0.07}_{-0.07}$&13.40$^{+0.07}_{-0.07}$\\
4C+17.41&a2769&201.13&16.42&1.89&0.73&12.81$^{+0.03}_{-0.04}$&13.04$^{+0.03}_{-0.04}$&12.80$^{+0.03}_{-0.04}$&13.03$^{+0.03}_{-0.04}$&13.54$^{+0.03}_{-0.04}$\\
4C+4.22&a2769&205.41&-4.43&11.92&0.71&12.72$^{+0.05}_{-0.06}$&12.95$^{+0.05}_{-0.06}$&12.72$^{+0.05}_{-0.06}$&12.94$^{+0.05}_{-0.06}$&13.45$^{+0.05}_{-0.06}$\\
4C+4.22&a2769&205.41&-4.43&13.33&0.92&12.75$^{+0.05}_{-0.06}$&12.99$^{+0.05}_{-0.06}$&12.74$^{+0.04}_{-0.06}$&12.98$^{+0.05}_{-0.06}$&13.48$^{+0.05}_{-0.06}$\\
J134217-040725&a3301&205.58&-4.14&9.19&1.20&12.99$^{+0.03}_{-0.03}$&13.22$^{+0.03}_{-0.03}$&12.99$^{+0.03}_{-0.03}$&13.21$^{+0.03}_{-0.03}$&13.72$^{+0.03}_{-0.03}$\\
4C+4.24&a2769&205.92&-3.57&9.39&1.25&13.01$^{+0.03}_{-0.03}$&13.24$^{+0.03}_{-0.03}$&13.00$^{+0.03}_{-0.03}$&13.23$^{+0.03}_{-0.03}$&13.74$^{+0.03}_{-0.03}$\\
\hline
\multicolumn{11}{l}{Table \ref{tab:N} continued.}\\
\hline
\end{tabular}
% \caption{Fitted column densities of features identified in this work. Columns give the background source of each sightline, project name, Galactic longitude and latitude, centroid velocity $v$, FWHM $\Delta v$, and column densities of the hyperfine levels of the OH ground-rotational state (where N$_1$~is the lowest level) and the total OH column density. The uncertainties are the 68\% credibility intervals.}
% \label{tab:N}
\end{table*}

\begin{figure*}
\begin{tabular}{cc}
\includegraphics[trim={0.7cm 0cm 1cm 0cm}, clip=true, width=0.45\linewidth]{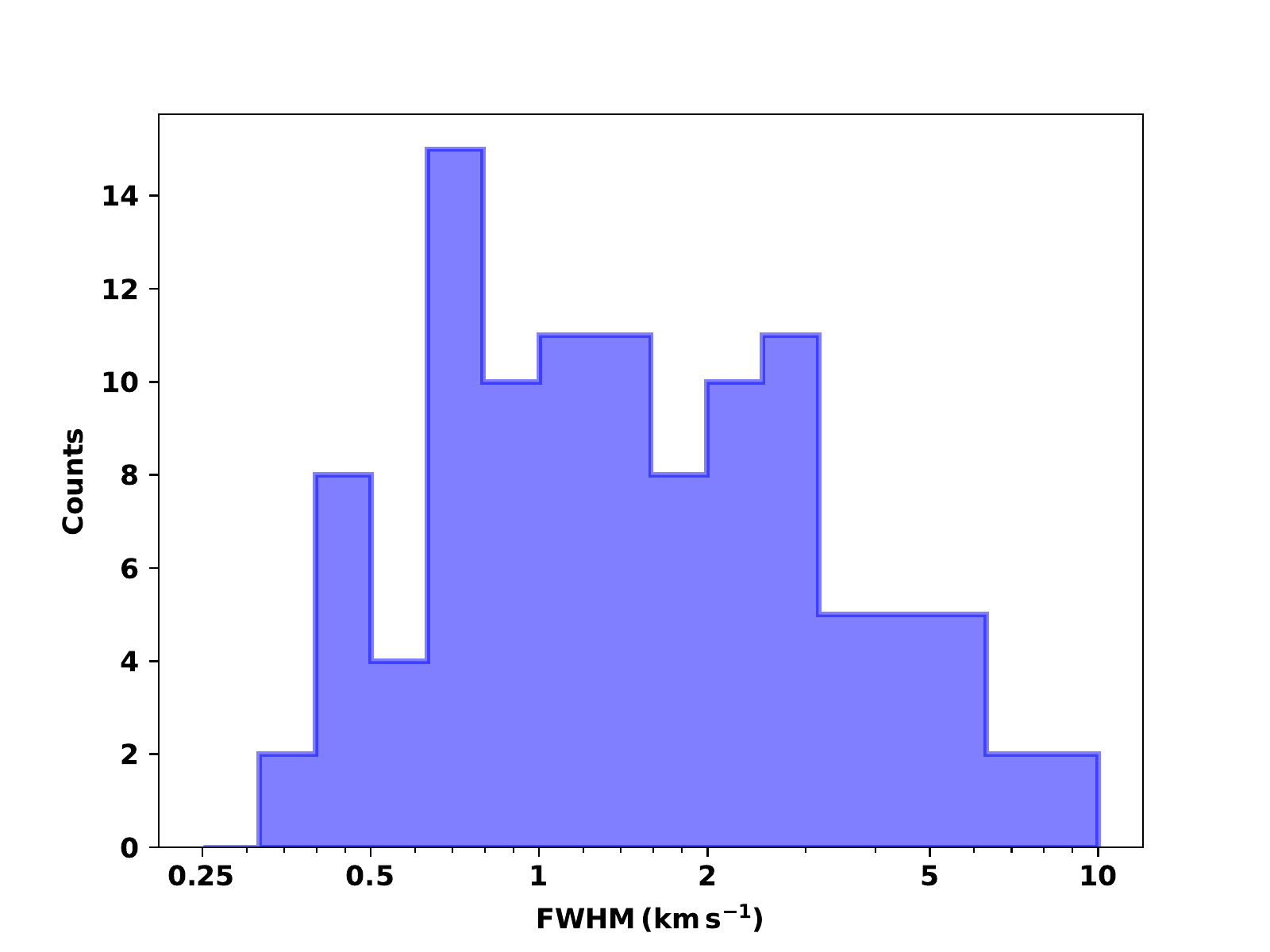}&
\includegraphics[trim={0.7cm 0cm 1cm 0cm}, clip=true, width=0.45\linewidth]{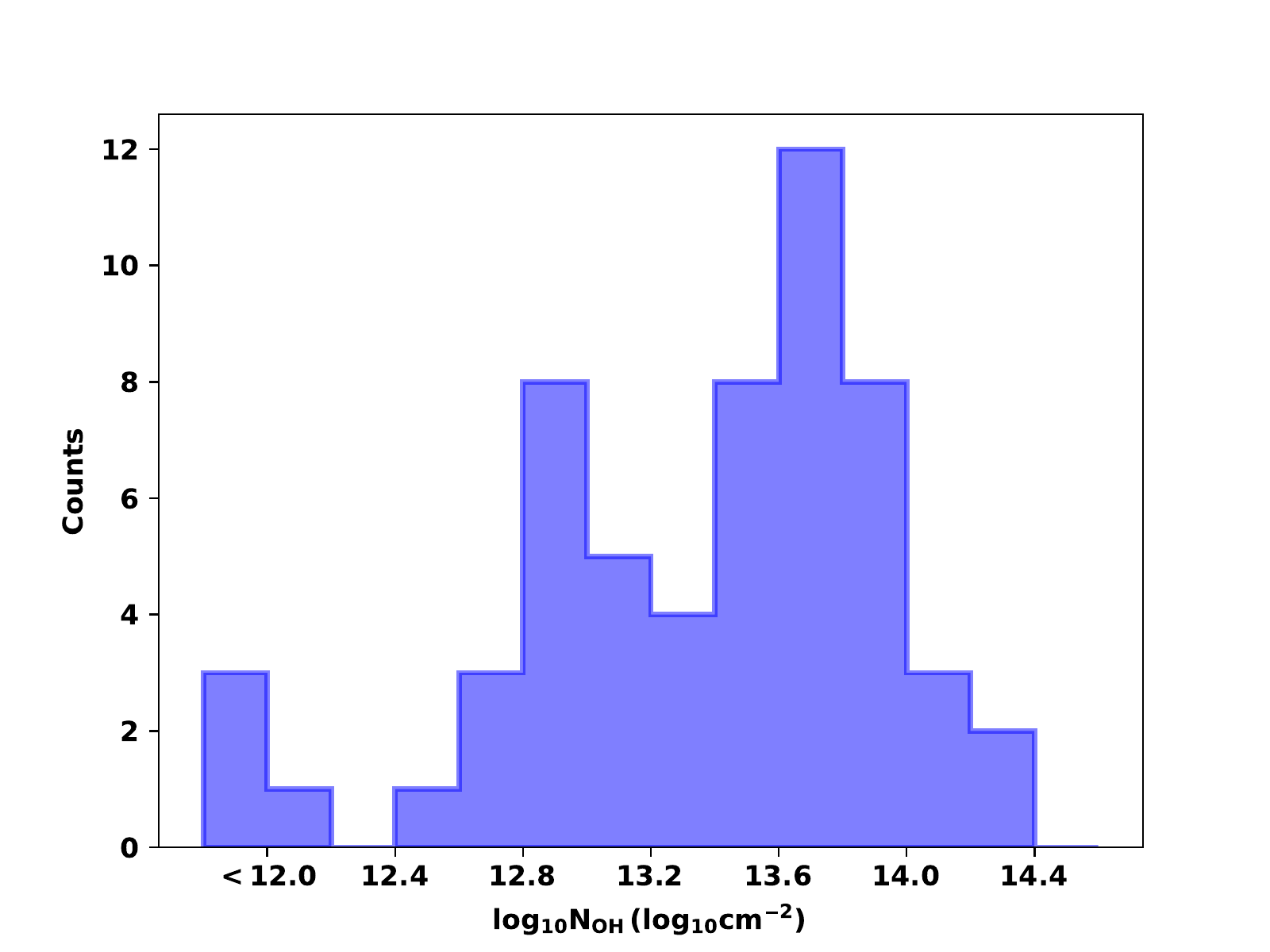}\\
\end{tabular}
\caption{Distribution of FWHM (left) and total OH column density (right) found from the sightlines examined in this paper. Note that the FWHM distribution has bin widths of equal $\log_{10}{\rm km\,s}^{-1}$. The leftmost bin in the column density plot contains all values below $N_{\rm OH}=10^{12}$ cm$^{-2}$. The vertical axes show counts.}
\label{fig:hist_ldv_lNOH}
\end{figure*}

\begin{figure*}
\begin{tabular}{cc}
\includegraphics[trim={0.9cm 0cm 1cm 0cm}, clip=true, width=0.45\linewidth]{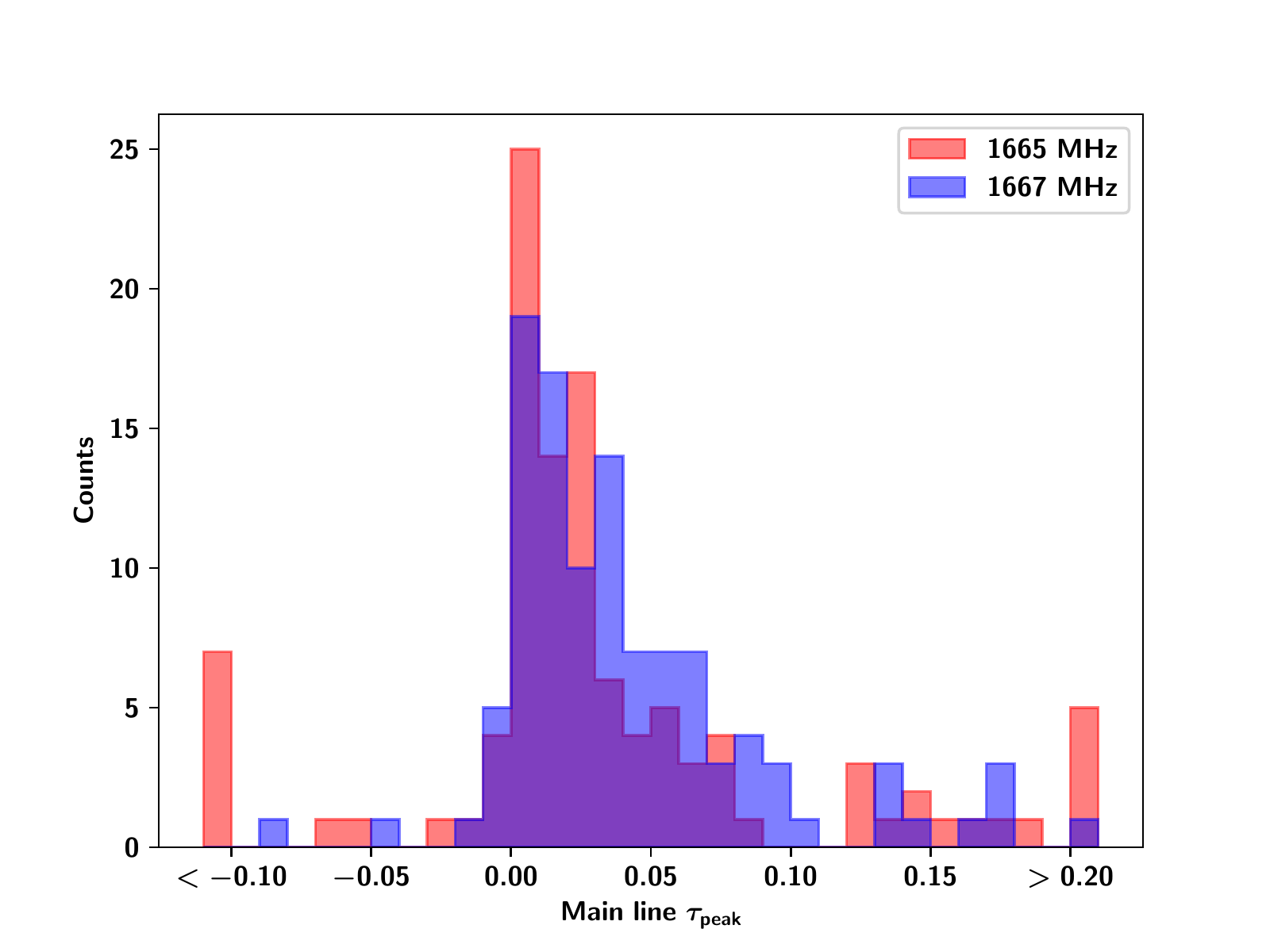}&
\includegraphics[trim={0.9cm 0cm 1cm 0cm}, clip=true, width=0.45\linewidth]{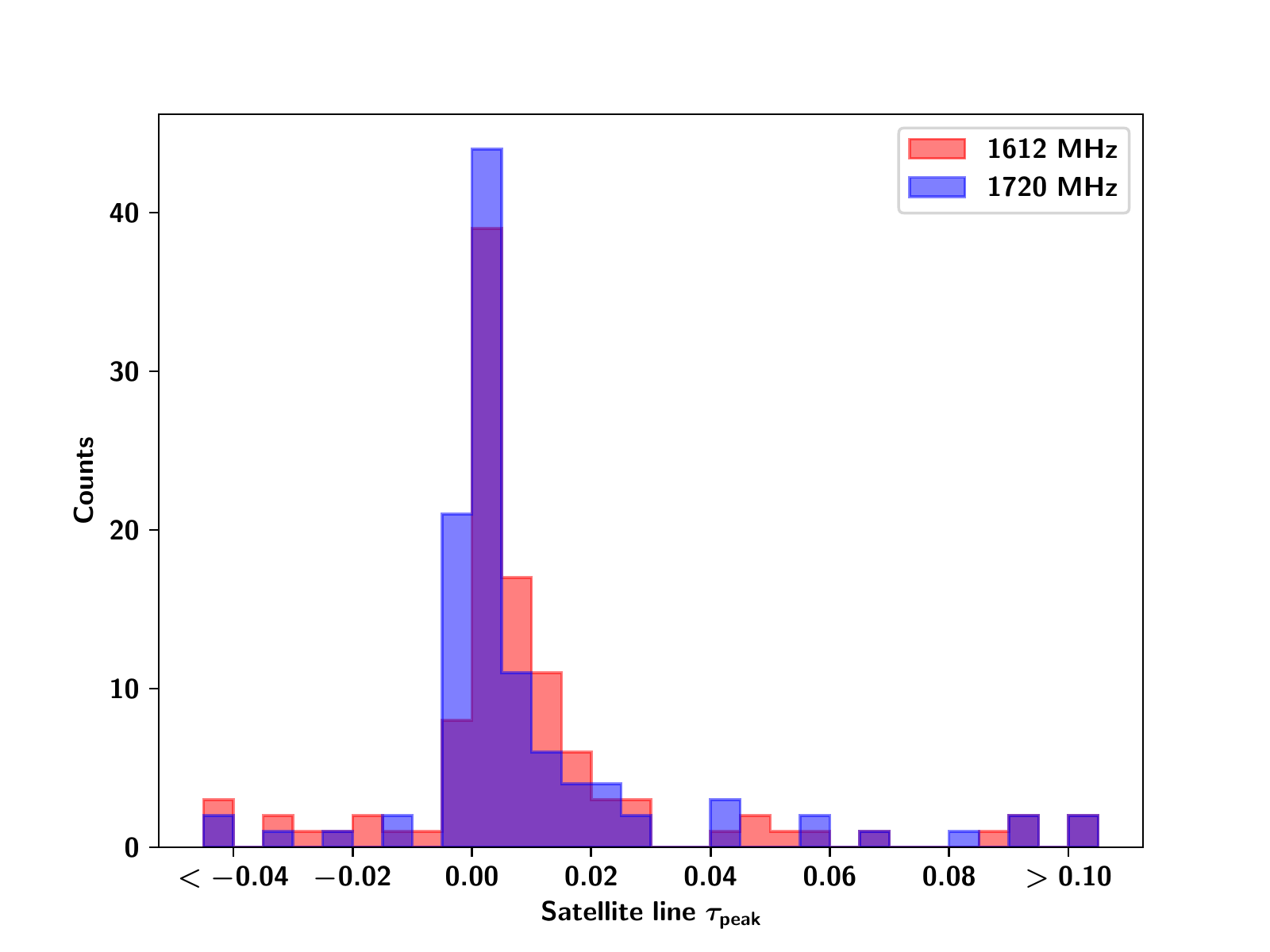}\\
\end{tabular}
\caption{Distribution of main-line (left) and satellite-line (right) peak optical depth found from the sightlines examined in this paper. The vertical axes show counts.}
\label{fig:hist_tau}
\end{figure*}

\begin{figure*}
\begin{tabular}{cc}
\includegraphics[trim={0.9cm 0cm 1cm 0cm}, clip=true, width=0.45\linewidth]{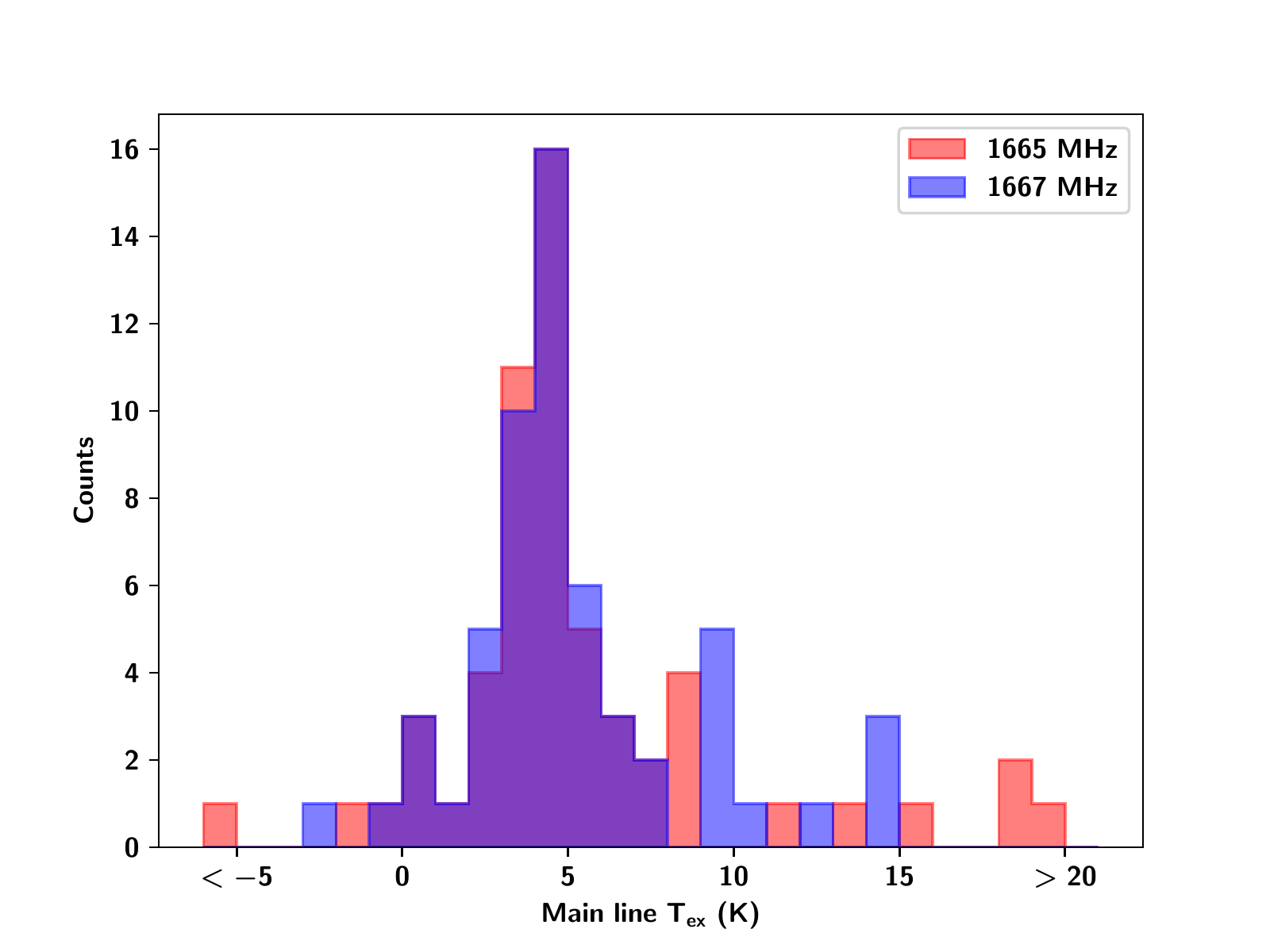}&
\includegraphics[trim={0.9cm 0cm 1cm 0cm}, clip=true, width=0.45\linewidth]{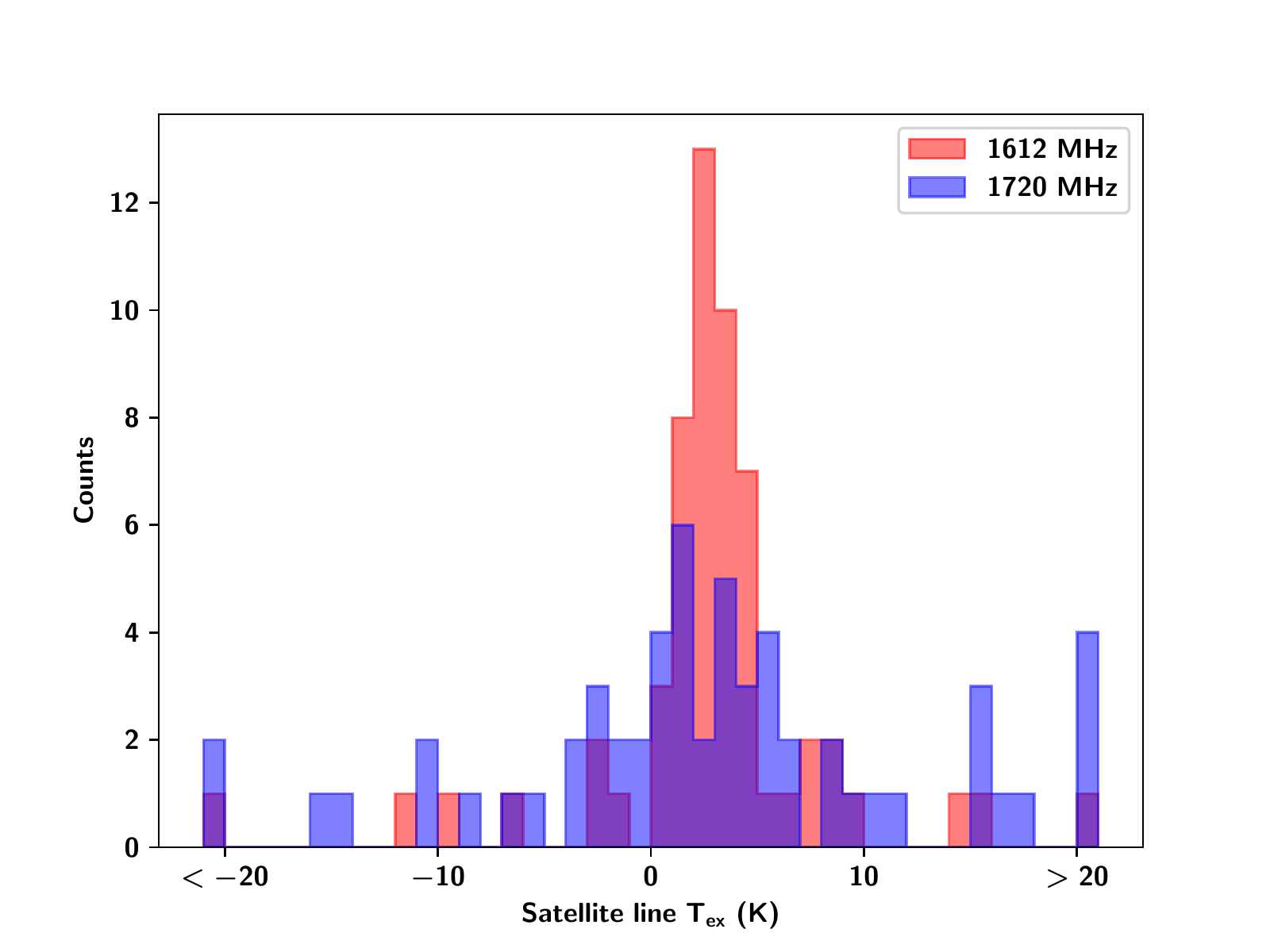}\\
\end{tabular}
\caption{Distribution of main-line (left) and satellite-line (right) excitation temperatures found from the sightlines examined in this paper. The vertical axes show counts.}
\label{fig:hist_Tex}
\end{figure*}

Figs. \ref{fig:hist_ldv_lNOH} to \ref{fig:hist_Tex} illustrate the distributions of the key parameters (FWHM, column density, optical depth and excitation temperature) of our fits. The distribution of FWHM (shown on a log scale in the left panel of Fig. \ref{fig:hist_ldv_lNOH}) suggests a log-normal distribution with a mean of 1.5\,km\,s$^{-1}$ and a 68\% confidence interval bound by $0.7 - 3.4$\,km\,s$^{-1}$. The distribution of total OH column density (shown in the right panel of Fig. \ref{fig:hist_ldv_lNOH}) suggests a typical OH column density of $\approx 10^{13.5}$cm$^{-2}$. The detection limit for OH column density is difficult to estimate with consistency as it depends not only on the noise level and channel width of the optical depth and expected brightness temperature spectra but on the excitation temperatures in the four transitions. However, with estimates of `typical' excitation temperatures $T_{\rm ex}\approx 2-5$\,K (see Fig. \ref{fig:hist_Tex}) we can estimate a detection limit of $N_{\rm OH}\approx 10^{12.5}-10^{13}{\rm cm}^{-2}$, which is consistent with the distribution in Fig. \ref{fig:hist_ldv_lNOH}. This therefore implies that our detections are incomplete and the typical column density of OH could be lower.

Fig. \ref{fig:hist_tau} shows the distribution of peak optical depths across the four OH ground-rotational state transitions. All detections are optically thin ($\tau_{\rm peak}\ll 1$) with approximately log-normal distributions. As would be expected from their relative transition strengths, the satellite lines have the lowest magnitude peak optical depths and the 1667\,MHz line has the highest. The trends in optical depth are examined more closely in the following section. Fig. \ref{fig:hist_Tex} shows the distribution of excitation temperature across the four transitions. The main-line excitation temperatures show a similar, roughly normal distribution centred at approximately 4\,K, while the satellite lines tend towards slightly lower values of about 3\,K. The satellite lines (and particularly the 1720\,MHz transition) are more often inverted (i.e. $T_{\rm ex}<0$) than the main lines. These trends are also examined more closely in the following section.

%% file: 5_Analysis.tex
As briefly outlined in the Observing OH subsection, this work represents an unprecedented analysis of OH in the diffuse ISM due primarily to our Gaussian decomposition algorithm \citep[\amoeba][]{Petzler2021a}. Generally speaking, other works tend to fit features in each transition separately \citep{NguyenQRieu1976,Dickey1981,Colgan1989,Liszt1996,Rugel2018,Li2018}, or solve all spectra simultaneously but channel-by-channel rather than component-by-component \citep[e.g.][]{Crutcher1977,Crutcher1979}. For this reason, we will discuss here the broad trends described by these earlier works, as a more detailed sightline-by-sightline comparison of measurements like optical depth, excitation temperature and column density is not strictly valid given the vast differences in our analyses. 

Figs. \ref{fig:results1} to \ref{fig:results7} (with representative examples shown in Fig. \ref{fig:example1}) show the results of the Gaussian decomposition of our spectra using \textsc{Amoeba}~\citep{Petzler2021a}. For sightlines observed with the ATCA (Figs. \ref{fig:results5} and \ref{fig:results6}) these plots show optical depth vs velocity for the four OH ground-rotational transitions in grey with the individual Gaussian components in red and the total fit in blue. The residuals of the total fits are shown in the fifth panel, and the sixth panel shows the residual of the optical depth sum rule:
\begin{equation*}
\tau_{\rm peak}(1612)+\tau_{\rm peak}(1720)-\tau_{\rm peak}(1665)/5-\tau_{\rm peak}(1667)/9
\end{equation*}
\noindent in black. The optical depth sum rule will hold when $|T_{\rm ex}|\gg h\nu_0/k_{\rm B}=0.08$, so features seen in the sum rule residuals indicate features for which $|T_{\rm ex}|\lesssim0.08\,$K or, more likely, places where maser sidelobes have contaminated the ATCA spectra. When analysing optical depth spectra only, \textsc{Amoeba}~includes an \textit{a priori}~distribution where deviations from the sum rule are expected to have a standard deviation of 0.5. This is intended as a weakly-informative prior, and is therefore much larger than the standard deviation of $\sim 10^{-3}$~that we found from our on-off observations (where \textsc{Amoeba}~does not assume that $|T_{\rm ex}|\gg h\nu_0/k_{\rm B}$). As a consequence of this prior, \textsc{Amoeba}~will tend not to fit signal caused by single-transition maser emission or other anomalies seen in only one transition, but will still be able to fit features that depart moderately from the optical depth sum rule. With this prior the fitted components from our ATCA observations yield a distribution of sum rule residuals with a standard deviation of 0.05. Any significant departures from the sum rule evident in our ATCA data are described in the Appendix.

The sightlines with on-off observations (Figs. \ref{fig:results1} to \ref{fig:results4} and \ref{fig:results7}) are generally well-fit, as evidenced by the lack of significant features in the residuals. Some minor exceptions can be seen in the observations towards SRC44 (Fig. \ref{fig:results7}) and 3C417 (Fig. \ref{fig:results1}) with residuals seen in the expected brightness temperature at 1720\,MHz, and PKS0528+134 (Fig. \ref{fig:results7}) and 4C+14.18 (Fig. \ref{fig:results2}) with residuals seen at 1612\,MHz. We note that all of these features are seen in the residuals of the expected brightness temperatures for the satellite-line transitions, which across all observations tend to have the lowest signal-to-noise ratios. \textsc{Amoeba}~assumes that the OH gas seen in the on-source and off-source positions have the same column densities in each of the four ground-rotational state levels as well as the same velocity dispersion. Therefore we interpret this lack of significant features in the residuals of the fits as validation of the underlying assumptions of \textsc{Amoeba}. We do not find evidence that OH has a significant `multi-phase' structure that would result in it having significantly different excitation in these positions, contrary to the finding of \citet{Liszt1996}.

We note the detection of four satellite-line `flips': two that have already been reported in \citet{Petzler2020} towards G340.79-1.02 (at $-29.22$ and $-26.44$\,km\,s$^{-1}$, see Fig. \ref{fig:results5}) and G353.41-0.30 (at $-19.10$ and $-14.40$\,km\,s$^{-1}$, see Fig. \ref{fig:results6}), and two that are new detections towards 4C+19.19 (at 1.12 and 2.70\,km\,s$^{-1}$, see Fig. \ref{fig:results3}) and 4C+14.18 (at 16.49 and 18.40\,km\,s$^{-1}$, see Fig. \ref{fig:results2}). \citet{Petzler2020} suggested that this profile type -- where the satellite lines show paired emission and absorption that then flip orientation across a closely blended feature -- generally indicates molecular gas on either side of a shock front. The flips towards G340.79-1.02 and G353.41-0.30 show the more common velocity orientation of the flip, with the 1720\,MHz stimulated emission seen at more negative velocities. These two sightlines are also associated on the sky and in velocity with known H\textsc{ii}~regions (G340.780-01.022 at $-25$\,km\,s$^{-1}$~ \citep{Caswell1987} and G353.408-00.381 at $-15.7$\,km\,s$^{-1}$~\citep{Quireza2006}, respectively), which \citet{Petzler2020} argue implies that an associated shock front is expanding from those H\textsc{ii}~regions towards the observer. In their picture, the 1720\,MHz-emitting gas is on the inside of the shock and collides with the 1612\,MHz-emitting gas in the surrounding molecular cloud: the enhanced radiation from the H\textsc{ii}~region and the surrounding dust inverts the 1612\,MHz line in the surrounding molecular cloud while the heating and compression from the shock switches off the 1612\,MHz emission and inverts the 1720\,MHz line.

On the other hand, the two new flips towards 4C+19.19 ($l^{\circ}=190.13$, $b^{\circ}=-1.64$) and 4C+14.18 ($l^{\circ}=196.98$, $b^{\circ}=1.10$) have the opposite velocity orientation and no clear H\textsc{ii}~association. \citet{Petzler2020} reported three such flips, all within the Taurus molecular cloud complex (and near to these two new detections though not in the same complex), towards G172.80-13.24 \citep[at 5.3 and 6.8\,km\,s$^{-1}$][]{Xu2016}, G173.40-13.26 \citep[at 5 and 8\,km\,s$^{-1}$][]{Ebisawa2019} and G175.83-9.36 (4C+27.14 from project a2600 at 7.1 and 7.8\,km\,s$^{-1}$, GNOMES collaboration). 
This third flip was observed twice in the data set examined in this paper, once in the a2600 project and once in a3301. The flip was visually apparent in the a2600 data, but this work fit the newer, higher signal-to-noise data from a3301 (which was not yet available at the time \citet{Petzler2020} was published) and a flip was not found. 
\citet{Petzler2020} propose that these flips, and by extension these two new detections towards 4C+19.19 and 4C+14.18, are not indicative of an enhanced radiation field or a shock, but may represent some other type of bulk motions such as the large shell proposed by \citet{Bialy2021}.

\subsection{Optical depth and excitation temperature relationships}
\begin{figure*}
\begin{tabular}{cc}
\includegraphics[width=0.495\linewidth]{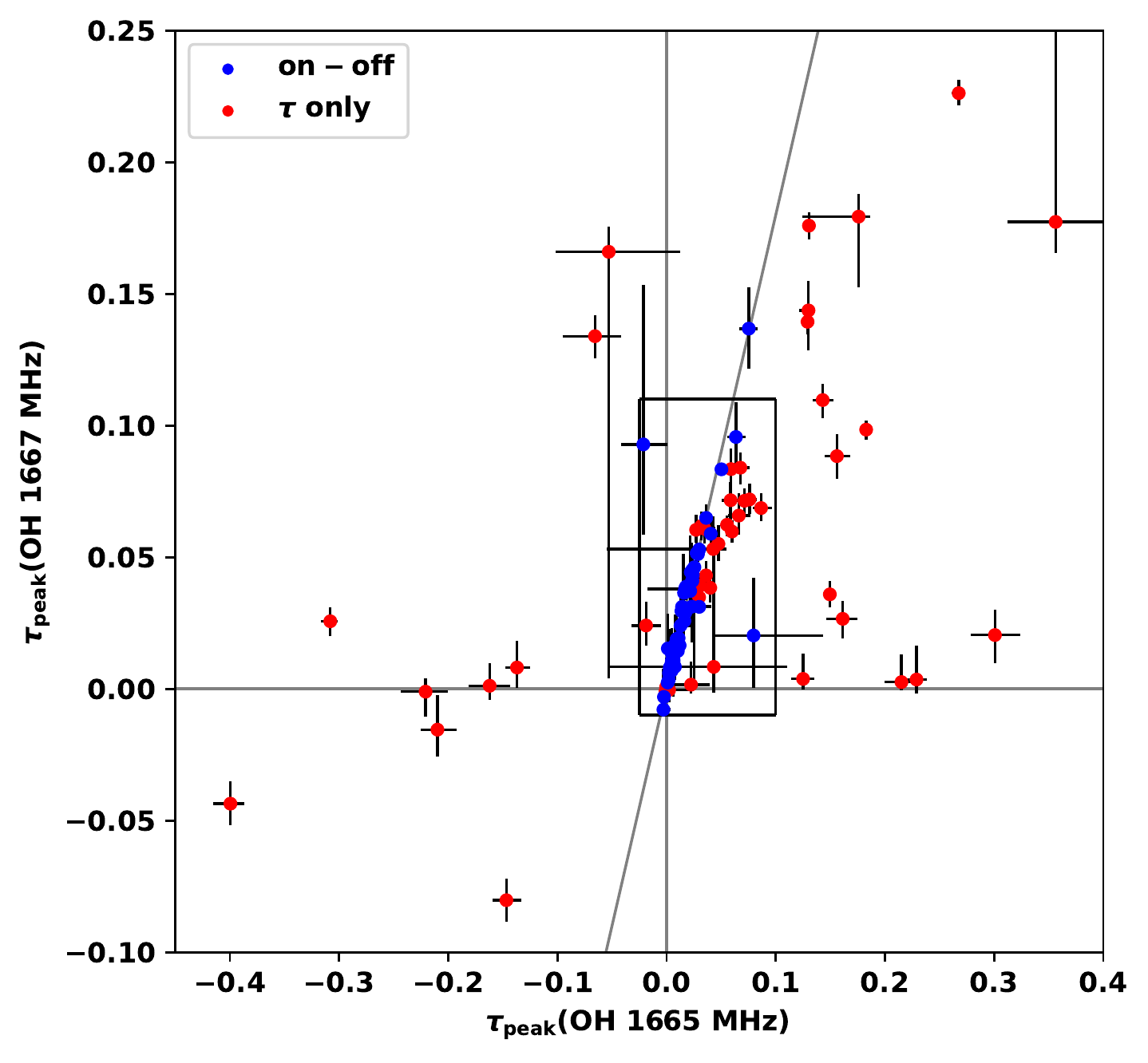}&
\includegraphics[width=0.33\linewidth]{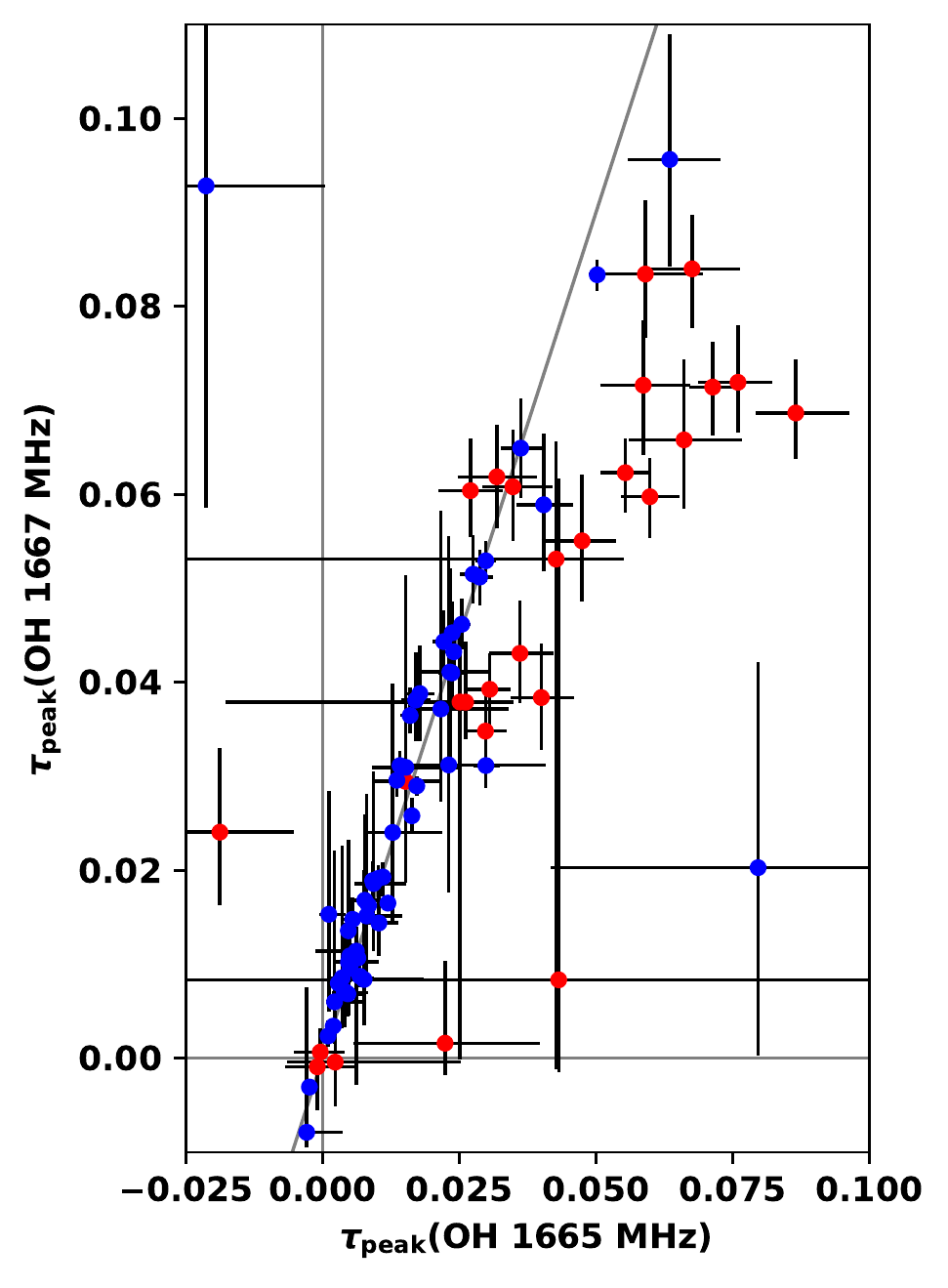}\\
\end{tabular}
\caption{Distribution of peak optical depths in the `main' lines at 1665 and 1667\,MHz. Features identified from our `on-off' data (from Arecibo) are shown in blue while our `optical depth only' data (from the ATCA) are shown in red. 
The rectangle in the left plot indicates the area enlarged in the plot on the right. The grey reference lines indicate the axes and where $\tau_{\rm peak}(1667)=\frac{9}{5}\tau_{\rm peak}(1665)$, which is the expected relationship between $\tau_{\rm peak}(1667)$~and $\tau_{\rm peak}(1665)$~when in local thermodynamic equilibrium (LTE), though adherence to this ratio is not sufficient evidence to conclude LTE. The error bars indicate the 68\% credibility intervals.}
\label{fig:tau_main}
\end{figure*}

\begin{figure}
\includegraphics[trim={0cm 0cm 0cm 0cm}, clip=true, width=\linewidth]{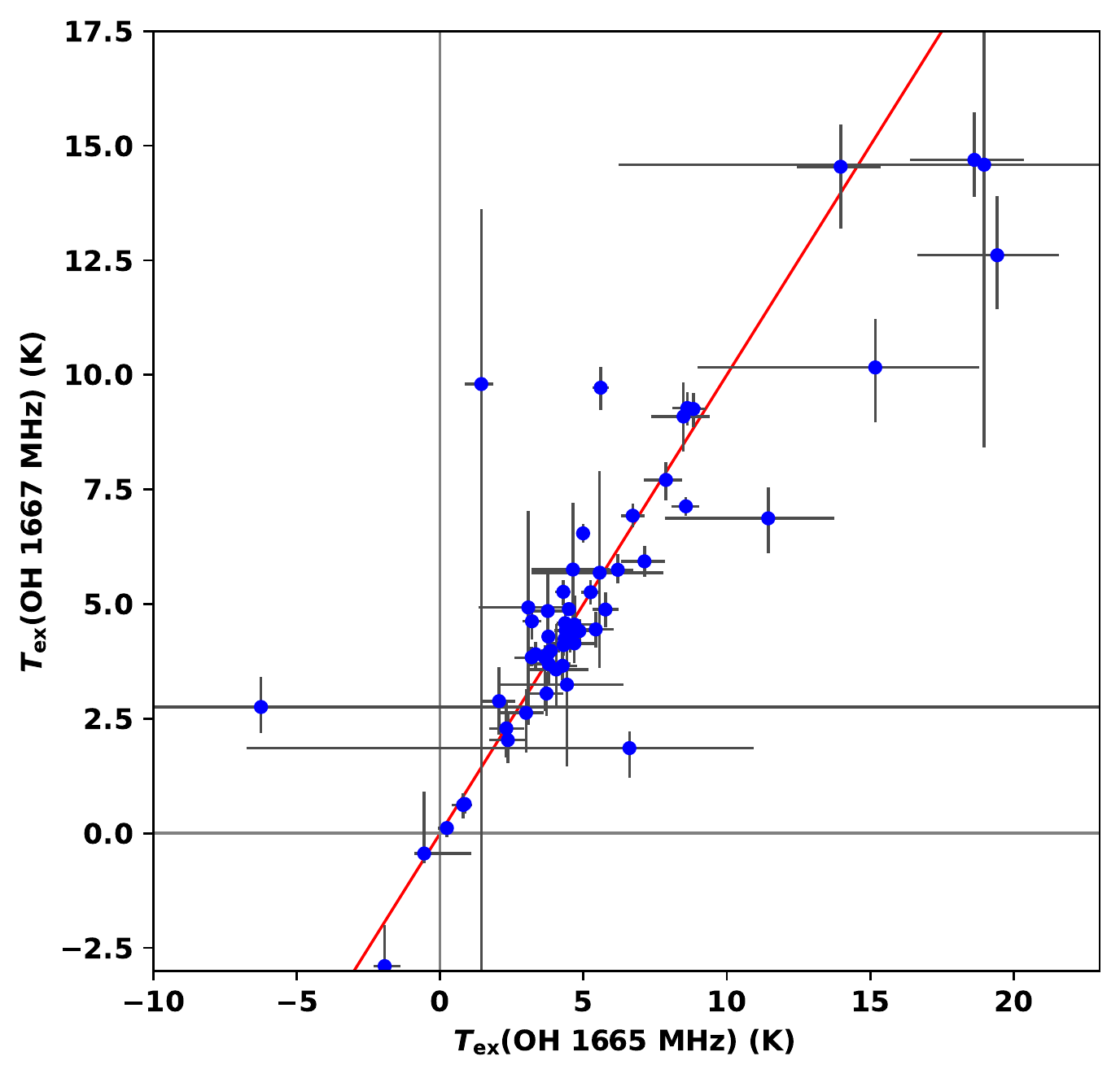}
\caption{Relationship between the OH `main-line' excitation temperatures found from the sightlines examined in this paper. The red reference line indicates where the two excitation temperatures are equal, and the error bars indicate the 68\% credibility intervals.}
\label{fig:Tex_main}
\end{figure}

\begin{figure*}
\begin{tabular}{cc}
\includegraphics[width=0.44\linewidth]{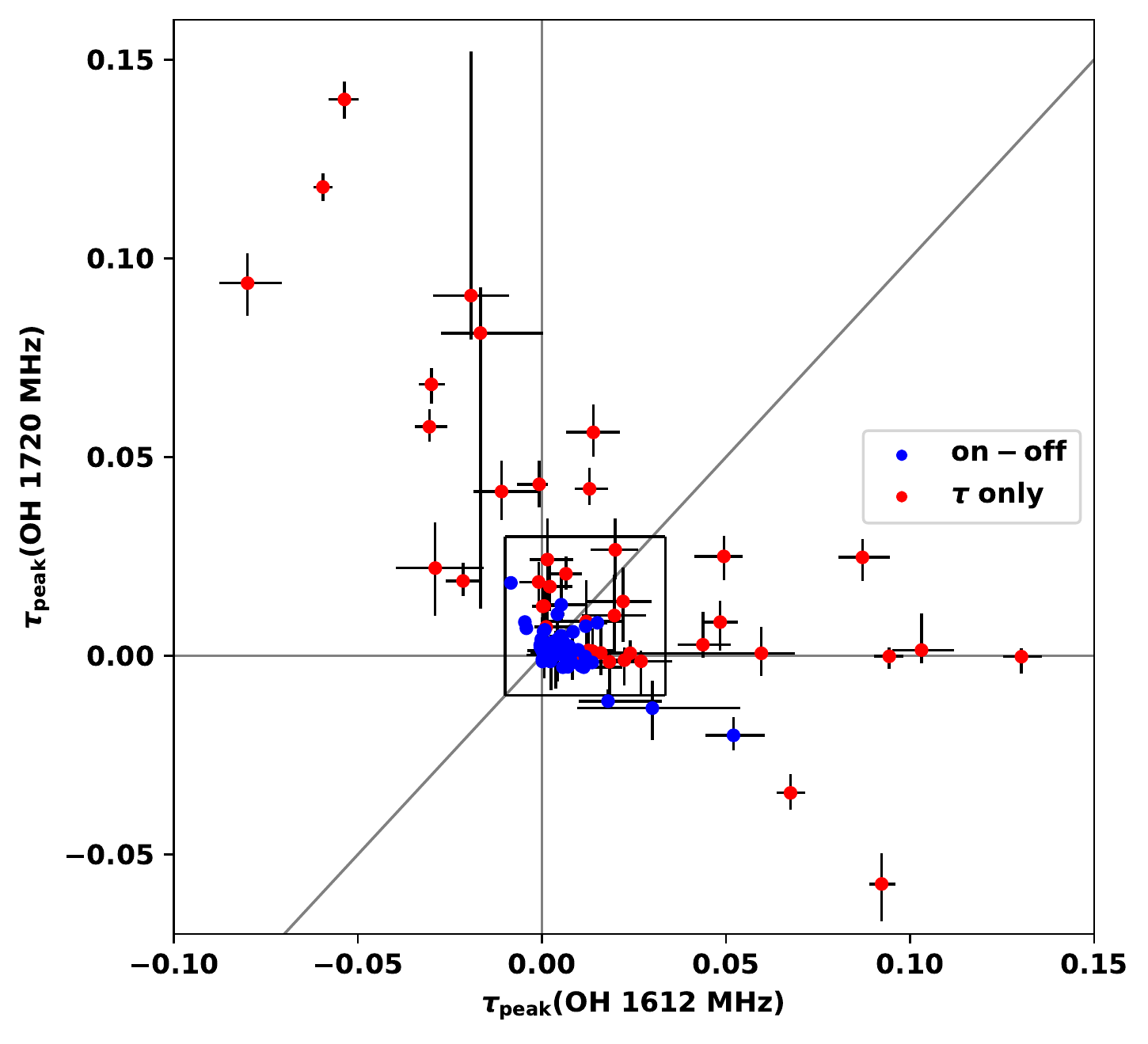}&
\includegraphics[width=0.4\linewidth]{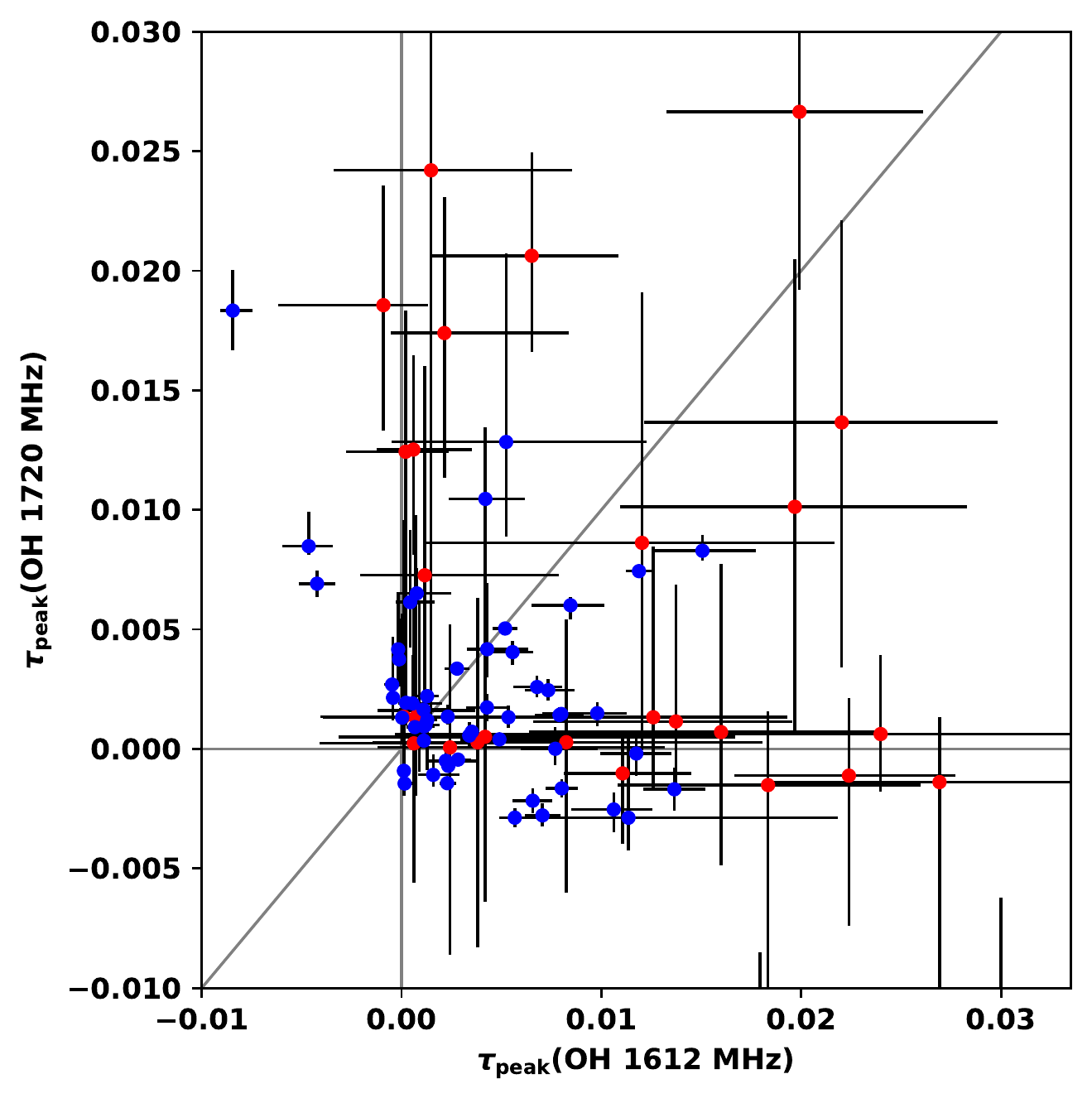}\\
\end{tabular}
\caption{Distribution of peak optical depths in the `satellite' lines at 1612 and 1720\,MHz. The rectangle in the left plot indicates the area enlarged in the plot on the right. The grey reference lines indicate the axes and where $\tau_{\rm peak}(1612)=\tau_{\rm peak}(1720)$, which is the expected relationship between $\tau_{\rm peak}(1612)$~and $\tau_{\rm peak}(1720)$~when in local thermodynamic equilibrium. The error bars indicate the 68\% credibility intervals.}
\label{fig:tau_sat}
\end{figure*}

\begin{figure}
\includegraphics[width=\linewidth]{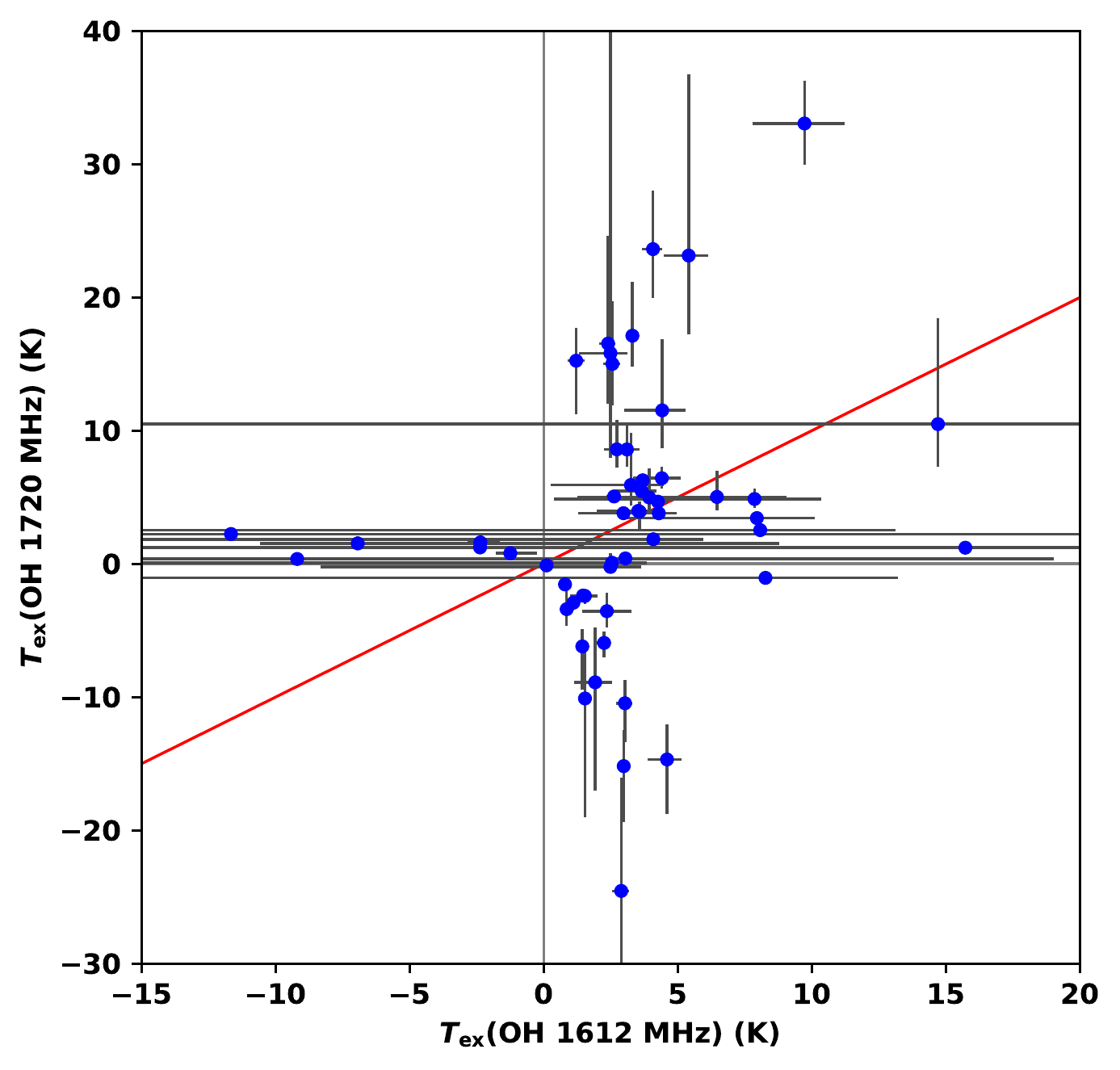}
\caption{Relationship between the OH `satellite-line' excitation temperatures found from the sightlines examined in this paper. The red reference line indicates where the two excitation temperatures are equal, and the error bars indicate the 68\% credibility intervals.}
\label{fig:Tex_sat}
\end{figure}

The relationships between main-line and satellite-line peak optical depths across the four OH ground-rotational transitions are shown in Figs. \ref{fig:tau_main} and \ref{fig:tau_sat} respectively. Similarly, the relationships between main-line and satellite-line excitation temperatures are shown in Figs. \ref{fig:Tex_main} and \ref{fig:Tex_sat}. Overall we find that while the excitation temperatures of the main lines are similar (median $|\Delta T_{\rm ex}({\rm main})|=0.6\,$K, 84\% show $|\Delta T_{\rm ex}({\rm main})|<2\,$K), those of the satellite lines show that the gas is generally not in LTE. In this subsection we will focus first on trends seen in the main lines, then on those seen in the satellite lines before commenting on the implications of both.

% Main lines

Fig. \ref{fig:tau_main} shows a significant difference in main-line peak optical depth relationship between our on-off data (shown in blue) and our optical-depth only data (shown in red). The features identified in our on-off data tend to have a main-line optical depth ratio of $5:9$~which is the expected ratio in the case of local thermodynamic equilibrium (LTE). LTE would also imply that the excitation temperatures of the main lines are equal, and they do tend towards similar values when we compare the main-line excitation temperatures in Fig. \ref{fig:Tex_main} (recall that we were unable to calculate excitation temperatures from our optical depth only data). The main-line excitation temperatures had a median difference of $|\Delta T_{\rm ex}({\rm main})|=0.6\,$K, and 84\% show $|\Delta T_{\rm ex}({\rm main})|<2\,$K.

On the other hand, features identified in our optical depth only data from the ATCA have main-line peak optical depths that show little discernible pattern aside from a slight tendency (seen in the right panel of Fig. \ref{fig:tau_main}) to have higher 1665\,MHz peak optical depth than that expected in LTE. Measurements of main-line optical depths from \citet{Li2018} show a pattern that is not inconsistent with this  -- there is a slight skew towards higher peak optical depth in the 1665\,MHz transition -- but the trend is much less pronounced. We note that \citet{Li2018} fit the main lines separately but did utilise on-off measurements. This trend, along with others noted in this subsection, are likely only apparent due to the large number of sightlines analysed in this work as well as our simultaneous fitting method, which is inherently more sensitive to lower optical depths.

Our data set from the ATCA differs from our on-off spectra both in the method by which features were identified (as described in the Method practicalities and limitations section), but also in the location of the lines-of-sight: our ATCA sightlines are in the Plane and towards the Galactic centre. It is therefore unclear which of these may be responsible for the differences seen in the main-line peak optical depth relationships. If we assume the latter case then we may conclude that deviations from the expected LTE ratio -- often referred to as `main-line anomalies' -- are more common in the Plane and towards the Galactic centre. Such main-line anomalies have been well-documented \citep[e.g.][and many others]{Crutcher1977,Crutcher1979} and indicate (as outlined in the Introduction) either a radiation field that differs significantly from a Planck distribution (such as from warm dust) or collisional excitations from particles that differ significantly from a Maxwellian distribution (such as from particle flows). Elaborating on the previous brief introduction, these conditions provide a significant difference in the energy budget between excitations into the upper and lower halves of the lambda doublets of the higher rotational states of OH. Then as these excited molecules cascade back into the ground-rotational state they remain on their respective side of the rotational ladder, but also remain on either the top or the bottom of the lambda doublet due to selection rules. Therefore any imbalance in the number of excitations into, say, the upper half of the lambda doublet in the infrared transitions into higher rotational states will result in a similar imbalance in the upper half of the lambda doublet in the ground-rotational state. This imbalance could be sufficient to invert one or both of the main-line transitions, but it could also result in the observed divergence from the LTE ratio. For example, in the presence of an infrared radiation field with sufficiently steep (negative) spectral profile, there will be fewer photons available at high energies compared to low energies. Therefore, transitions into the lower half of the lambda doublets of excited rotational states will be more common than transitions into the upper halves. As these OH molecules cascade back into the ground-rotational state they will tend to over-populate the lower levels of the ground-rotational state, thus sub-thermally exciting all four ground-rotational state transitions, and more particularly the 1665\,MHz transition (and the 1720\,MHz transition, though we discuss this later) as its lower level has fewer sub-levels \citep{Elitzur1976etal}. This sub-thermal excitation could then lead to the systematically higher peak optical depths in the 1665\,MHz transition as seen in the right panel of Fig. \ref{fig:tau_main}. We also note that we have identified 16 features for which either one or both of the main lines have a negative optical depth, implying that those lines are inverted, and all but 3 of these are from our sightlines observed with the ATCA and are therefore located in the Plane and towards the Galactic centre. From these it appears that inversions of the 1665\,MHz line are more common than those of the 1667\,MHz line, and in cases where the 1667\,MHz line is inverted it is more common for the 1665\,MHz line to also be inverted, though the small sample size is insufficient to draw significant conclusions from these trends.

% Satellite lines

Turning our attention now to the satellite lines, we see no tendency towards the expected LTE ratio of $1:1$, and the satellite-line excitation temperatures (see the right panel of Fig. \ref{fig:hist_Tex}, noting that these represent on-off observations only) are clearly unequal. This is consistent with the findings of nearly all previous works that have measured satellite-line optical depths. 

Looking closer at the relationship between satellite-line optical depths (see Fig. \ref{fig:tau_sat}), we note that while it is most common for both to be positive (63/109 points are found in the first quadrant of Fig. \ref{fig:tau_sat}), it is more common for the 1720\,MHz transition to have a negative optical depth (27/109) than it is for the 1612\,MHz line (19/109). Negative optical depths imply a population inversion. This is consistent with works such as \citet{Turner1982} and \citet{Dawson2022} who note that inversions of the 1720\,MHz line are ubiquitous in the ISM. However, we note that in the case of our sightlines observed with the ATCA (in the Plane and towards the Galactic centre) this trend disappears and it is marginally more common for the 1612\,MHz line to have a negative optical depth (12/51) compared to the 1720\,MHz line (9/51). We also note that in these cases the 1612\,MHz line tends towards more negative optical depths than the 1720\,MHz, a possible indication that it is more strongly inverted (though not conclusively as this could be a column density effect).

Continuing from the previous brief introduction, satellite line inversions are caused by an imbalance in cascades into the ground-rotational state from the first and second excited rotational states \citep{Elitzur1976}. Collisions can selectively excite into just the first excited rotational level but not the second, which then leads to an enhancement of cascade pathways into the $F=2$~levels of the ground-rotational state and inversion of the 1720\,MHz line \citep{Elitzur1976}. On the other hand, an enhanced radiation field can excite OH into both the first and second excited rotational states, which will cascade into the $F=1$~and $F=2$~levels of the ground-rotational state equally. In this case, since the $F=1$~levels have a lower degeneracy ($g=3$) than the $F=2$~levels ($g=5$), this mechanism can invert the 1612\,MHz line \citep{Elitzur1976etal}. However, at the low column densities identified in this work ($N_{\rm OH}\lesssim 10^{15}{\rm cm}^{-2}$) this mechanism is generally disrupted because the cascade from the second-excited rotational level becomes optically thin. This disruption can then allow the 1720\,MHz line to invert but only weakly \citep{Elitzur1992}. At these low column densities the 1612\,MHz line is able to weakly invert ($|\tau_{\rm peak}|\lesssim 0.02$) in gas with low number density \citep[$n_{\rm OH}\lesssim 10^{3}$cm$^{-3}$]{Petzler2020}. Therefore while we may speculate that the gas hosting the 1612\,MHz inversion has a low number density, the cause of the 1720\,MHz inversions is less clear.

Overall, as noted by nearly all works who have measured all four OH ground-rotational state transitions, it is much more likely for the satellite lines to be inverted than the main lines. However, this trend becomes much less significant when we consider only our sightlines observed with the ATCA. As previously mentioned, these observations differ from our Arecibo observations in two key ways: the locations of the sightlines were in the Plane towards the Galactic centre, but were also analysed differently as they only consisted of optical depth spectra. In addition, when we sub-divide our data set in this way we become increasingly limited in our conclusions due to small sample size effects. We therefore cautiously summarise that while non-LTE excitations of OH (as primarily evidenced by the behaviour of the satellite line peak optical depths) are clearly the norm in the diffuse ISM, these trends appear more pronounced along sightlines toward the Galactic centre. Further, the precise excitation mechanisms that dominate this non-LTE behaviour also appear to be different towards the Galactic centre.

\subsection{Comparison of OH and H\textsc{i}~CNM component parameters}
A selection of the sightlines in this work with OH detections had previously been observed in H\textsc{i}~absorption as part of the Millennium survey \citep{Heiles2003}. The CNM components from these sightlines (a total of 327 components) were identified by \citet{Nguyen2019} as part of the GNOMES collaboration, and these are compared to our OH fits in Figs. \ref{fig:CNM1} to \ref{fig:CNM4} in the Appendix. In this work we wish to draw comparisons between the properties of OH as obtained from our fits and any associated CNM gas. We therefore attempted to match our OH features (in velocity) to the CNM components identified by \citet{Nguyen2019} for each sightline. This was done via a by-eye comparison of the OH feature centroid velocities to those of the CNM components identified by \citet{Nguyen2019}. 

In all cases the FWHM of OH detections in this work overlapped in velocity with the FWHM of H\textsc{i}~absorption features identified by \citet{Nguyen2019}. In many cases (i.e. 4C+17.23 in Fig. \ref{fig:CNM2}, 4C+28.11 in Fig. \ref{fig:CNM3}) there is a clear association between a given OH feature and an individual CNM component (i.e. the two components line up in velocity with no other nearby features). However, in other cases (e.g. 3C092 in Fig. \ref{fig:CNM1}, 4C+11.15 in Fig. \ref{fig:CNM2}) the association with an individual CNM component is more ambiguous. In addition, the process by which \citet{Nguyen2019} fit the CNM components was restricted by consideration of the complimentary H\textsc{i}~emission data and physical constraints (i.e. spin temperature) on the resulting components \citep[for details see][]{Heiles2003}. Therefore at times the H\textsc{i}~CNM fits may be too conservative for a feature-by-feature comparison with OH. For example the very high signal-to-noise of the H\textsc{i}~data towards 3C131 (see Fig. \ref{fig:CNM1}) may justify a more complex fit to the feature at 5\,km\,s$^{-1}$~which may yield better matches to the complex OH fit from this work. Given the CNM fits as they are, we have several instances where we must choose between one or more potential CNM components for a given OH feature (e.g. 4C+17.41 in Fig. \ref{fig:CNM2}), in which case we used our judgement to match either the closest component in velocity, or the more narrow CNM component. Additionally, there were several instances where we matched one or more OH component to the same CNM component (e.g. 4C+04.22 in Fig. \ref{fig:CNM1}). This process resulted in a total of 43 matches between 43 OH components and 26 H\textsc{i}~CNM components. These matches are summarised in Table \ref{tab:match}.

\begin{figure*}
    \centering
    \begin{tabular}{cc}
    \includegraphics[width=0.45\linewidth]{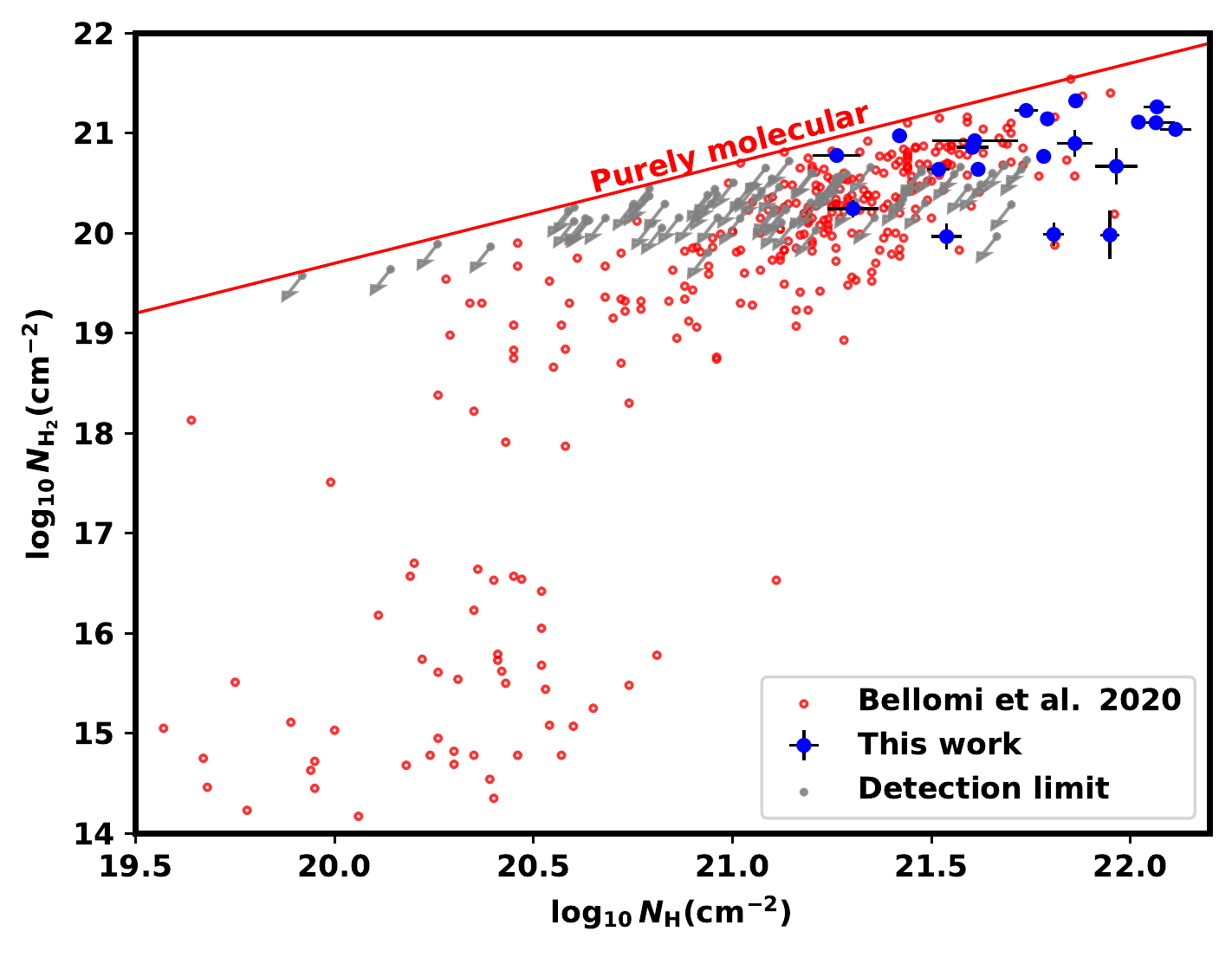}&
    \includegraphics[width=0.45\linewidth]{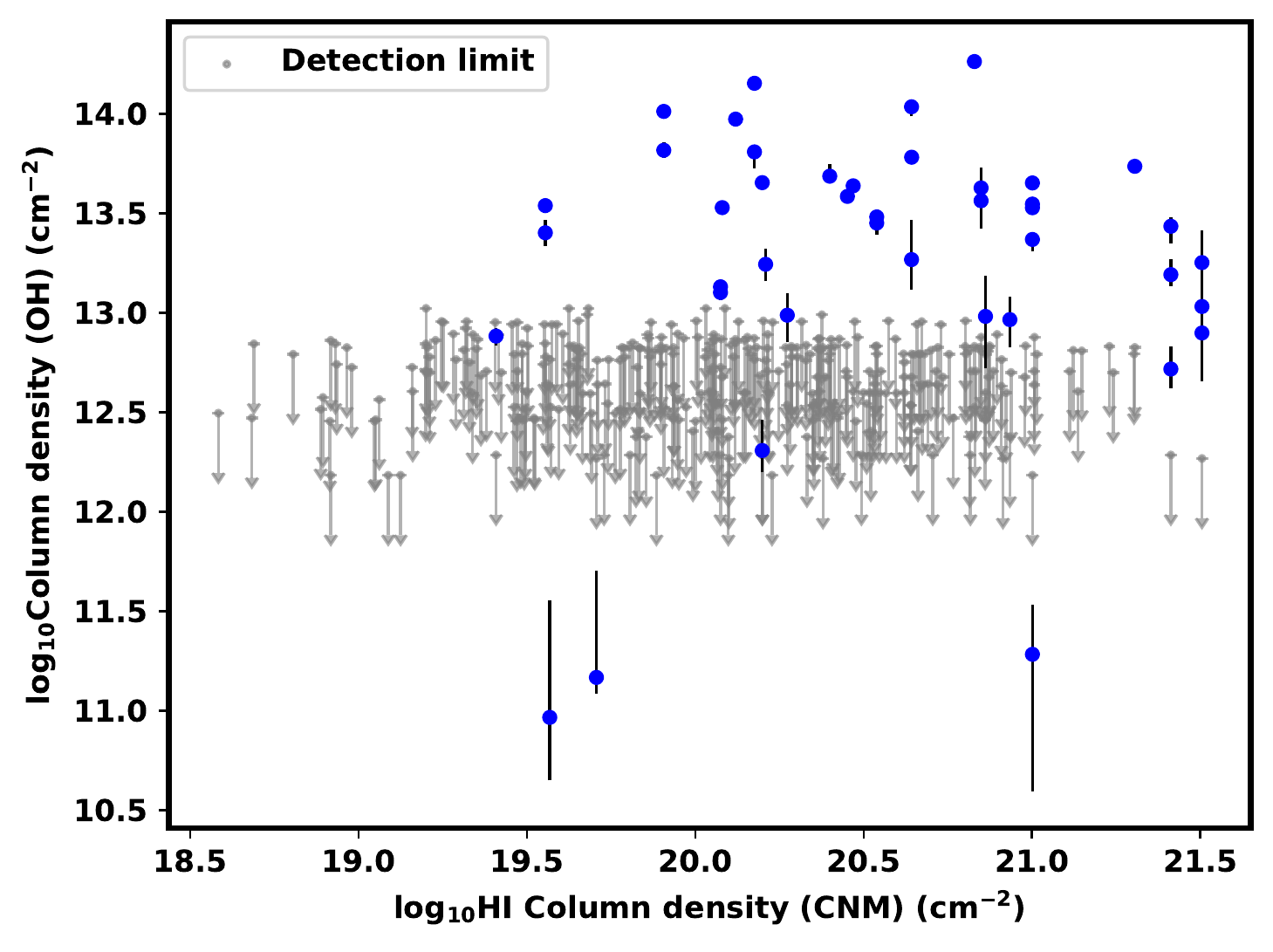}\\
    \end{tabular}
    \caption{Left: relationship between line-of-sight integrated H$_2$~column density (found from $N_{\rm H_2}=10^7N_{\rm OH}$) and total H column density (found from $N_{\rm H}=N_{\rm H\textsc{i}}({\rm CNM})+N_{\rm H\textsc{i}}({\rm WNM})+2N_{\rm H_2}$) for each sightline with both OH and H\textsc{i} observations from this work (blue), and from \citet[][red]{Bellomi2020}. Right: relationship between OH column density and H\textsc{i} CNM column density for matching OH and H\textsc{i}~features. The detection limits in both plots are estimated from the $2\times$\,rms noise in our optical depth data, the median excitation temperatures determined from our sightlines with detections and a feature width of 0.3\,km\,s$^{-1}$~(i.e. three times our typical channel width): grey arrows indicate the upper limit, under which detections may be missing.}
    \label{fig:logNOH_v_NHI}
\end{figure*}

UV studies of H$_2$~indicate that the molecular gas fraction $f_{\rm H_2}=2N_{\rm H_2}/(N_{\rm H\textsc{i}}+2N_{\rm H_2})$~sharply increases at a total gas column density of $N_{\rm H}=N_{\rm H\textsc{i}}+2N_{\rm H_2}\approx 10^{21}$\,cm$^{-2}$~\citep{Savage1977,Rachford2002,Gillmon2006}, at which point the total H\textsc{i} column density is expected to saturate in the Milky Way galaxy \citep{Reach1994,Meyerdierks1996,Douglas2007,Barriault2010,Lee2012,Liszt2014} and in other galaxies \citep{Wong2002,Blitz2006,Leroy2008,Wong2009}. At this H\textsc{i}~column density (in solar metallicity environments) there is sufficient dust shielding for H$_2$~to persist. Beyond this limit any additional H\textsc{i} will be converted to H$_2$.

\citet{Bellomi2020} illustrated this transition in what they term a `kingfisher' diagram, shown in the left panel of Fig. \ref{fig:logNOH_v_NHI}. Their data was a selection of those included in \citet{Gudennavar2012}, and included direct measurements of H$_2$~from UV absorption lines, thus was able to probe much lower molecular column densities than this work. In Fig. \ref{fig:logNOH_v_NHI} we show the data from \citet{Bellomi2020} in red which illustrate the atomic-to-molecular transition evident from N$_{\rm H}\approx 10^{20}$~to $10^{21}$cm$^{-2}$. Also included in this plot (in blue) are the results from this work. The total molecular column density per sightline from this work was found from the sum of the column densities of individual OH components along each sightline, then converted to $N_{\rm H_2}$~using the relative abundance of OH to H$_2$~of 10$^{-7}$~\citep[][and references therein]{Nguyen2018}, against total hydrogen column density found from the sum of the column densities of all WNM and CNM components (taken from \citealt{Nguyen2018}) along each sightline plus twice the computed H$_2$~column density. The detection limit of our data shown in Fig. \ref{fig:logNOH_v_NHI} was estimated from the $2\times$\,rms noise in our optical depth data, the median excitation temperatures determined from our sightlines with detections and a feature width of 0.3\,km\,s$^{-1}$~(i.e. three times our typical channel width).  

We can see from Fig. \ref{fig:logNOH_v_NHI} that our detections represent lines of sight with total $N_{\rm H_2}$~much higher than that at which the atomic-to-molecular transition is seen to occur. Our detections also fall in a region of the kingfisher plot where there is not a strong relationship between the molecular and total column density. Therefore it is not surprising that we do not see a relationship between $N_{\rm H_2}$~and $N_{\rm H}$~in our data. We also do not see a relationship when we compare individual matched features' H\textsc{i} and OH column densities, as illustrated on the right panel in Fig. \ref{fig:logNOH_v_NHI}. 

\begin{figure*}
    \centering
    \begin{tabular}{cc}
    \includegraphics[trim={0cm 0cm 1cm 0cm}, clip=true, width=0.45\linewidth]{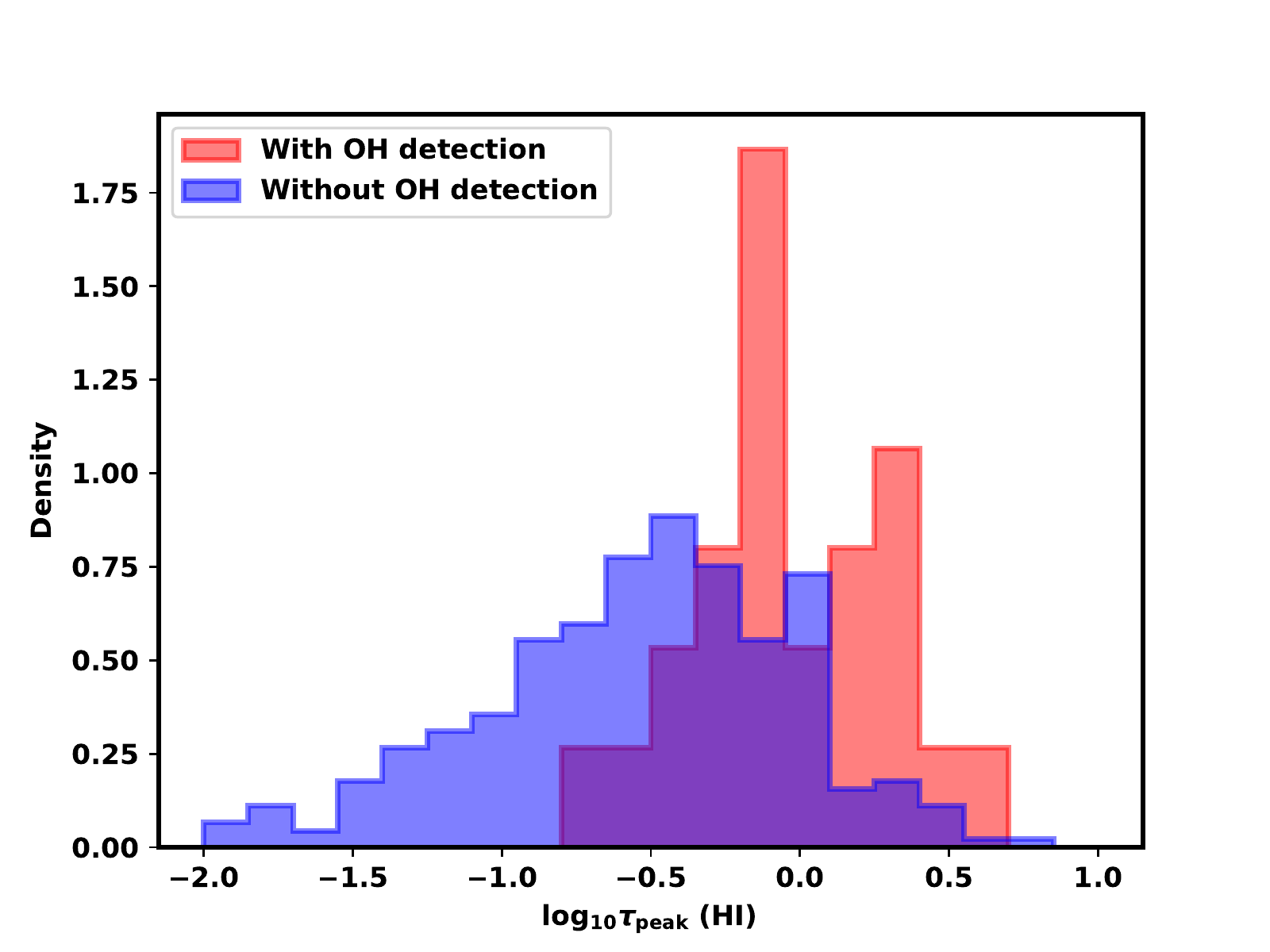}&
    \includegraphics[trim={0cm 0cm 1cm 0cm}, clip=true, width=0.45\linewidth]{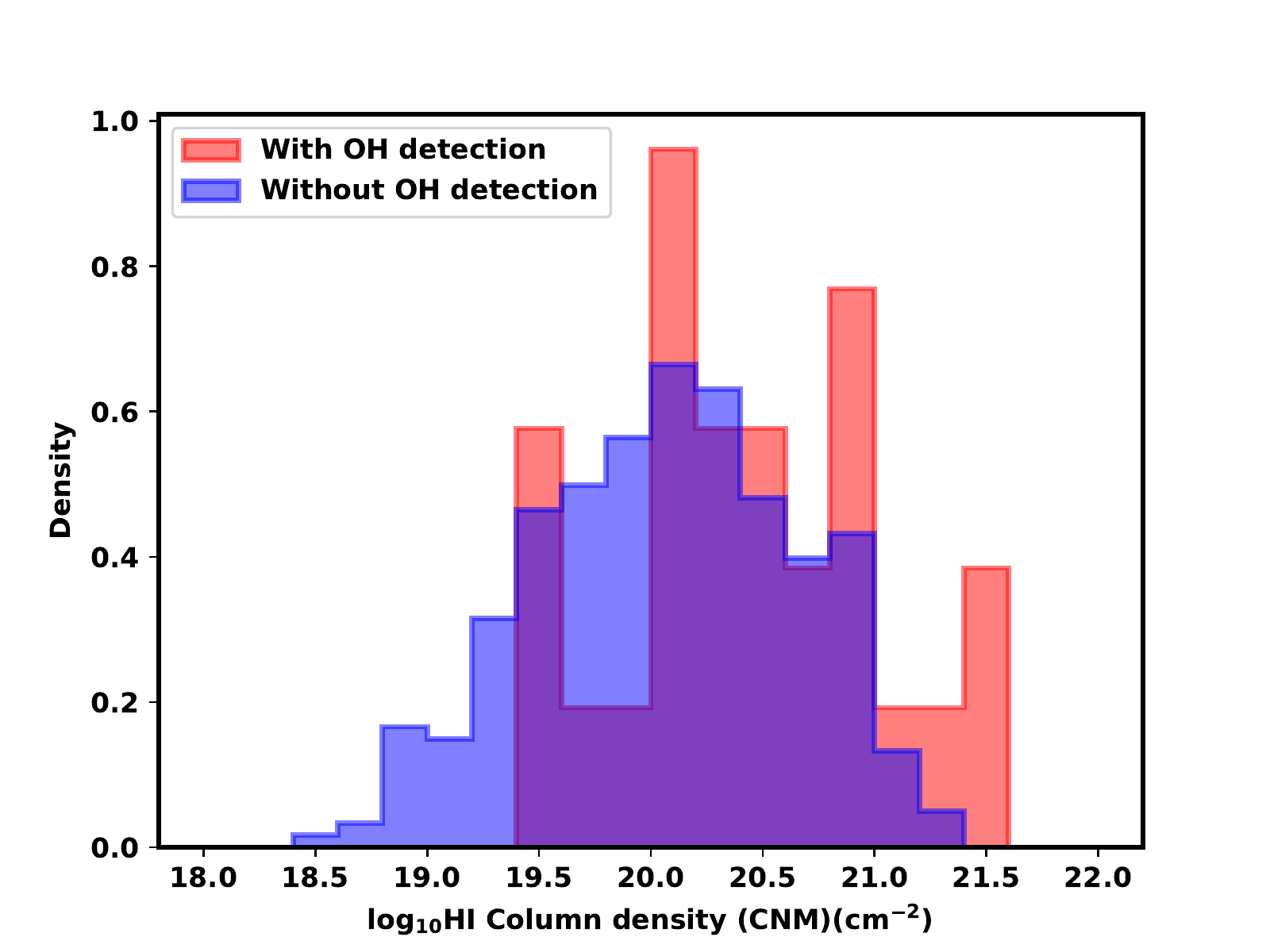}\\
    \end{tabular}
    \caption{Normalised histograms showing the distribution of $\log_{10}$\,peak H\textsc{i} CNM optical depth (left) and $\log_{10}$\,H\textsc{i} CNM column density (right) of H\textsc{i}~CNM features found by \citet{Nguyen2019} both with (red) and without (blue) a matching OH component. Both sets of distributions differ significantly, with a Kolmogorov-Smirnov p-value of $3\times 10^{-6}$~for peak H\textsc{i} CNM optical depth, and 0.02 for H\textsc{i} CNM column density.}
    \label{fig:matchvno_TAU0_NHI}
\end{figure*}

There are some significant differences between H\textsc{i}~CNM components with associated OH and those without. The histograms shown in Fig. \ref{fig:matchvno_TAU0_NHI} compare the distributions of the H\textsc{i}~CNM components with an associated OH component (red) and without such an association (blue) across H\textsc{i} CNM peak optical depth and column density. In both cases these two distributions differ significantly, with components associated with OH tending towards higher values of both parameters. The significance of these different distributions was measured via the Kolmogorov-Smirnov test, which resulted in a p-value for the CNM peak optical depth distributions of $3\times 10^{-6}$, and 0.02 for the column density distributions. 

Though Fig. \ref{fig:matchvno_TAU0_NHI} implies that CNM clouds with higher peak optical depth or column density are more likely to contain detectable OH, the lack of a clear linear relationship between OH and CNM column density (see right panel of Fig. \ref{fig:logNOH_v_NHI}) does not imply that CNM clouds with higher peak optical depth or column density contain \textit{more} OH. Instead, these data suggest the existence of a threshold CNM optical depth or column density under which any OH will not be detected, but over which there is not then a linear relationship between how much OH (and by extension, H$_2$) will form. Again, this is consistent with the findings of \citet{Bellomi2020}.

\begin{figure*}
\begin{tabular}{cc}
    \includegraphics[trim={0cm 0cm 1cm 0cm},clip=true,width=0.45\linewidth]{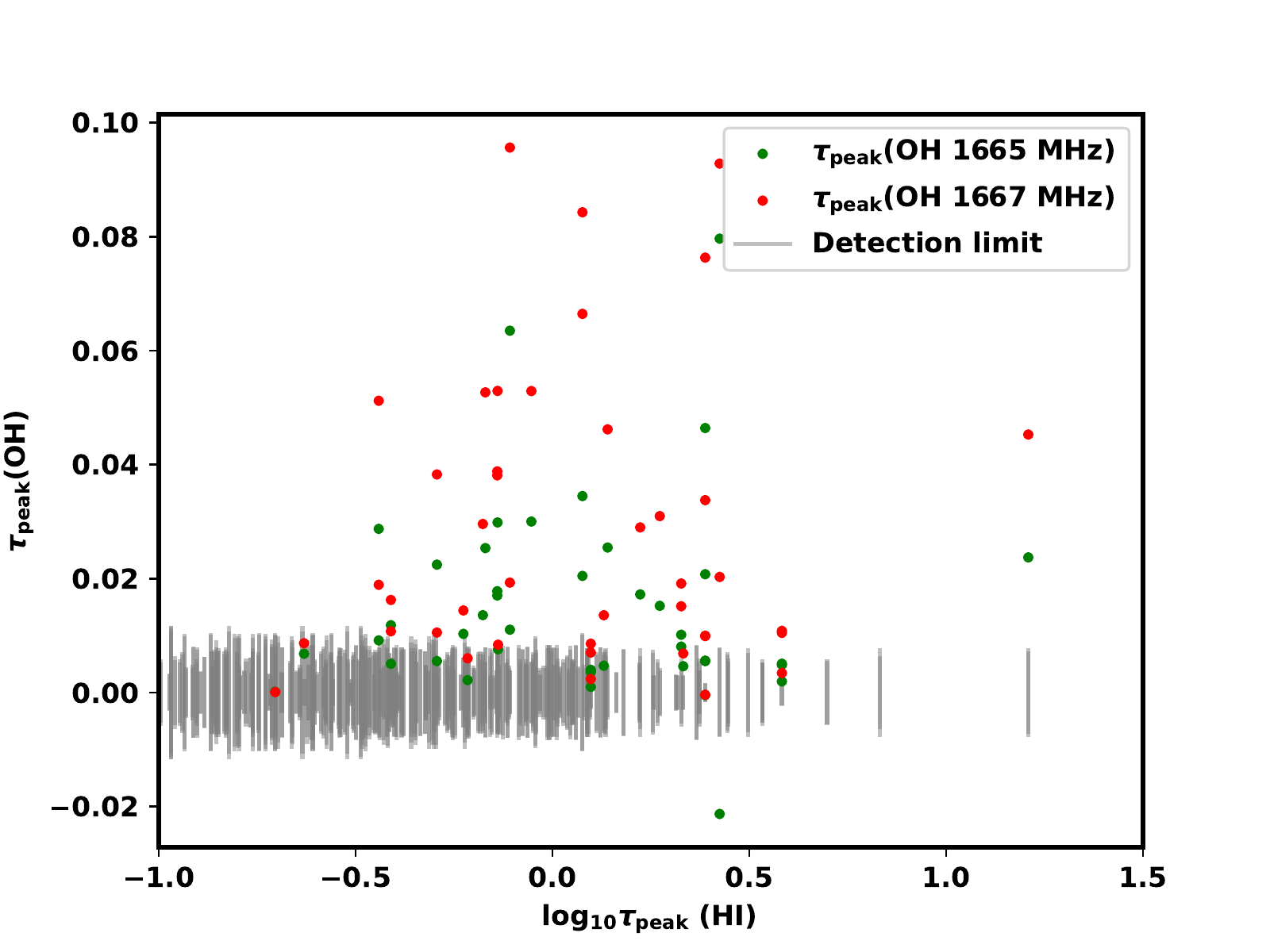}&
    \includegraphics[trim={0cm 0cm 1cm 0cm},clip=true,width=0.45\linewidth]{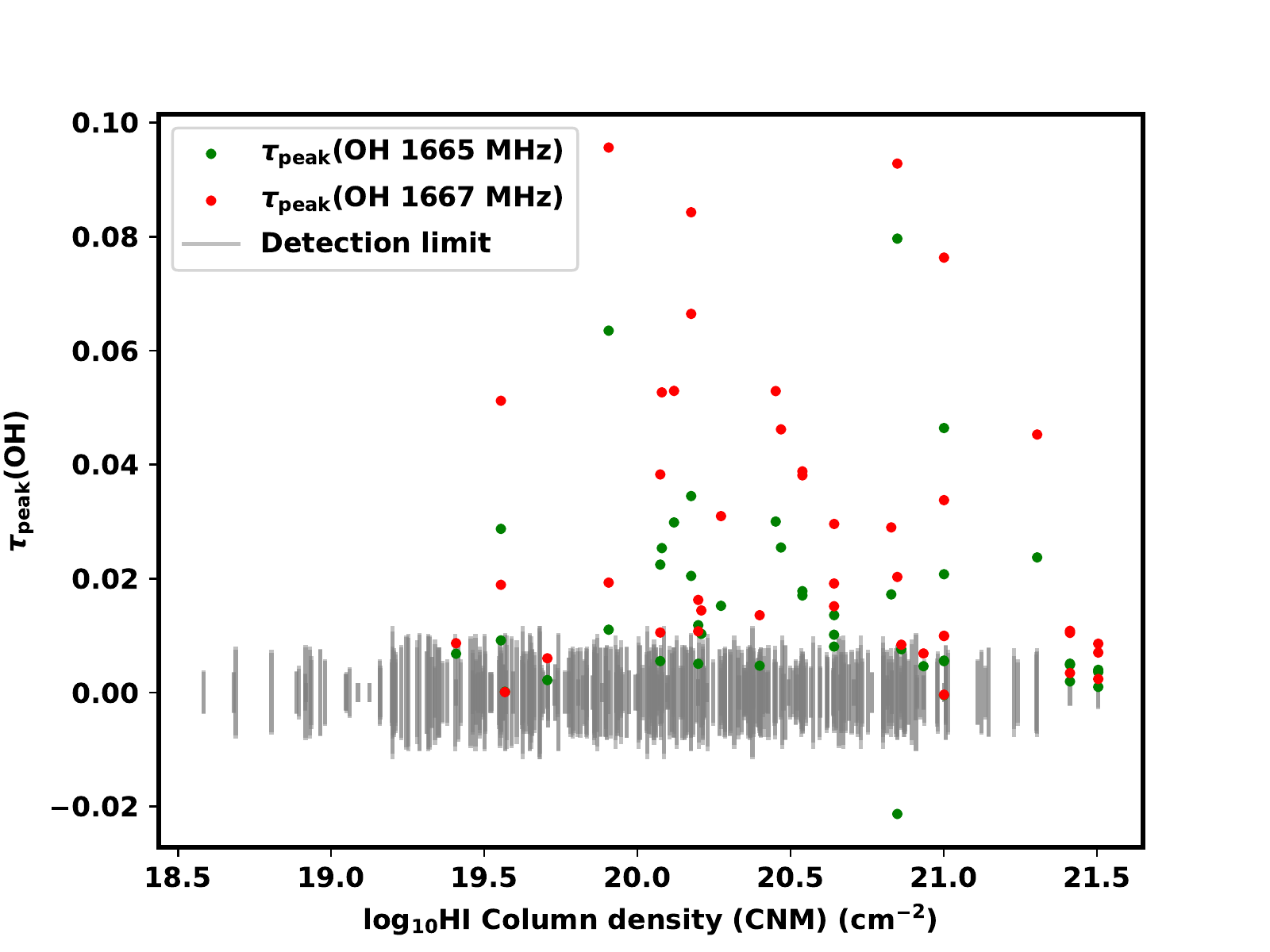}\\
\end{tabular}
    \caption{Relationship between OH peak main line optical depth and H\textsc{i}~optical depth (left), and H\textsc{i}~CNM column density (right) for matching OH and H\textsc{i}~features. The detection limit is estimated to be twice the standard deviation of the noise in spectra for which a match for a CNM component was not found: vertical grey lines connect these $\pm 2 \sigma$~values as an indication of the range for which detections may be missing.}
    \label{fig:maintau_v_logTAU0_NHI}
\end{figure*}

Focusing on this apparent tendency of OH to be more readily detectable in clouds with higher peak H\textsc{i} CNM optical depth or column density, we naturally would like to establish if this is due to OH at lower peak H\textsc{i} CNM optical depth or column density being undetectable or whether it is due to it being absent. In other words, is the apparent lack of OH `real' or a symptom of our sensitivity? Bearing in mind the complexities of detectibility discussed in previous sections, generally speaking the strongest influence on whether or not an OH feature is detected is the signal to noise ratio of the 1667\,MHz peak optical depth (which will in turn generally depend on the brightness of the background continuum) as it tends to have the highest signal-to-noise ratio of the 8 spectra comprising each sightline, followed by that of the 1665\,MHz peak optical depth. Fig. \ref{fig:maintau_v_logTAU0_NHI} shows the relationships between these key parameters that drive detectability (main-line optical depth) and H\textsc{i} CNM peak optical depth and column density. Detection limits are indicated by grey vertical lines that connect the $\pm 2 \sigma$~values for spectra for which a match for a CNM component was not found (thus indicating the range for which detections may be missing). We estimated the OH peak optical depth detection limits to be approximately equal to twice the standard deviation of the noise in the optical depth spectra based on the findings of \citet{Petzler2021a} that for spectra with a signal-to-noise ratio of 2 \textsc{Amoeba}~is able to recover 90\% of features present in on-off data. 

From Fig. \ref{fig:maintau_v_logTAU0_NHI} there does not appear to be a trend of decreasing main-line peak optical depths (and therefore decreasing detectablility) at lower peak H\textsc{i} CNM optical depth or column density. This is of course not definitive evidence that the OH is absent as it is still possible that the pattern we see in Fig. \ref{fig:matchvno_TAU0_NHI} is a reflection of the detectability of the OH due to the complex nature of the relationship between the abundance of OH (i.e. its column density) and its optical depth. However, if we were to go so far as to assume that the differences in the distributions with and without OH detections seen in Fig. \ref{fig:matchvno_TAU0_NHI} are real, we may attribute this to the shielding of the H\textsc{i} gas: at higher H\textsc{i} CNM peak optical depth and column density molecular gas will be shielded from dissociating UV radiation, allowing the molecular gas to accumulate such that there is sufficient OH to be detected.

\begin{figure*}
    \centering
    \begin{tabular}{cc}
    \includegraphics[trim={0cm 0cm 1cm 0cm}, clip=true, width=0.45\linewidth]{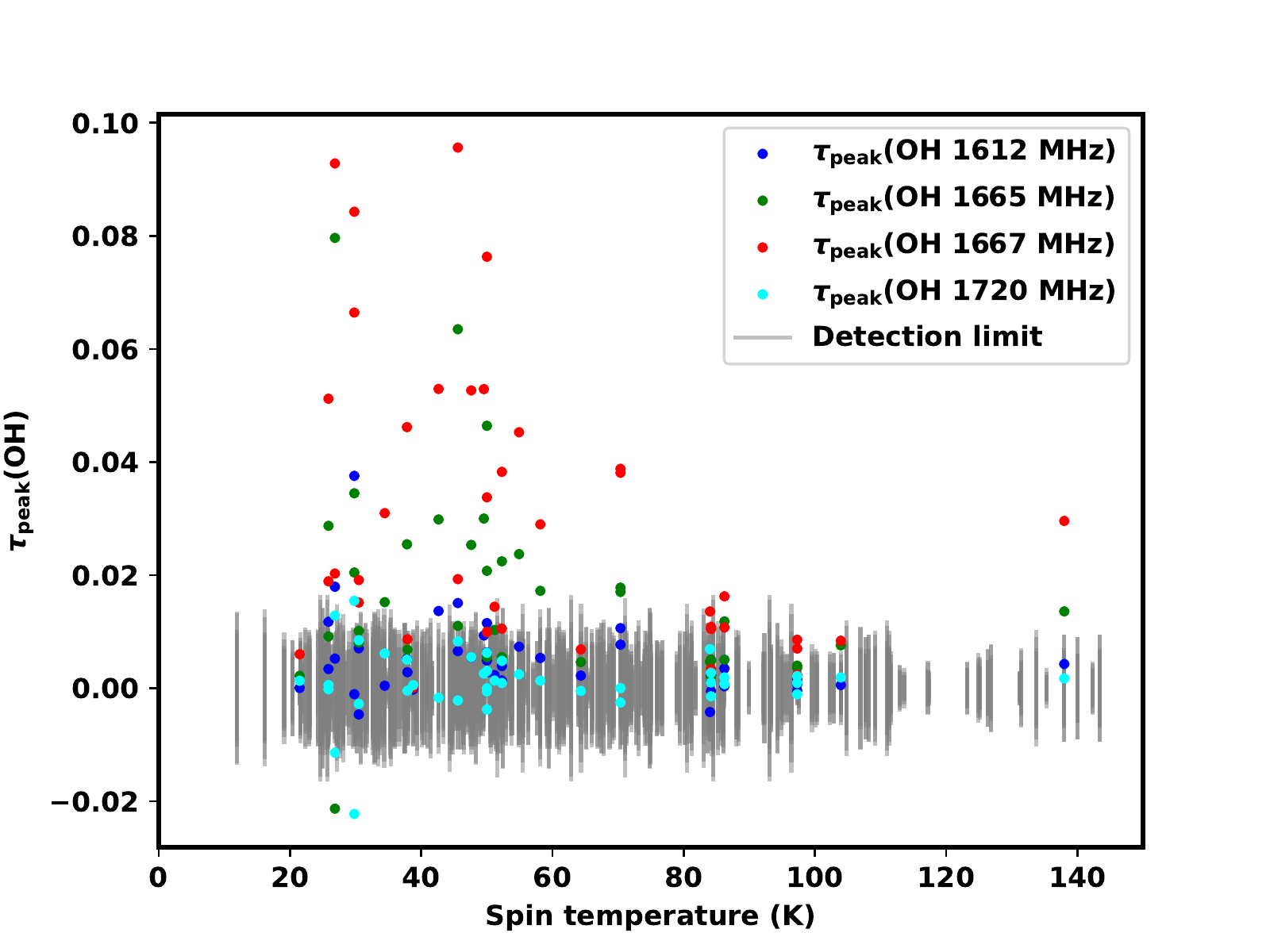}&\includegraphics[trim={0cm 0cm 1cm 0cm}, clip=true, width=0.45\linewidth]{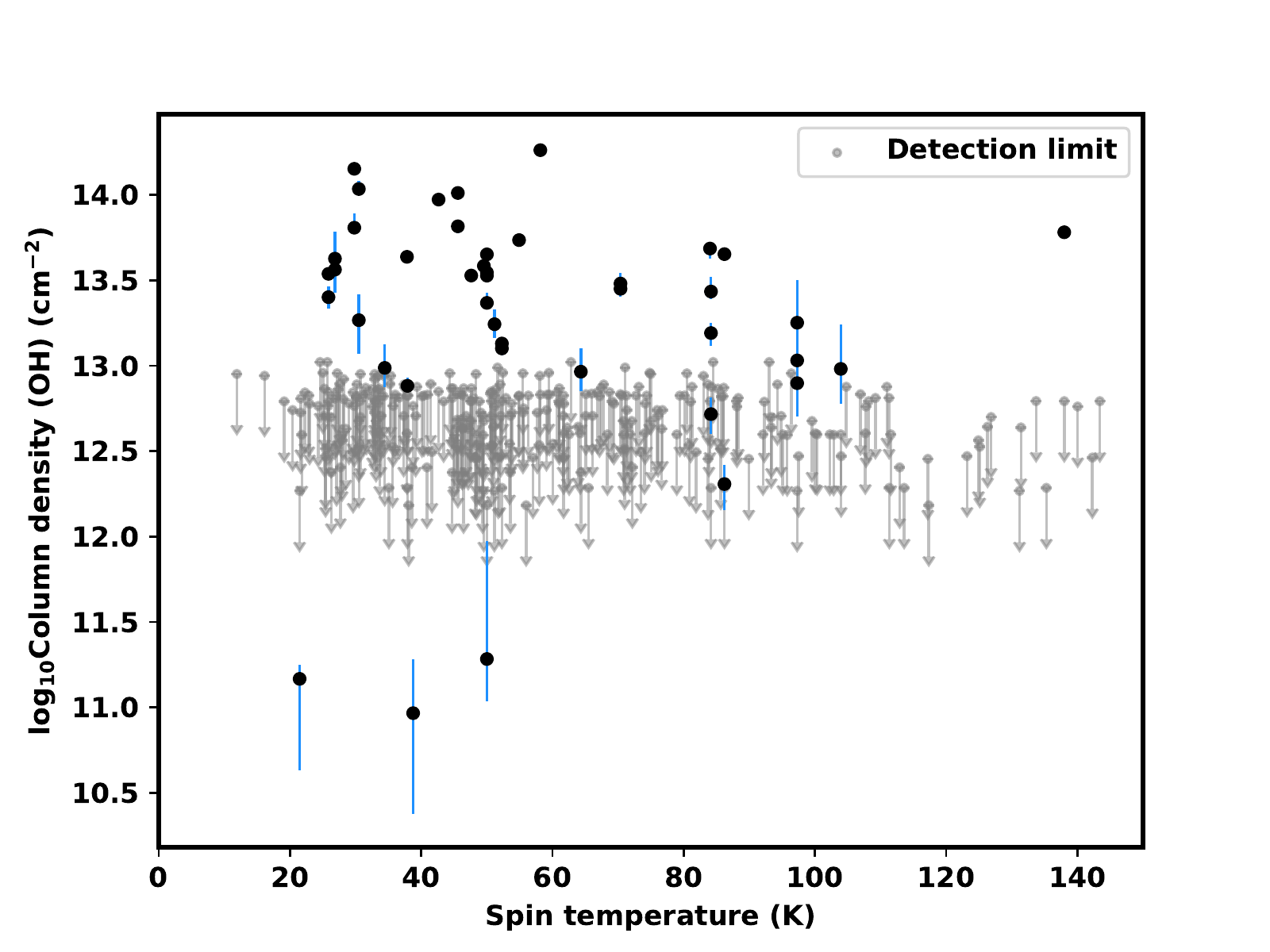}\\
    \end{tabular}
    \caption{Relationships between (left) OH peak optical depth and (right) $\log_{10}$~OH column density, and H\textsc{i}~CNM spin temperature for matching OH and H\textsc{i}~features. The detection limit for optical depth is estimated to be twice the standard deviation of the noise in spectra for which a match for a CNM component was not found: vertical grey lines connect these $\pm 2 \sigma$~values as an indication of the range for which detections may be missing. The detection limit for column density was estimated from the $2\times$\,rms noise in our optical depth data, the median excitation temperatures determined from our sightlines with detections and a feature width of 0.3\,km\,s$^{-1}$~(i.e. three times our typical channel width).}
    \label{fig:tau_v_Ts}
\end{figure*}

We then looked for other relationships between the parameters of the OH fits and those of the H\textsc{i}~CNM fits. Very few pairs of parameters show notable trends, and of these none are strong enough to be predictive. Some of these weak relationships are however interesting, such as the relationships between OH optical depths and H\textsc{i}~CNM spin temperature, illustrated in the left panel of Fig. \ref{fig:tau_v_Ts}. In each of the four ground-rotational state transitions there is a significantly wider range of OH optical depths (and more so for the main-line transitions at 1665 and 1667\,MHz) for components matched with H\textsc{i}~CNM components with a low spin temperature. At higher spin temperatures the optical depths in all four OH transitions approach zero. Since this trend is strongest in the main lines, which tend not to exhibit anomalous excitation we may cautiously associate optical depth with total column density. Indeed, when we compare our fitted OH column density to spin temperature (see the right panel in Fig. \ref{fig:tau_v_Ts}) we do see this same trend where higher OH column densities are seen at lower H\textsc{i} CNM spin temperatures, though the trend is less pronounced. This is consistent with a scenario where more molecular gas is able to accumulate in H\textsc{i} CNM gas with low spin temperature. 

Unfortunately, none of these trends hint at a direct relationship between the parameters of the molecular and CNM gas. Indeed, this may be the more interesting result as it is consistent with a scenario where the molecular gas is effectively decoupled from the cold atomic phase. It may therefore be the case that the molecular gas traced in these observations is not mixed significantly with the CNM, in contrast (for example) to the suggestion of \citet{Stanimirovic2014} for sightlines towards Perseus.

%% file: 6_Conclusions.tex
We have presented observations of the four ground-rotational state transitions of hydroxyl towards 107 sightlines: 92 sets of `on-off' observations in and out of the Galactic plane from the Arecibo telescope, and 15 sets of optical depth spectra from the ATCA. Using the Bayesian Gaussian decomposition algorithm \amoeba~we identify 109 features across 38 of these sightlines (27 from Arecibo, 11 from the ATCA). We find significant departures from LTE which are more apparent in the satellite lines (at 1612 and 1720\,MHz) than in the main lines (at 1665 and 1667\,MHz). These departures are more pronounced along sightlines through the Galactic centre, though these were observed with the ATCA and only consist of optical depth spectra rather than the on-off spectra obtained for sightlines observed with Arecibo. Assuming these differences are real, we attribute non-LTE behaviour of the main lines to non-Planckian radiation fields or non-Maxwellian collisional distributions in this region. We attribute non-LTE behaviour of the satellite lines to collisional excitations or enhanced radiation fields along with low number density.

We compare our OH fit parameters to H\textsc{i} CNM parameters published by \citet{Nguyen2019}. No direct relationships are found between these parameters, though some trends are evident. First, we identify a tendency for CNM features with an associated OH feature to have higher H\textsc{i} peak optical depth and higher CNM column density than those without, which is naturally explained by the shielding of the molecular gas by the CNM from dissociating UV radiation that would prevent the accumulation (and therefore detection) of molecular gas. Second, higher H\textsc{i} spin temperature components host only low optical depth OH, whereas lower spin temperature components host a wider range of OH optical depths. Since this trend was more apparent in the main lines, we associate the optical depth with column density and interpret this as an indication that more molecular gas can accumulate in CNM gas of lower spin temperature. We do not believe that any of these trends indicate a direct interaction between the molecular and CNM gas, and we speculate that this may indicate a decoupling of the molecular gas from the CNM once it accumulates. However, more complex fits to CNM features might alter these conclusions significantly.

With the currently accepted limitations of using CO to probe the molecular content of the ISM, we will continue to rely on other tracers of this regime such as OH. The sensitivity of OH excitations to its environment -- and particularly the readiness of its lines to invert -- provides an invaluable probe of the conditions of the molecular ISM. Though this work represents an unprecedented number of features identified in all four OH ground-rotational state transitions, our analysis was at many times limited by the small number of features displaying a given behaviour. Such analyses would therefore benefit from a significant increase in the number of examined sightlines. OH of course also has its own limitations, namely the weakness of its transitions. This is a limitation we can resolve if we seek more integration time in our observations \citep{Busch2021}. Hopefully the future study of OH will include wide range, deep observations with which we can unravel some of the current mysteries of the atomic to molecular transition in the ISM.

%% file: 7_Appendix.tex
\section{Gaussian models}

% Amoeba fits plots:
% These are plotted using 'Display_final_data_fits.py' and are stored in 'amoeba/Final_fits'

\begin{figure*}
    \centering
    \begin{tabular}{cc}
    \includegraphics[trim={0cm 0.6cm 0cm 1cm}, clip=true, width=0.45\linewidth]{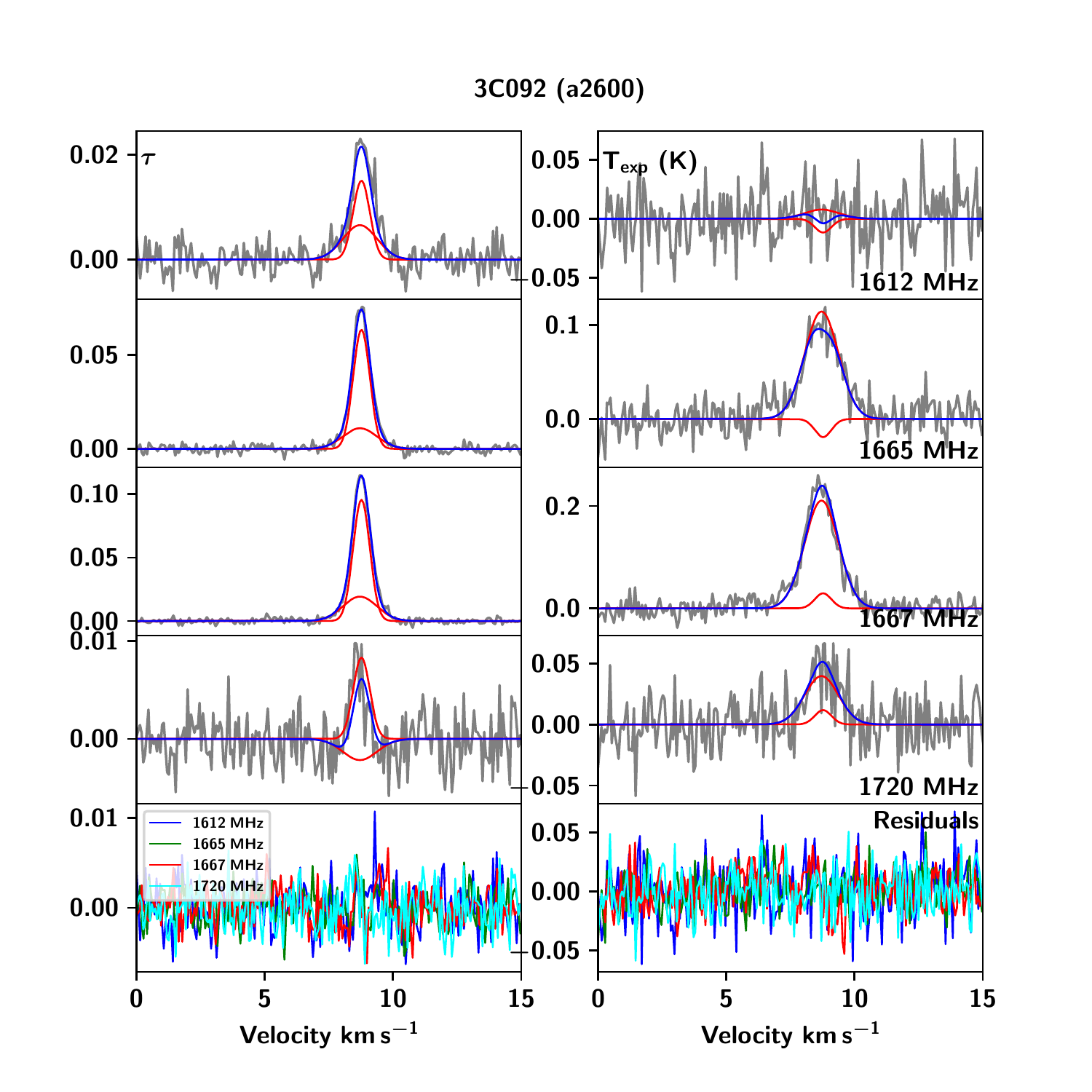}&
    \includegraphics[trim={0cm 0.6cm 0cm 1cm}, clip=true, width=0.45\linewidth]{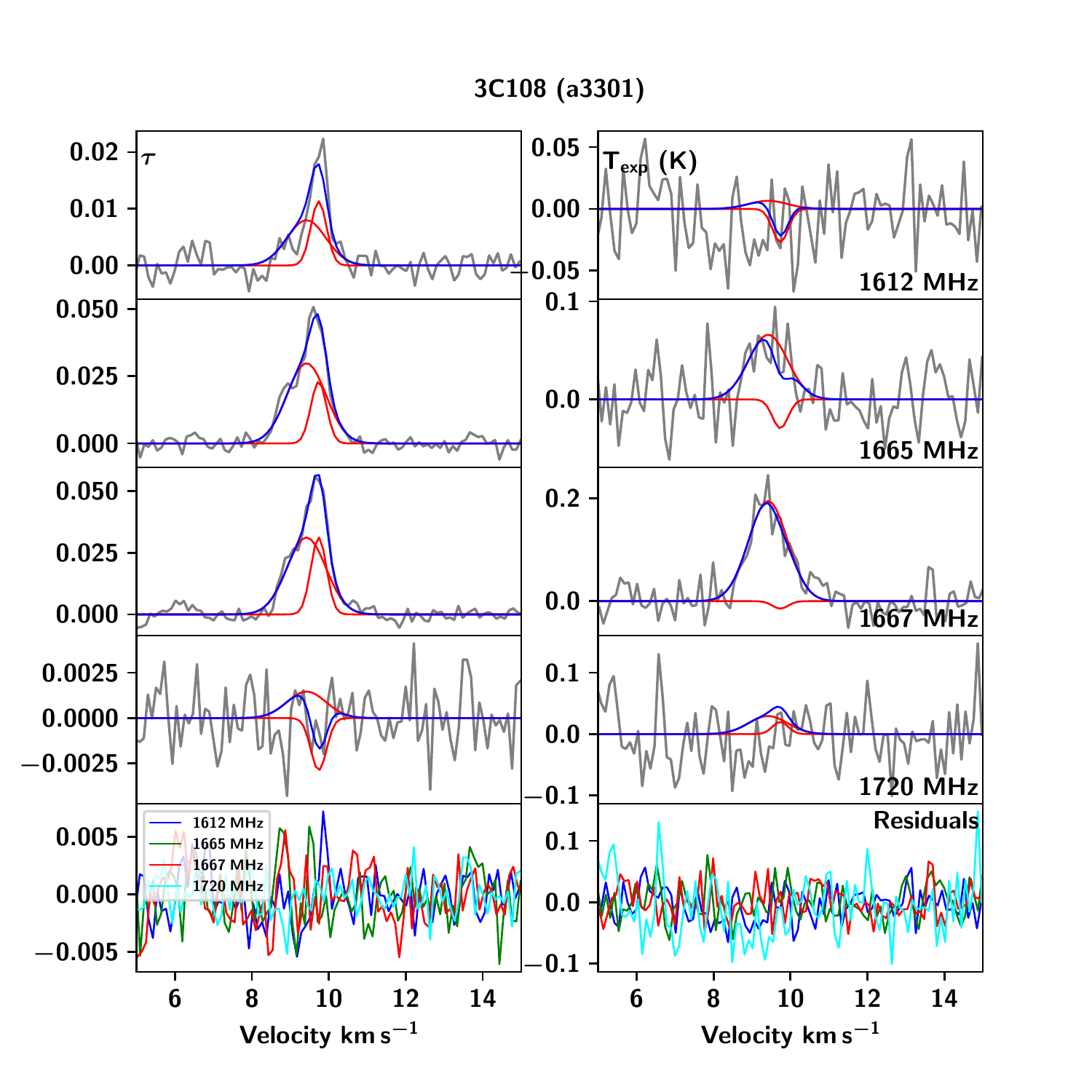}\\
    \includegraphics[trim={0cm 0.6cm 0cm 1cm}, clip=true, width=0.45\linewidth]{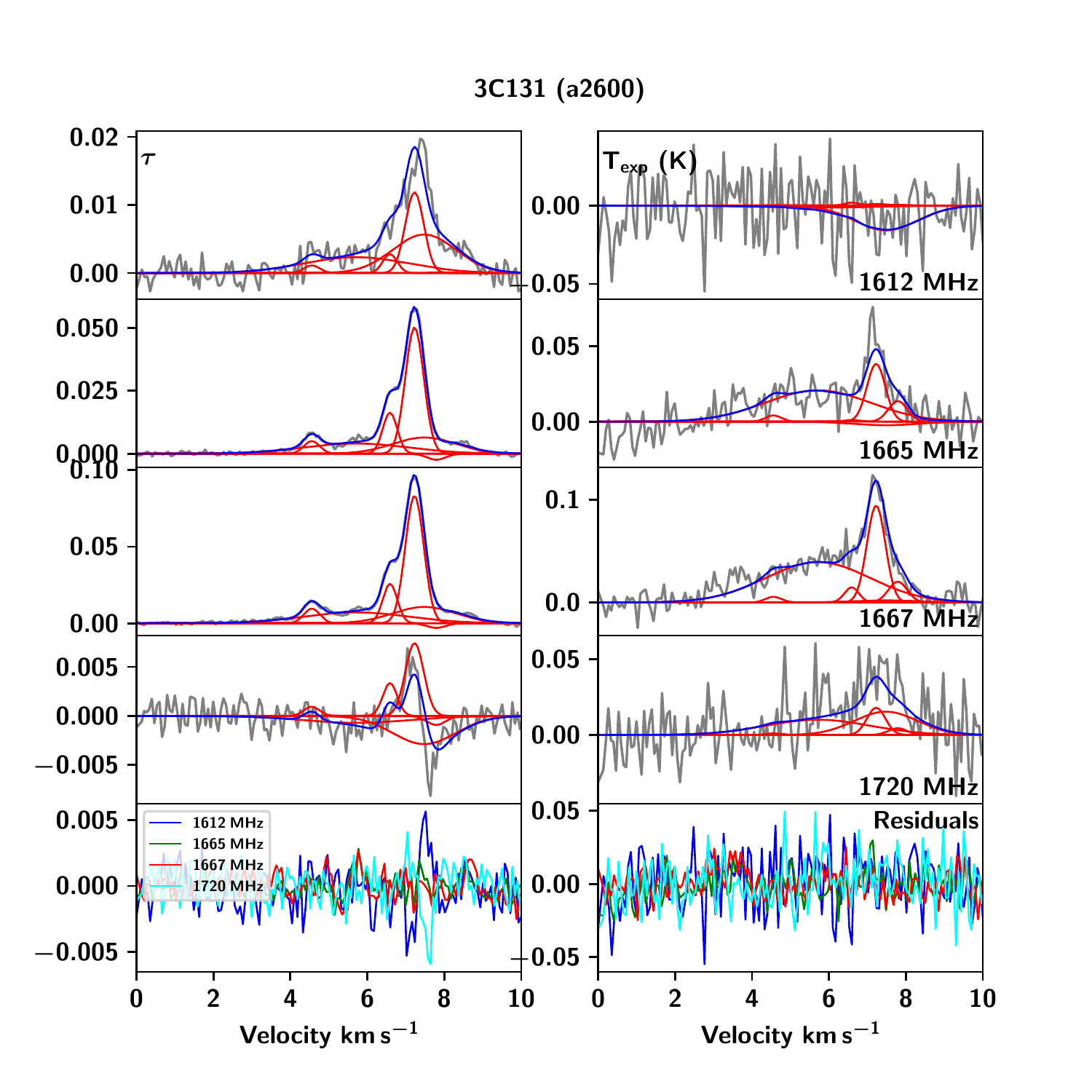}&
    \includegraphics[trim={0cm 0.6cm 0cm 1cm}, clip=true, width=0.45\linewidth]{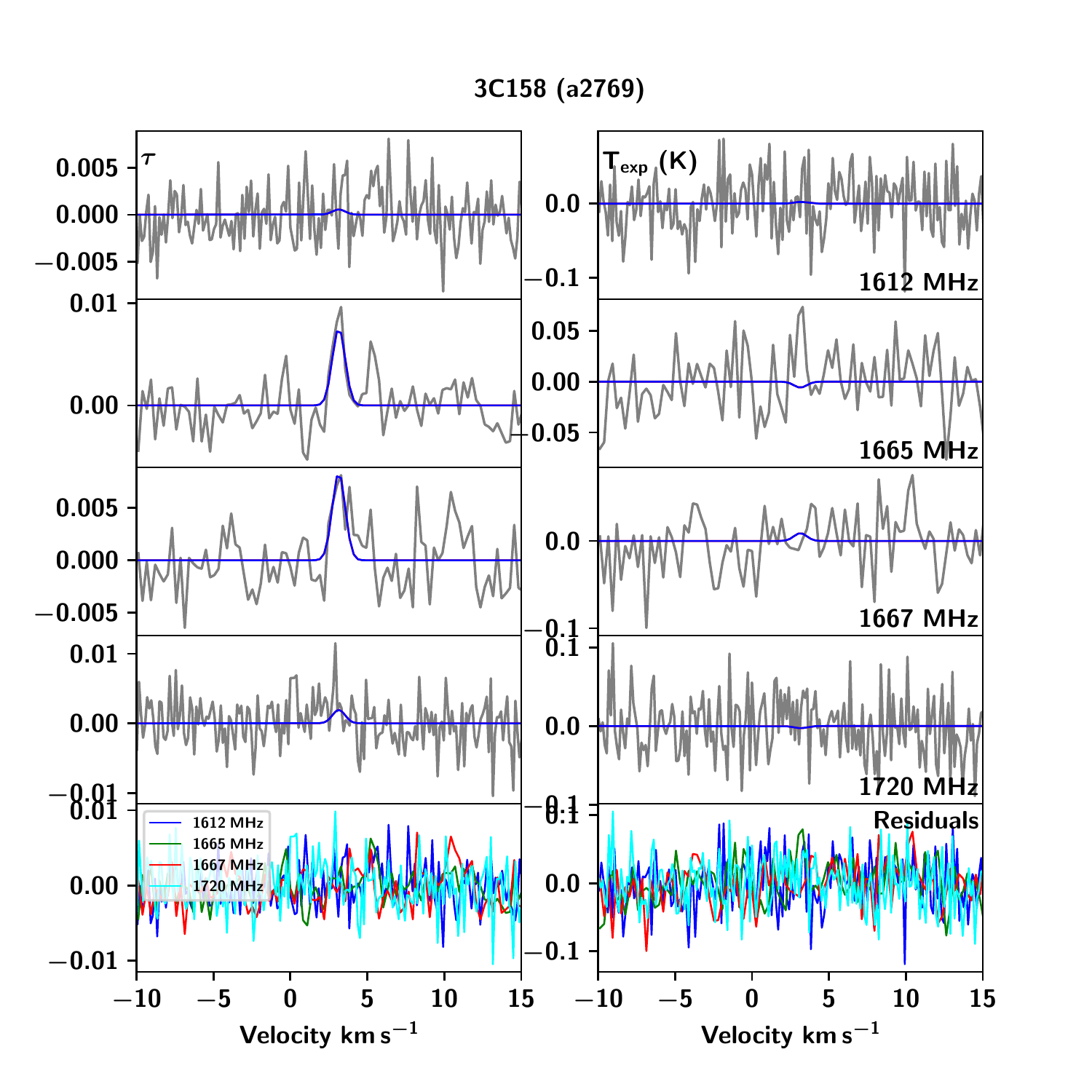}\\
    \includegraphics[trim={0cm 0.6cm 0cm 1cm}, clip=true, width=0.45\linewidth]{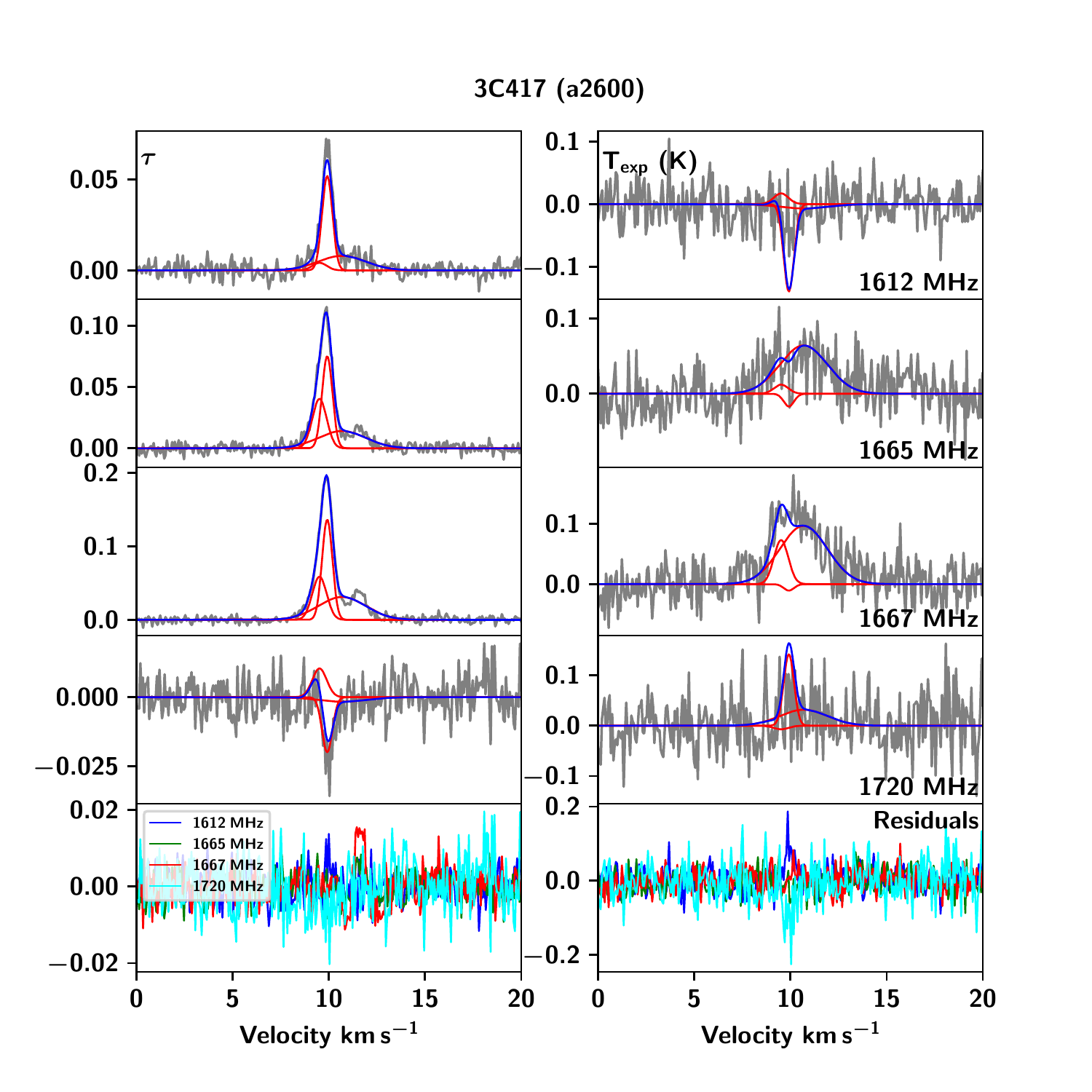}&
    \includegraphics[trim={0cm 0.6cm 0cm 1cm}, clip=true, width=0.45\linewidth]{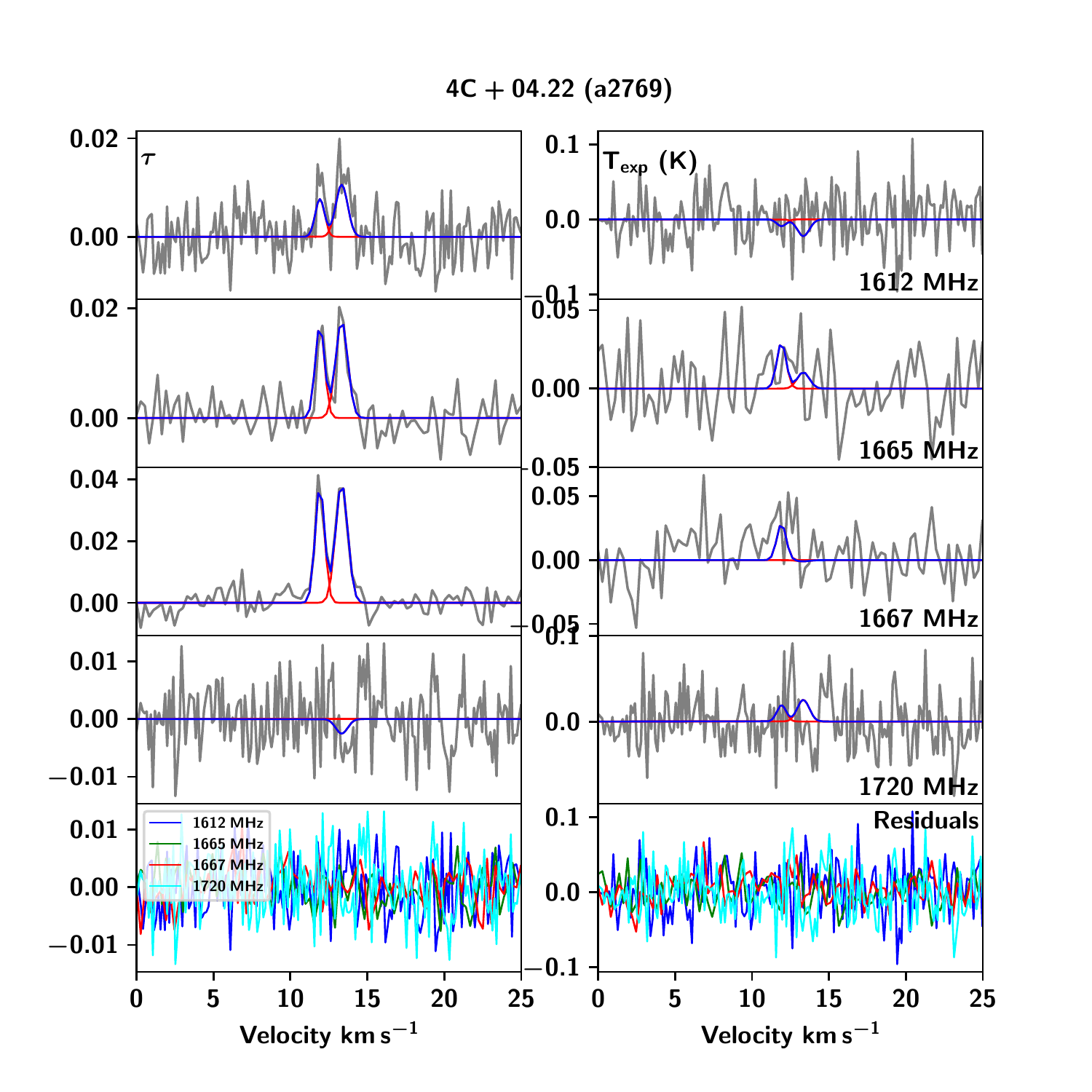}\\
    \end{tabular}
    \caption{The top four left hand panels of each plot show optical depth observations in grey, and the top four right hand panels show the expected brightness temperature data in grey. These panels show the individual fitted components in red and the total fit for each spectrum in blue. The bottom panels show the residuals of these total fits in each of the four ground-rotational state transitions of OH. This figure shows the sightlines (left to right, top to bottom) towards 3C092, 3C108, 3C131, 3C158, 3C417 and 4C+04.22.}
    \label{fig:results1}
\end{figure*}

\begin{figure*}
    \centering
    \begin{tabular}{cc}
    \includegraphics[trim={0cm 0.6cm 0cm 1cm}, clip=true, width=0.45\linewidth]{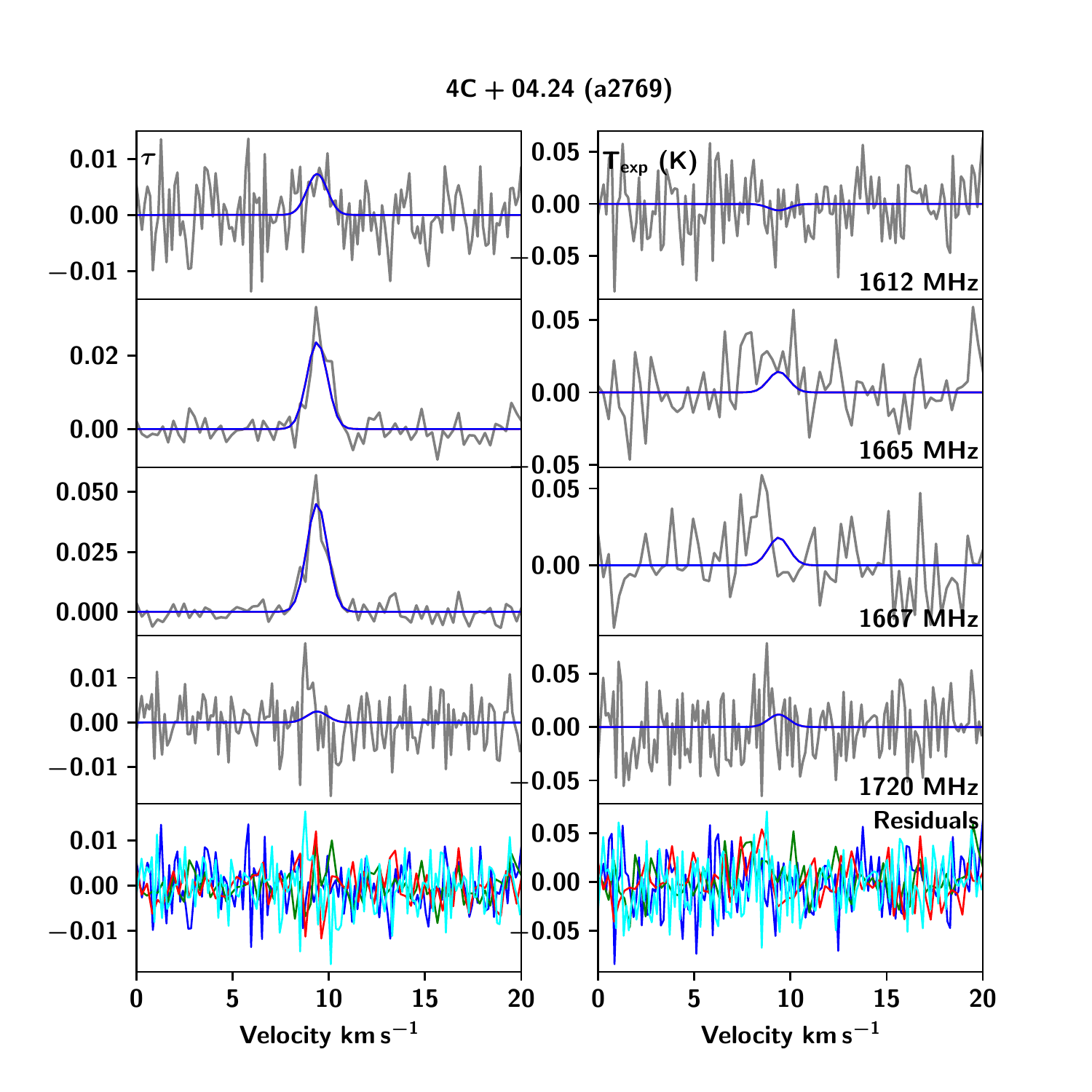}&
    \includegraphics[trim={0cm 0.6cm 0cm 1cm}, clip=true, width=0.45\linewidth]{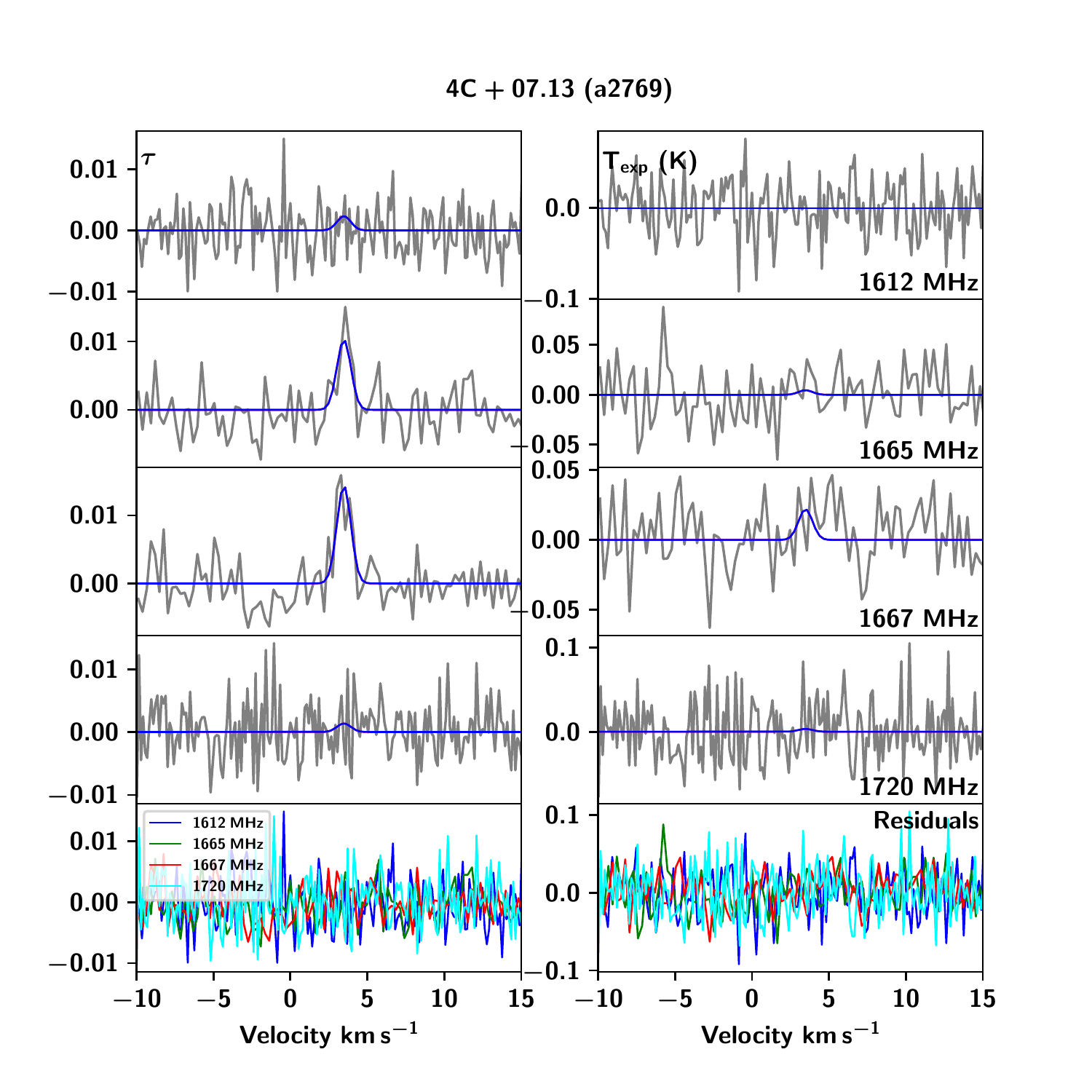}\\
    \includegraphics[trim={0cm 0.6cm 0cm 1cm}, clip=true, width=0.45\linewidth]{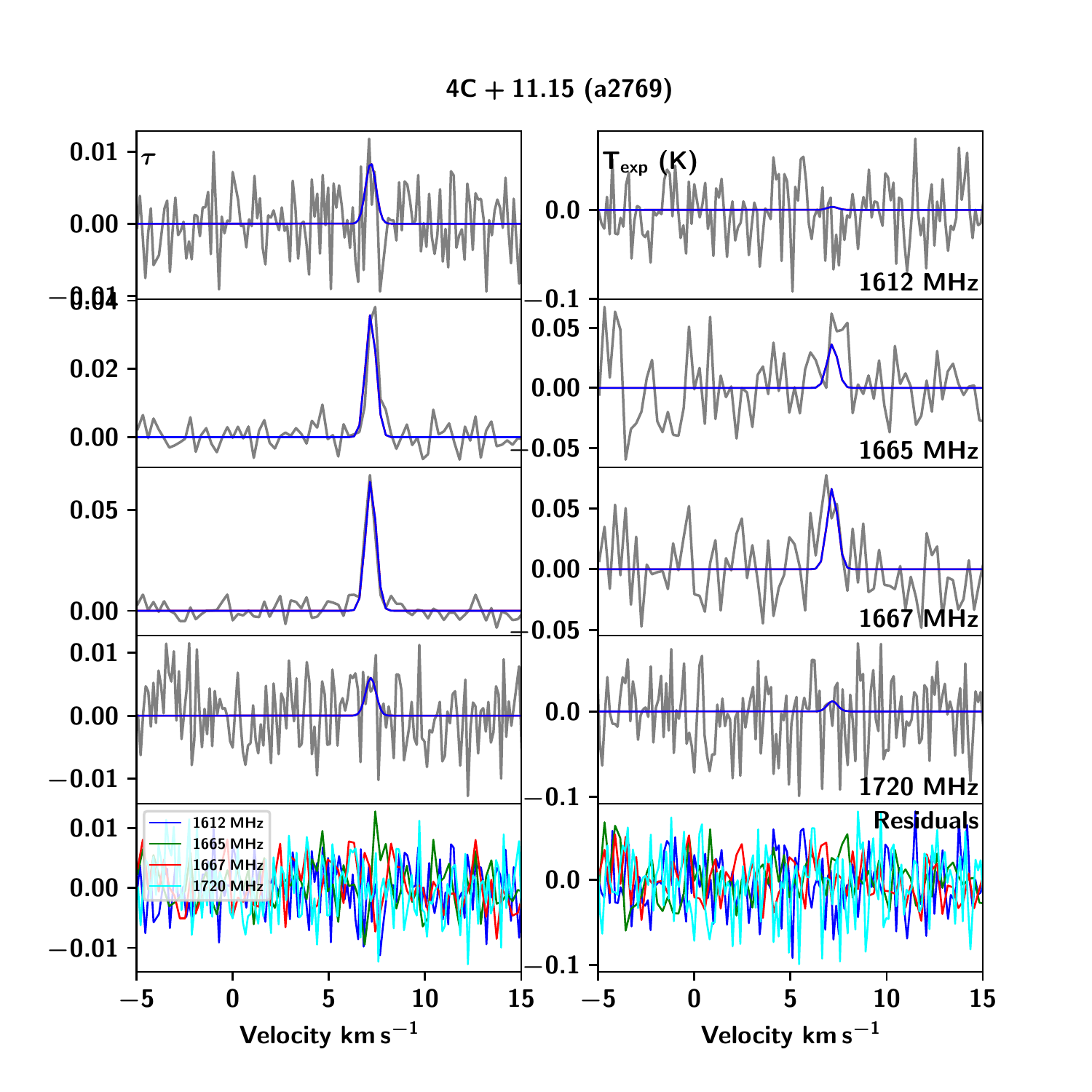}&
    \includegraphics[trim={0cm 0.6cm 0cm 1cm}, clip=true, width=0.45\linewidth]{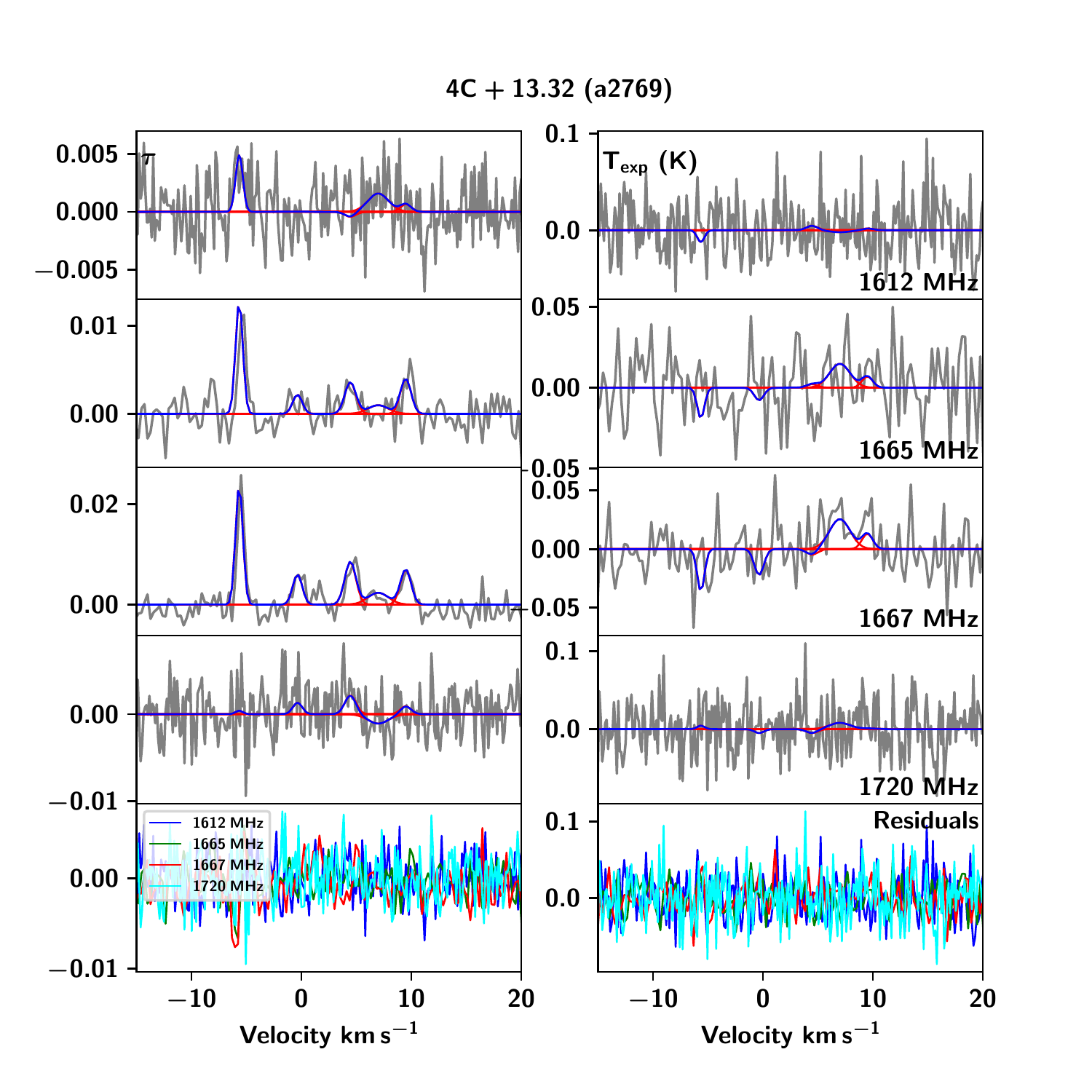}\\
    \includegraphics[trim={0cm 0.6cm 0cm 1cm}, clip=true, width=0.45\linewidth]{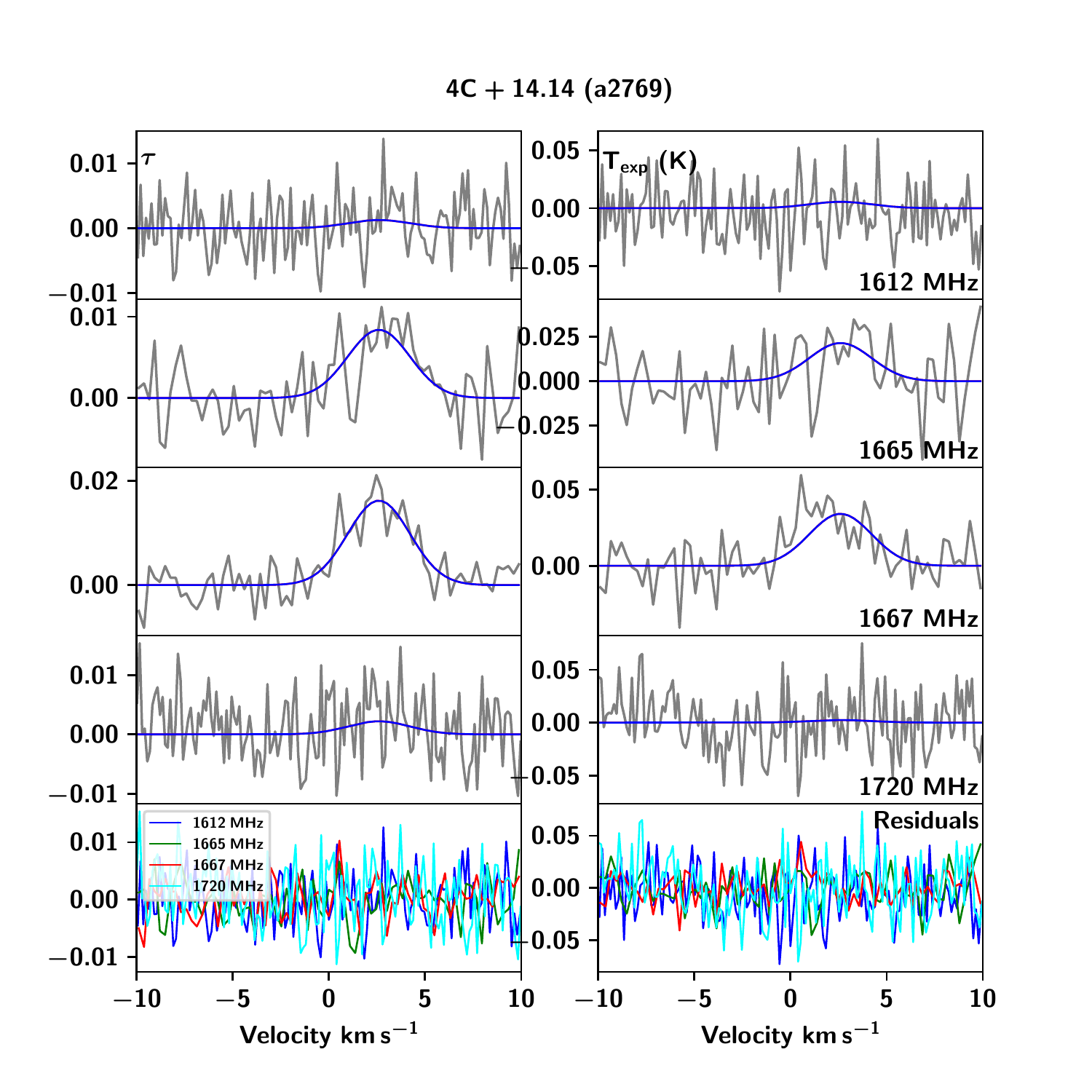}&
    \includegraphics[trim={0cm 0.6cm 0cm 1cm}, clip=true, width=0.45\linewidth]{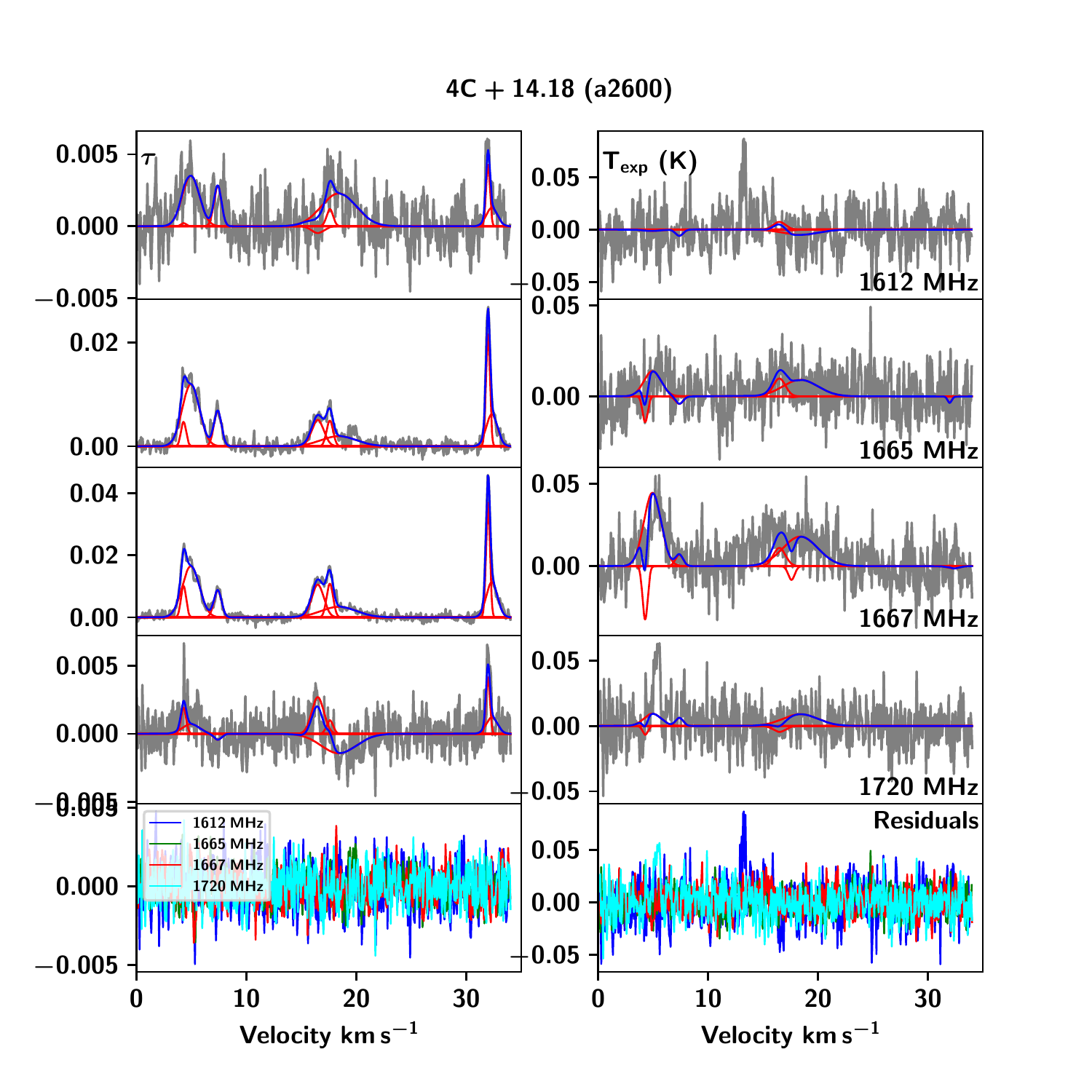}\\
    \end{tabular}
    \caption{Same as Fig. \ref{fig:results1} for 4C+04.24, 4C+07.13, 4C+11.15, 4C+13.32, 4C+14.14 and 4C+14.18.}
    \label{fig:results2}
\end{figure*}

\begin{figure*}
    \centering
    \begin{tabular}{cc}
    \includegraphics[trim={0cm 0.6cm 0cm 1cm}, clip=true, width=0.45\linewidth]{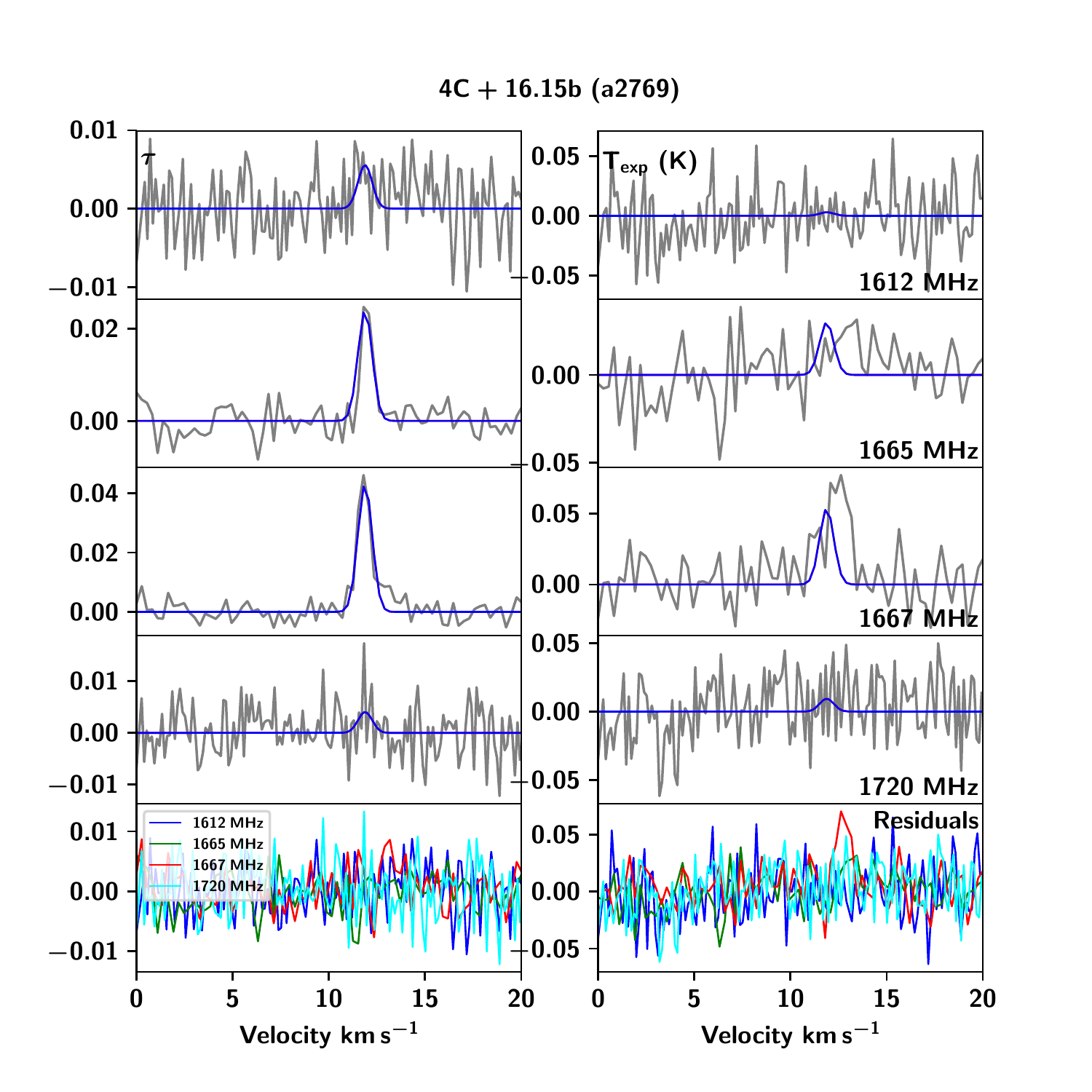}&
    \includegraphics[trim={0cm 0.6cm 0cm 1cm}, clip=true, width=0.45\linewidth]{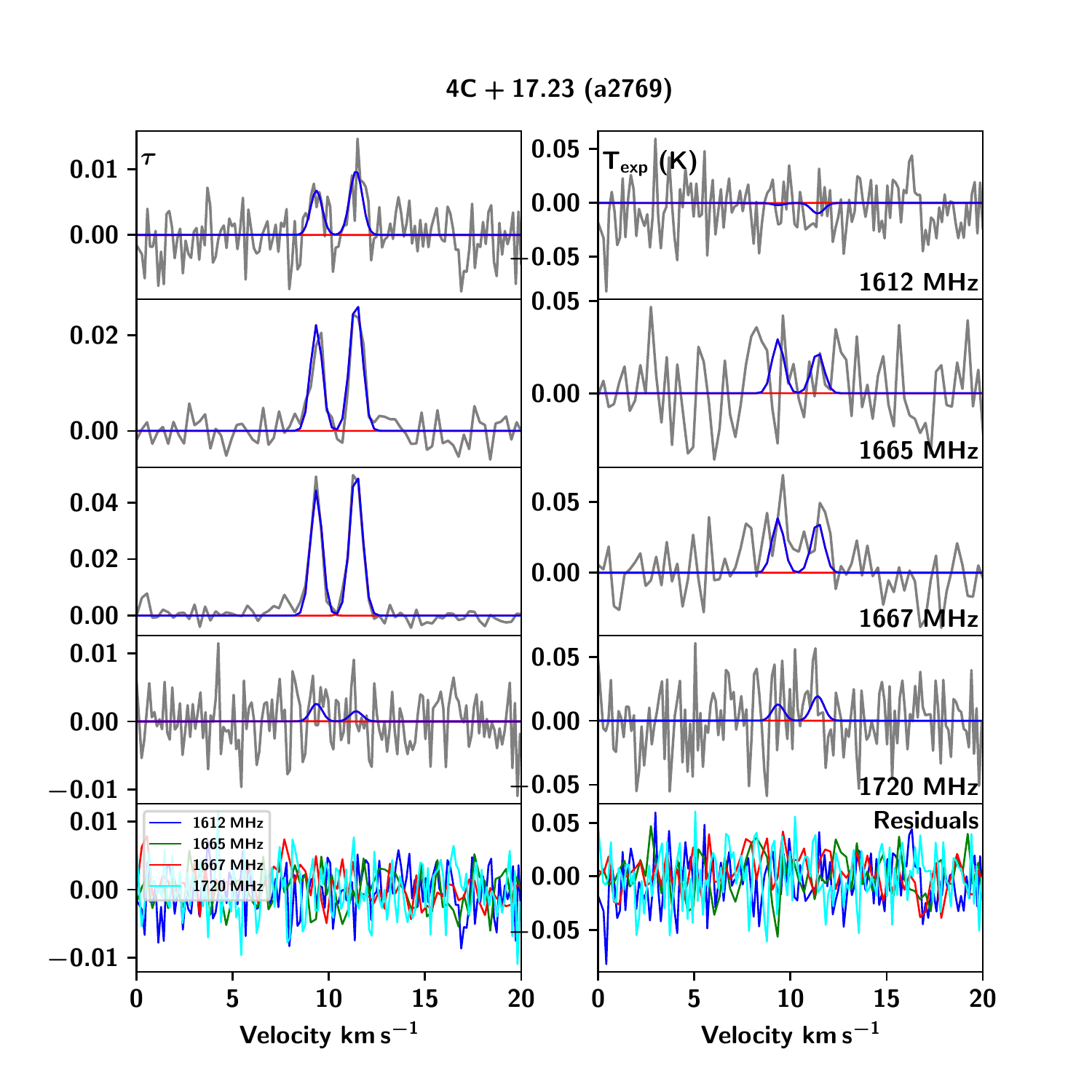}\\
    \includegraphics[trim={0cm 0.6cm 0cm 1cm}, clip=true, width=0.45\linewidth]{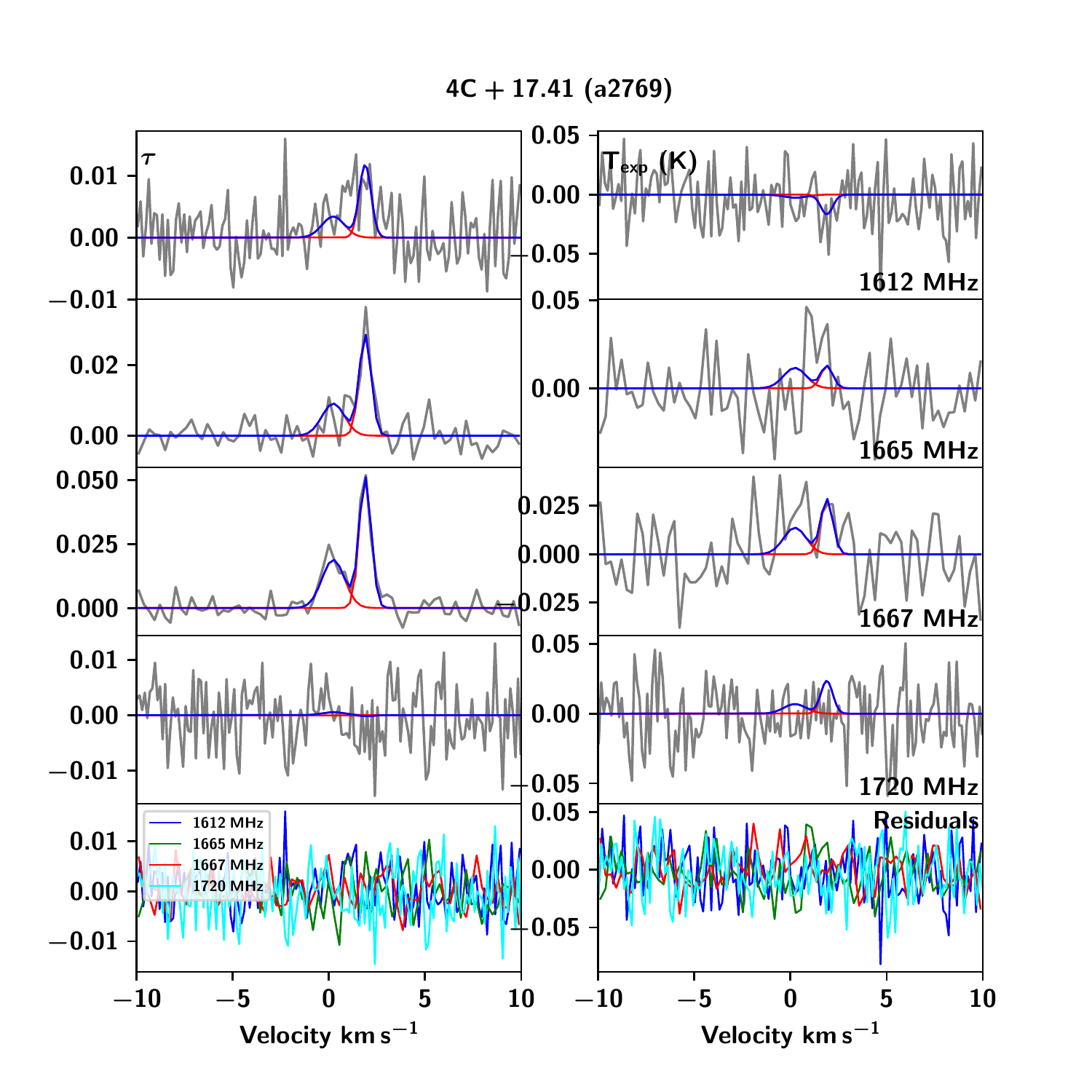}&
    \includegraphics[trim={0cm 0.6cm 0cm 1cm}, clip=true, width=0.45\linewidth]{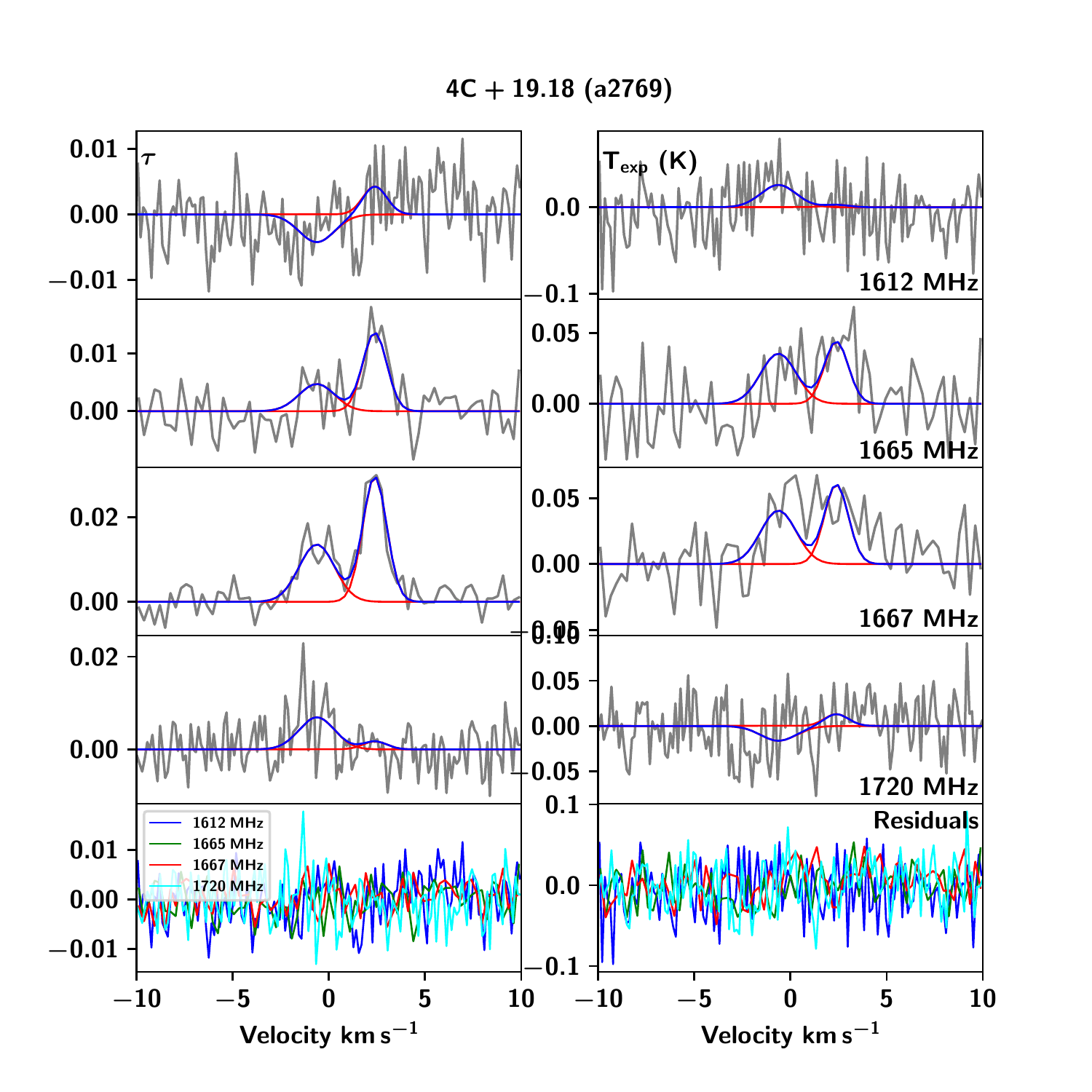}\\
    \includegraphics[trim={0cm 0.6cm 0cm 1cm}, clip=true, width=0.45\linewidth]{Figures/a2769_4C+1919.pdf}&
    \includegraphics[trim={0cm 0.6cm 0cm 1cm}, clip=true, width=0.45\linewidth]{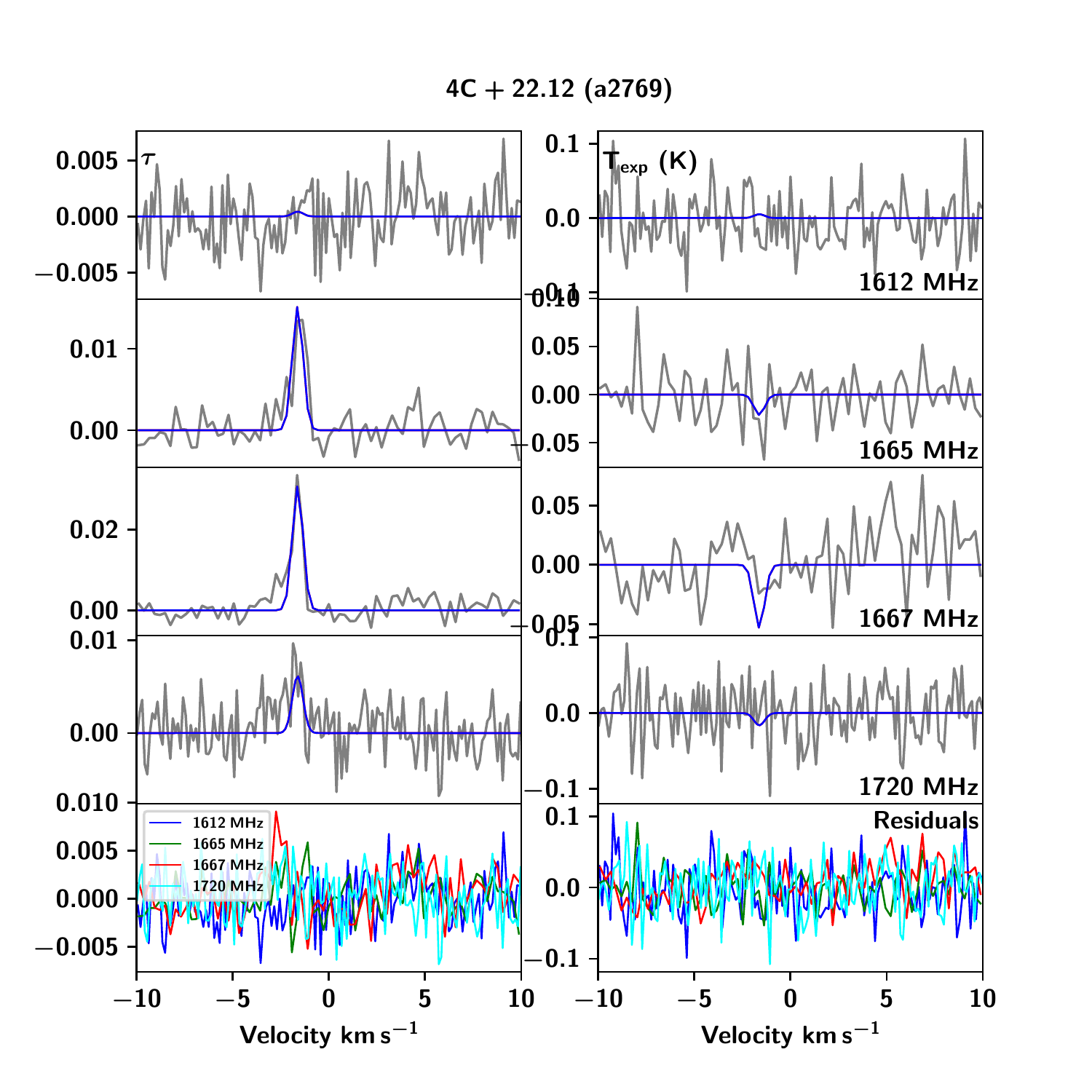}\\
    \end{tabular}
    \caption{Same as Fig. \ref{fig:results1} for 4C+16.15b, 4C+17.23, 4C+17.41, 4C+19.18, 4C+19.19 and 4C+22.12.}
    \label{fig:results3}
\end{figure*}

\begin{figure*}
    \centering
    \begin{tabular}{cc}
    \includegraphics[trim={0cm 0.6cm 0cm 1cm}, clip=true, width=0.45\linewidth]{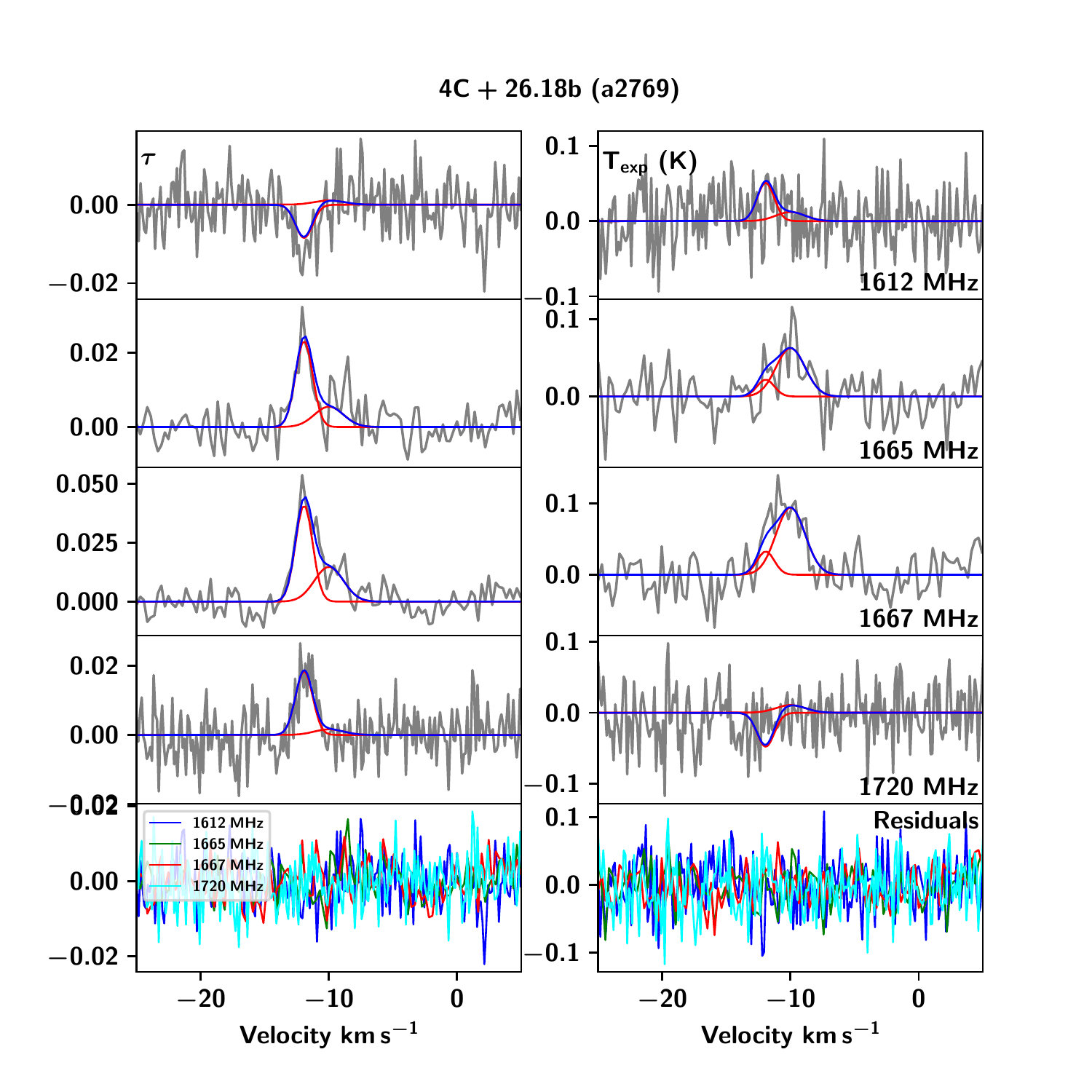}&
    \includegraphics[trim={0cm 0.6cm 0cm 1cm}, clip=true, width=0.45\linewidth]{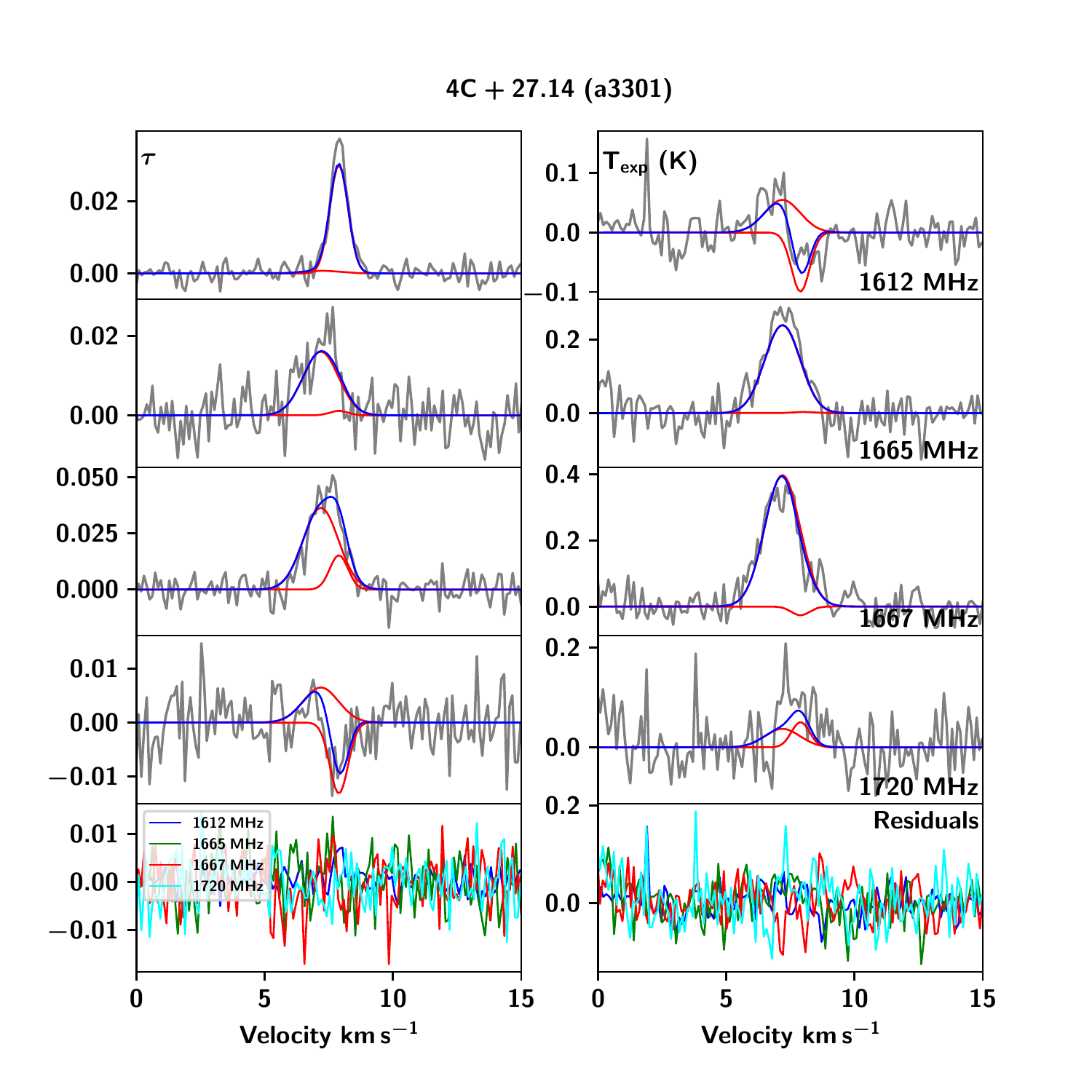}\\
    \includegraphics[trim={0cm 0.6cm 0cm 1cm}, clip=true, width=0.45\linewidth]{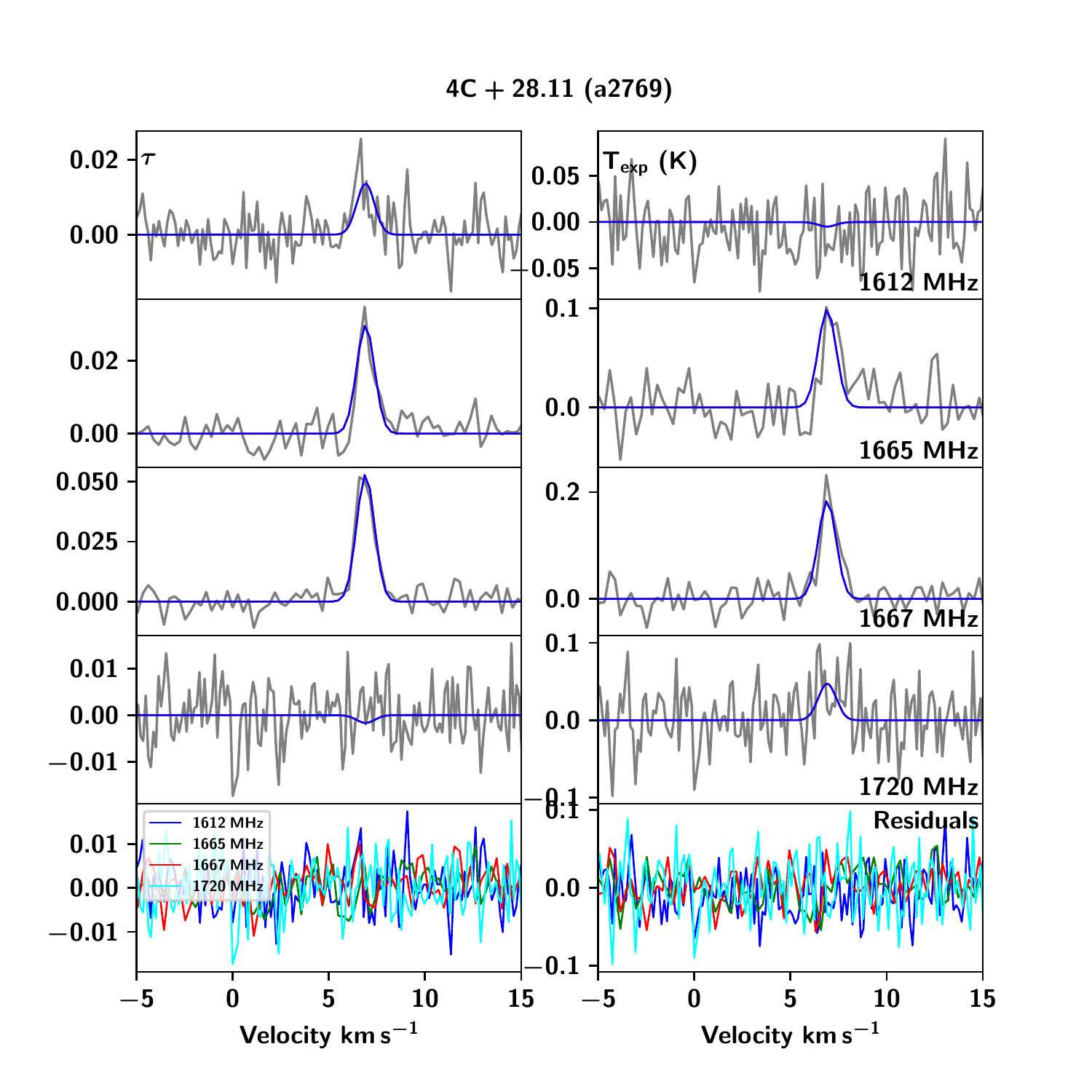}&
    \includegraphics[trim={0cm 0.6cm 0cm 1cm}, clip=true, width=0.45\linewidth]{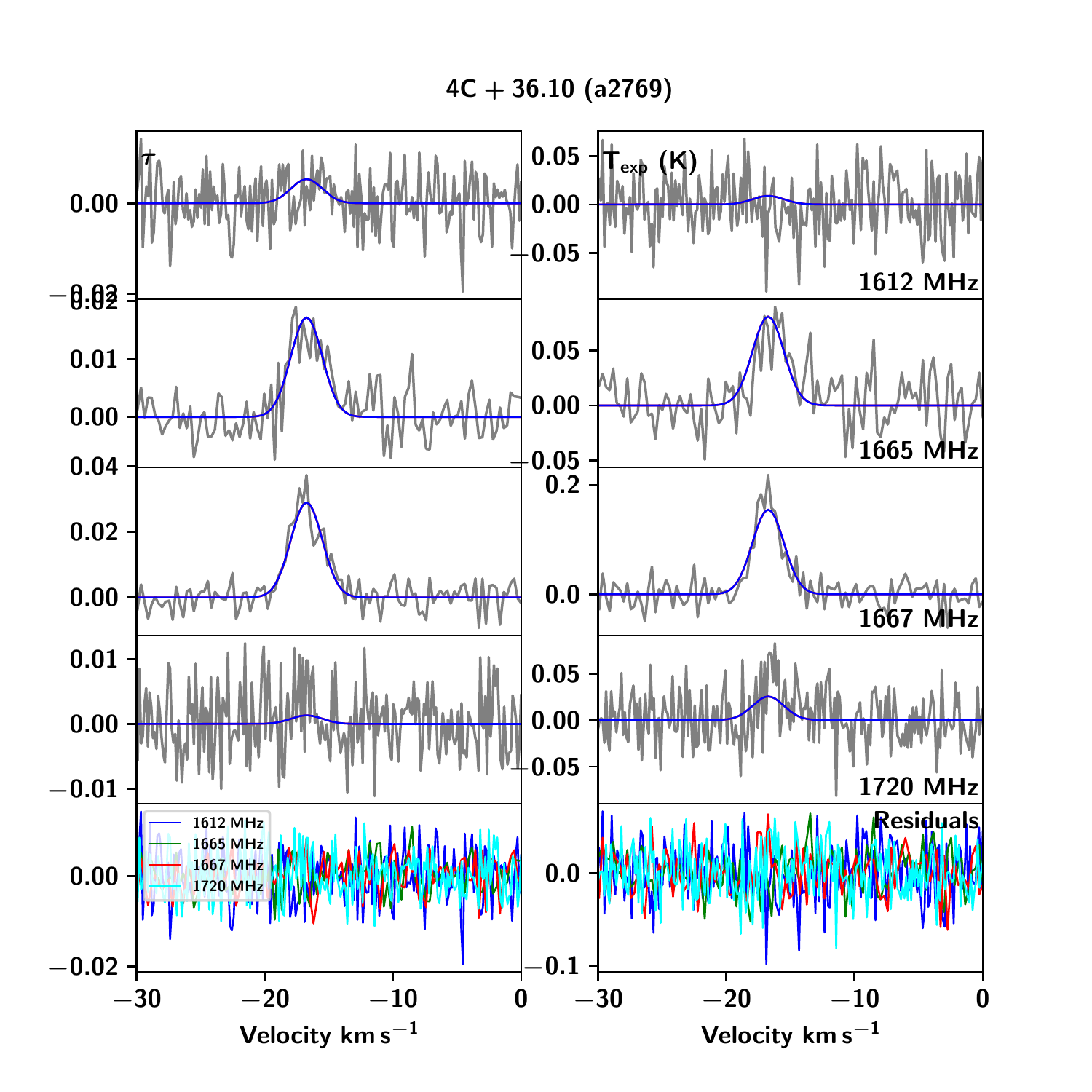}\\
    \multicolumn{2}{c}{\includegraphics[trim={0cm 0.6cm 0cm 1cm}, clip=true, width=0.45\linewidth]{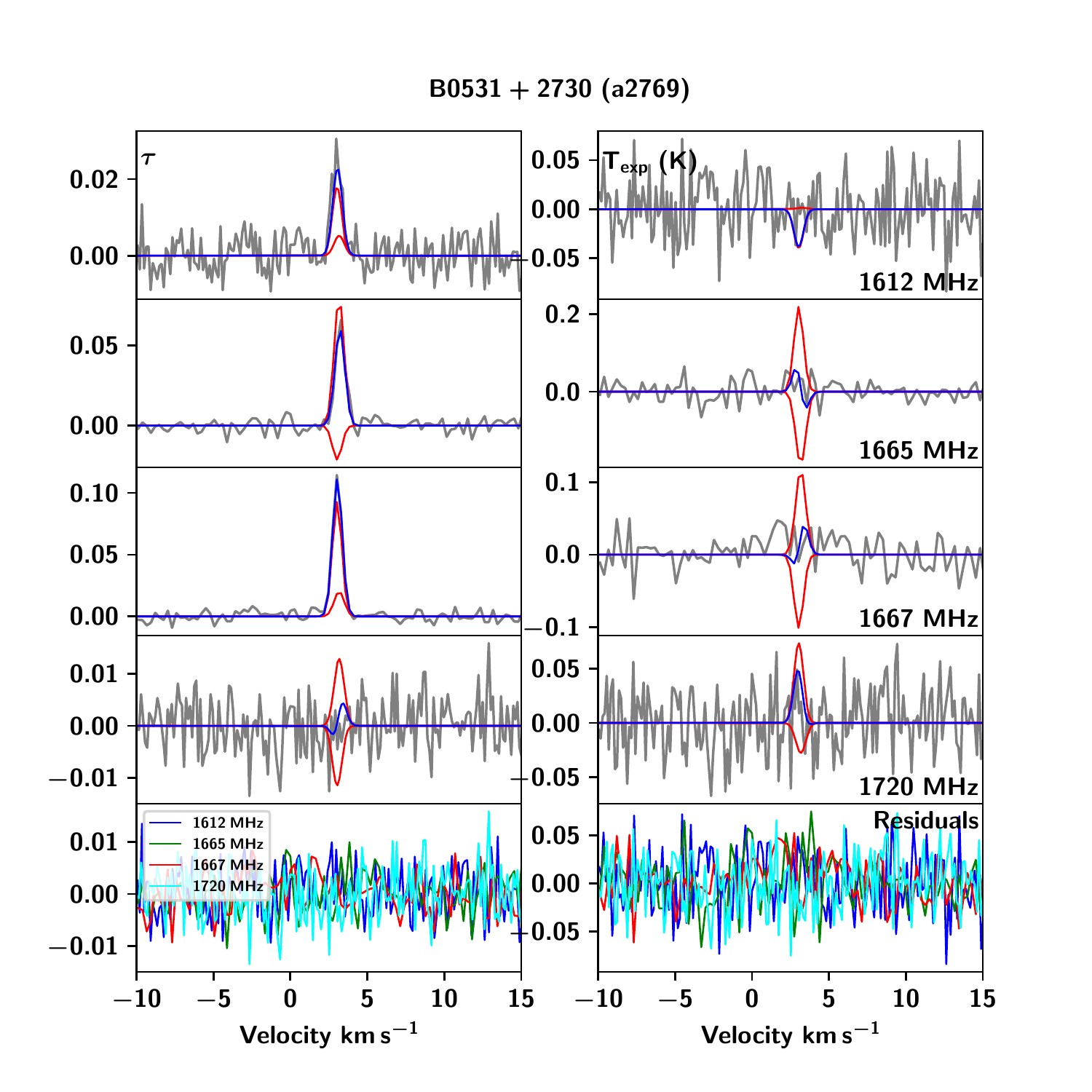}}\\
    \end{tabular}
    \caption{Same as Fig. \ref{fig:results1} for 4C+26.18b, 4C+27.14, 4C+28.11, 4C+36.10 and B0531+2730.}
    \label{fig:results4}
\end{figure*}

% reword this to be shorter, it's already mentioned in the text
Figs. \ref{fig:results1} to \ref{fig:results7} show the results of the Gaussian decomposition of our spectra using \textsc{Amoeba}~\citep{Petzler2021a}. For sightlines observed with the ATCA (Figs. \ref{fig:results5} and \ref{fig:results6}) these plots show optical depth vs velocity for the four OH ground-rotational transitions in grey with the individual Gaussian components in red and the total fit in blue. The residuals of the total fits are shown in the fifth panel, and the sixth panel shows the residual of the optical depth sum rule ($\tau_{\rm peak}(1612)+\tau_{\rm peak}(1720)-\tau_{\rm peak}(1665)/5-\tau_{\rm peak}(1667)/9$) in black.

\subsection{Comments on individual sightlines}

\begin{figure*}
    \centering
    \begin{tabular}{cc}
    \includegraphics[trim={0cm 0.6cm 0cm 1cm}, clip=true, width=0.45\linewidth]{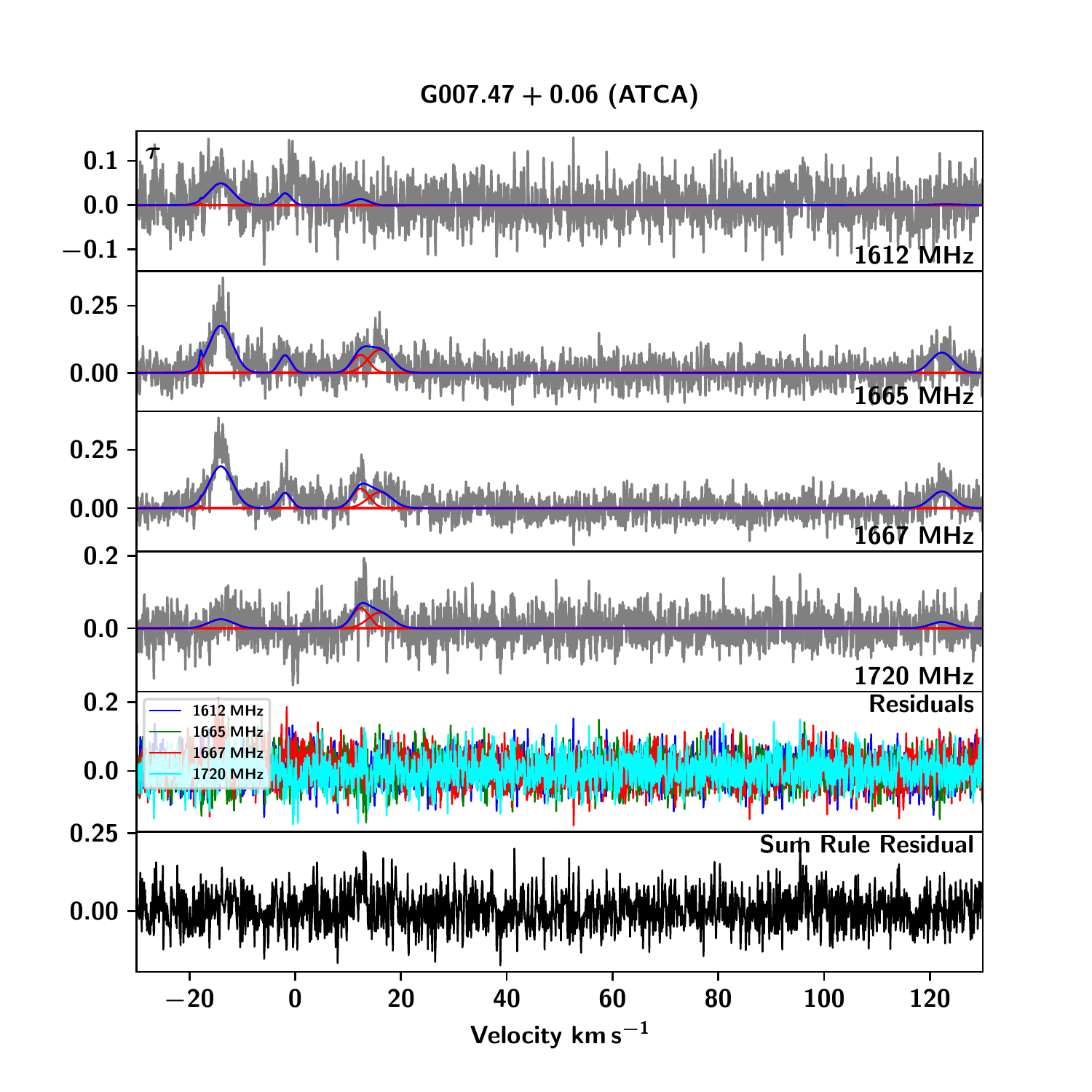}&
    \includegraphics[trim={0cm 0.6cm 0cm 1cm}, clip=true, width=0.45\linewidth]{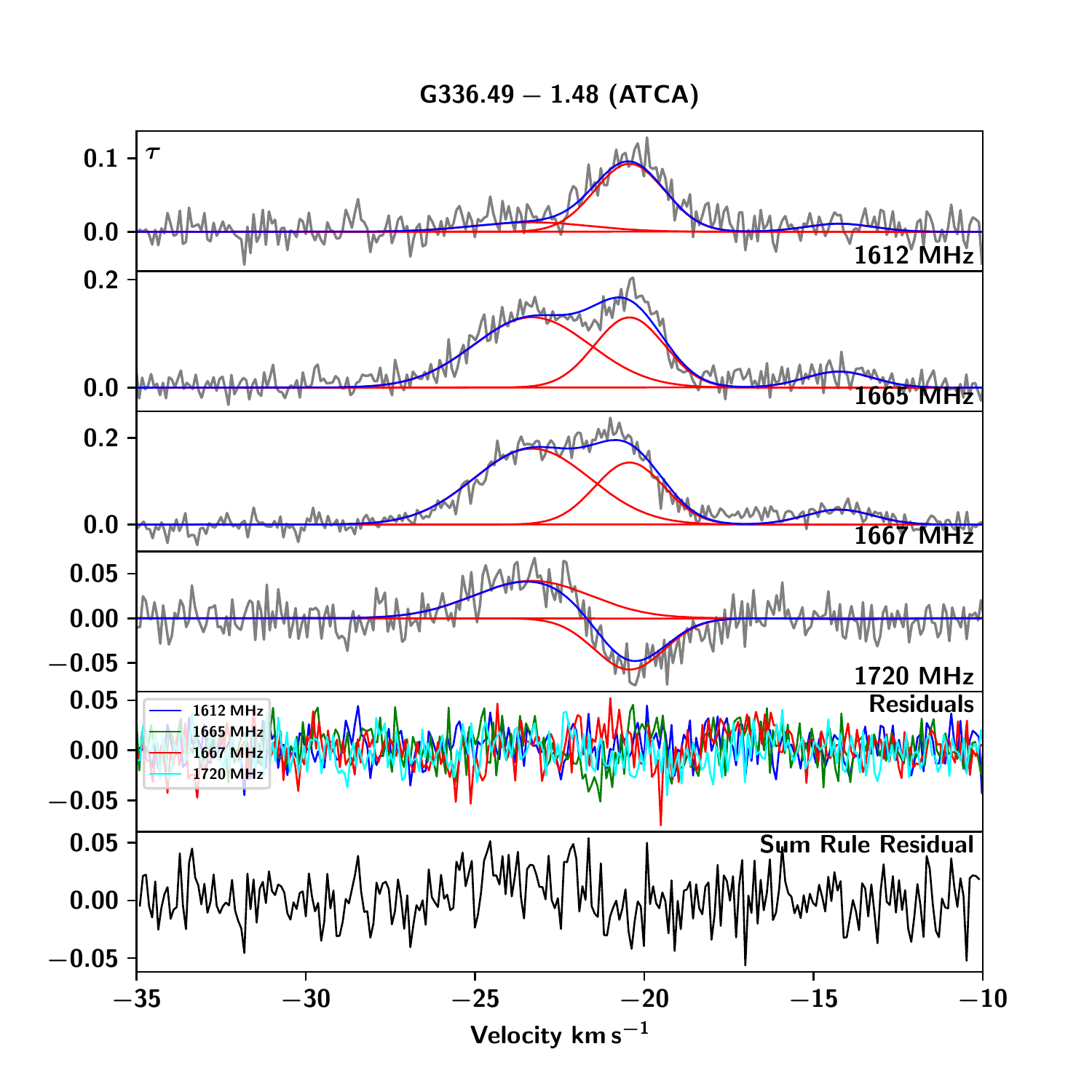}\\
    \includegraphics[trim={0cm 0.6cm 0cm 1cm}, clip=true, width=0.45\linewidth]{Figures/ATCA_g340b.pdf}&
    \includegraphics[trim={0cm 0.6cm 0cm 1cm}, clip=true, width=0.45\linewidth]{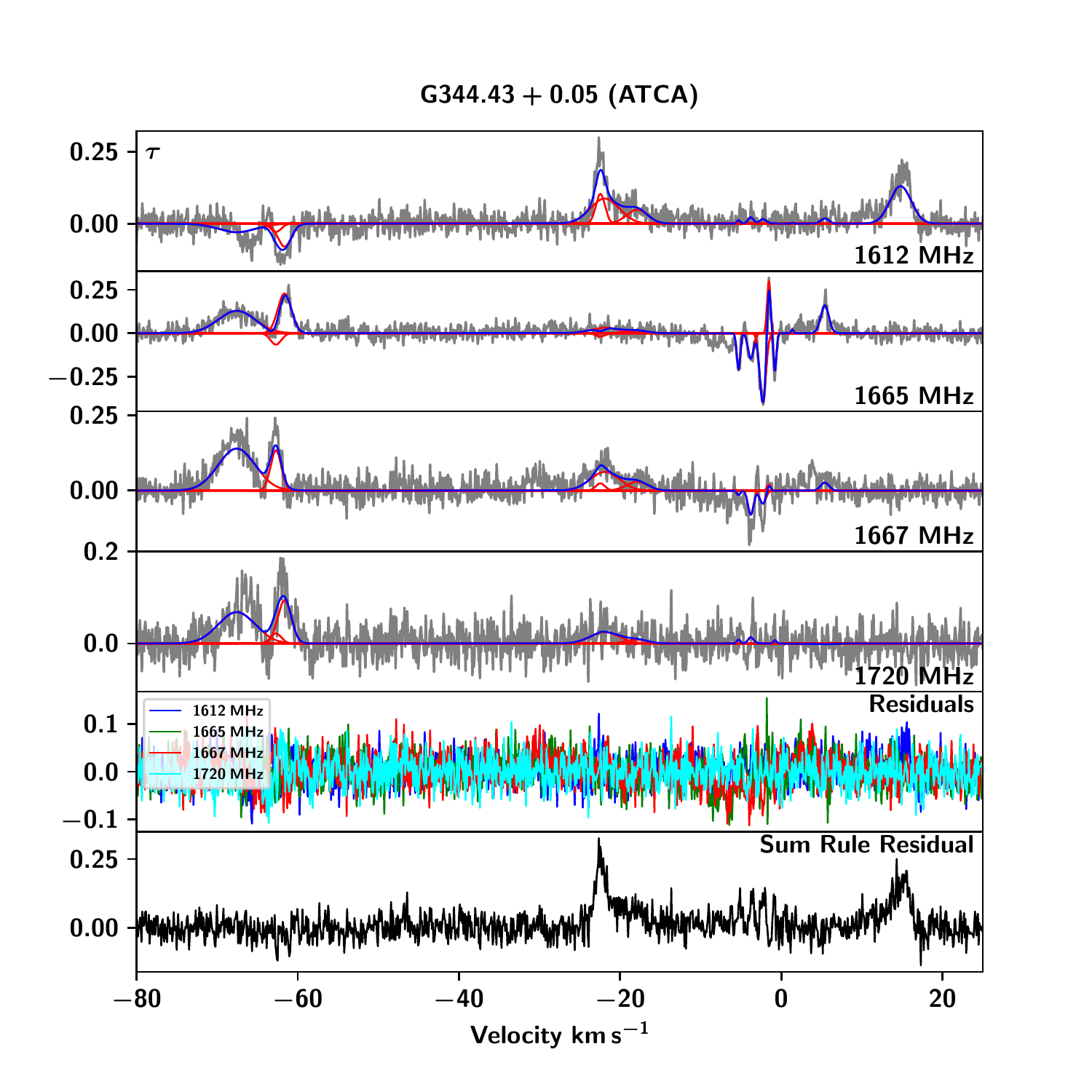}\\
    \includegraphics[trim={0cm 0.6cm 0cm 1cm}, clip=true, width=0.45\linewidth]{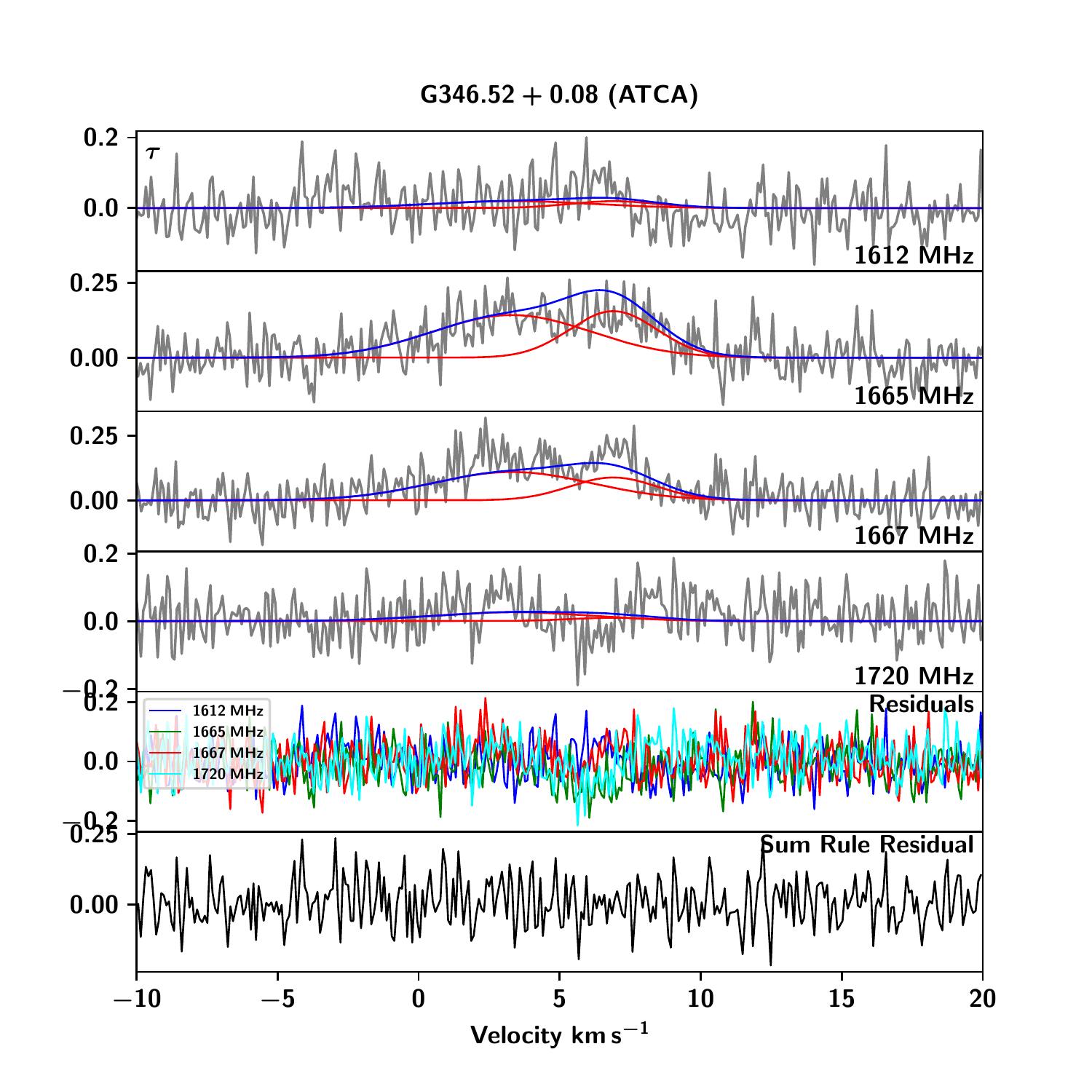}&
    \includegraphics[trim={0cm 0.6cm 0cm 1cm}, clip=true, width=0.45\linewidth]{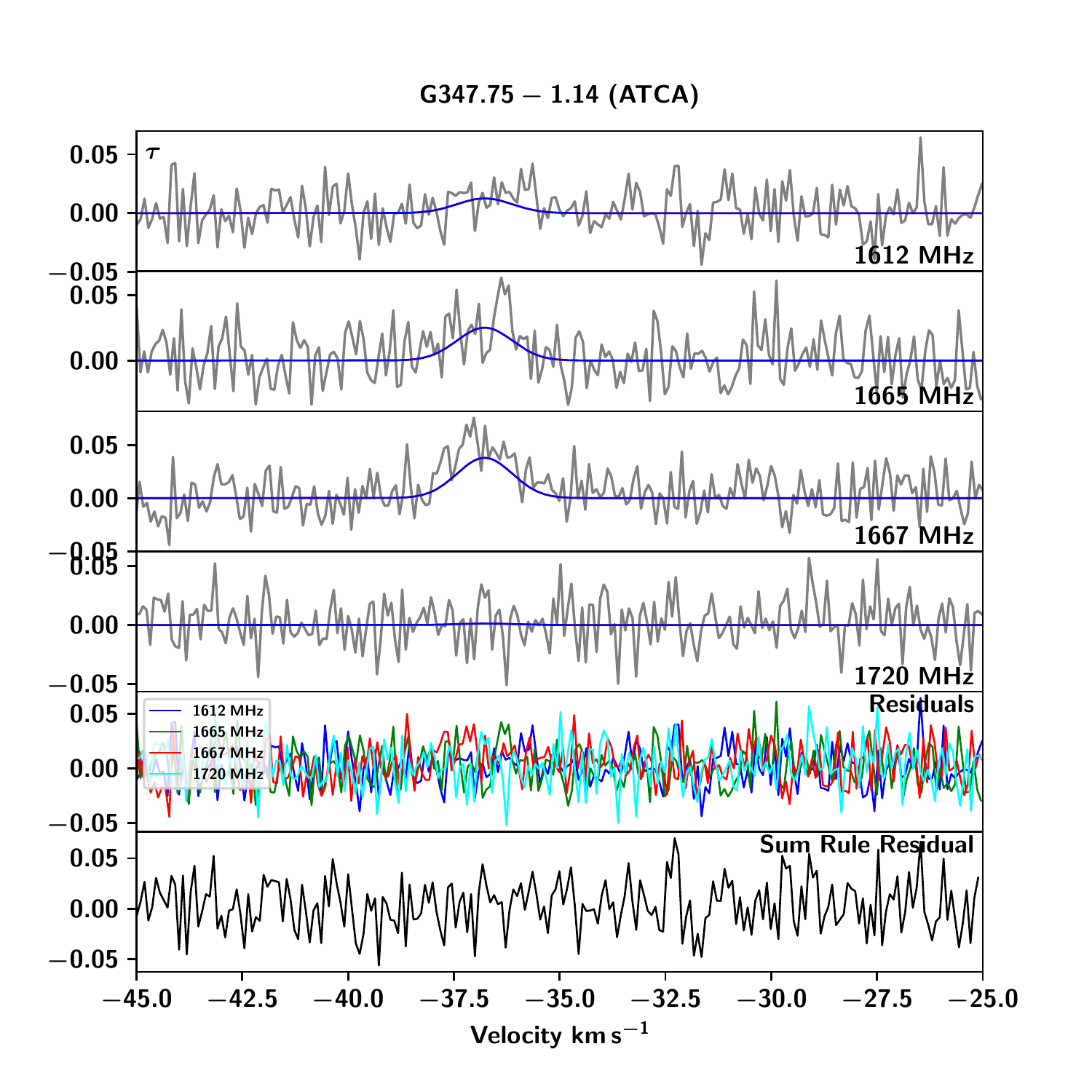}\\
    \end{tabular}
    \caption{The top four panels of each plot show optical depth data in grey, individual features in red and total fit in blue for the 1612, 1665, 1667 and 1720\,MHz transitions. The fifth panel shows the residuals of the total fits with 1612\,MHz in blue, 1667\,MHz in green, 1667\,MHz in red, and 1720\,MHz in cyan. The bottom panel shows the residual of the OH optical depth sum rule: $\tau_{\rm peak}(1612)+\tau_{\rm peak}(1720)-\tau_{\rm peak}(1665)/5-\tau_{\rm peak}(1667)/9$. This figure shows the sightlines toward G007.47+0.06, G336.49-1.48, G340.79-1.02, G344.43+0.05, G346.52+0.08 and G347.75-1.14.}
    \label{fig:results5}
\end{figure*}

\begin{figure*}
    \centering
    \begin{tabular}{cc}
    \includegraphics[trim={0cm 0.6cm 0cm 1cm}, clip=true, width=0.45\linewidth]{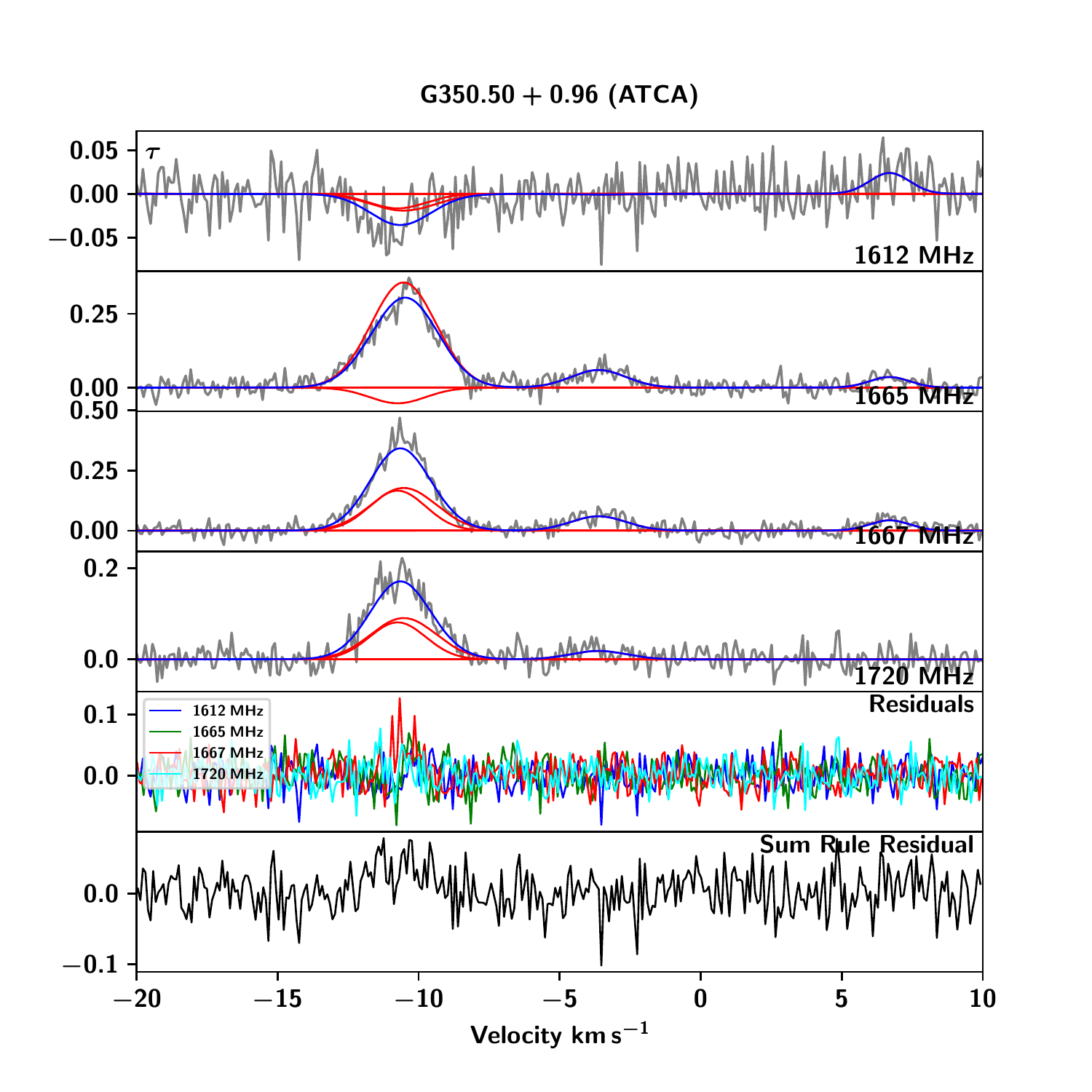}&
    \includegraphics[trim={0cm 0.6cm 0cm 1cm}, clip=true, width=0.45\linewidth]{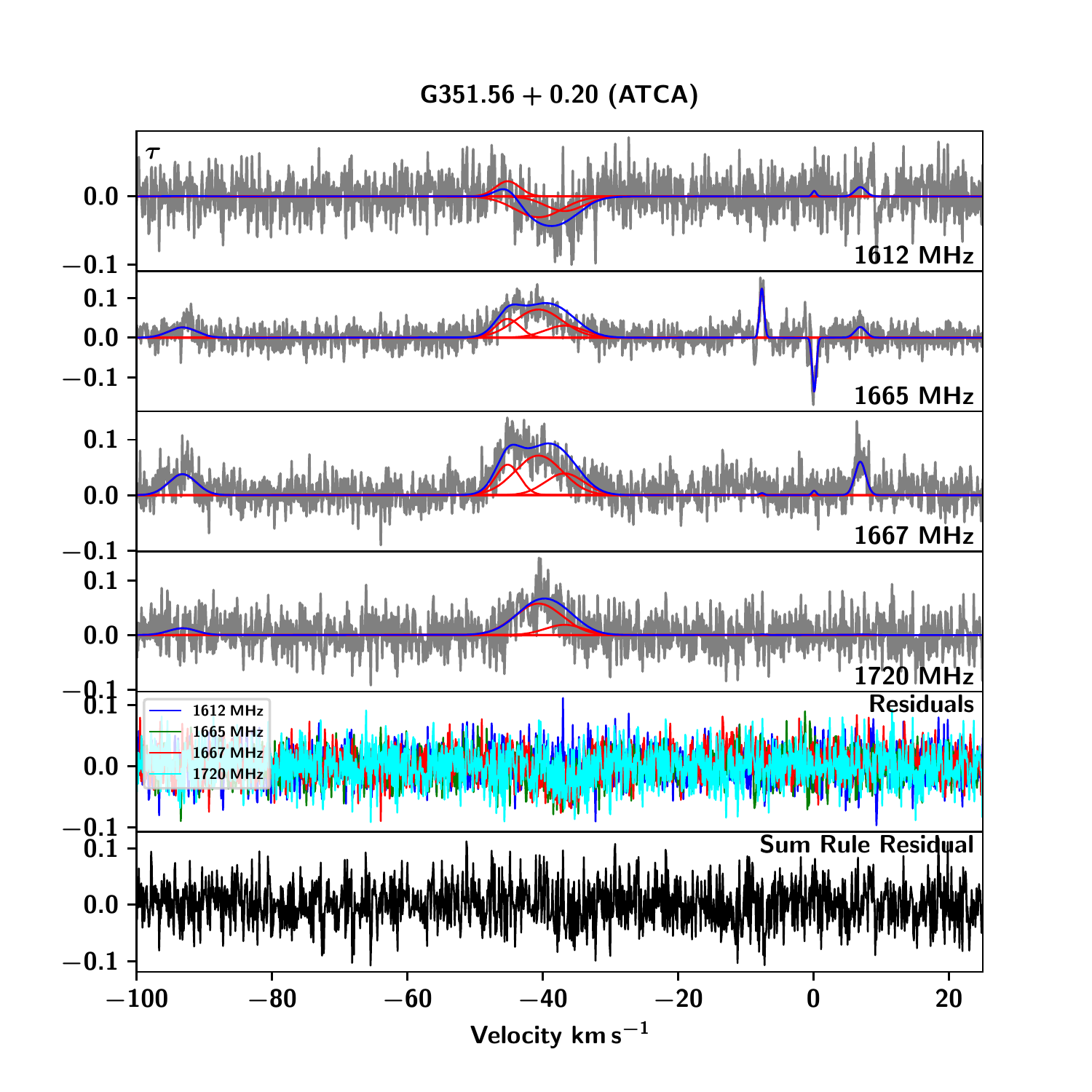}\\
    \includegraphics[trim={0cm 0.6cm 0cm 1cm}, clip=true, width=0.45\linewidth]{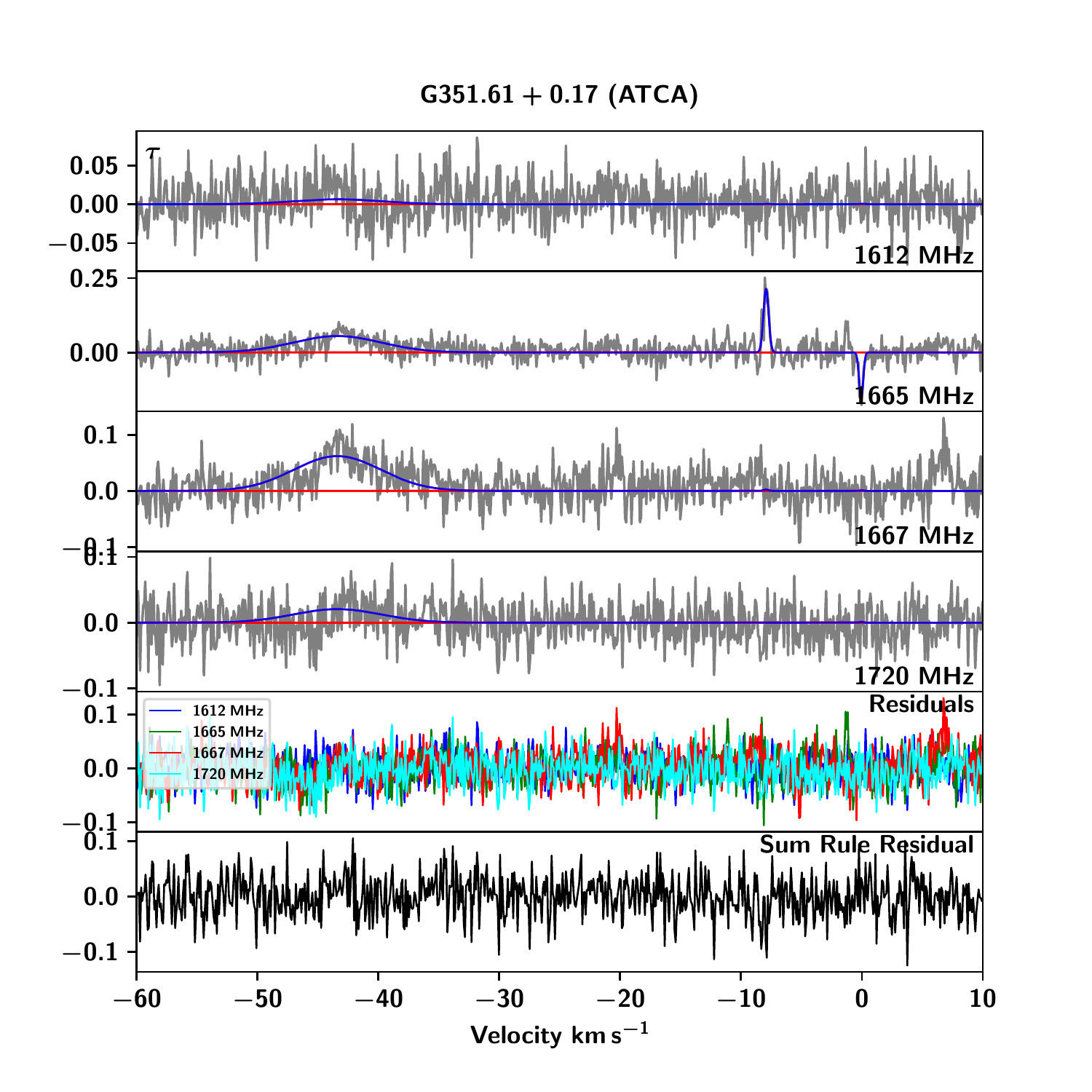}&
    \includegraphics[trim={0cm 0.6cm 0cm 1cm}, clip=true, width=0.45\linewidth]{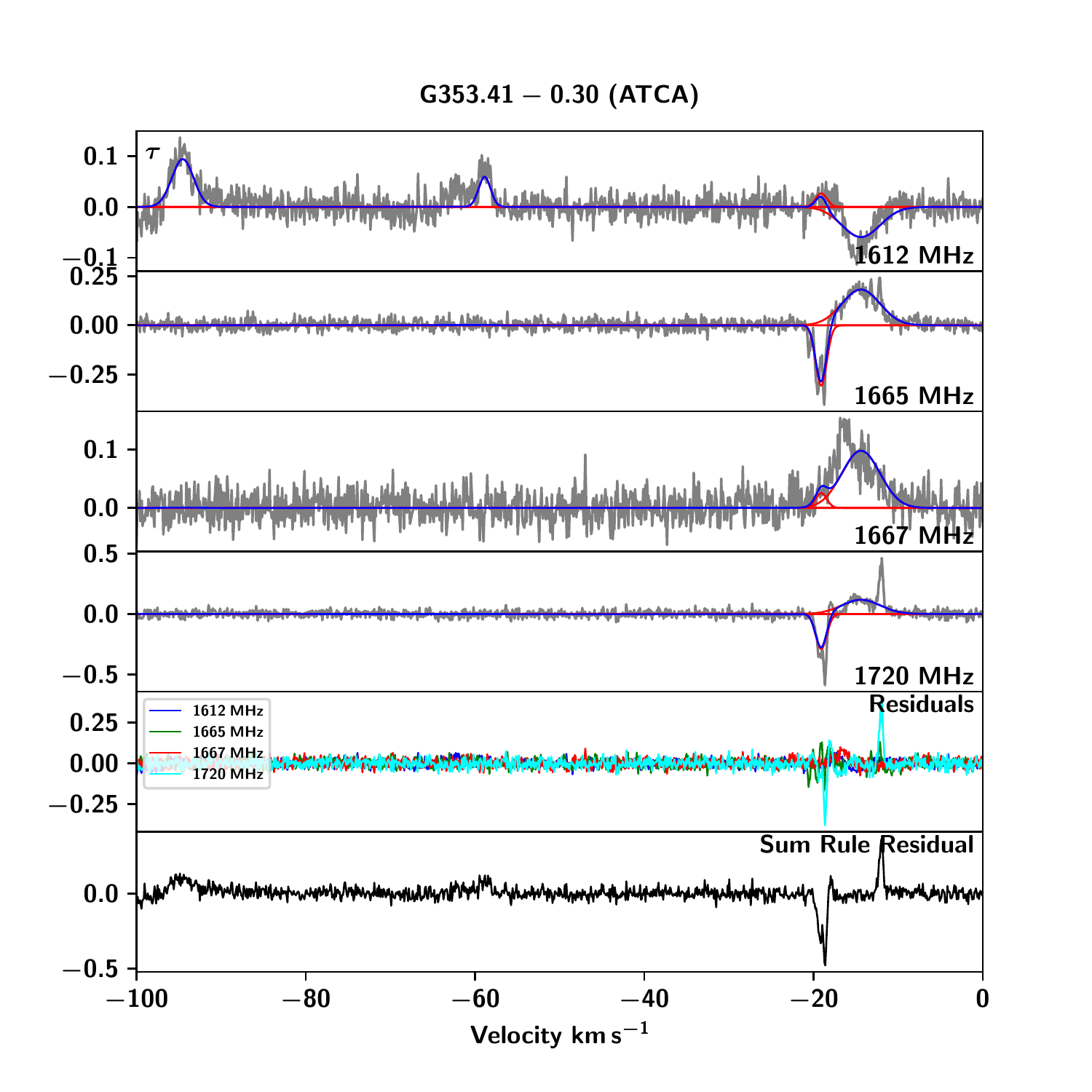}\\
    \multicolumn{2}{c}{\includegraphics[width=0.45\linewidth]{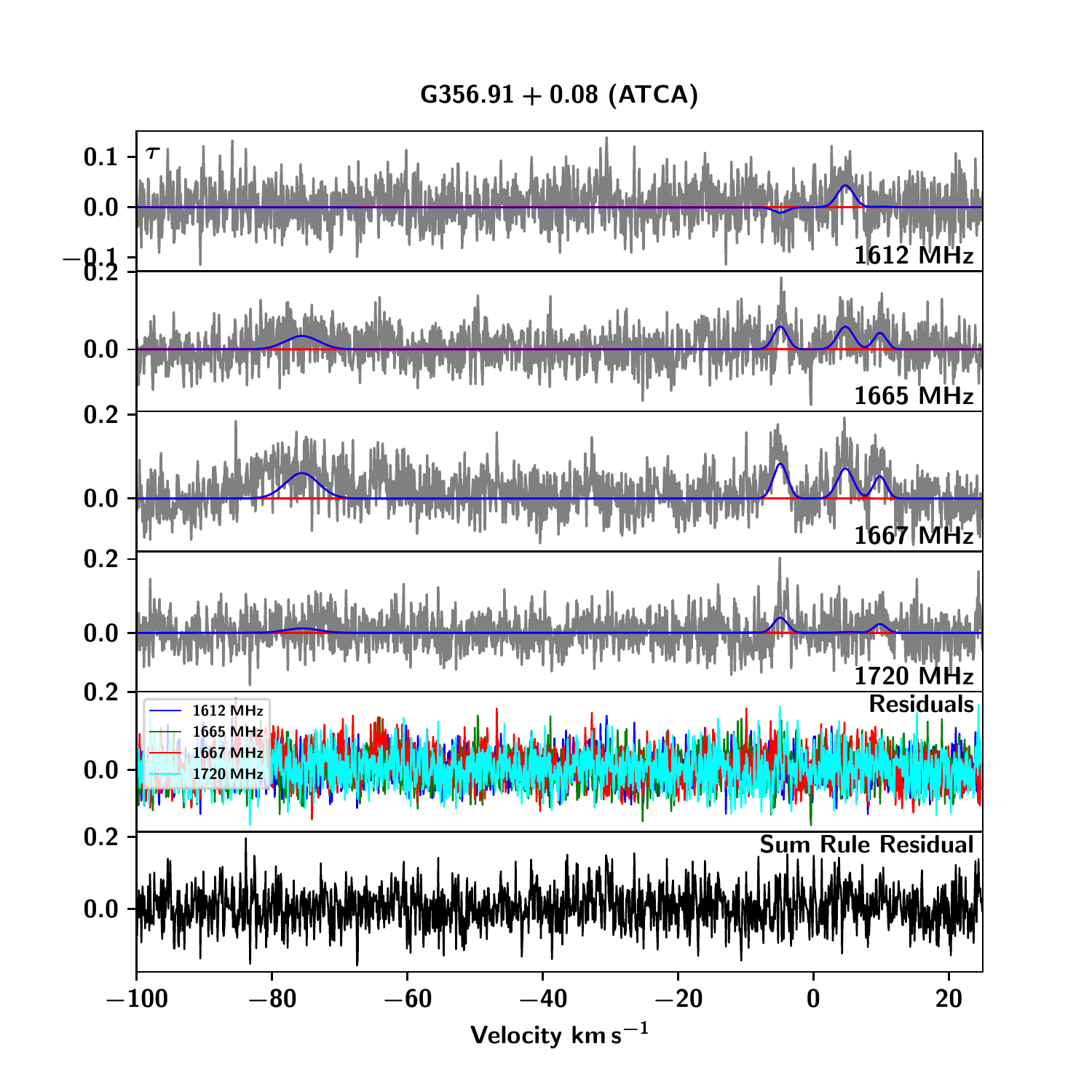}}\\
    \end{tabular}
    \caption{Same as Fig. \ref{fig:results5} for G350.50+0.96, G351.56+0.20, G351.61+0.17, G353.41-0.30 and G356.91+0.08.}
    \label{fig:results6}
\end{figure*}

\begin{figure*}
    \centering
    \begin{tabular}{cc}
    \includegraphics[trim={0cm 0.6cm 0cm 1cm}, clip=true, width=0.45\linewidth]{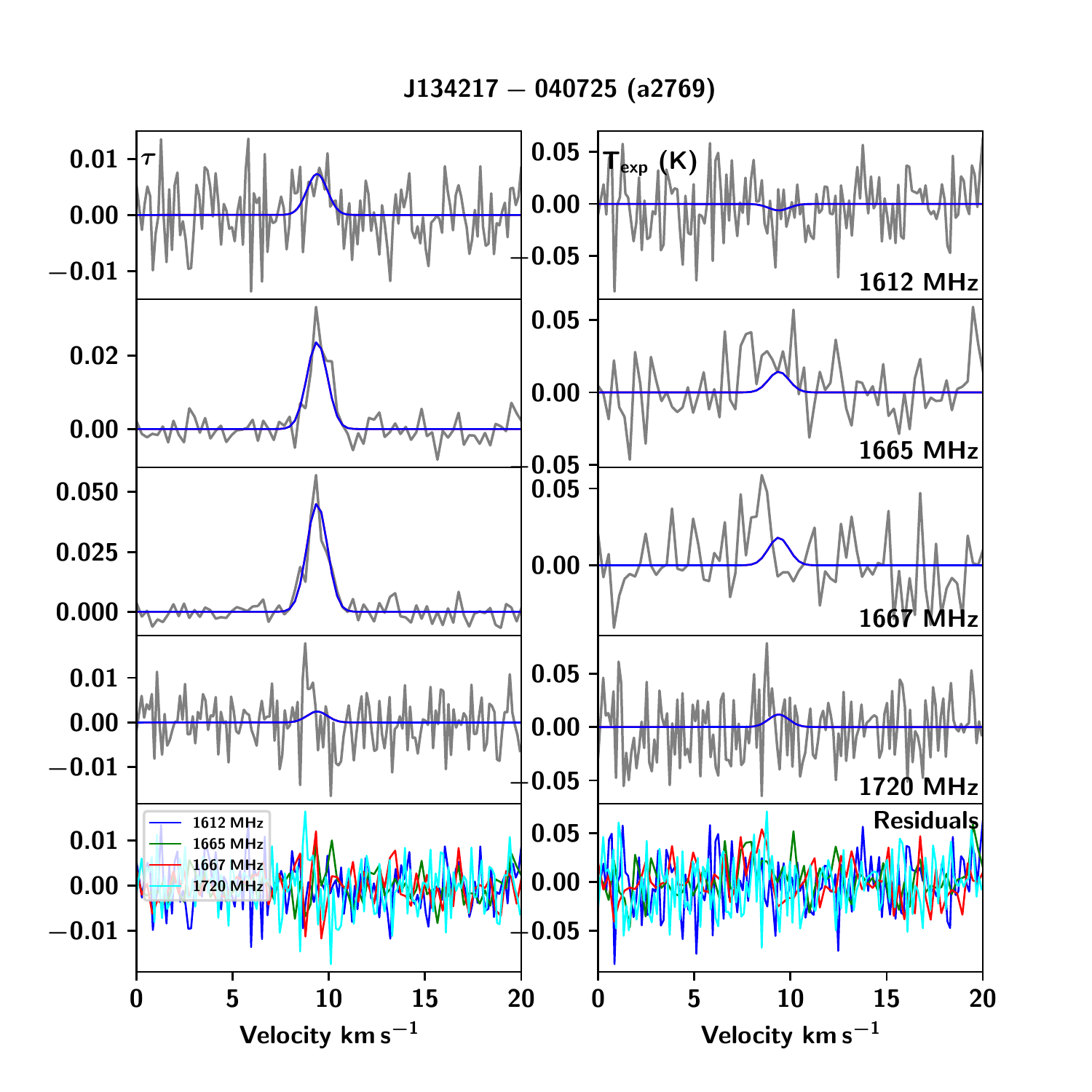}&
    \includegraphics[trim={0cm 0.6cm 0cm 1cm}, clip=true, width=0.45\linewidth]{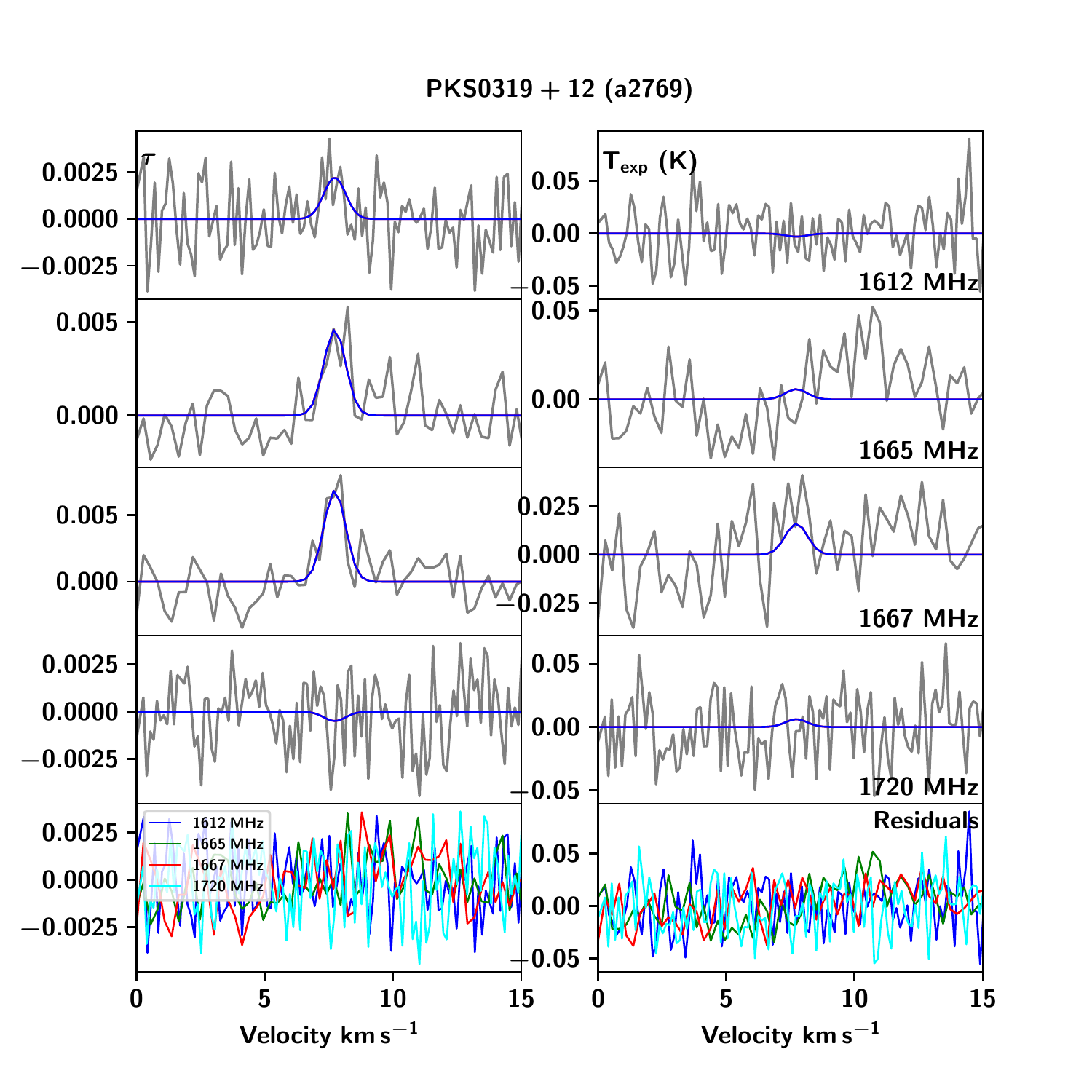}\\
    \includegraphics[trim={0cm 0.6cm 0cm 1cm}, clip=true, width=0.45\linewidth]{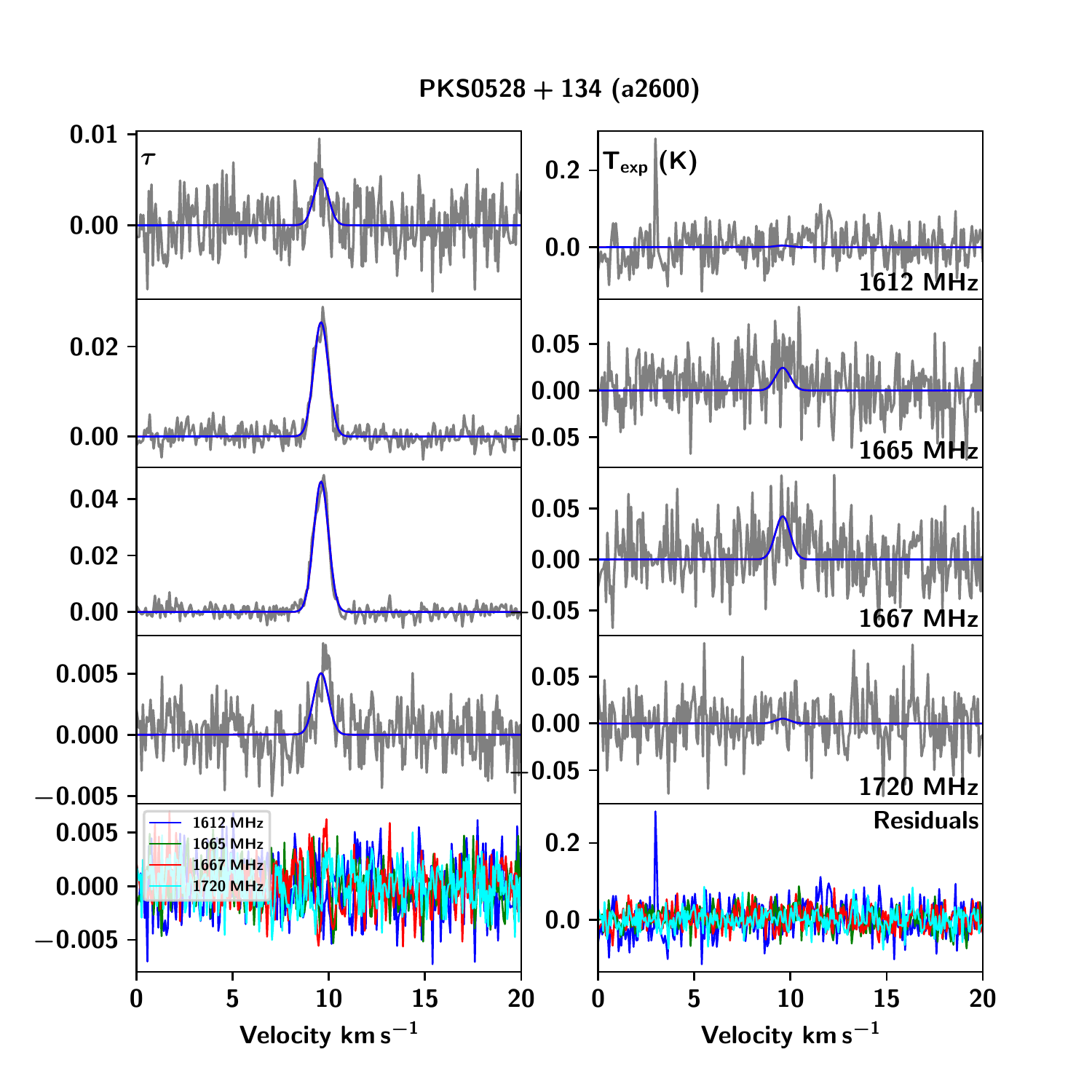}&
    \includegraphics[trim={0cm 0.6cm 0cm 1cm}, clip=true, width=0.45\linewidth]{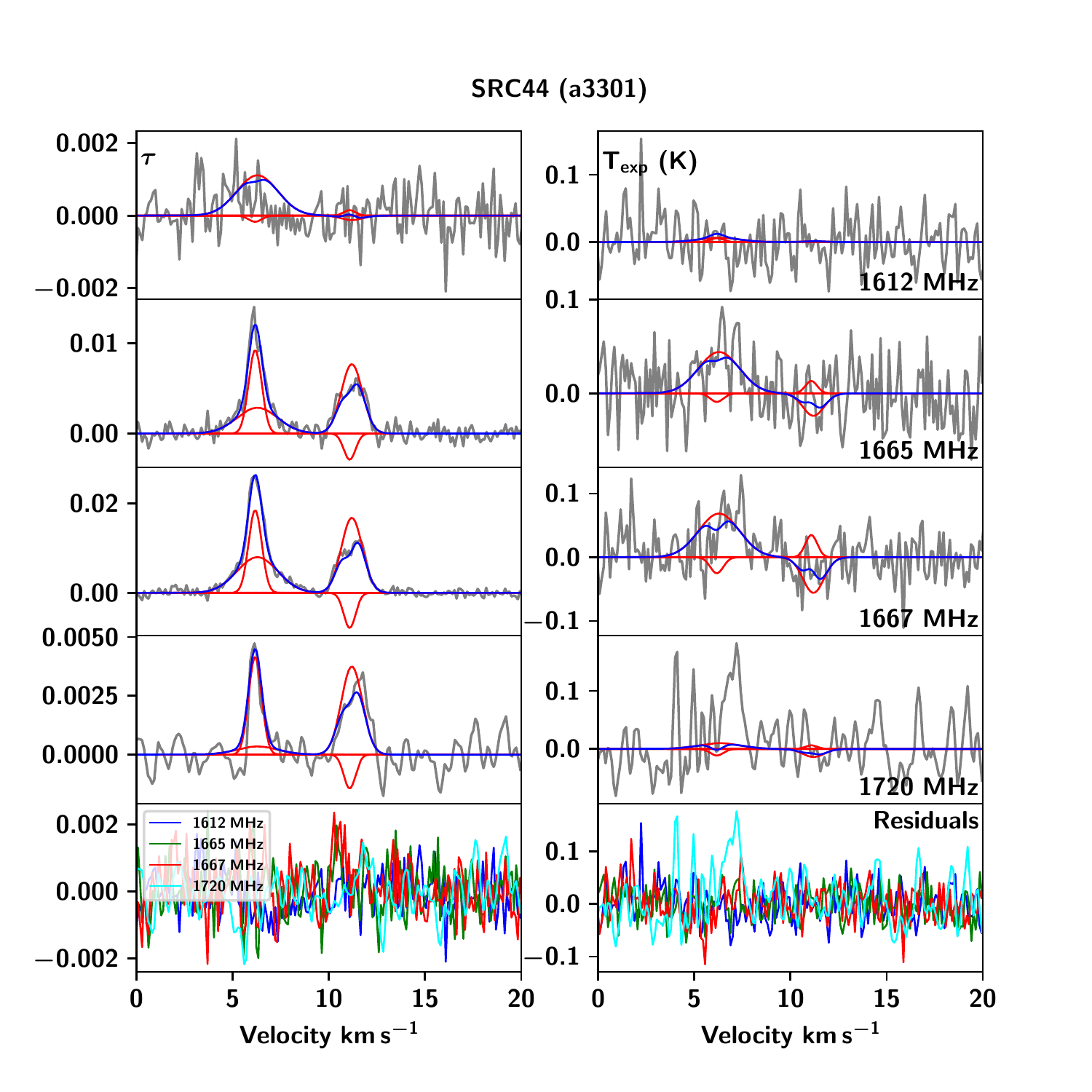}\\
    \end{tabular}
    \caption{Same as Fig. \ref{fig:results1} for J134217-040725, PKS0319+12, PKS0528+134 and SRC44.}
    \label{fig:results7}
\end{figure*}

Significant departures from the sum rule are evident for the sightlines towards G344.43+0.05 (at $\approx-22$ and 15\,km\,s$^{-1}$, see Fig. \ref{fig:results5}) and G353.41-0.30 (at $\approx -95$, $-59$, $-19$ and $-12$\,km\,s$^{-1}$, see Fig. \ref{fig:results6}). Both features in the sum rule residuals towards G344.43+0.05 and those at $-95$ and $-59$\,km\,s$^{-1}$~towards G353.41-0.30 are due primarily to features seen in the 1612\,MHz line and resemble the profile of a `double-horned' maser \citep[see e.g. Fig. 2 in][for representative examples]{Caswell1999}. These double-horn masers arise in evolved stellar envelopes \citep[e.g.][]{deJong1983,Werner1980,Hyland1972}, that due to their expansion are observed as two Doppler-shifted components. Such masers of course have negative optical depths, and we attribute these features in our data to the presence of 1612\,MHz masers in the negative sidelobes of the sightline.

The deviation seen towards G353.41-0.30 at $-12$\,km\,s$^{-1}$~is dominated by a feature in the 1720\,MHz line and was not fit by \textsc{Amoeba}. This feature is likely not fit because its deviation from the sum rule of nearly 0.5 is penalised by our previously mentioned weak prior. As a comparison, the maximum deviation from the sum rule across our ATCA data set for features that \textsc{Amoeba}~\textit{did} fit is the neighbouring feature at $-19$\,km\,s$^{-1}$~along this same sightline, which had a sum rule deviation of $-0.2$. This feature at $-19$\,km\,s$^{-1}$~also has significant optical depth at 1665\,MHz, where the feature at $-12$\,km\,s$^{-1}$~only had a marginal feature at 1665\,MHz. 
There is a 1720\,MHz maser towards this background source at $-19.4$\,km\,s$^{-1}$~\citep{Caswell2004,Ogbodo2020}, as well as a 1667\,MHz maser at $-19.7$\,km\,s$^{-1}$~\citep{Caswell1998}. On the other hand we were not able to identify any known 1720\,MHz masers at $-12$\,km\,s$^{-1}$~along this sightline or nearby. 
This is a good demonstration of \textsc{Amoeba}'s hesitancy to fit single-transition features that violate the optical depth sum rule, as the lack of significant signal in the other transitions lead to preference of the null model (i.e. the absence of a feature).



\section{Comparing OH to H\textsc{i} CNM}

\onecolumn
\begin{table}
    \centering
    \begin{tabular}{llrrrrrrc}
\hline
&&&&\multicolumn{2}{c}{OH fits}&\multicolumn{3}{c}{Matched H\textsc{i}~CNM fits}\\
Source&Project&$l^{\circ}$&$b^{\circ}$&\multicolumn{1}{c}{$v$}&\multicolumn{1}{c}{$\Delta v$}&\multicolumn{1}{c}{$v$}&\multicolumn{1}{c}{$\Delta v$}&$\exp(-\tau_{\rm peak})$\\
&&&&\multicolumn{2}{c}{km\,s$^{-1}$}&\multicolumn{2}{c}{km\,s$^{-1}$}&\\
\hline
3C092&a2600&159.74&-18.41&8.71&1.56&9.37&1.17&0.5\\
3C092&a2600&159.74&-18.41&8.77&0.76&9.37&1.17&0.5\\
3C131&a2600&171.44&-7.80&4.56&0.48&5.15&4.23&0.1\\
3C131&a2600&171.44&-7.80&5.71&3.21&5.15&4.23&0.1\\
3C131&a2600&171.44&-7.80&6.59&0.44&5.15&4.23&0.1\\
3C131&a2600&171.44&-7.80&7.23&0.56&5.15&4.23&0.1\\
3C131&a2600&171.44&-7.80&7.48&1.93&5.15&4.23&0.1\\
3C131&a2600&171.44&-7.80&7.79&0.57&5.15&4.23&0.1\\
3C158&a2769&196.64&0.17&3.14&0.98&4.41&4.96&0.5\\
4C+04.22&a2769&205.41&-4.43&11.92&0.71&12.30&3.50&0.5\\
4C+04.22&a2769&205.41&-4.43&13.33&0.92&12.30&3.50&0.5\\
4C+04.24&a2769&205.92&-3.57&9.39&1.25&9.07&1.17&0.0\\
4C+07.13&a2769&178.87&-36.27&3.48&1.07&3.23&2.74&0.6\\
4C+13.32&a2769&197.15&-0.85&-0.35&1.02&-0.68&2.00&0.5\\
4C+13.32&a2769&197.15&-0.85&4.46&1.20&8.75&13.53&0.3\\
4C+13.32&a2769&197.15&-0.85&6.99&2.21&8.75&13.53&0.3\\
4C+13.32&a2769&197.15&-0.85&9.50&1.15&8.75&13.53&0.3\\
4C+14.18&a2600&196.98&1.10&4.28&0.55&4.77&2.43&0.7\\
4C+14.18&a2600&196.98&1.10&4.94&1.84&4.77&2.43&0.7\\
4C+14.18&a2600&196.98&1.10&7.39&0.81&7.47&1.49&0.8\\
4C+14.18&a2600&196.98&1.10&16.49&1.29&17.48&4.13&0.0\\
4C+14.18&a2600&196.98&1.10&17.59&0.70&17.48&4.13&0.0\\
4C+14.18&a2600&196.98&1.10&18.40&3.76&17.48&4.13&0.0\\
4C+14.18&a2600&196.98&1.10&31.98&0.42&32.81&2.30&0.6\\
4C+14.18&a2600&196.98&1.10&32.33&1.19&32.81&2.30&0.6\\
4C+17.23&a2600&176.36&-24.24&9.35&0.72&9.11&1.93&0.5\\
4C+17.23&a2600&176.36&-24.24&11.42&0.77&11.30&3.33&0.4\\
4C+17.41&a2769&201.13&16.42&0.23&1.38&0.94&1.97&0.7\\
4C+17.41&a2769&201.13&16.42&1.89&0.73&0.94&1.97&0.7\\
4C+19.18&a2769&190.09&-2.17&-0.62&2.18&0.20&1.14&0.3\\
4C+19.18&a2769&190.09&-2.17&2.39&1.47&2.87&2.46&0.5\\
4C+19.19&a2769&190.13&-1.64&1.12&1.42&2.26&3.49&0.1\\
4C+19.19&a2769&190.13&-1.64&2.70&3.15&2.26&3.49&0.1\\
4C+22.12&a2769&188.07&0.04&-1.62&0.66&-2.29&1.50&0.2\\
4C+27.14&a2600&175.83&-9.36&-0.80&1.05&-0.69&2.49&0.8\\
4C+27.14&a2600&175.83&-9.36&7.01&1.36&7.13&2.17&0.3\\
4C+27.14&a2600&175.83&-9.36&7.83&0.80&7.13&2.17&0.3\\
4C+28.11&a2769&166.06&-17.22&6.91&1.10&6.81&2.19&0.5\\
4C+36.10&a2769&172.98&2.44&-16.74&2.89&-17.93&3.58&0.2\\
B0531+2730&a2769&179.87&-2.83&3.04&0.72&1.36&5.09&0.1\\
B0531+2730&a2769&179.87&-2.83&3.17&0.78&1.36&5.09&0.1\\
PKS0319+12&a2769&170.59&-36.24&7.73&1.01&7.88&3.20&0.1\\
PKS0528+134&a2600&191.37&-11.01&9.60&0.90&9.68&2.90&0.3\\
\hline
    \end{tabular}
    \caption{OH features identified in this work matched with corresponding H\textsc{i}~CNM components identified by \citet{Nguyen2019} (see text for criteria used to match components). Columns give the targeted background source of each sightline, the project name, Galactic longitude and latitude, centroid velocity $v$, FWHM $\Delta v$, (repeated without uncertainties from Table \ref{tab:tau} for identification) and the centroid velocity $v$, FWHM $\Delta v$~and $\exp (-\tau_{\rm peak})$~found by \citet{Nguyen2019}.}
    \label{tab:match}
\end{table}
\twocolumn

% Comparing my fits to Hiep:
% These plots are all made using 'Plot_Hiep_amoeba.py' and are stored in 'amoeba/plots/'

\begin{figure*}
    \centering
    \begin{tabular}{cc}
    \includegraphics[width=0.45\linewidth]{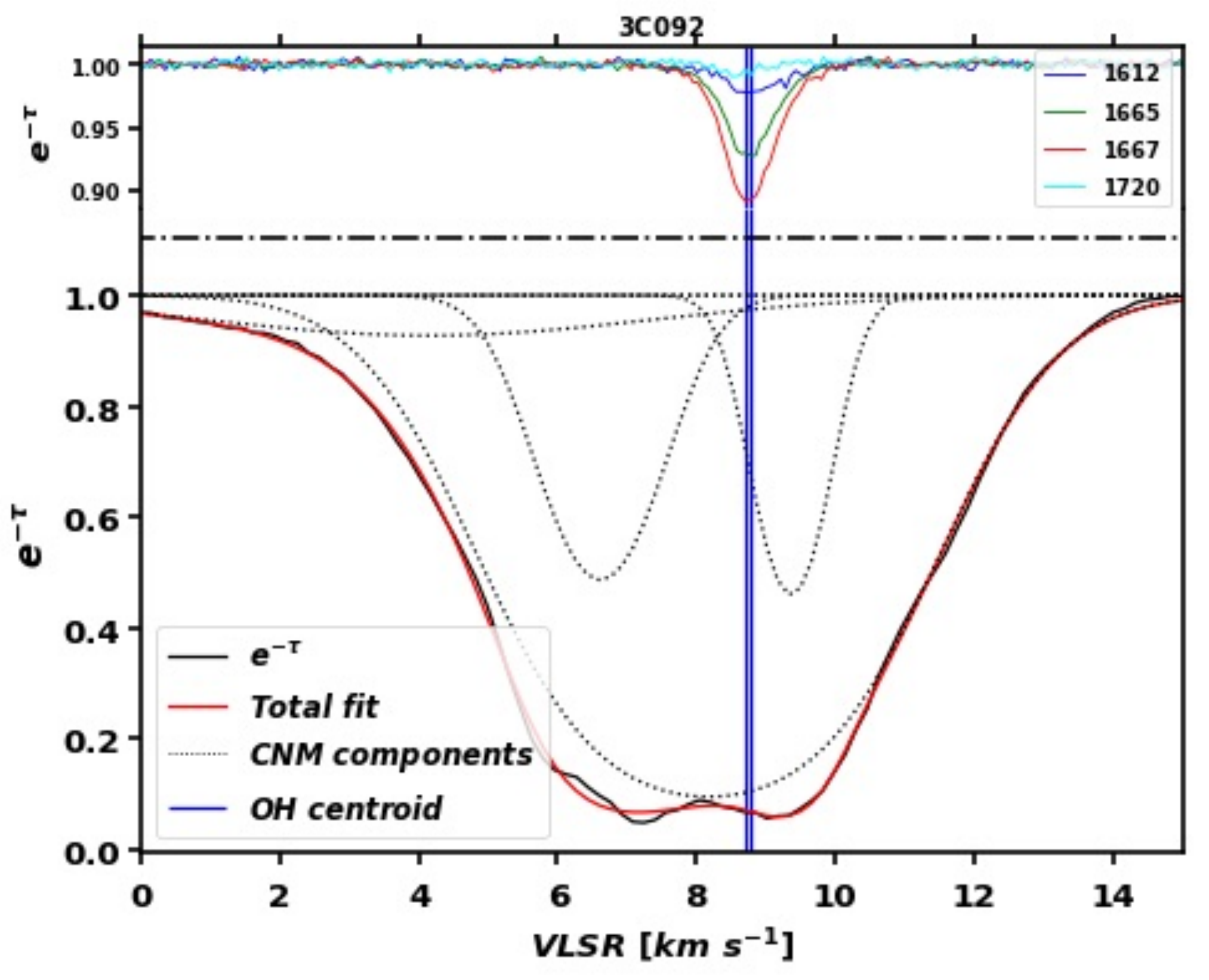}&
    \includegraphics[width=0.45\linewidth]{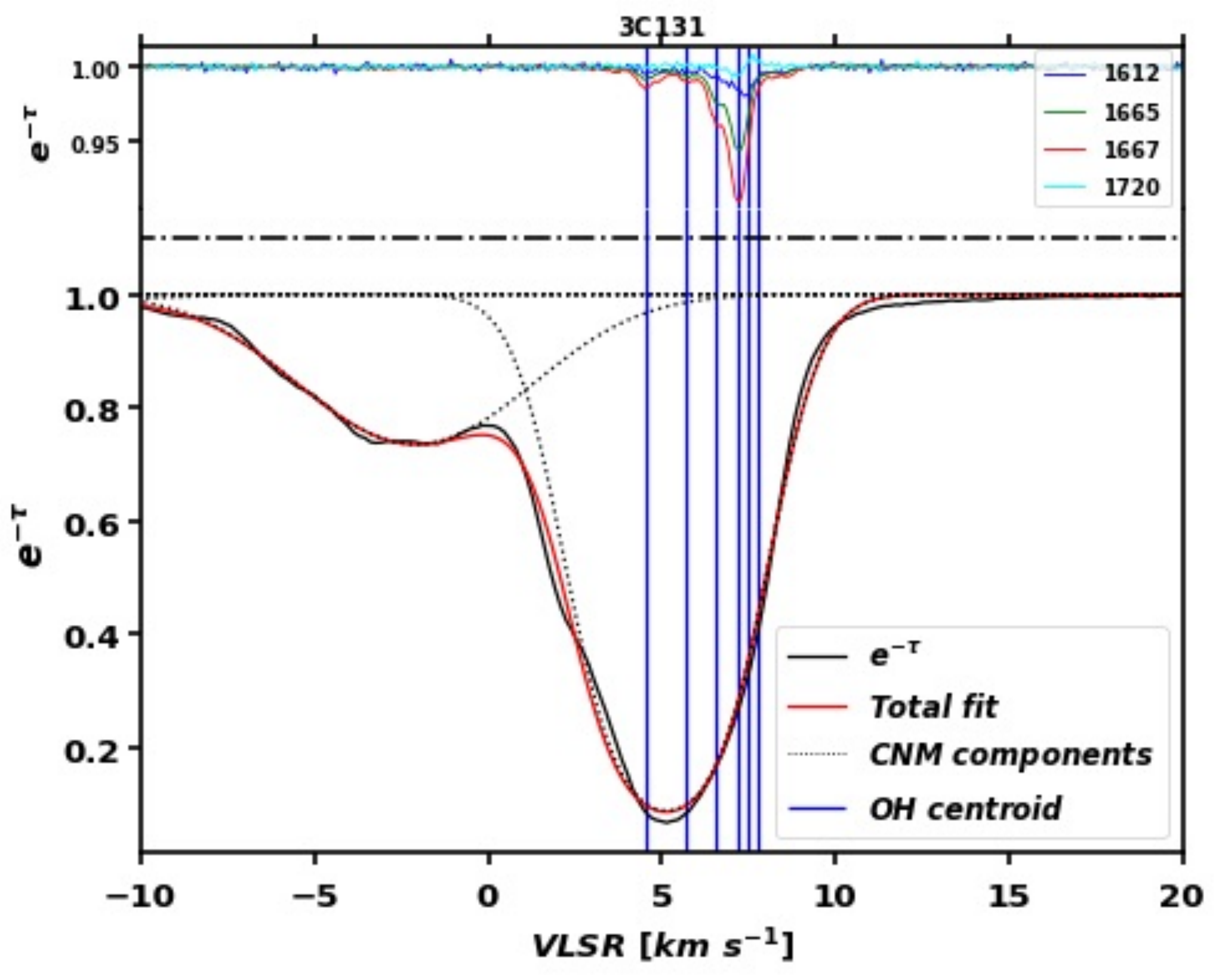}\\
    \includegraphics[width=0.45\linewidth]{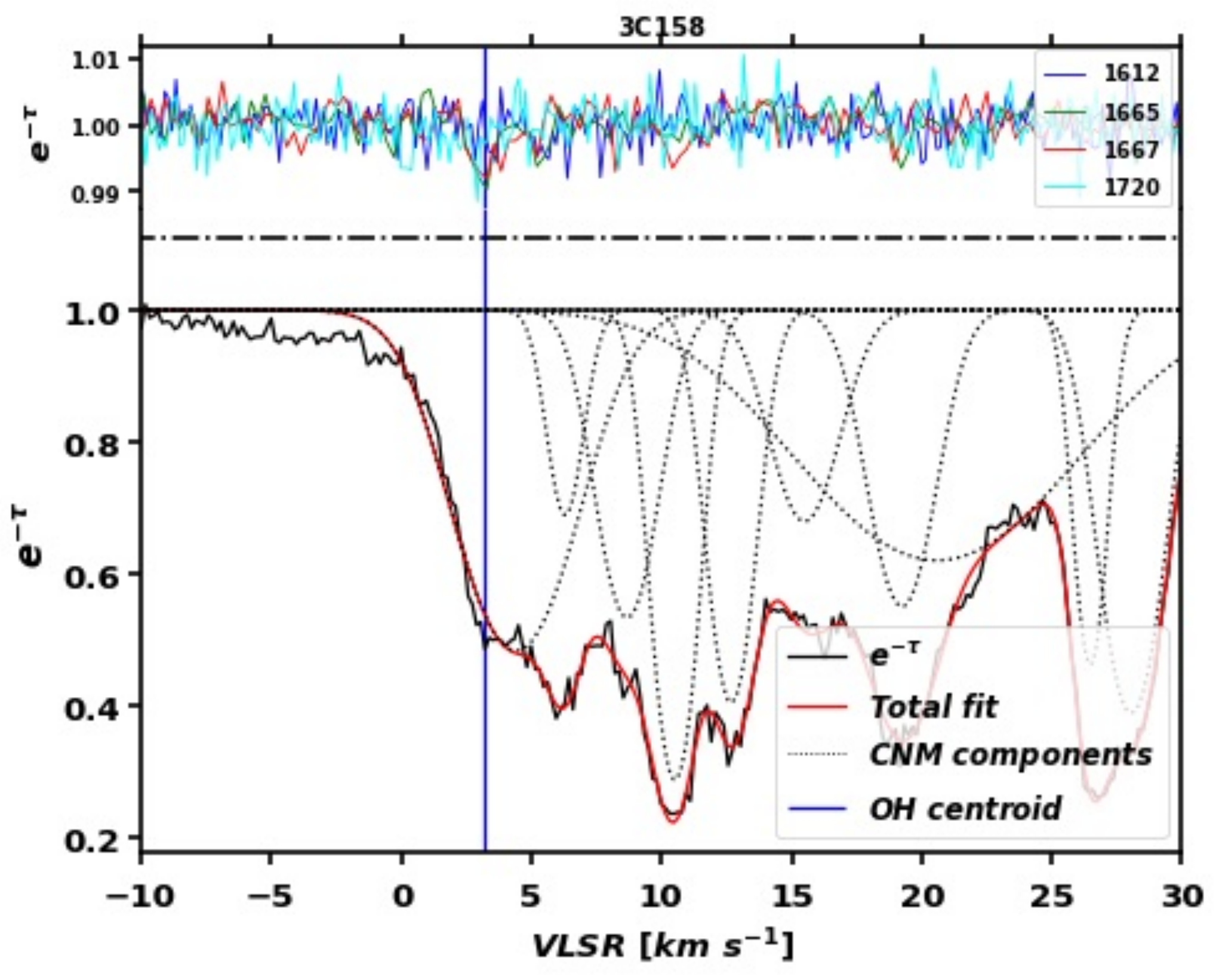}&
    \includegraphics[width=0.45\linewidth]{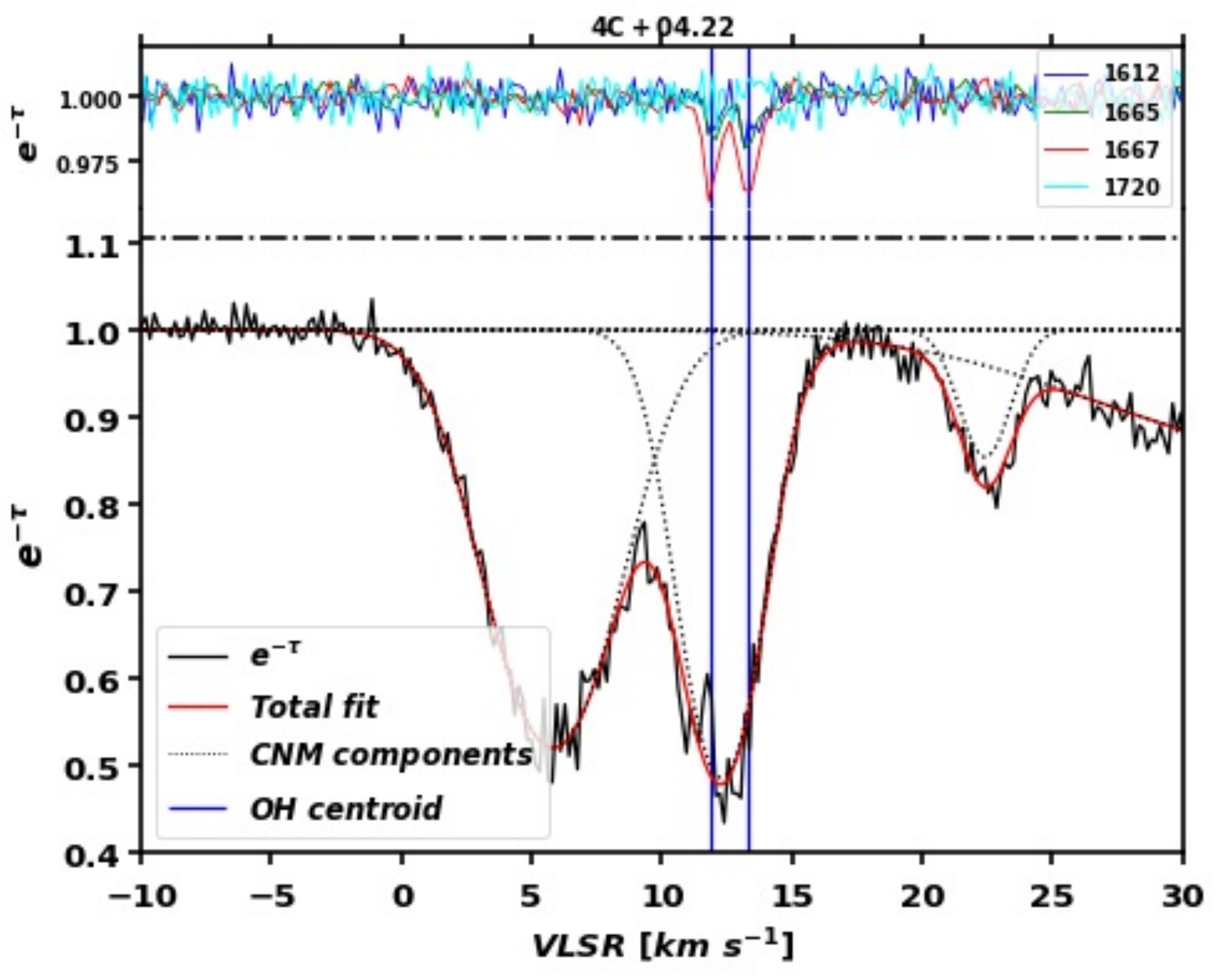}\\
    \includegraphics[width=0.45\linewidth]{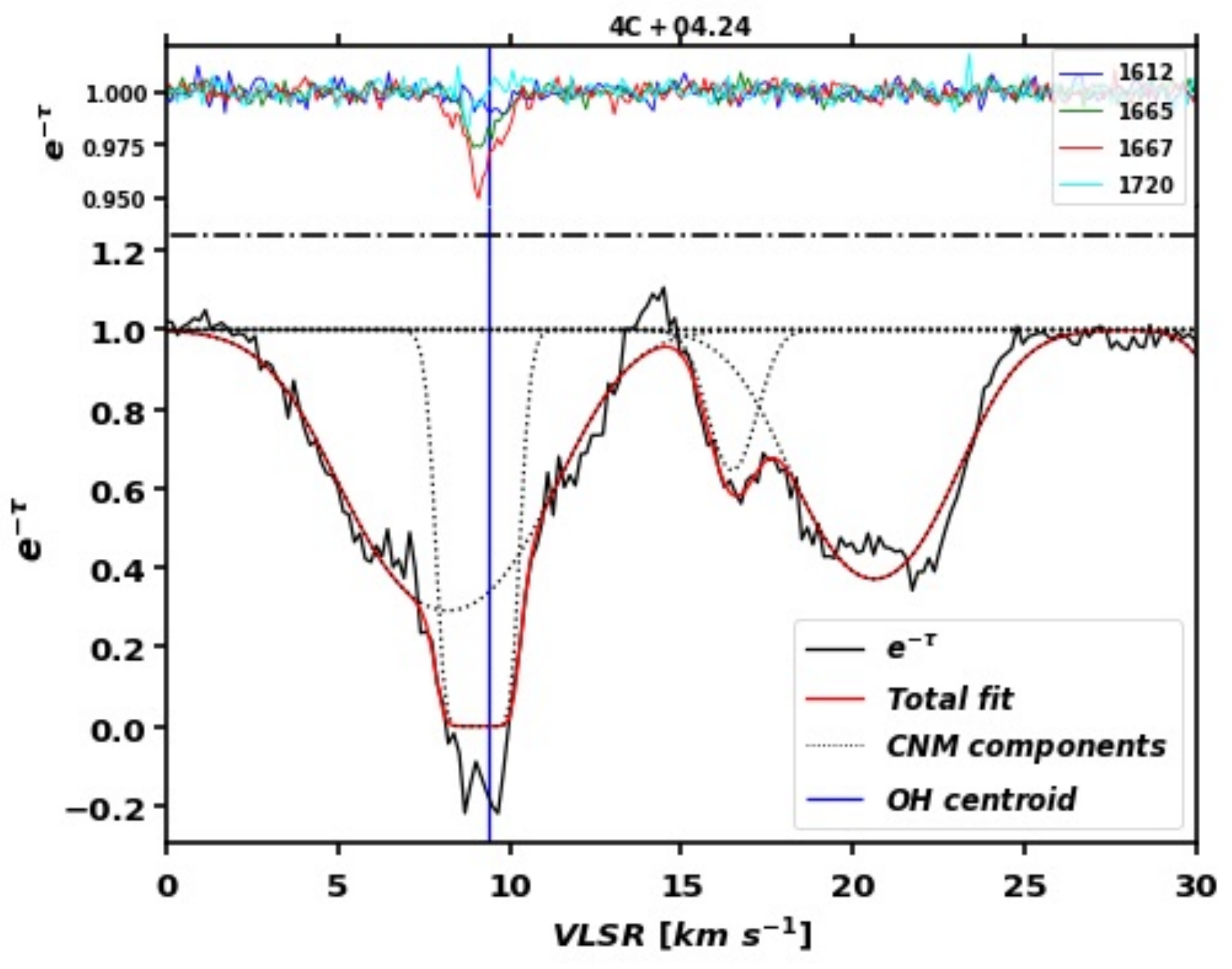}&
    \includegraphics[width=0.45\linewidth]{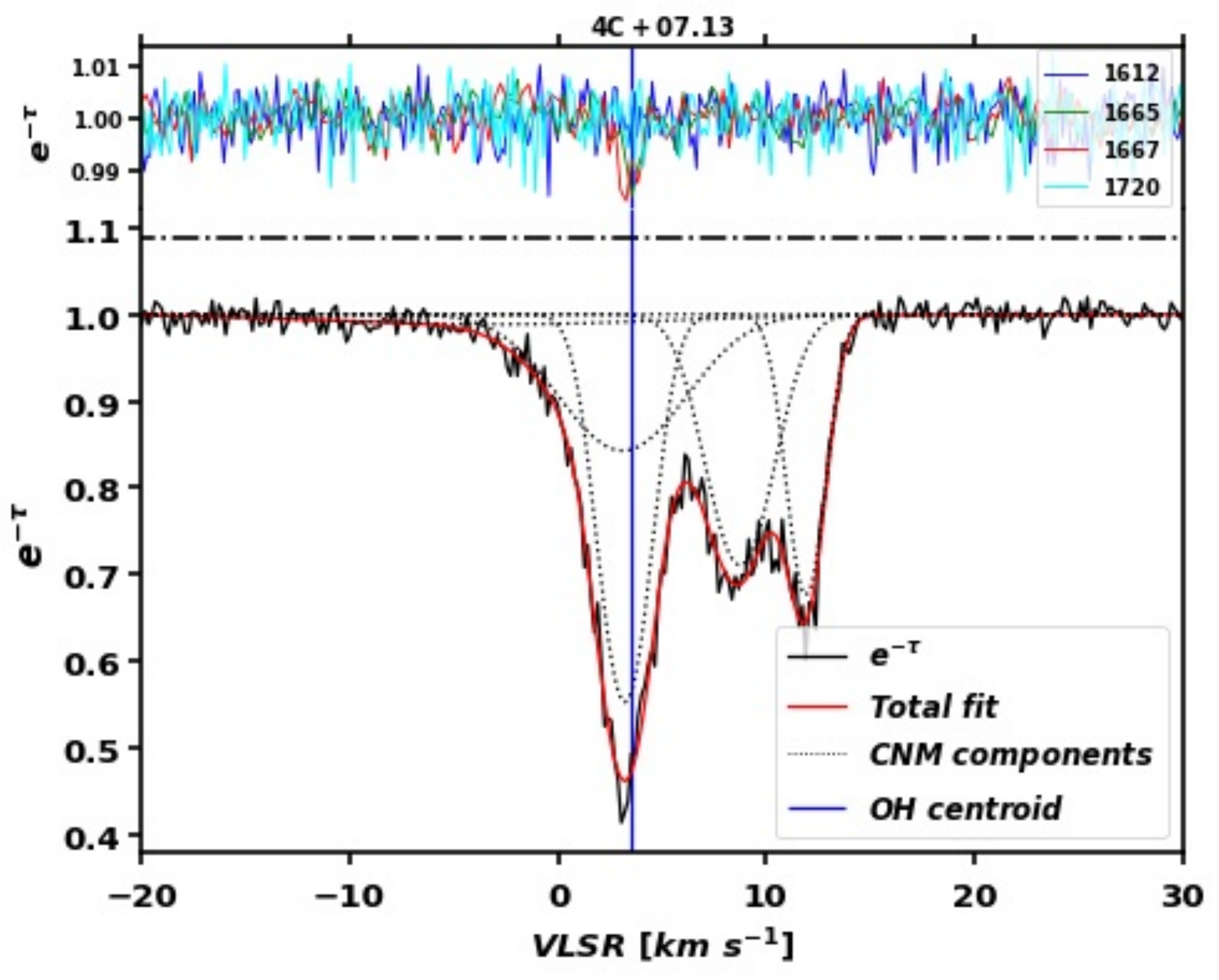}\\
    \end{tabular}
    \caption{Each plot shows OH spectra from this work in the top panel with fitted centroid velocities indicated by the vertical blue lines. The bottom panel shows H\textsc{i}~absorption data (black) with fitted CNM components (black dotted lines) and total CNM fit (red) as reported by \citet{Nguyen2019}. From right to left, top to bottom this figure shows 3C092, 3C131, 3C158, 4C+04.22, 4C+04.24 and 4C+07.13.}
    \label{fig:CNM1}
\end{figure*}
\begin{figure*}
    \centering
    \begin{tabular}{cc}
    \includegraphics[width=0.45\linewidth]{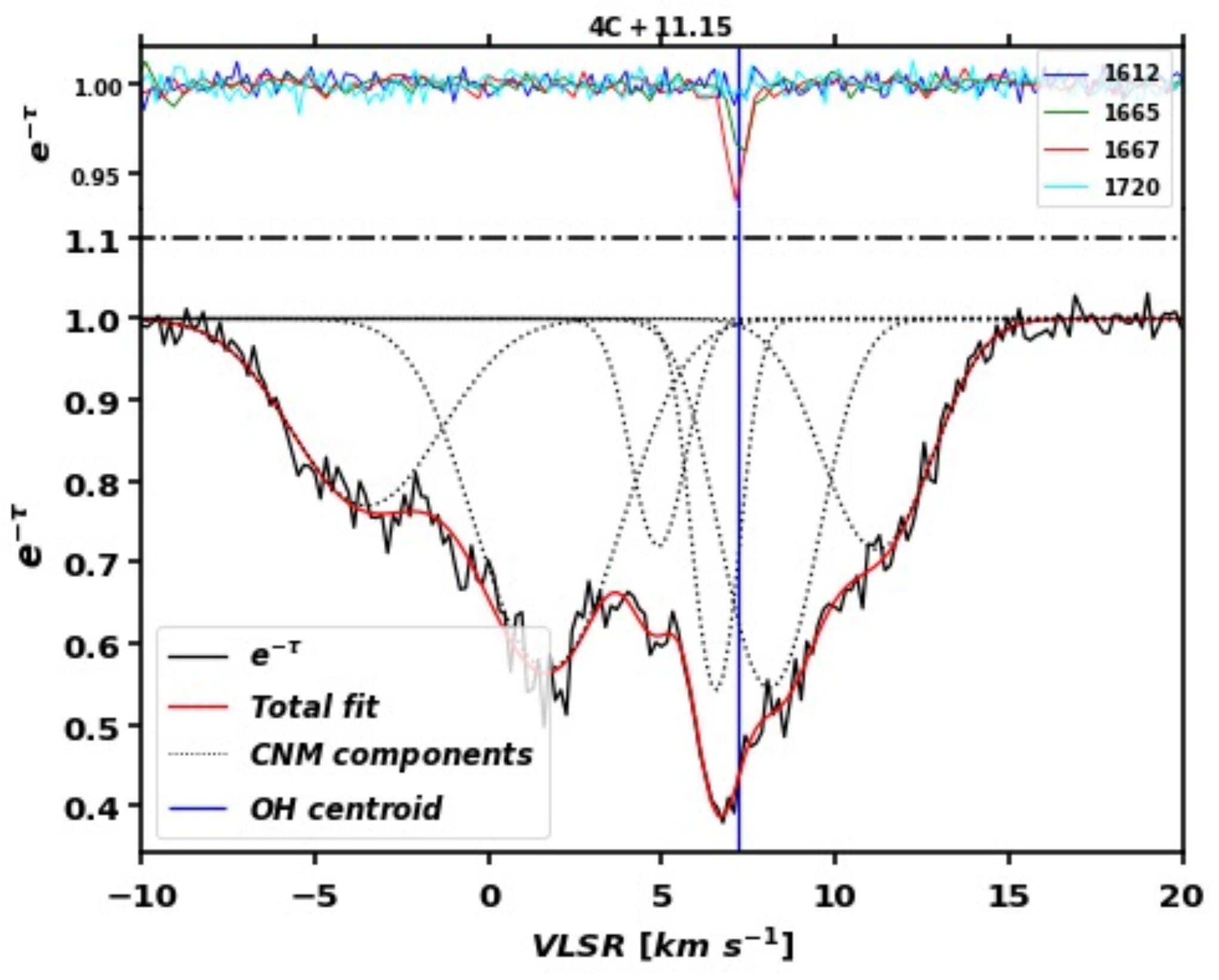}&
    \includegraphics[width=0.45\linewidth]{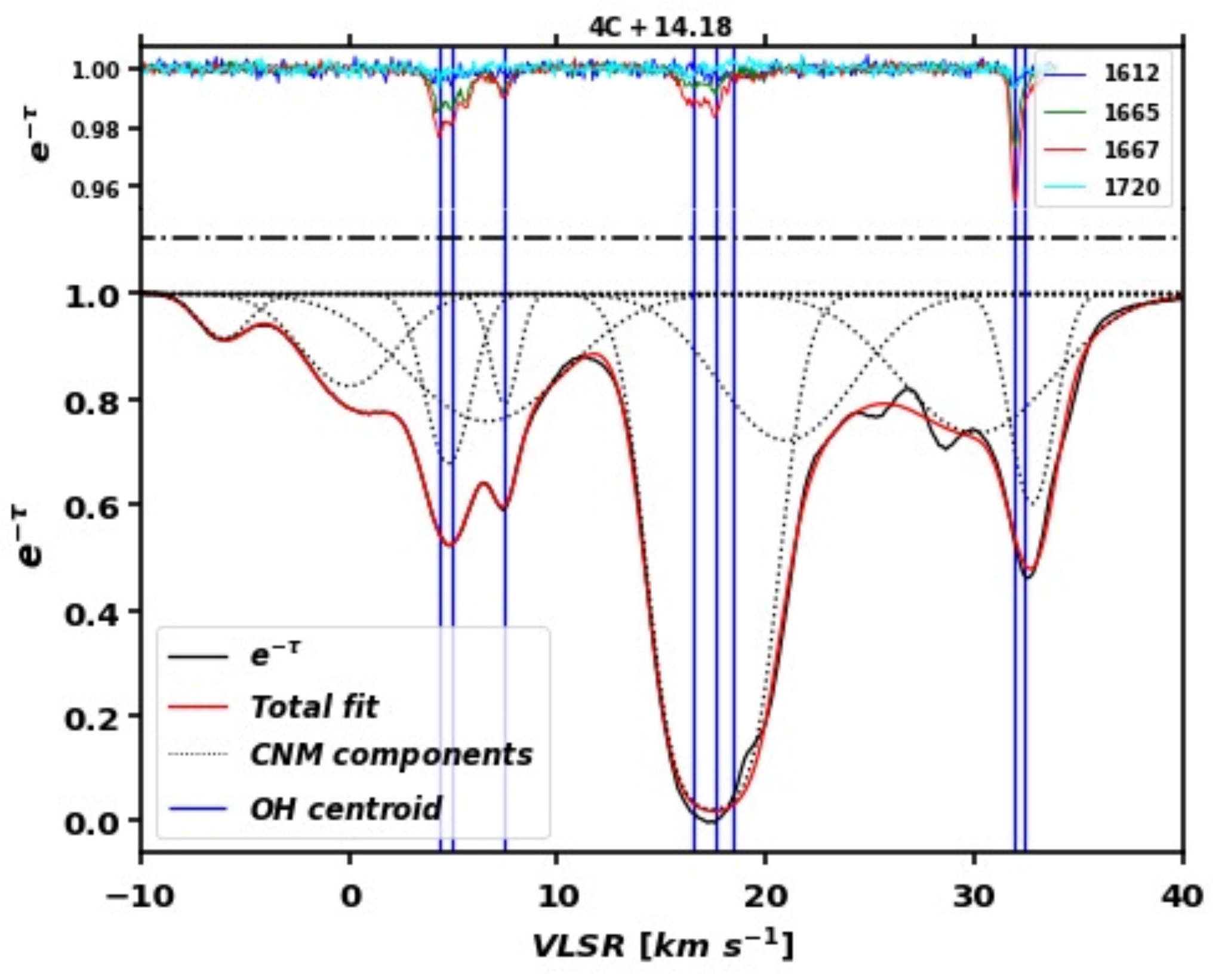}\\
    \includegraphics[width=0.45\linewidth]{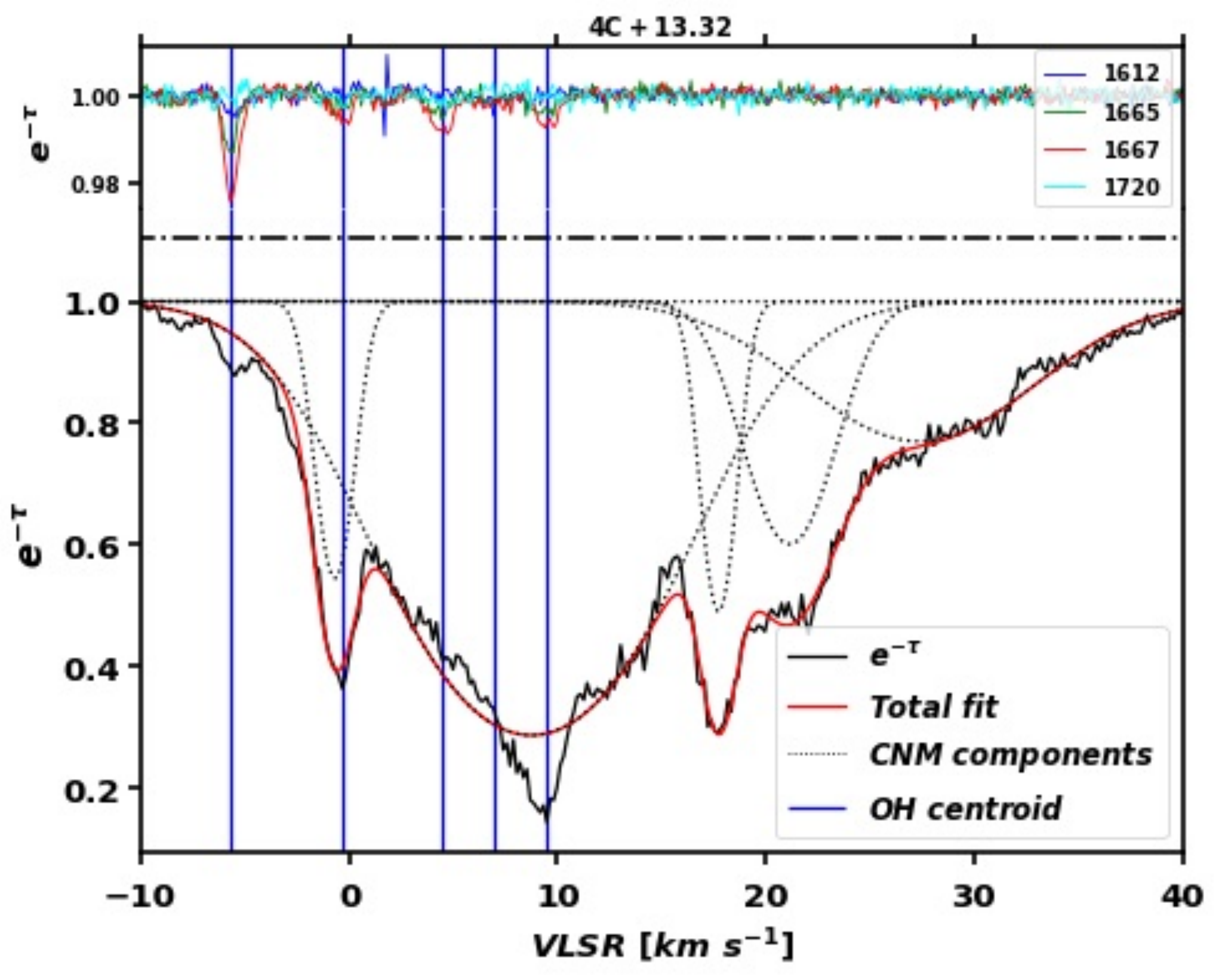}&
    \includegraphics[width=0.45\linewidth]{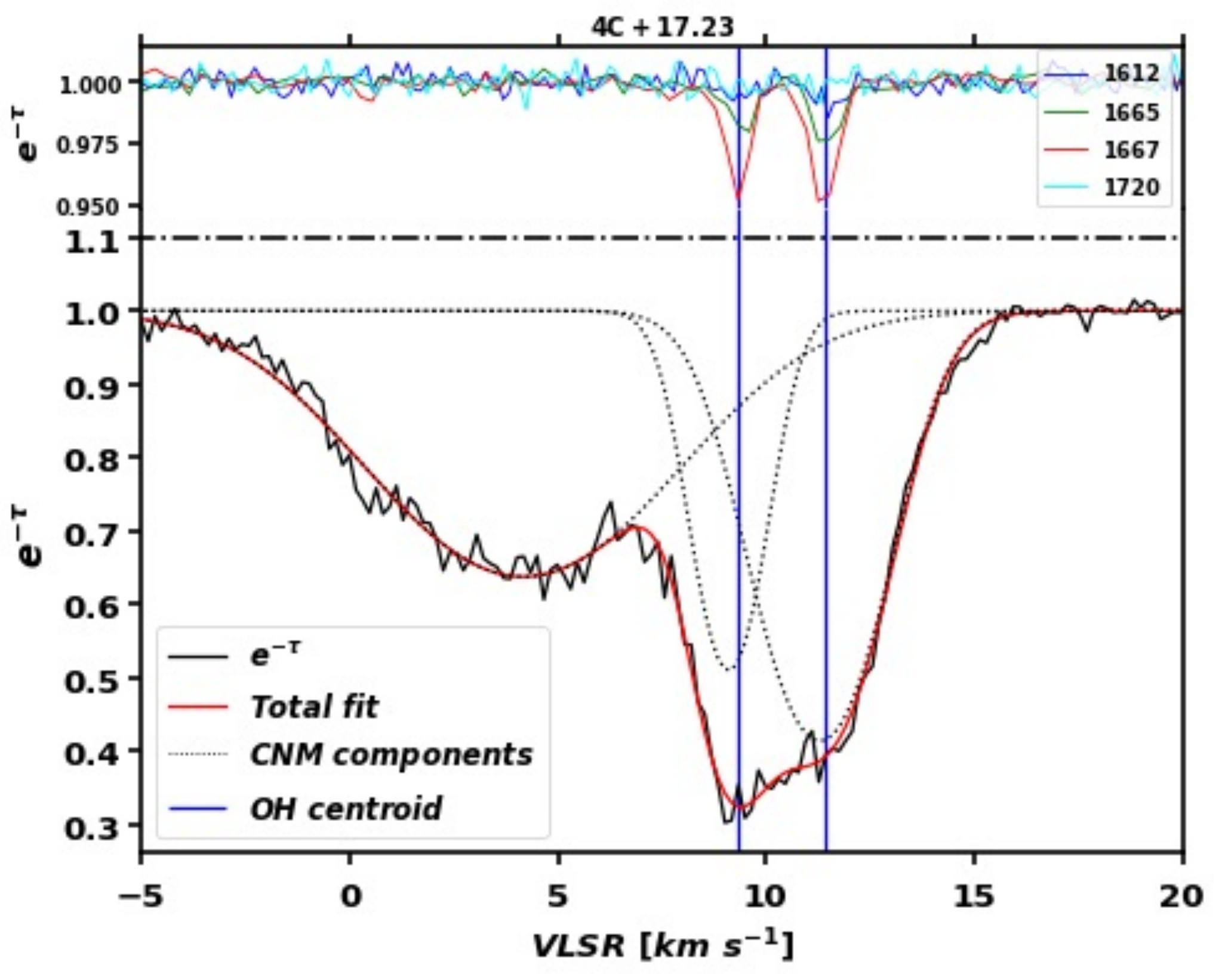}\\
    \includegraphics[width=0.45\linewidth]{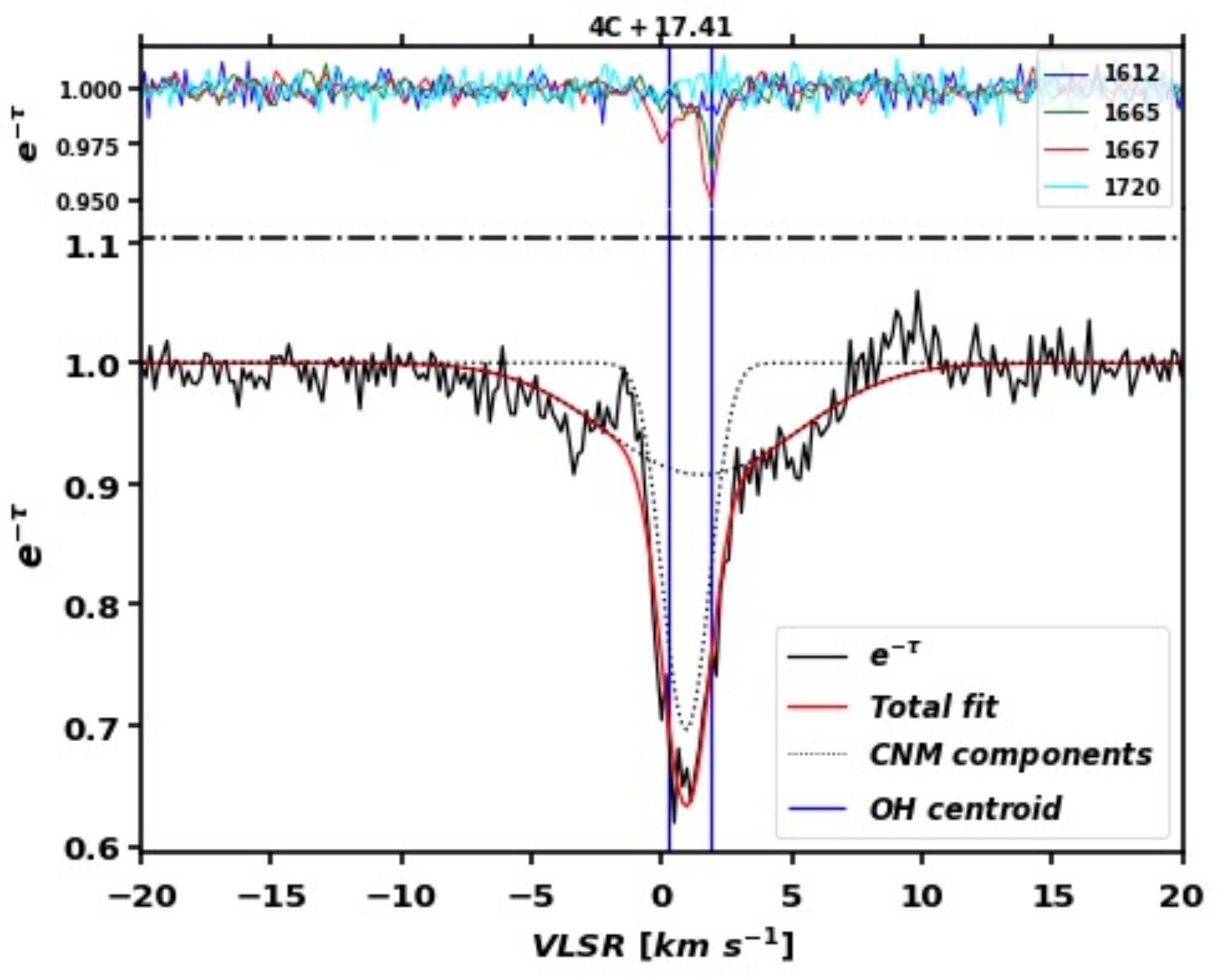}&
    \includegraphics[width=0.45\linewidth]{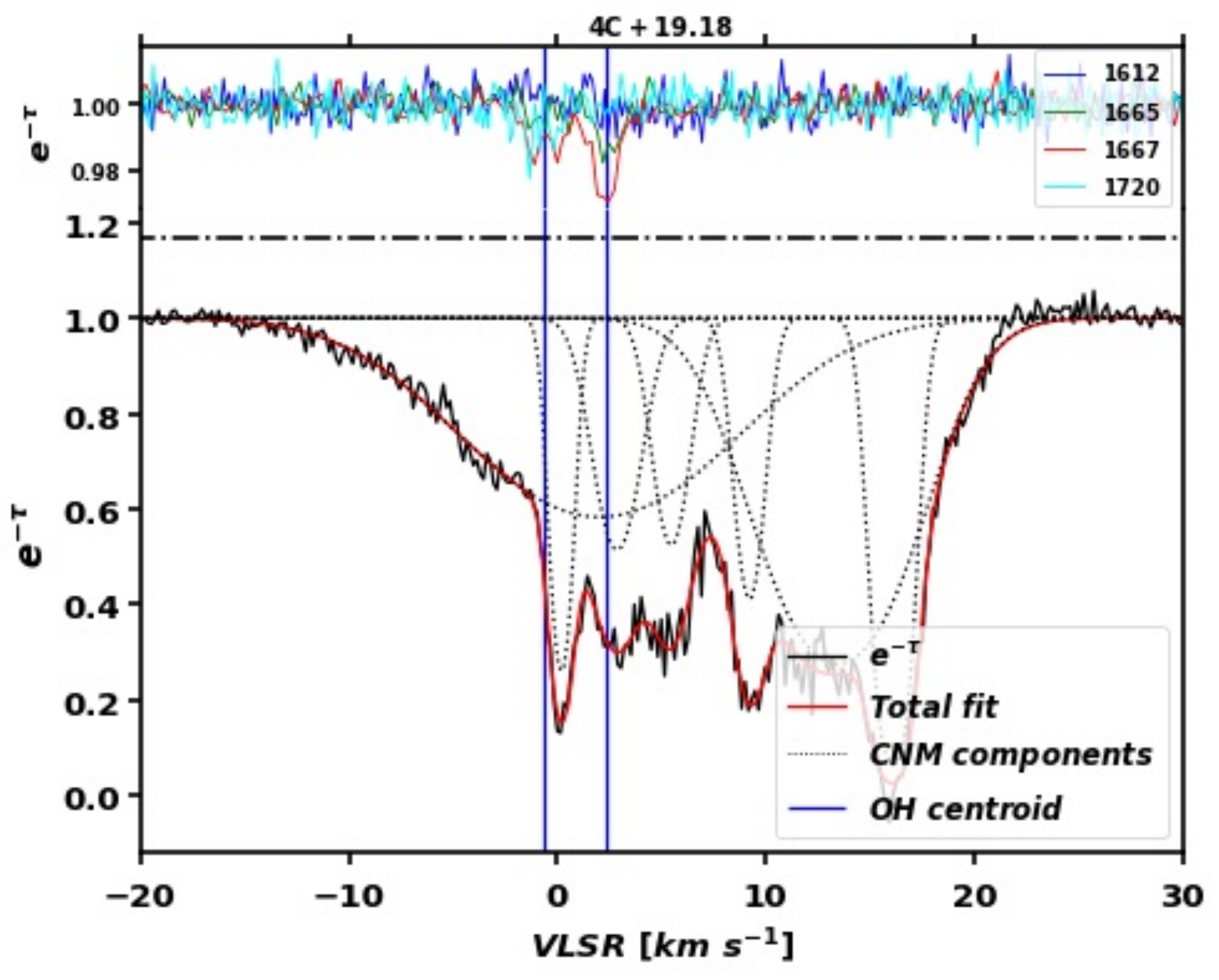}\\
    \end{tabular}
    \caption{Same as Fig. \ref{fig:CNM1} for 4C+11.15, 4C+14.18, 4C+13.32, 4C+17.23, 4C+17.41 and 4C+19.18.}
    \label{fig:CNM2}
\end{figure*}
\begin{figure*}
    \centering
    \begin{tabular}{cc}
    \includegraphics[width=0.45\linewidth]{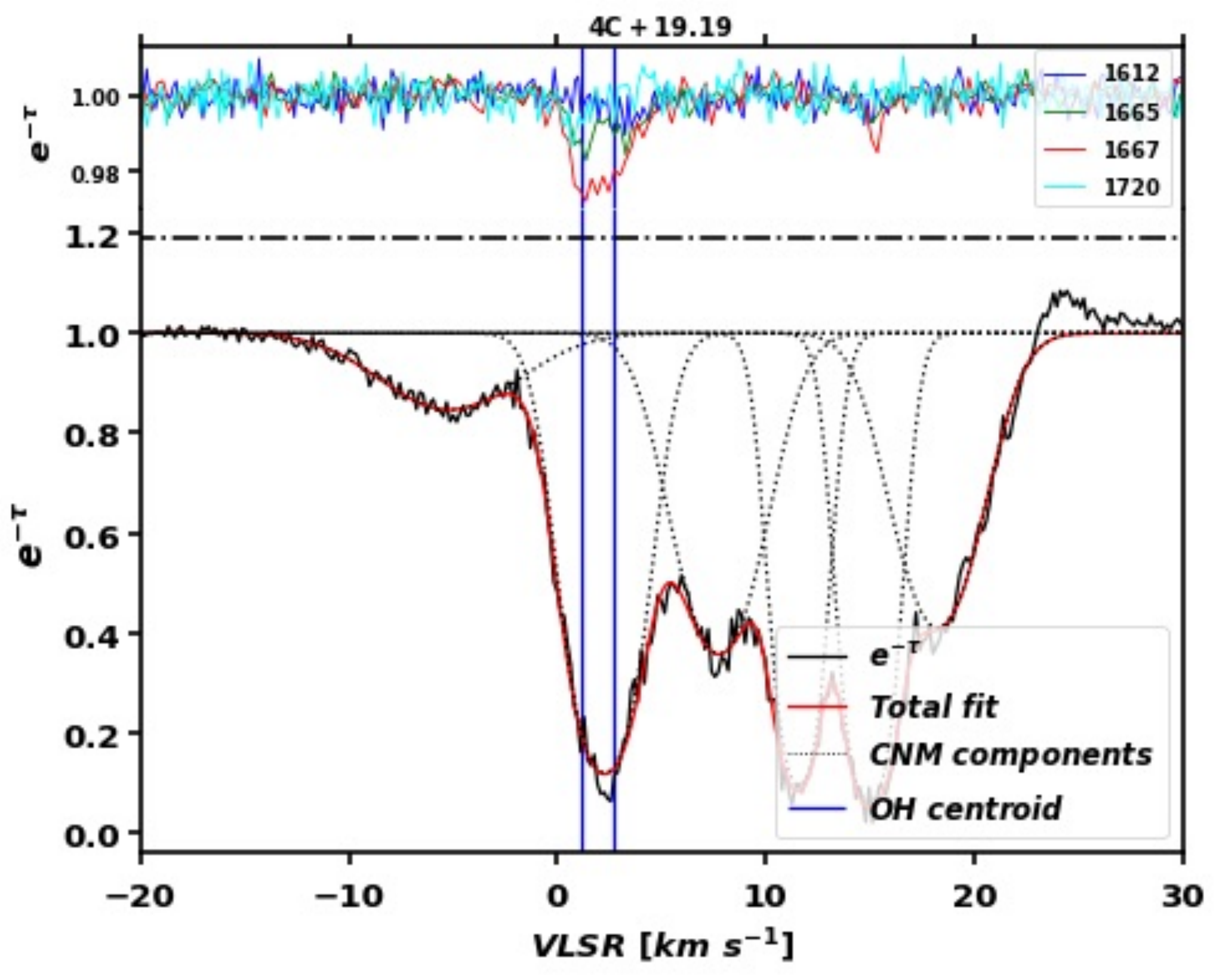}&
    \includegraphics[width=0.45\linewidth]{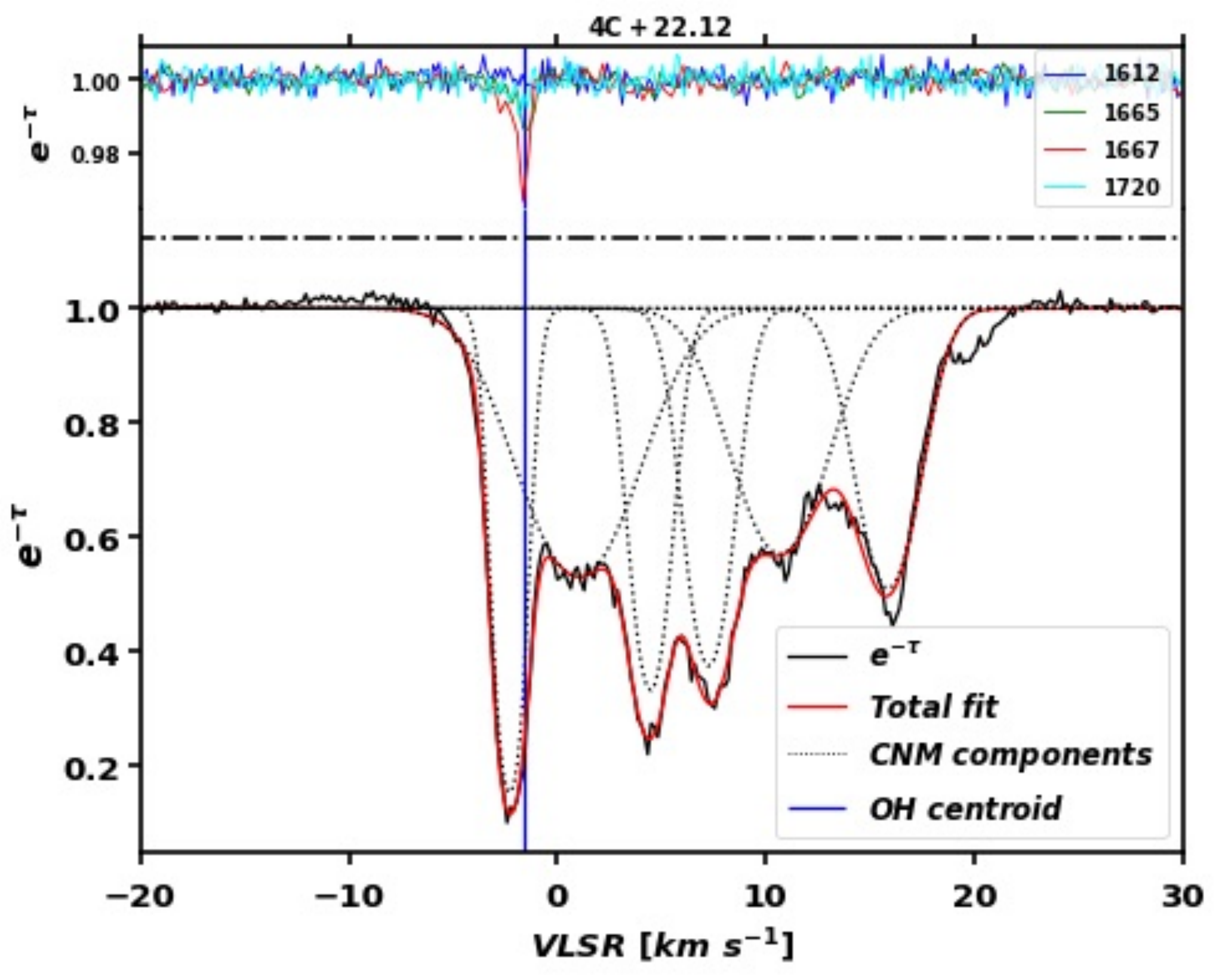}\\
    \includegraphics[width=0.45\linewidth]{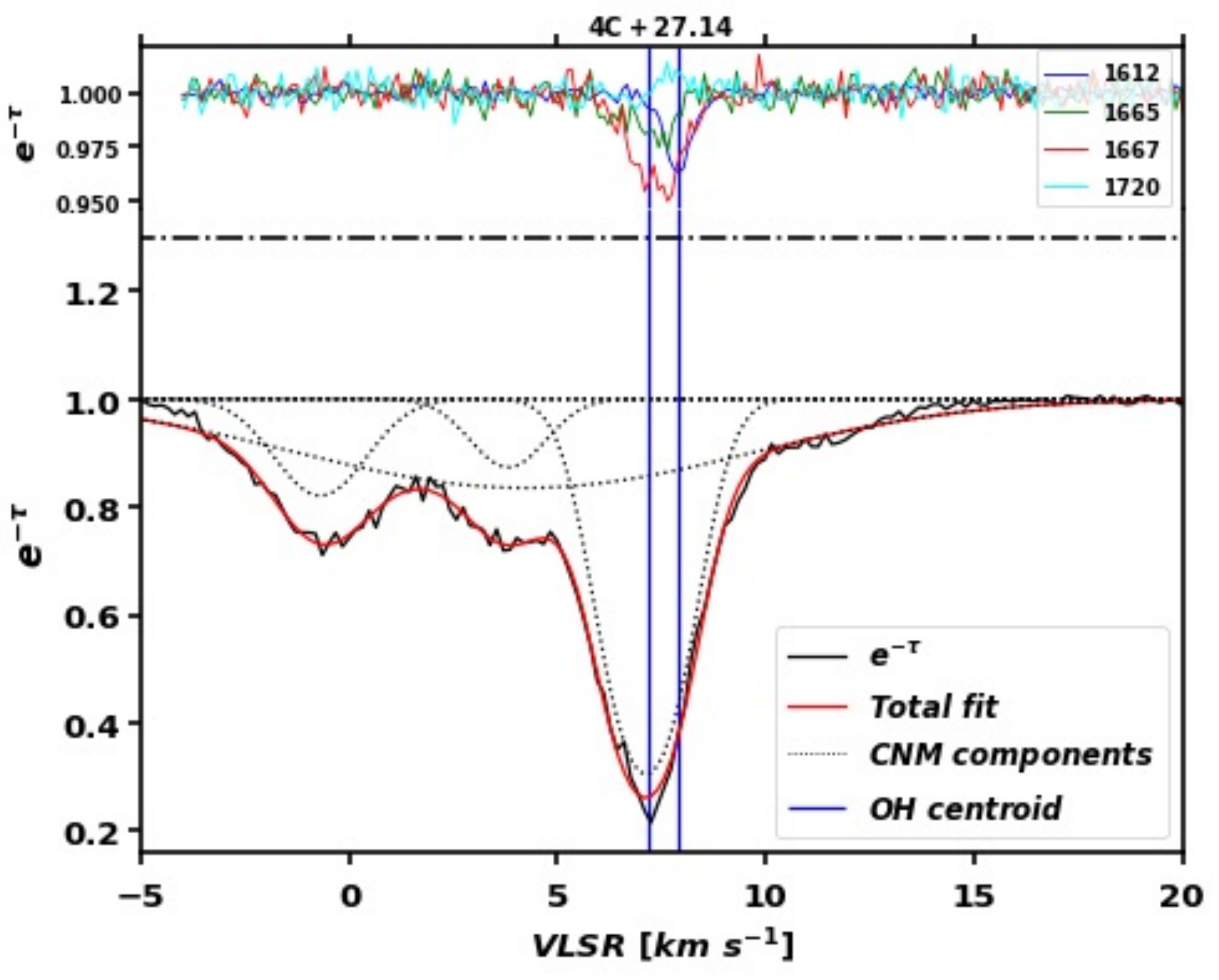}&
    \includegraphics[width=0.45\linewidth]{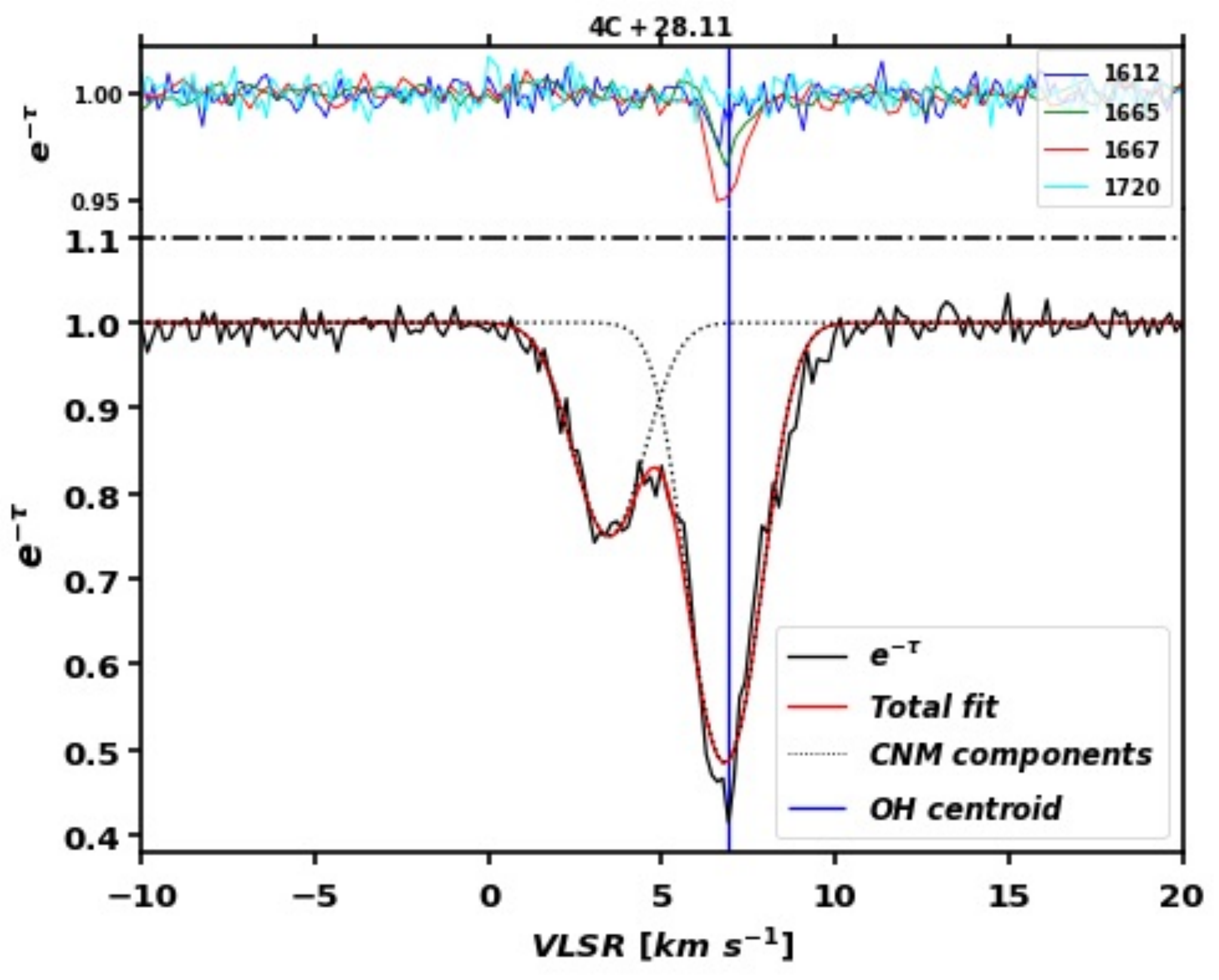}\\
    \includegraphics[width=0.45\linewidth]{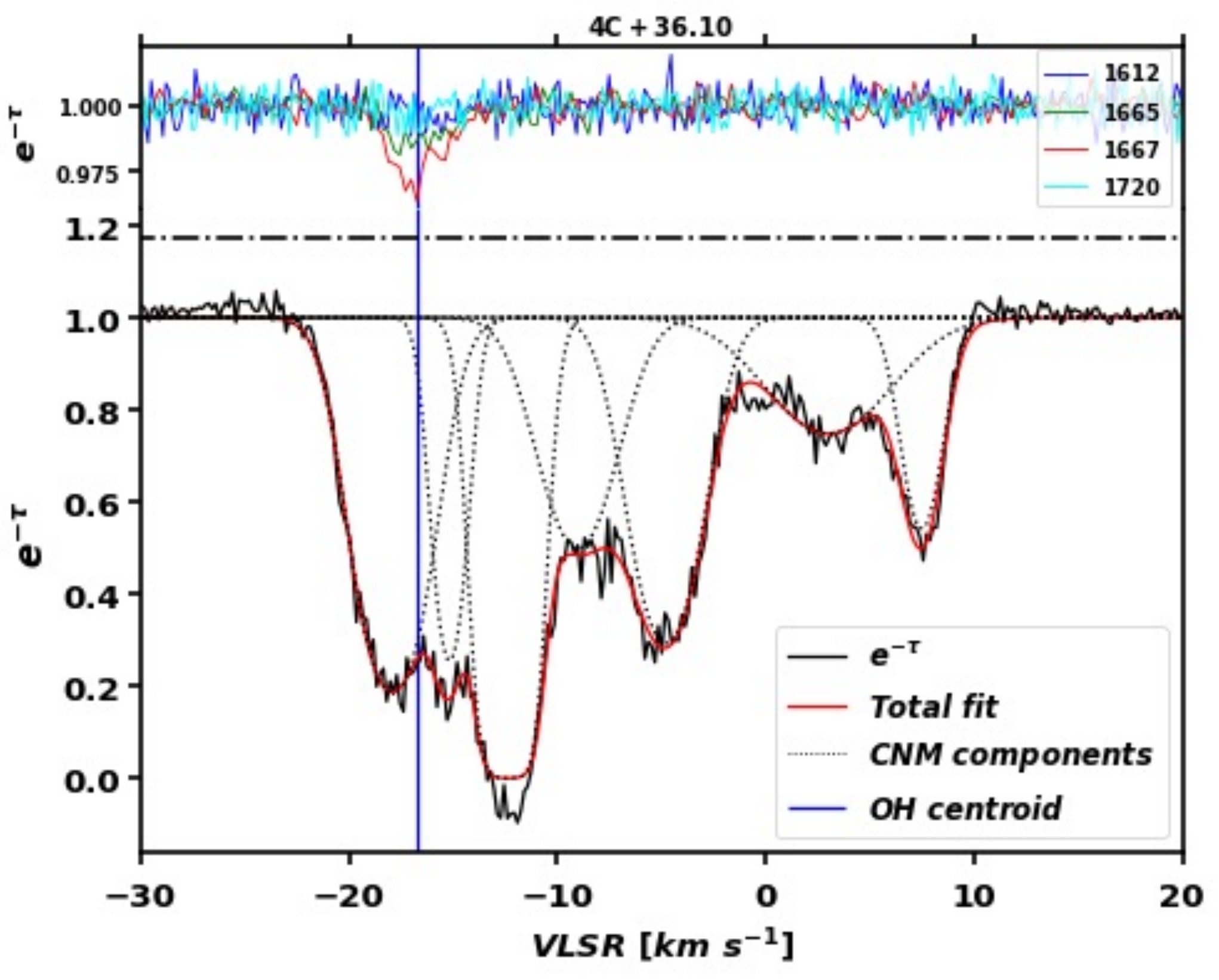}&
    \includegraphics[width=0.45\linewidth]{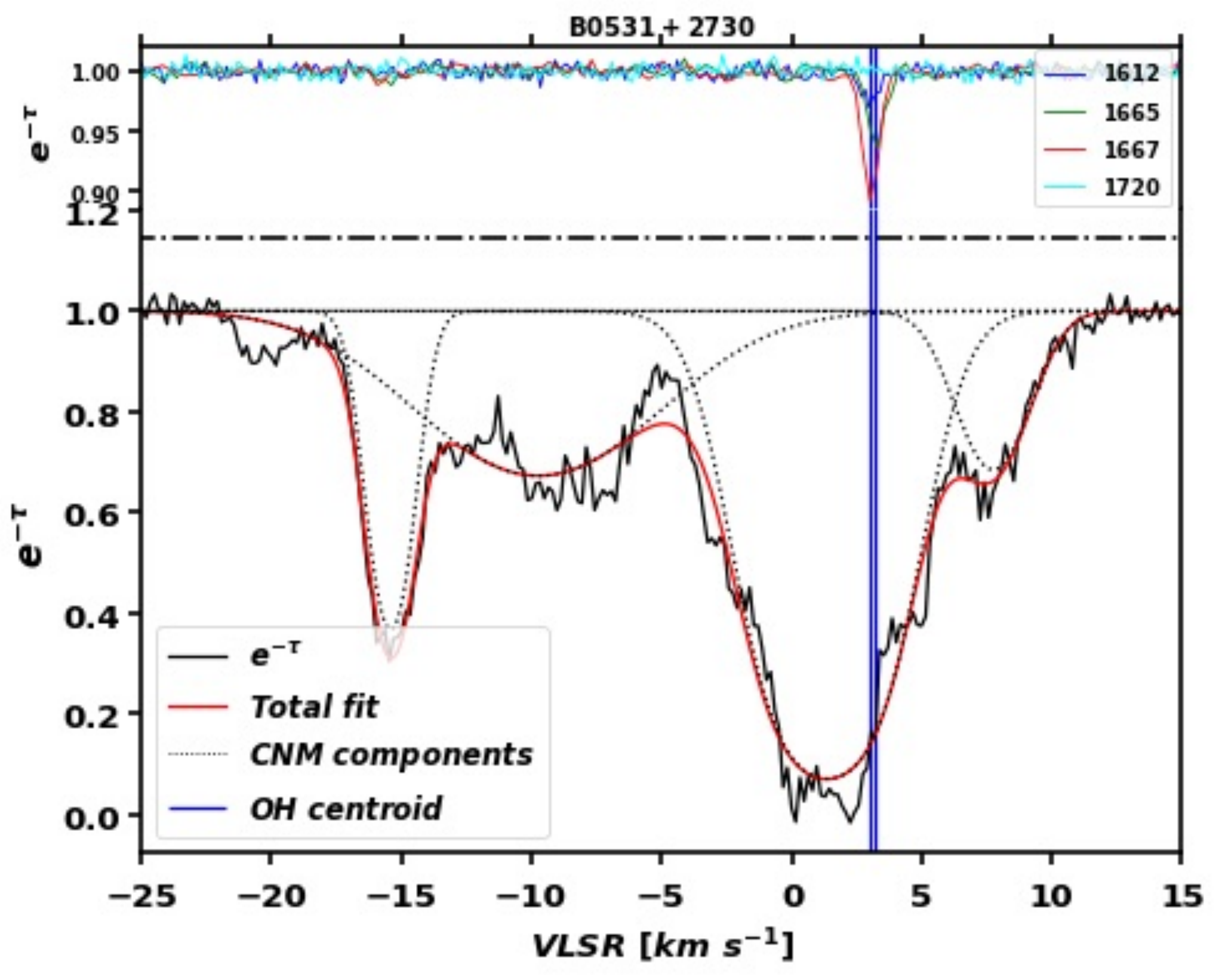}\\
    \end{tabular}
    \caption{Same as Fig. \ref{fig:CNM1} for 4C+19.19, 4C+22.12, 4C+27.14, 4C+28.11, 4C+36.10 and B0531+2730.}
    \label{fig:CNM3}
\end{figure*}
\begin{figure*}
    \centering
    \begin{tabular}{cc}
    \includegraphics[width=0.45\linewidth]{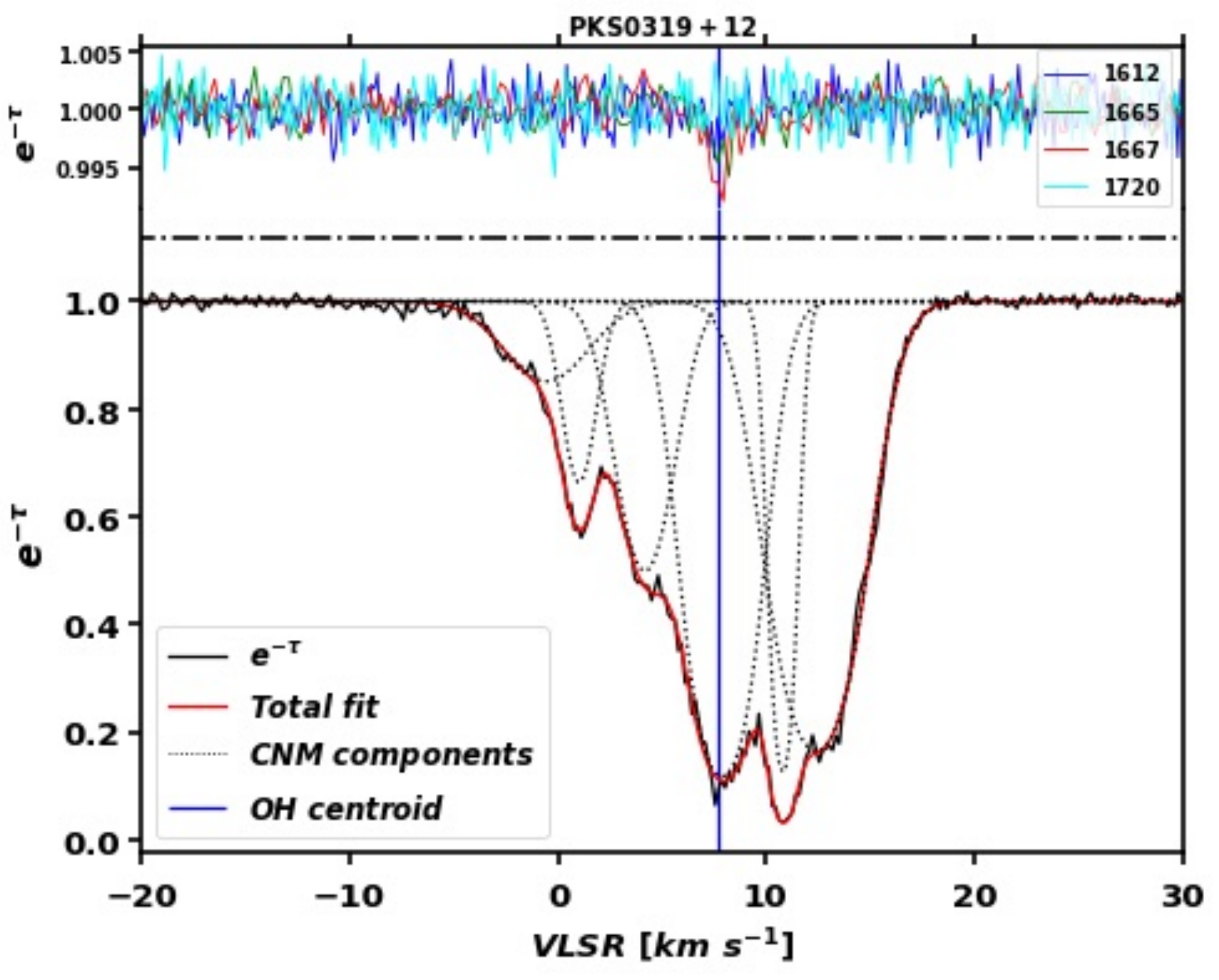}&
    \includegraphics[width=0.45\linewidth]{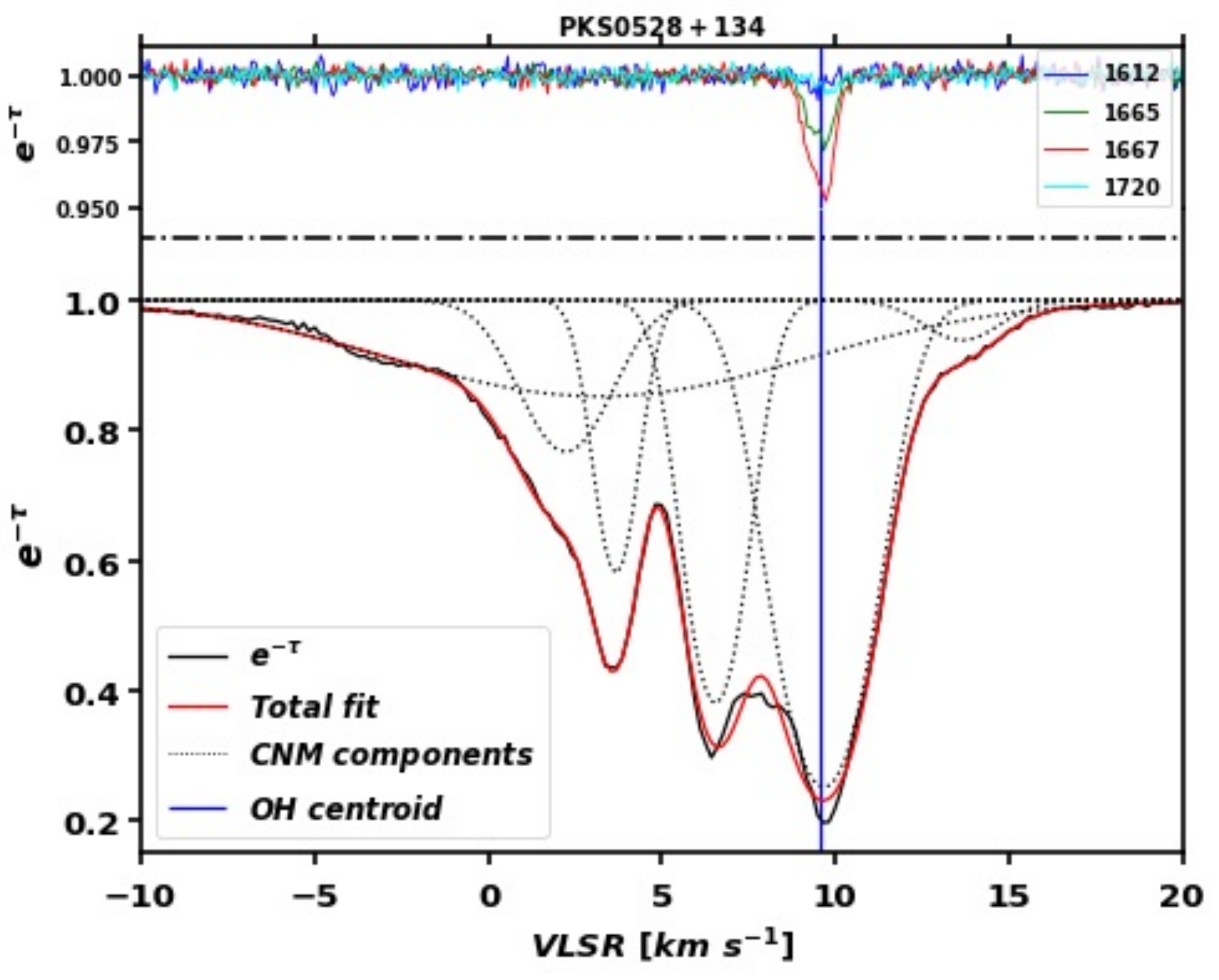}\\
    \end{tabular}
    \caption{Same as Fig. \ref{fig:CNM1} for PKS0319+12 and PKS0528+134.}
    \label{fig:CNM4}
\end{figure*}